\documentclass[12pt]{article}

\usepackage{amsmath}
\usepackage{amssymb}

\countdef\cpart=1
\def\part#1{%
\newpage
\advance\cpart by 1
\section*{Part \the\cpart: #1}
\addcontentsline{toc}{section}{Part \the\cpart: #1}
}

\usepackage[nottoc]{tocbibind}
\usepackage[usenames,dvipsnames,svgnames,table]{xcolor}
\usepackage[colorlinks,citecolor=DarkGreen,linkcolor=FireBrick,urlcolor=FireBrick,linktocpage]{hyperref}
\numberwithin{equation}{subsection}
\usepackage{graphicx}
\usepackage{float}
\usepackage[height=8.8in,width=6.45in]{geometry}
\usepackage{tgtermes}
\usepackage{tgadventor}
\usepackage[T1]{fontenc}
\usepackage[mathscr]{euscript}
\usepackage[export]{adjustbox}

\usepackage{tikz}
\def\boo{0.0}
\def\xlattice#1#2#3{
\begin{tikzpicture}[scale=.5]
\filldraw[color=black!5!white](-.5,-.5) rectangle (1.5,1.5);
\draw[->] (-2,0) -- (3,0);
\draw[->] (0,-2) -- (0,3);
\filldraw[fill=white,draw=gray](0,2) circle (.8em);
\foreach \x in {-1,0,1,2} {
	\foreach \y in {-1,0,1,2}{
		\pgfmathsetmacro\a{mod(#1 * \x - #2 * \y,2)}
		\ifx\a\boo
			\filldraw[color=#3] (\x,\y) circle (.5em);
		\else
			\filldraw[fill=white,draw=gray] (\x,\y) circle (.5em);
		\fi
	}
}
\end{tikzpicture}
}
\def\zlattice#1#2#3{
\begin{tikzpicture}[scale=.5]
\filldraw[color=black!5!white](-.5,-.5) rectangle (1.5,1.5);
\draw[->] (-2,0) -- (3,0);
\draw[->] (0,-2) -- (0,3);
\foreach \x in {-1,0,1,2} {
	\foreach \y in {-1,0,1,2}{
		\pgfmathsetmacro\a{mod(#1 * \x - #2 * \y,2)}
		\ifx\a\boo
			\filldraw[color=#3] (\x,\y) circle (.5em);
		\else
			\filldraw[fill=white,draw=gray] (\x,\y) circle (.5em);
		\fi
	}
}
\end{tikzpicture}
}

\def\lattice#1#2#3{
\begin{tikzpicture}[scale=.4]
\filldraw[color=black!5!white](-.5,-.5) rectangle (3.5,3.5);
\draw[->] (-3,0) -- (5,0);
\draw[->] (0,-3) -- (0,5);
\foreach \x in {-2,-1,0,1,2,3,4} {
	\foreach \y in {-2,-1,0,1,2,3,4}{
		\pgfmathsetmacro\a{mod(#1 * \x - #2 * \y,4)}
		\ifx\a\boo
			\filldraw[color=#3] (\x,\y) circle (.5em);
		\else
			\filldraw[fill=white,draw=gray] (\x,\y) circle (.5em);
		\fi
	}
}
\end{tikzpicture}
}

\def\slattice#1{
\begin{tikzpicture}[scale=.4]
\filldraw[color=black!5!white](-.5,-.5) rectangle (3.5,3.5);
\draw[->] (-3,0) -- (5,0);
\draw[->] (0,-3) -- (0,5);
\foreach \x in {-2,-1,0,1,2,3,4} {
	\foreach \y in {-2,-1,0,1,2,3,4}{
		\pgfmathsetmacro\a{mod(\x,2)*mod(\x,2)+mod(\y,2)*mod(\y,2)}
		\ifx\a\boo
			\filldraw[color=#1] (\x,\y) circle (.5em);
		\else
			\filldraw[fill=white,draw=gray] (\x,\y) circle (.5em);
		\fi
	}
}
\end{tikzpicture}
}
\usetikzlibrary{decorations.pathmorphing,fit}
\tikzset{my/.style={scale=1.2,>=stealth,thick,baseline=(A.center)}}
\tikzset{snake it/.style={decorate,decoration={snake,segment length=1.5mm, amplitude=.3mm}}}
\tikzset{curve/.style={relative,out=-30,in=210}}

\usepackage{xcolor}
\usepackage{mdframed}
\newenvironment{claim}{  \begin{mdframed}[linecolor=black!0,backgroundcolor=black!10]\noindent\ignorespaces}{\end{mdframed}}

\renewenvironment{figure}[1][]{
  \begin{originalfigure}[#1]
    \begin{mdframed}[linecolor=black!0,backgroundcolor=black!1]
}{
    \end{mdframed}
  \end{originalfigure}
}

\renewenvironment{table}[1][]{
  \begin{originaltable}[#1]
    \begin{mdframed}[linecolor=black!0,backgroundcolor=black!1]
}{
    \end{mdframed}
  \end{originaltable}
}

\def\Nequals#1{$\mathcal{N}{=}\,#1$}

\def\bar#1{\overline{#1}{}}

\def\cA{\mathcal{A}}

\def\cD{\mathcal{D}}

\def\cH{\mathcal{H}}

\def\cL{\mathcal{L}}

\def\cN{\mathcal{N}}
\def\cO{\mathcal{O}}
\def\cP{\mathcal{P}}
\def\cQ{\mathcal{Q}}

\def\cW{\mathcal{W}}

\def\bC{\mathbb{C}}
\def\bH{\mathbb{H}}
\def\bO{\mathbb{O}}
\def\bR{\mathbb{R}}
\def\bZ{\mathbb{Z}}
\def\sJ{\mathsf{J}}

\def\sM{\mathsf{M}}

\def\sQ{\mathsf{Q}}
\def\sq{\mathsf{q}}

\def\fg{\mathfrak{g}}

\def\so{\mathop{\mathfrak{so}}}
\def\sp{\mathop{\mathfrak{sp}}}
\def\slash#1{\ooalign{{\text{$#1$}}\crcr \hss\big/\hss}}

\def\tr{\mathop{\mathrm{tr}}\nolimits}

\def\diag{\mathop{\mathrm{diag}}}

\def\Pf{\mathop{\mathrm{Pf}}}
\def\rank{\mathop{\mathrm{rank}}}
\def\Im{\mathop{\mathrm{Im}}}

\def\Ind{\mathop{\mathrm{Ind}}\nolimits}
\def\Res{\mathop{\mathrm{Res}}\nolimits}

\def\Spin{\mathop{\mathrm{Spin}}}

\def\SO{\mathop{\mathrm{SO}}}

\def\SU{\mathop{\mathrm{SU}}}
\def\SL{\mathop{\mathrm{SL}}}
\def\GL{\mathop{\mathrm{GL}}}
\def\su{\mathop{\mathfrak{su}}}

\def\Sp{\mathop{\mathrm{Sp}}}

\def\UU{{\mathrm{U}}}
\def\U{{\mathrm{U}}}

\def\CP{{\mathbb{CP}}}

\def\vev#1{\langle#1\rangle}
\def\ket#1{|#1\rangle}
\def\ii{\mathrm{i}}
\def\LambdaRG{E}
\usepackage{mathtools}

\let\oldstar\star
\def\star{{\oldstar}}
\def\Vol{\mathop{\mathrm{Vol}}}
\def\Re{\mathop{\mathrm{Re}}}
\def\Im{\mathop{\mathrm{Im}}}
\def\CS{\mathrm{CS}}
\def\cc{c.c.}

\def\exercise#1{%
\begin{mdframed}[linecolor=black!0,backgroundcolor=black!10]
\textbf{Exercise.} #1 
\end{mdframed}
}
\def\answer{\paragraph{Answer.} }

\usepackage{bxcjkjatype} 

\begin{document}

\begin{titlepage}

\begin{flushright}
IPMU-18-0182 \\
version 2
\end{flushright}

\vskip 4cm

\begin{center}

{\Large Lectures on 4d \Nequals1 dynamics and related topics}

\vskip 1cm
 Yuji Tachikawa 
\vskip 1cm

\begin{tabular}{ll}
  & Kavli Institute for the Physics and Mathematics of the Universe (WPI), \\
& University of Tokyo,  Kashiwa, Chiba 277-8583, Japan\\
\end{tabular}

\vskip 1cm

\end{center}

\noindent 
This lecture series consists of mostly independent three parts:

\smallskip

\noindent\textbf{Part 1:} a  tour of basic 4d \Nequals1 supersymmetric dynamics,
covering the pure super Yang-Mills, the Seiberg duality for classical groups,
the supersymmetric index on $S^3 \times S^1$, and the $a$-maximization.

\smallskip

\noindent\textbf{Part 2:} a brief look at 2d \Nequals{(2,2)} and \Nequals{(0,2)} 
dynamics,
covering the Landau-Ginzburg and  Calabi-Yau models,
the elliptic genera on $T^2$, and the triality of Gadde-Gukov-Putrov.

\smallskip

\noindent\textbf{Appendix:} a review of basic properties of supersymmetries in various dimensions.

\smallskip

\noindent 
It is based on lectures at Tohoku University, Osaka University, Kyoto University and Nagoya University in the last several years.
Part of the content is based on an unpublished collaboration with F. Yagi and another with K. Kikuchi.

\end{titlepage}

\setcounter{tocdepth}{3}
\tableofcontents
\newpage

\setcounter{section}{-1}
\section{Introduction}
\label{sec:introduction}

The study of 4d \Nequals1 supersymmetric theories is a rich and deep subject.
In addition to  possible applications to elementary particle phenomenology,
it is now an integral part of the thriving field of the study of supersymmetric dynamics in arbitrary dimensions and their inter-relationships. 

There are so many things to be learned in this field, even when restricted to very important results.
The depth of the history of the field is also significant: it is almost 45 years\footnote{%
This introduction was written in November 2018.} since 4d \Nequals1 unbroken supersymmetry was first introduced by Wess and Zumino in \cite{Wess:1974tw},
and it is almost 25 years since Seiberg duality was found \cite{Seiberg:1994pq}.
One might feel at a loss where to start, when  s/he would like to enter this vast scientific territory.

Thankfully, there are already many excellent reviews on this subject, e.g.~%
the classic lecture notes by Intriligator and Seiberg \cite{Intriligator:1995au},
two versions by Argyres \cite{ArgyresReview1996,ArgyresReview2001},
the TASI lecture by Peskin \cite{Peskin:1997qi},
the lecture by Shifman \cite{Shifman:1995ua},
an unorthodox one by Strassler \cite{Strassler:2003qg},
the book by Terning \cite{TerningBook} and the accompanying set of slides \cite{TerningSlides},
a recent review by Tanedo \cite{TanedoReview},
and the book by Dine \cite{DineBook}.

When the author was asked to give a series of lectures on this subject, therefore, 
he was not sure if there was much reason to give yet another set of lectures.
Cannot the supposed audience simply read one of these great reviews?
There seemed to be even less reason to write up the author's version of lectures.
Still, the author realized that most of the available reviews do not adequately cover two fundamental techniques of the field introduced more recently than the others, 
namely the method of the supersymmetric index on $S^3\times S^1$ and the principle of the $a$-maximization.

The emphasis in this set of lecture notes is, then, on these more recent techniques,
and to present a number of examples where their power can be felt. 
That said, any set of lecture notes on 4d \Nequals1 dynamics cannot lack a section on Seiberg duality.
Here, the author aimed to present a somewhat different take on this important subject.
For example, we treat the case $N_f \sim 3N$ first and lowering $N_f$ next, rather than the more historical approach of starting with low $N_f$ and raising it up.
We also study all three series of classical groups, $\SU$, $\Sp$ and $\SO$ in this set of lecture notes,
so that the readers can find which part of the story holds universally and which part does not.
The author does \emph{not} think  that this different take is in fact better, 
but at least the readers can find a different light shed on this by now familiar subject.

Another aim in this set of lecture notes is to try to show to the reader that the modern analysis of the supersymmetric dynamics is not confined in one specific choice of spacetime dimensions $d$ and the number $\cN$ of supersymmetry.
Rather, in the last few years, the supersymmetric dynamics in the entire possible choice of $d$ and $\cN$ became interrelated and intertwined.
The author tried to illustrate this by treating the case of 2d \Nequals{(2,2)} theories and 2d \Nequals{(0,2)} theories, following the pattern of the analysis of the 4d \Nequals1 case.
The reader should note however that this presentation is ahistorical: 
rather, various developments and techniques in 2d supersymmetric dynamics in the 1980s were an important undertone which allowed the study of 4d supersymmetric theories to blossom in the 1990s.

\bigskip\bigskip

\paragraph{Organization of the notes:}

\begin{figure}
\centering
\includegraphics[width=.7\textwidth]{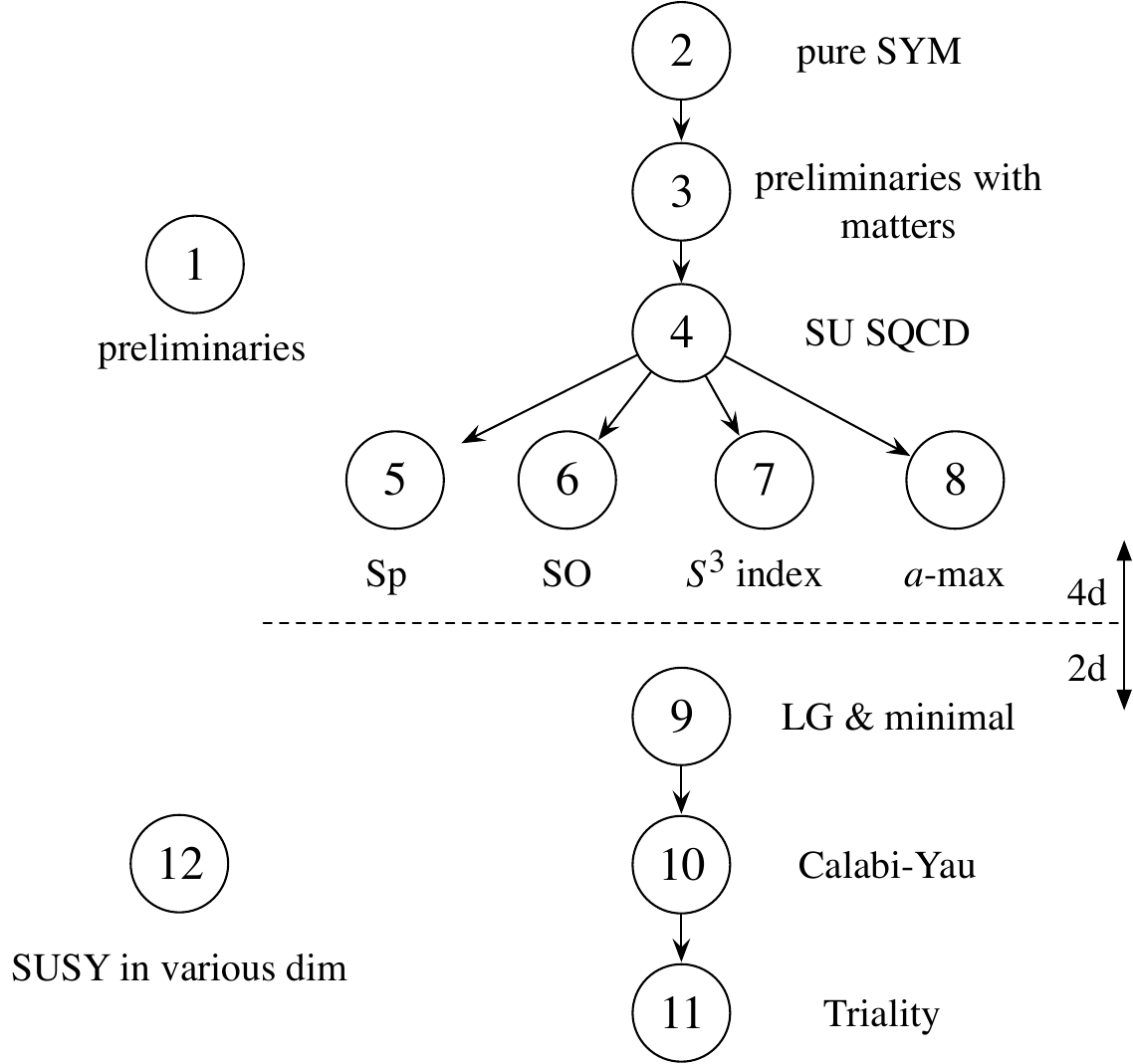}
\caption{Interrelationships of the contents of this set of lecture notes.
\label{fig:diagram}}
\end{figure}

This set of lectures is divided into mostly independent three parts, of rather unequal lengths.
It consists of twelve sections, whose interrelationships are summarized in Fig.~\ref{fig:diagram}.

\bigskip
\noindent\textbf{Part 1} has about 90 pages and contains eight sections.
There, we will give a general overview of 4d \Nequals1 supersymmetric dynamics.

We start by  reviewing preliminary materials in Sec.~\ref{sec:preliminaries}, including one-loop running, anomalies, line operators and \Nequals1 superspace.
The main content begins in Sec.~\ref{sec:pure}, where we study the behavior of pure super Yang-Mills.
We count the number of vacua both in the infrared limit and in the ultraviolet limit.

Next in Sec.~\ref{sec:matters}, we will study supersymmetric gauge theories with matters from perturbative points of view.
In Sec.~\ref{sec:SUSQCD}, we will analyze the infrared behavior of the $\SU(N)$ gauge theory with $N_f$ flavors, for $0<N_f< 3N$, in a non-perturbative manner.
An essential ingredient is the conserved $\U(1)$ R-symmetry, which allows us to fix many of the properties of the infrared limit.
Another essential ingredient in the analysis is the Seiberg duality, which will also be introduced there.
In Sec.~\ref{sec:SpSQCD} and in Sec.~\ref{sec:SOSQCD}, the analysis will be generalized to $\Sp(N)$ and $\SO(N)$.
The $\Sp(N)$ case goes without any surprises, and is in fact simpler than the $\SU$ case.
The $\SO(N)$ case turns out to be far more subtle and interesting.

In Sec.~\ref{sec:SCI}, we will introduce and study the supersymmetric index on $S^3\times S^1$.
We will provide a few case studies, one of which is a detailed check of the Seiberg duality.
Finally in Sec.~\ref{sec:amax}, we will learn the $a$-maximization, which often allows us to determine the $\U(1)$ R symmetry in the infrared superconformal algebra using the ultraviolet description of the theory.
Again we will provide a few case studies.

The author  hopes that a large part of the techniques necessary to read papers on 4d dynamics with unbroken \Nequals1 symmetry is covered by this general overview.
Three notable omissions are: the detailed explanation of the superfield formalism,  the analysis of the moduli space of supersymmetric vacua, and the study of the exactly-marginal operators.
The author thinks that the first two topics can be learned from existing textbooks and reviews,
and that the third can be easily learned from \cite{Leigh:1995ep,Green:2010da} after the reader goes through this review.
We do not discuss any aspects of supersymmetry breaking or any connection to particle phenomenology either; they need to be studied somewhere else.

\bigskip
\noindent\textbf{Part 2} is of about 30 pages, consisting of three sections.
There, we will study a few examples of 2d \Nequals{(2,2)} and \Nequals{(0,2)} dynamics,
in particular the ones which can be analyzed quite analogously to the 4d \Nequals1 cases we will study in Part 1.
In Sec.~\ref{sec:LGmin}, we will analyze 2d \Nequals{(2,2)} Landau-Ginzburg models and their correspondence with the 2d minimal models.
In Sec.~\ref{sec:CY}, we will study geometric Calabi-Yau sigma models and their realizations as Landau-Ginzburg orbifolds and as $\U(1)$ gauge theories.
We will halve the number of supersymmetry in Sec.~\ref{sec:triality}, where we will discuss the triality of Gadde, Gukov and Putrov, and its relation to the 4d Seiberg duality.

This part 2 is not and is not intended to be a general overview of 2d \Nequals{(2,2)} and \Nequals{(0,2)} dynamics;
a proper treatment will require an entirely new set of lectures.
Rather, it is meant to illustrate that many of the techniques we learn in the 4d \Nequals1 case can be adapted and used in this different setting,
and that dualities in lower dimensions (this time in 2d) can be understood by a compactification of a higher-dimensional theory (this time in 4d).

We will see again that the conserved $\U(1)$ R-symmetry fixes many of the properties of the infrared limit;
we will use the elliptic genus, i.e.~the supersymmetric index of the theory on $T^2=S^1\times S^1$;
and we will encounter the $c$-extremalization, which allows us to fix the $\U(1)$ R-symmetry in the infrared superconformal symmetry in terms of the ultraviolet description.

\bigskip
\noindent\textbf{Part 3} is an appendix of about 10 pages.
In its only section, Sec.~\ref{sec:various},
we briefly review the structure of the supersymmetry and the superconformal symmetry in various dimensions.
Hopefully this short summary is of some use to those readers who would like to enter this intricate world that is the study of supersymmetric dynamics in various dimensions.
As a final topic, we also provide a short summary of known theories with 16 supercharges.

\newpage
\section*{Acknowledgements}
\addcontentsline{toc}{section}{Acknowledgements}

This lecture note is based on the author's lectures at various Japanese universities, including Tohoku University (Oct. 2011), Kyoto University (Oct. 2014), Osaka University (Feb. 2015), and Nagoya University (Mar. 2017).
The lectures contain some unpublished results:
The content of Sec.~\ref{sec:yagi} is based on the author's unpublished collaboration with F. Yagi,
and a part of the content of Sec.~\ref{sec:E7surprise} is based on the author's unpublished collaboration with K. Kikuchi.

The author would like to thank first all the participants of those lectures.
The author would also like to thank many people who made valuable comments to the earlier versions of the manuscript:
P. Agarwal,
V. Dobrev,
R. Eager,
D. Gang,
S. Giacomelli,
S. M. Hosseini,
K. Intriligator,
K. Kanno,
Z. Komargodski,
N. Maru,
E. Nardoni,
S. Razamat,
J. Song,
and G. Zafrir.
The author would like to thank 
Y. Enoki,
Y. Lee, 
S. Nawata,
V. P. Spiridonov,
J. Yagi,
and K. Yonekura in particular, who read the manuscript carefully and found many errors and inaccuracies. 
The remaining errors are of course all the fault of the author.
This set of lecture notes would have never been written up without the repeated encouragement from Mr. Nakamura of Springer. 
The author would like to thank him, and any reader who finds this set of lecture notes useful should thank him too.

The author is partially supported  by 
JSPS KAKENHI Grant-in-Aid, Wakate-A, No.17H04837 
and Kiban-S, No.16H06335 and 
by WPI Initiative, MEXT, Japan at IPMU, the University of Tokyo.

\part{A tour of 4d \Nequals1 dynamics}
In this Part 1, we will give a tour of fundamental techniques to study 4d \Nequals1 supersymmetric dynamics.
After briefly reviewing the perturbative analysis, we will study the Seiberg duality,
not only in the $\SU$  case but also in the $\Sp$ and $\SO$ cases.
We will then learn how to compute supersymmetric indices on $S^3\times S^1$,
and how to find the superconformal R-charge by performing the $a$-maximization.

\section{Preliminaries}
\label{sec:preliminaries}
In this first section, we collect various preliminary results about 4d quantum field theories, which are not necessarily supersymmetric. 
Sec.~\ref{subsec:oneloop} contains the one-loop running of the gauge coupling,
mainly to explain our notations and conventions.
Sec.~\ref{subsec:anomalies} discusses the perturbative and global anomalies of fermions,
again mainly to explain our notations and conventions of the anomaly polynomial.
Sec.~\ref{sec:groupglobal} introduces the physical difference of gauge groups sharing the same Lie algebras, paying particular attentions to the spectra of line operators.
Sec.~\ref{sec:4dsuperfields} is a very brief summary of 4d \Nequals1 superfield formalism.
Except Sec.~\ref{sec:groupglobal}, the content is totally standard, and can be found in many other places.

For detailed expositions on somewhat advanced aspects of non-supersymmetric quantum field theory, the readers are referred to e.g.~a classic account by Coleman \cite{Coleman} or more recent books by Shifman \cite{Shifman} and by Mari\~no \cite{Marino}.
On anomalies, the TASI lecture note by Harvey \cite{Harvey:2005it} can be highly recommended.
A recent paper \cite{Cordova:2018cvg} also contains an excellent overview of the traditional anomalies in Sec.~2 and Sec.~5.2.

\subsection{One-loop running}
\label{subsec:oneloop}
In this book we use the convention that the covariant derivatives \emph{do not} contain the gauge coupling.
For example, 
the covariant derivative of an $\UU(N)$ gauge field has the form $D_\mu=\partial_\mu \delta^a_b + i (A_\mu){}^a_b$ ($a,b=1,\ldots,N$).
The gauge kinetic term contains the gauge coupling instead: \begin{equation}
S \supset -\int d^4 x \frac{1}{2g^2} \tr F_{\mu\nu}F_{\mu\nu}, \label{nonsusygaugekin}
\end{equation} where we always use the Wick-rotated, Euclidean form in this book.

The one-loop renormalization of the gauge coupling in the presence of Weyl fermions in the representation $R_f$
and complex scalars in the representation $R_s$
is then given by the following standard form: \begin{equation}
\LambdaRG\frac{d}{d\LambdaRG}g=-\frac{g^3}{(4\pi)^2}\left[
\frac{11}{3}C(\text{adj})-\frac23  C(R_f)-\frac13   C(R_s)
\right]+O(g^4).\label{nonsusyRG}
\end{equation}
Here, $\LambdaRG$ is the energy scale at which $g$ is measured, and we use the convention that all fermions are written in terms of left-handed Weyl fermions.
The quantity $C(R)$ is defined by the relation \begin{equation}
\tr (T_R)^u (T_R)^v =C(R) \delta^{uv}
\label{CR}
\end{equation} where $(T_R)^u$ for $u=1,\ldots, \dim G$ is the $u$-th  generator of the gauge algebra in the representation $R$,
where the generators are normalized so that $C(\text{adj})$ is equal to the dual Coxeter number of the gauge algebra,
where $\text{adj}$ stands for the adjoint representation.
In this lecture note, we only need the following: \begin{align}
\SU(N):\quad & C(\text{adj})=N,& C(\text{fund})& =\frac12,\\
\SO(N):\quad & C(\text{adj})=N-2, & C(\text{vec})&=1,\\
\Sp(N):\quad & C(\text{adj})=N+1, & C(\text{fund})&=\frac12.
\end{align}
Here, \text{fund} of $\SU(N)$, $\text{vec}$ of $\SO(N)$, $\text{fund}$ of $\Sp(N)$ have dimensions $N$, $N$, $2N$, respectively.
The reader might not be familiar with the $\Sp$ group; a detailed introduction will be given in Sec.~\ref{sec:sp}.

The renormalization group equation \eqref{nonsusyRG} can be also written as \begin{equation}
\LambdaRG\frac{d}{d\LambdaRG}\frac{8\pi^2}{g^2}=\left[
\frac{11}{3}C(\text{adj})-\frac23  C(R_f)-\frac13   C(R_s)
\right] + O(g). \label{b}
\end{equation}
This form will be more natural in view of our convention \eqref{nonsusygaugekin} of having $1/g^2$ as one of the coefficients of terms in the Lagrangian.

\subsection{Anomalies}
\label{subsec:anomalies}
\subsubsection{Chiral anomaly}
\label{sec:chiral}
Non-abelian gauge theories have an important source of non-perturbative effects, called instantons. This is a nontrivial classical field configuration in the Euclidean $\mathbb{R}^4$ with nonzero integral of \begin{equation}
16\pi^2 k :=\int_{\mathbb{R}^4} \tr F_{\mu\nu}\tilde F^{\mu\nu}.
\end{equation} In the standard normalization of the trace for $\SU(N)$, $k$ is automatically an integer, and is called the instanton number.
The theta term in the Euclidean path integral appears as \begin{equation}
\exp\left[i\frac{\theta}{16\pi^2} \int \tr F_{\mu\nu} \tilde F^{\mu\nu}\right],
\label{theta}
\end{equation} 
and a configuration with the instanton number $k$ has a nontrivial phase $e^{i\theta k}$.
Note that a shift of $\theta$ by $2\pi$ does not change this phase. 
The shift $\theta\to\theta+2\pi$ does not change the physics on $\bR^4$.

Using \begin{equation}
\tr F_{\mu\nu}F_{\mu\nu} = \frac12 \tr(F_{\mu\nu}\pm \tilde F_{\mu\nu})^2 \mp  \tr F_{\mu\nu} \tilde F_{\mu\nu} 
\ge \mp \tr F_{\mu\nu}\tilde F_{\mu\nu}, 
\end{equation} we find that \begin{equation}
\int d^4 x \tr F_{\mu\nu} F_{\mu\nu} \ge 16\pi^2 |k| \label{bound}
\end{equation}
 which is saturated only when \begin{equation}
F_{\mu\nu} + \tilde F_{\mu\nu}=0 \quad
\text{or}
\quad F_{\mu\nu} - \tilde F_{\mu\nu}=0 
\label{SD}
\end{equation} depending on the sign of $k$.
Therefore, within configurations of fixed $k$, those satisfying relations \eqref{SD} give the dominant contributions to the path integral.  The solutions to \eqref{SD} are called instantons or anti-instantons, depending on the sign of $k$. 

In an instanton background, the weight in the path integral coming from the gauge kinetic term is \begin{equation}
\exp\left[ 
-\frac{1}{2g^2}\int \tr F_{\mu\nu}F^{\mu\nu}
+
i\frac{\theta}{16\pi^2}\int \tr F_{\mu\nu}\tilde F^{\mu\nu}
\right]
=e^{2\pi i \tau k}.\label{instanton_contri}
\end{equation} We similarly have the contribution $e^{2\pi i\bar\tau |k|}$ in an anti-instanton background. 

Let us consider  a charged left-handed fermion $\psi_\alpha$ is in the representation $R$ of the gauge group. 
Its complex conjugate, $\bar\psi_{\dot\alpha}$ is automatically in the representation $\bar R$.
It is known that the number of zero modes in $\psi_\alpha$ minus the number of zero modes in $\bar\psi_{\dot\alpha}$ 
is $2C(R)k$. 
In particular, the fermion path integral in a background gauge field configuration $A$ with positive instanton number $k$
\begin{equation}
\vev{O_1O_2\cdots}_A =  \int [D\psi] [D\bar\psi] O_1O_2 \cdots e^{-S} \label{bgf}
\end{equation}
 vanishes unless the product of the operators $O_1 O_2\cdots$ contains $2C(R)k$ more $\psi$'s than $\bar\psi$'s,
 to absorb these excess zero modes.
 
Loosely speaking, the path integral measures $[D\psi]$ and  $[D\bar\psi]$ contain both infinite number of integrations variables. However, there is a certain sense that there is a finite difference by $2C(R)k$ in the number of integration variables. 
Equivalently, under the spacetime-independent rotation \begin{equation}
\psi \to e^{i\varphi} \psi,\quad
\bar\psi \to e^{-i\varphi} \bar\psi,
\label{U1rot}
\end{equation} the fermionic path integration measure is transformed as \begin{equation}
\begin{aligned}\relax
[D\psi] & \to [D\psi]e^{+\infty i\varphi + 2C(R)k i\varphi  }, \\
[D\bar\psi] & \to [D\bar\psi]e^{-\infty i\varphi }. 
\end{aligned}
\end{equation} 
When combined, we have  \begin{equation}
[D\psi][D\bar\psi] \to [D\psi][D\bar\psi] e^{2C(R)k i\varphi} = 
[D\psi][D\bar\psi] \exp\left[
2C(R)\varphi \frac{i}{16\pi^2}\int \tr F_{\mu\nu}\tilde F^{\mu\nu}
\right].\label{phaseshift}
\end{equation}
For \eqref{bgf} to be non-vanishing, then, we need to have $O_1O_2\cdots $ to contain $2C(R)k$ additional $\psi$'s than $\bar \psi$'s.

Note that the shift \eqref{phaseshift}  can be compensated by a shift of the $\theta$ angle, $\theta\to \theta+2C(R)\varphi$. 
As we recalled before, the shift $\theta\to\theta+2\pi$ does not change the physics.
Therefore, the rotation of the field $\psi$ by $\exp(\frac{2\pi i}{2C(R)})$ is a genuine, unbroken symmetry.

\subsubsection{General perturbative anomaly}
\label{sec:anomalypoly}
In Sec.~\ref{sec:chiral}, we considered the change of the phase under the $\UU(1)$ rotation \eqref{U1rot} of the path integral \eqref{bgf} of fermions charged under $G$.
Let us generalize the discussion.

\paragraph{The anomaly polynomial:}
Say we have a Weyl fermion $\psi$ in a representation $R$ of $G$ in even spacetime dimension $n$ on a spacetime manifold $M_n$.
Let us consider the fermion path integral under a given $G$ gauge field $A_\mu$: \begin{equation}
Z[A_\mu] = \int [D\psi][D\bar\psi] e^{-\int \bar\psi D_\mu \sigma^\mu \psi}.
\end{equation}  
We ask what is the relation between $Z[A^g_\mu]$ and $Z[A_\mu]$,
where\begin{equation}
A^g_\mu = g^{-1} A_\mu g  + g^{-1} \partial_\mu g.
\end{equation} is the gauge connection transformed by a gauge transformation $g:M_n \to G$.

It is known that, for an infinitesimal gauge transformation $g=1+ \epsilon\chi +O(\epsilon^2)$, 
there is an anomalous phase transformation of the form \begin{equation}
Z[A^g] = \exp( 2\pi i  \epsilon\int_{M_n} F_R(\chi,A))  Z[A]
\end{equation}
given by the integral of a degree-$n$ differential form $F_R(\chi,A)$ depending on the gauge parameter $\chi$ and the gauge field $A$.

Let us provide a general formula for $F_R(\chi,A)$.
The formula might look scary when encountered for the first time.
To reduce the worry, we will give an explicit example shortly.

The degree-$n$ form $F_R(\chi,A)$ is given in terms of the gauge variation $\delta_\chi$ of the Chern-Simons term $\CS(A)$ which is a degree-$(n+1)$ form
\begin{equation}
d F_R[\chi, A] = \delta_\chi \CS_R(A).
\end{equation}
The Chern-Simons term $\CS_R(A)$ is in turn given in terms of the degree-$(n+2)$ form $\cA_R(A)$ known as the anomaly polynomial:
\begin{equation}
d\CS_R(A) = \cA_R(A).
\end{equation} 
The explicit form of  $\cA_R(A)$ is given by 
\begin{equation}
\cA_R(A) =\left[ 
(1+ \frac{p_1}{24} + \frac{7p_1^2-4p_2}{5760}+\cdots ) \tr_R e^{ \frac{1}{2\pi} F^a (T_R)^a  }
\right]_\text{$(n+2)$-form part}
\end{equation}
where $p_1$ and $p_2$ are certain degree-4 and degree-8 forms composed from the spacetime curvature called the Pontryagin classes
and \begin{equation}
F^a :=  \frac12 F_{\mu\nu}^a dx^\mu dx^\nu
\end{equation} is the  $a$-th gauge curvature of $A_\mu$ expressed as a $2$-form,
and finally $(T^a)_R$ is the $a$-th generator of the gauge algebra in the representation $R$,
which we take to be Hermitean.
We often use the anomaly polynomial $\cA_R$ directly, instead of computing the actual gauge variation $F_R$.

\paragraph{Chiral anomaly:}
As an explicit example, let us reproduce the chiral anomaly using this general framework.
Take $G=\UU(1)\times \SU(N)$,
and consider a 4d chiral fermion $\psi$ which has charge $+1$ under $\UU(1)$ and transforms in the $N$-dimensional fundamental representation under $\SU(N)$.
We choose the spacetime to be flat, so that the gravitational terms $p_1$ and $p_2$ vanish.
Then the anomaly polynomial is \begin{equation}
\cA(A) = \left[\tr e^{\frac{1}{2\pi} ( F_{\UU(1)}+ F_G )} \right]_\text{$6$-form part}
=\frac1{3!}\tr [\frac{1}{2\pi} ( F_{\UU(1)}+ F_G )]^3
\supset \frac12 \frac{1}{(2\pi)^3} F_{\UU(1)}  \tr (F_G )^2
\end{equation}
where in the last expression we extracted the only term relevant to the $\UU(1)$-$G$-$G$ anomaly.
We then have \begin{equation}
\CS_{\UU(1)G^2}(A)= \frac12 \frac{1}{(2\pi)^3} A_{\UU(1)} \tr (F_G)^2
\end{equation} and then \begin{equation}
F(\chi,A) = \frac12 \frac{1}{(2\pi)^3} \chi_{\UU(1)} \tr (F_G)^2.
\end{equation} 
We now convert the differential form expression $\int \tr (F_G)^2 $ 
to a more explicit spacetime integral $\int d^4x \tr F_{\mu\nu} \tilde F_{\mu\nu}$: 
the coefficient turns out to be \begin{equation}
\int \tr(F_G)^2 = \frac12\int d^4 x F_{\mu\nu}\tilde F_{\mu\nu}.
\end{equation}
Then indeed we find that under the $\U(1)$ gauge transformation by $\psi\to e^{i\epsilon\chi} \psi$,
we get the phase rotation \begin{equation}
Z[A^{g}]=\exp(i \epsilon\chi \frac{1}{16\pi^2} \int d^4x \tr F_{\mu\nu}\tilde F_{\mu\nu}) Z[A],
\end{equation}
reproducing \eqref{phaseshift} when $R$ is the fundamental of $\SU(N)$.

\paragraph{Gauge anomaly:}
When we consider $G$ as a dynamical gauge group,
we need to perform the path integral over the gauge field $A$.
In this case, we need to require that the fermion path integral is gauge invariant $Z[A^g]=Z[A]$.
This is guaranteed if and only if the anomaly polynomial $\cA$ vanishes.

We remind the reader that the pure gauge part of the anomaly polynomial of
a chiral fermion in representation $R$ of $G$ is given by the formula \begin{equation}
\cA=\frac16\tr_R (\frac{F}{2\pi})^3.
\end{equation}

Since $\tr_{\bar R} F^3/6  = -\tr_{R} F^3/6$, if a chiral fermion appears in a real representation, 
the anomaly automatically cancels.
The case of the Standard Model is more subtle.
The chiral fermions contained in a single generation are given as follows\footnote{%
Why did  nature choose such a strange combination of gauge charges to cancel the gauge anomaly?
The author's take is that our universe is being simulated by a PhD student trying to complete her/his thesis.
Many of the more natural looking representations were already treated in papers from the past,
and therefore the student needed to come up with a contrived spectrum to write a new paper.
}: \begin{equation}
\begin{array}{c|cccccccccc}
&Q_L & \bar u_R & \bar d_R & \ell_L & \bar e_R \\ 
\hline
\SU(3) & 3 & \bar 3 & \bar 3 & 1 & 1  \\
\SU(2) & 2 & 1 & 1 & 2 & 1 \\
\UU(1) & 1/6 & -2/3 & 1/3 & -1/2 & 1
\end{array}.
\end{equation}
Let us check the $\U(1)$ part of the cancellation: 
 \begin{equation}
\cA=\frac{1}6 (\frac{F_{\UU(1)}}{2\pi})^3(3\cdot 2\cdot \left(\frac16\right)^3 
+ 3\cdot \left(-\frac23\right)^3
+ 3\cdot \left(\frac13\right)^3
+ 2\cdot \left(-\frac12\right)^3
+ 1\cdot 1^3)=0.
\end{equation}

\exercise{Check the cancellation of the full anomaly polynomial of the Standard Model.}

\paragraph{On terminologies:}

Before proceeding, we would like to make a remark on the terminology.
Let us fix the spacetime dimension to $n=2d$.
When a theory $Q$, such as a theory of free fermions, has a symmetry $G$, we can introduce the background $G$-gauge field  $A_G$ to the system.
The anomaly refers to a controllable change in the phase of the partition function under the $G$ gauge transformation.

Suppose $G=G_0\times F$. 
We can try to perform the path integral over $A_{G_0}$.
If possible, the result is a $G_0$-gauge theory, where a dynamical $A_{G_0}$ gauge field couples to $Q$.
Whether this is possible or not is controlled by the pure $G_0$ anomaly of $Q$.
This part is usually called as  the \emph{gauge anomaly}.

The part $F$ often remains as the flavor symmetry of the $G_0$ gauge theory.
The $F^{d+1}$ part of the anomaly is usually called as the \emph{'t Hooft anomaly}, after its use in the anomaly matching in \cite{tHooft:1979rat}.

If $F=\U(1)$ and the $(G_0)^d F$ part of the anomaly is nonzero, the current of $F$ is no longer conserved.
This was what happened in the original Adler-Bell-Jackiw anomaly \cite{Adler:1969gk,Bell:1969ts}
when $G_0=\U(1)$ is the Maxwell field, and the effect  is called as the \emph{axial anomaly}.
This was also the source of the mass of the $\eta'$ \cite{tHooft:1976rip}, for which $G_0$ is a non-Abelian  gauge group. 
The effect is often called as the \emph{chiral anomaly}.

There are indeed many names. But all these phenomena come from the same underlying mechanism, and are controlled by the anomaly polynomial.

\subsubsection{Non-renormalization of the anomaly}
\label{sec:nonrenanom}
Let us consider an $n$-dimensional quantum field theory $Q$ with flavor symmetry $G$.
We can then consider its partition function $Z_Q[A]$ in the presence of the background $G$ gauge field $A_\mu$ on the $n$-dimensional manifold $M_n$.
Its phase in general depends on the gauge choice: we have \begin{equation}
Z_Q[A^g] = \exp( 2\pi i  \epsilon\int_{M_n} F_Q(\chi,A))  Z_Q[A]
\end{equation} 
for an infinitesimal gauge transformation $g=1+\epsilon\chi+O(\epsilon^2)$.

The discussion in the previous subsection was for the particular case when the theory $Q$ in question is simply a free fermion field in the representation $R$ of $G$.
Even for the general case, the anomalous phase $F_Q(\chi,A)$ is controlled by the Chern-Simons term $\CS_Q(A)$,
which is then determined by the anomaly polynomial $\cA_Q(A)$ via the descent relations
\begin{equation}
d F_Q[\chi, A] = \delta_\chi \CS_Q(A),\qquad
d\CS_Q(A) = \cA_Q(A). 
\label{generaldescent}
\end{equation}

\begin{figure}
\centering
\includegraphics[scale=.3,valign=b]{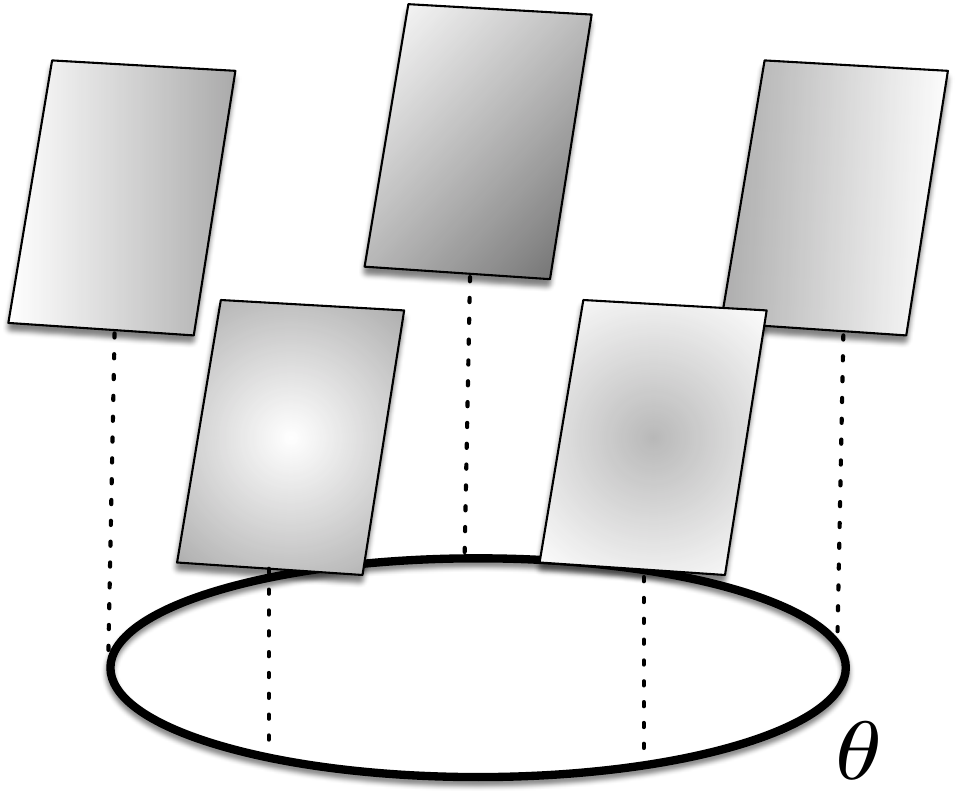}\qquad\qquad
\includegraphics[scale=.3,valign=b]{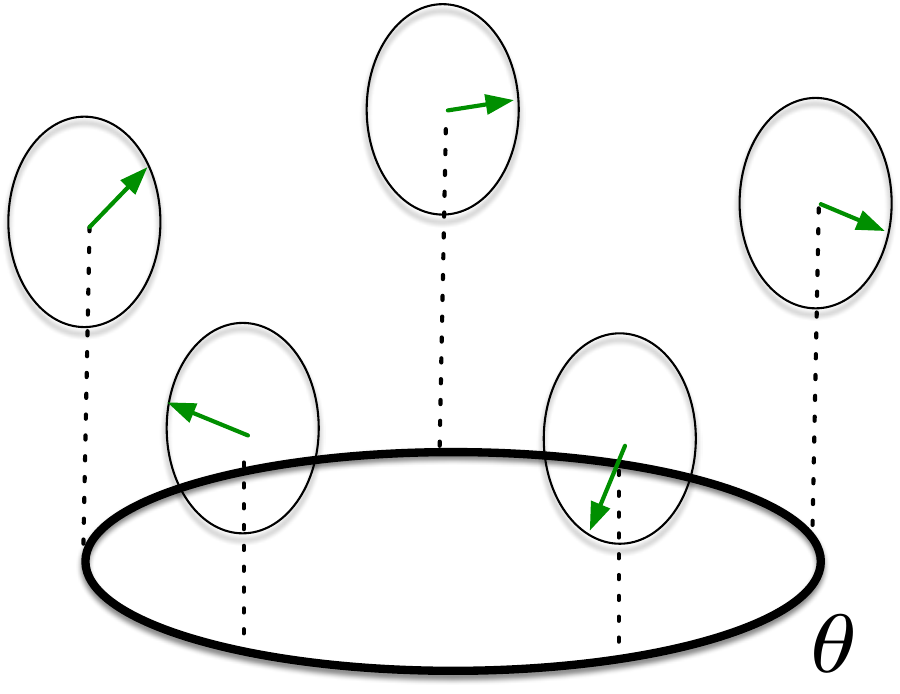}
\caption{Left: gauge configurations $A^{g(\theta)}$ obtained by a $\theta$-dependent gauge transformation $g(\theta)$ from a given $A$; the fiber is supposed to be $\bR^n$ and the gradation on it tries to depict a gauge field.
Right: the corresponding partition function $Z_Q[A^{g(\theta)}]$. The absolute value stays constant but the phase varies.
\label{fig:one}}
\end{figure}

We can give a more direct significance for the anomaly polynomial $\cA_Q(A)$.
Consider a one-parameter family of gauge transformations \begin{equation}
g(\theta;x)\in G 
\end{equation} for $\theta\in[0,2\pi]$ and $x\in M_n$, where we require $g(0;x)=g(2\pi;x)$.
Since $Z_Q[A^{g(0)}]=Z_Q[A^{g(2\pi )}]$, the total change of the phase of $Z_Q[A^{g(\theta)}]$ as we continuously change $\theta$ from $0$ to $2\pi$ is of the form  $2\pi c$ for an integer $c$.
See Fig.~\ref{fig:one} for an illustration.

\begin{figure}
\centering
\includegraphics[scale=.3,valign=b]{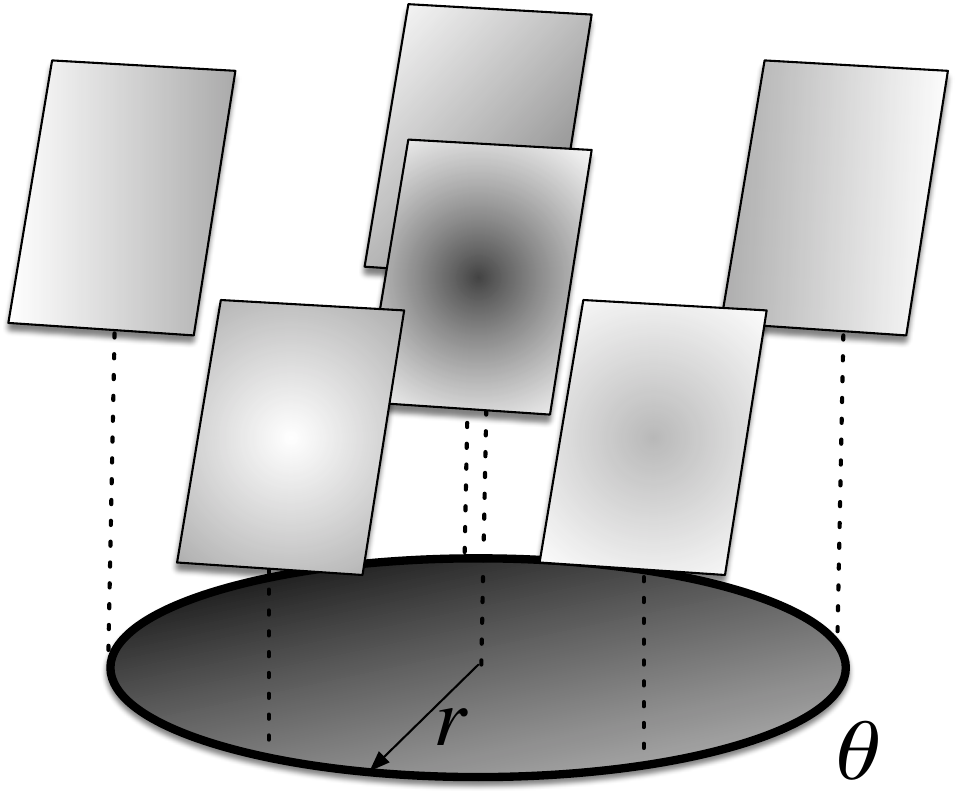}\qquad\qquad
\includegraphics[scale=.3,valign=b]{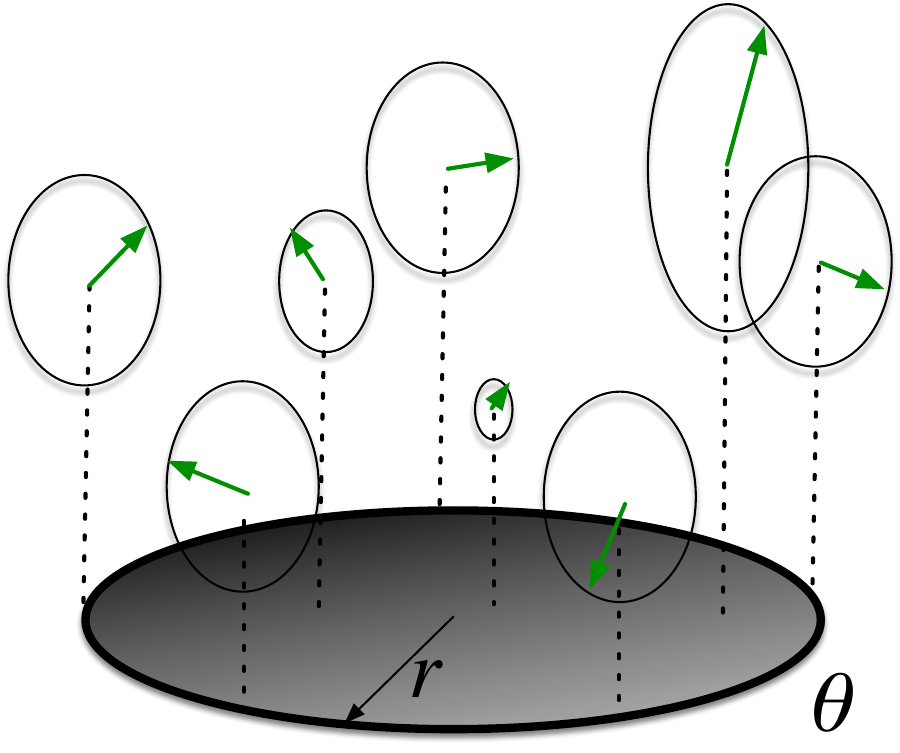}
\caption{Left: gauge configurations are now filled inside the disk. In the interior, it is not generally a gauge transform of the originally given $A$.
Right: the corresponding partition function $Z_Q[A(r,\theta)]$. 
The absolute value also changes in the interior.
\label{fig:two}}
\end{figure}

We now consider a two-parameter family of gauge fields \begin{equation}
A(r,\theta; x )
\end{equation} where $(r,\theta)$ parameterize a disk $D^2$ of radius 1, such that its value at the boundary circle at $r=1$ is given by  \begin{equation}
A(r=1,\theta) = A^{g(\theta)}.
\end{equation}
Note that in the interior $r<1$, the gauge configuration is not usually a gauge transform of the originally given $A$.
This defines a gauge field on $D^2\times M_n$, see Fig.~\ref{fig:two}.
The descent equation \eqref{generaldescent} means that the integer $c$ characterizing the total phase rotation is given by \begin{equation}
c=\int_{D^2\times M_n} \cA_Q(A).
\end{equation}

This geometric consideration tells us that the anomaly polynomial $\cA_Q(A)$ cannot be renormalized, since the renormalization is a continuous process, whereas an integer cannot be continuously changed.

This non-renormalization statement, when applied to a weakly-coupled system of fermions, scalars and gauge fields, implies the original Adler-Bardeen theorem \cite{Adler:1969er}.
The same statement, when applied to the ultraviolet limit and the infrared limit of a strongly-coupled system, is usually known as the 't Hooft anomaly matching condition, originally introduced in \cite{tHooft:1979rat}.\footnote{%
In passing, we note that the original argument in \cite{tHooft:1979rat} using a spectator fermion to cancel the anomaly and making the flavor symmetry weakly dynamical is not quite satisfactory
for a general quantum field theory $Q$ with symmetry $G$,
since there is no guarantee that there is actually a spectator fermion which can cancel the anomaly of $Q$.
For example, in four dimensions, the $E_8$-symmetric theory of Minahan and Nemeschansky \cite{Minahan:1996cj} is known to have 1/5 of the smallest possible $\U(1)$-$E_8$-$E_8$ anomaly of an $E_8$-charged fermion \cite{Cheung:1997id}.
In this case, we can take five decoupled copies of Minahan-Nemeschansky's $E_8$ theory and apply the original argument,
but there is no a priori guarantee that this trick of considering multiple decoupled copies should work.

There is a different justification of the 't Hooft anomaly matching condition by Coleman and Grossman \cite{Coleman:1982yg} which appeared a few years after the original reference \cite{tHooft:1979rat}.
There, the relevant correlator of current operators was analyzed using analyticity and unitarity,
and the anomalous part was shown to be equal when compared between the ultraviolet limit and the infrared limit.
}

\subsubsection{Global anomaly}
Let us next consider Witten's global anomaly \cite{Witten:1982fp}.
Take a chiral fermion in the doublet of gauge $\SU(2)$.
There is no perturbative anomaly, since $\tr( F_a \sigma^a)^3=0$ 
just because a cube of a traceless $2\times2$ matrix is traceless.
This means that a continuous $\SU(2)$ gauge transformation does not change the phase of $[D\psi][D\bar \psi]$. 
What matters then is 
  maps $g:\mathbb{R}^4\to \SU(2)$ identified up to continuous change.
Mathematically, they are characterized by the homotopy group $\pi_4(\SU(2))$, 
and it is known that \begin{equation}
\pi_4(\SU(2))=\pi_4(S^3)=\mathbb{Z}_2.
\end{equation} Let $g_0:\mathbb{R}^4\to \SU(2)$ be the one corresponding to the nontrivial element in this $\bZ_2$.  
It is known that under this $[D\psi][D\bar \psi]$ gets a minus sign.
So, one cannot have an odd number of Weyl fermions in the doublet representation in an $\SU(2)$ gauge theory.

\exercise{Check that there are even number of $\SU(2)$ doublets in the Standard Model.}

It is known that $\pi_4(G)$ is nontrivial only for $G=\SU(2)$ or $G=\Sp(n)$; 
$\SU(2)$ is a special case of $\Sp(n)$ when $n=1$.
In this case $\pi_4(G)=\bZ_2$.\footnote{%
For the computation of $\pi_4(G)$, see \cite{BottSamelson}.
The computation boils down to the fact that only for $G=\Sp(n)$ there is a root which is a nontrivial multiple of a weight. 
Indeed, the long root of $\Sp(n)$ is twice a weight vector.
See also \url{https://mathoverflow.net/questions/259487/computation-of-pi-4-of-simple-lie-groups}.
}
A chiral fermion in the fundamental $2n$ dimensional representation of $\Sp(n)$ is known to produce the minus sign under the topologically nontrivial gauge transformation.
Therefore, one cannot have an odd number of Weyl fermions in the fundamental representation in an $\Sp(n)$ gauge theory.

Before proceeding, we note that the global anomaly we saw here can in fact be computed using the perturbative anomaly we reviewed in Sec.~\ref{sec:anomalypoly}.
In the case of the doublet fermion of $\SU(2)$,
we embed it into a triplet fermion of $\SU(3)$, and use the fact that 
a topologically nontrivial gauge transformation $g:\bR^4\to \SU(2)$ 
can be continuously connected to a trivial gauge transformation if considered as a map
$g:\bR^4\to \SU(3)$ via the embedding $\SU(2)\subset \SU(3)$.
For details, see  the original paper by Elitzur and Nair \cite{Elitzur:1984kr}.

\subsection{Gauge groups vs gauge algebras}
\label{sec:groupglobal}
In physics literature, we often do not make a distinction between two gauge groups sharing the same Lie algebra,
such as $\SU(2)$ and $\SO(3)$ sharing $\su(2)=\so(3)$
or $\SU(4)$ and $\SO(6)$ sharing $\su(4)=\so(6)$.
This distinction does not matter if we only consider gauge theories formulated on a flat $\bR^4$ without any line operators.
But there are situations where this distinction does matter
and can be used to a great effect, allowing us to obtain more information about the systems under consideration.

For more details about this subsection, the readers are referred to \cite{Aharony:2013hda,Kapustin:2014gua}.
Before proceeding, we also mention an interesting paper \cite{Tong:2017oea} where the issues explained here were studied in the case of the Standard Model.

\subsubsection{The 't Hooft magnetic flux and the Stiefel-Whitney class}
\label{sec:StiefelWhitney}
Let us first consider an $\SO(3)$ gauge configuration on $T^2$ given as follows.
We start from a flat space $\bR^2$ parameterized by $x$ and $y$, 
and identity $x\sim x+L$ and $y\sim y+L$.
When making this identification,
we perform  an $\SO(3)$ gauge transformation by $u=\diag(-1,-1,+1)$ in the $x$-direction
and by $v=\diag(+1,-1,-1)\in \SO(3)$ in the $y$-direction.
Since $uv=vu$, these two identifications are compatible, and indeed define an $\SO(3)$ gauge bundle on $T^2$.

This configuration does not lift to an $\SU(2)$ configuration, however. 
The element $u$ lifts to $U=i\sigma_z \in \SU(2)$
and $v$ to $V=i\sigma_x\in \SU(2)$,
and they do not commute: $UV=-VU$. 

It is easy to generalize this to $\SU(N)$ and $\SU(N)/\bZ_N$:
we take \begin{equation}
U=c\diag(1,\omega,\omega^2,\ldots,\omega^{N-1}),\qquad
V=c' \begin{pmatrix}
0 & & & 1 \\
1& 0 &  \\
& \ddots & \ddots & \\
 & & 1 & 0
\end{pmatrix}
\end{equation}
where $\omega=e^{2\pi i/N}$  and we choose the phase $c$ and $c'$ so that $U,V\in \SU(N)$.
They satisfy $ UV=\omega VU$ and therefore do not define a consistent $\SU(N)$ bundle on $T^2$.
They still define an $\SU(N)/\bZ_N$ bundle.

Such gauge configurations were first considered in the physics literature in the late 70s by 't Hooft \cite{tHooft:1979uj}, and is known under the name of the 't Hooft magnetic flux.
In mathematics such bundles are said to have a non-trivial  second Stiefel-Whitney class\footnote{%
The phrase `the second Stiefel-Whitney class' is, in the mathematics literature,
usually restricted to the obstructions related to lifting an $\SO$ bundle to a $\Spin$ bundle.
In physics the phrase is often used in a generalized sense, associated to any group covering another.
}; 
the work of Stiefel and Whitney goes back to late 1930s and early 1940s.

A configuration with a nontrivial second Stiefel-Whitney class can be obtained in a different way.
Consider now a two-sphere $S^2$ surrounding a point in the space $\bR^3$.
We consider a $\U(1)$ Dirac monopole background on this $S^2$,
and regard it as an $\SO(3)$ or an $\SU(2)$ configuration.
The charge of the Dirac monopole is specified by the rotation number of the $\U(1)$ gauge transformation around the equator of $S^2$.
Suppose this $\U(1)$ rotation around the equator corresponds to a $360^\circ$ rotation in the gauge $\SO(3)$.
This is a valid $\SO(3)$ monopole configuration.
However, this does not lift to a consistent $\SU(2)$ configuration, since one needs a $720^\circ$ rotation in $\SU(2)$ in this normalization.

This $\SO(3)$ magnetic monopole configuration therefore has a 't Hooft magnetic flux, or equivalently has a nontrivial second Stiefel-Whitney class.
We see that the smallest possible magnetic charge of an $\SU(2)$ configuration is twice the smallest possible magnetic charge of an $\SO(3)$ configuration.

\subsubsection{Line operators}
We define an $\SU(2)$ gauge theory to be a quantum theory obtained by the path integral over all possible $\SU(2)$ gauge configurations.
Similarly, we define an $\SO(3)$ gauge theory by considering all possible $\SO(3)$ configurations.
There can be $\SO(3)$ configurations which do not lift to $\SU(2)$.
Two distinct $\SU(2)$ configurations might descend to the same $\SO(3)$ configurations.
Therefore, they give rise to different theories.\footnote{%
In the physics literature, we sometimes see a misguided statement that the choice of the charged matter fields determines the global structure of the group, that a pure $\su(2)$ gauge theory without any charged matter will have the gauge group $\SO(3)$, for example.
Our point of view is different. Even for a pure gauge theory without charged matter fields, we can make a choice of the gauge configurations to path integrate over.
}

Even on the flat $\bR^4$, an $\SO(3)$ gauge theory and an $\SU(2)$ gauge theory behave differently,
once we start considering line operators. 
Let us first introduce different types of line operators.

\paragraph{'t Hooft line operators:}
Let us first consider a 't Hooft line operator on a contour $C\subset \bR^4$ of a $G$ gauge theory.
This is defined e.g.~by demanding that the gauge field configuration very close to $C$ is given by a Dirac monopole of a $\U(1)$ subgroup of the gauge group $G$.
As we saw, in the unit where the allowed 't Hooft operators of an $\SO(3)$ gauge theory have integer magnetic charges,
the allowed 't Hooft operators of an $\SU(2)$ gauge theory have even integer charges.

We can also use a rougher classification of 't Hooft operators, by specifying the 't Hooft magnetic flux or equivalently the Stiefel-Whitney class of the bundle as restricted on $S^2$ around the contour $C$.
Then, an $\SO(3)$ gauge theory has 't Hooft operators distinguished by its $\bZ_2$ quantum number,
whereas an $\SU(2)$ gauge theory does not have any topologically nontrivial 't Hooft operator.
This $\bZ_2$ quantum number is the modulo 2 reduction of the charge of the $\U(1)$ Dirac monopoles embedded in $\SU(2)$ or $\SO(3)$ discussed above.

\paragraph{Wilson line operators:}
Wilson line operators might be more familiar to the readers: they are specified by giving a contour $C\subset \bR^4$ and providing a representation $R$ of the gauge group.
We then insert in the path integral the path-ordered exponential $P\exp\int A$ of the gauge field $A$ in the representation $R$.

For an $\SU(2)$ gauge theory, an irreducible representation $R$ is specified by its maximal spin, which can either be integer or half-integer.
Among them, only those with integer spin are allowed in an $\SO(3)$ gauge theory.

We can make a rougher classification of Wilson line operators, by specifying how the central element  $-1\in \SU(2)$ acts on them.
It is well-known that this element acts as $+1$ on an integer-spin representation and as $-1$ on a half-integer-spin representation.
This provides a $\bZ_2$ classification of Wilson line operators of an $\SU(2)$ gauge theory.

\paragraph{Dyonic line operators:}
We can also consider a dyonic line operator, obtained by placing a Wilson line operator and a 't Hooft line operator on neighboring contours $C$ and $C'$, and bringing $C'$ on top of $C$.
For a moment, let us only specify the gauge algebra to be $\su(2)$ and remain agnostic about whether the gauge group is $\SO(3)$ or $\SU(2)$.
In this case, the electric/magnetic charge of a line operator is specified by $(\lambda_e,\lambda_m)\in \bZ\times \bZ$.
Here, we use the normalization that $\lambda_e$ is twice the spin of the $\SU(2)$ representation so that it is even for an integer-spin representation and odd for a half-integer-spin representation,
and that $\lambda_m$ is the integer magnetic charge of the Dirac monopole  for $\U(1)\subset \SO(3)$.
A rougher $\bZ_2\times\bZ_2$ classification is obtained by reducing these integer charges modulo 2.

\subsubsection{Full spectrum of line operators: $\SU(2)$ vs $\SO(3)$}
\label{sec:su2so3}
\begin{figure}
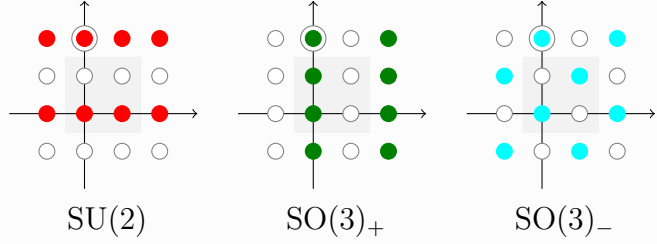

\centering
\begin{tabular}{ccc}
\xlattice01{Red} & \xlattice10{Green} & \xlattice11{Cyan} \\
$\SU(2)$ & $\SO(3)_+$ & $\SO(3)_-$
\end{tabular}
\caption{The weights of line operators of gauge theories with the Lie algebra $\mathfrak{su}(2)$.
There, the horizontal axis is for $\lambda_e$ and the vertical axis is for $\lambda_m$. The shaded regions in the figure give the $\bZ_2$ charges.
The symbol $\odot$  marks the magnetic charge of the condensate in a pure Yang-Mills theory.
\label{fig:su2lines}
}
\end{figure}

As we already saw, an $\SU(2)$ theory has line operators with even $\lambda_m$,
while an ordinary $\SO(3)$ theory has line operators with even $\lambda_e$.
For an illustration, see the entries $\SU(2)$ and $\SO(3)_+$ in Fig.~\ref{fig:su2lines}.

There is in fact another consistent choice of line operators.
To see this, we note that the failure of the Dirac quantization law between an electric line operator of charge $\lambda_e$ 
and a magnetic line operator of charge $\nu_m$ is given by $(-1)^{\lambda_e \nu_m}$.
Therefore, the phase inconsistency between two line operators with charges $(\lambda_e,\lambda_m)$ and $(\nu_e,\nu_m)$ is $(-1)^{\lambda_e \nu_m-\lambda_m\nu_e}$.
Therefore, for any consistent choice $S$ of charges of line operators, we require \begin{equation}
\lambda_e \nu_m-\lambda_m\nu_e \in 2\bZ\qquad\text{for any}\quad (\lambda_e,\lambda_m), (\nu_e,\nu_m) \in S.
\end{equation}
We also require that \begin{equation}
 (\lambda_e+\nu_e,\lambda_m+\nu_m)\in S
\qquad\text{for any}\quad (\lambda_e,\lambda_m), (\nu_e,\nu_m) \in S.
\end{equation}

For an $\SU(2)$ theory, we start with purely electric line operators, which has a charge of the form $(\lambda_e,\lambda_m)=(\text{arbitrary},0)$.
It is easy to see that the only maximal set $S$ of consistent charges we can add is of the form
$(\lambda_e,\lambda_m)=(\text{arbitrary},\text{even})$.
In contrast, for an $\SO(3)$ theory, we start with purely electric line operators whose charge is given by $(\lambda_e,\lambda_m)=(\text{even},0)$.
Then there are two maximal sets $S$ of consistent charges.
We already saw one, which takes $(\lambda_e,\lambda_m)=(\text{even},\text{arbitrary})$.
We also find another one, which takes $(\lambda_e,\lambda_m)=(\text{even},\text{even})$ or $=(\text{odd},\text{odd})$. 
These two choices are known as an $\SO(3)_\pm$ gauge theories, see Fig.~\ref{fig:su2lines}.

The distinction between $\SO(3)_+$ and $\SO(3)_-$ can be given in terms of the Lagrangian.
To see this, we use the Witten effect \cite{Witten:1979ey} under the change of the theta angle $\theta$ \eqref{theta} from $0$ to $2\pi$.
In our normalization, this sends a line operator with charge $(\lambda_e,\lambda_m)$ to a line operator with charge $(\lambda_e+\lambda_m,\lambda_m)$
This will map an $\SO(3)_+$ theory to an $\SO(3)_-$ theory and vice versa:
\begin{equation}
\SO(3)_{+}^{\theta}=\SO(3)_{-}^{{\theta+2\pi}}.
\label{SO3+-}
\end{equation}

This also means that when $G=\SO(3)$ the  true periodicity of $\theta$ is $4\pi$.
This is due to the fact that on spin manifolds\footnote{On non-spin manifolds  there can be ``quarter instantons'' and the periodicity of $\theta$ is $8\pi$. }, the instanton number of $\SO(3)$ gauge theories is a multiple of $1\over 2$ in our normalization.
Naively, the shift of $\theta$ by $2 \pi$ does not change the local physics.  But since the insertion of the line operators in $\bR^4$ creates a nontrivial topology, it allows us to distinguish $\theta$ from $\theta+2\pi$ locally on $\bR^4$.

\subsubsection{Full spectrum of line operators: $\SU(4)$ vs $\SO(6)=\SU(4)/\bZ_2$ vs $\SU(4)/\bZ_4$}
It is not difficult to repeat the analysis for other Lie algebras and corresponding Lie groups.
For concreteness, let us discuss the case of the algebra $\su(4)\simeq \so(6)$ and the corresponding Lie groups $\SU(4)$, $\SO(6)=\SU(4)/\bZ_2$ and $\SU(4)/\bZ_4$.

The center of $\SU(4)$ is $\{1,i,-1,-i\} \in \SU(4)$,
and this allows us to classify its irreducible representation by $\bZ_4$.
Conversely, the homotopy group of $\SU(4)/\bZ_4$ is $\bZ_4$,
and the 't Hooft magnetic flux or equivalently the generalized second Stiefel-Whitney class is labeled by $\bZ_4$.
Therefore, a charge of a line operator is given by $(\lambda_e,\lambda_m)\in \bZ_4\times \bZ_4$,
and two line operators with charges $(\lambda_e,\lambda_m)$ and $(\nu_e,\nu_m)$ are consistent with each other only when \begin{equation}
\lambda_e \nu_m-\lambda_m\nu_e =0 \mod 4.
\end{equation}

\begin{figure}
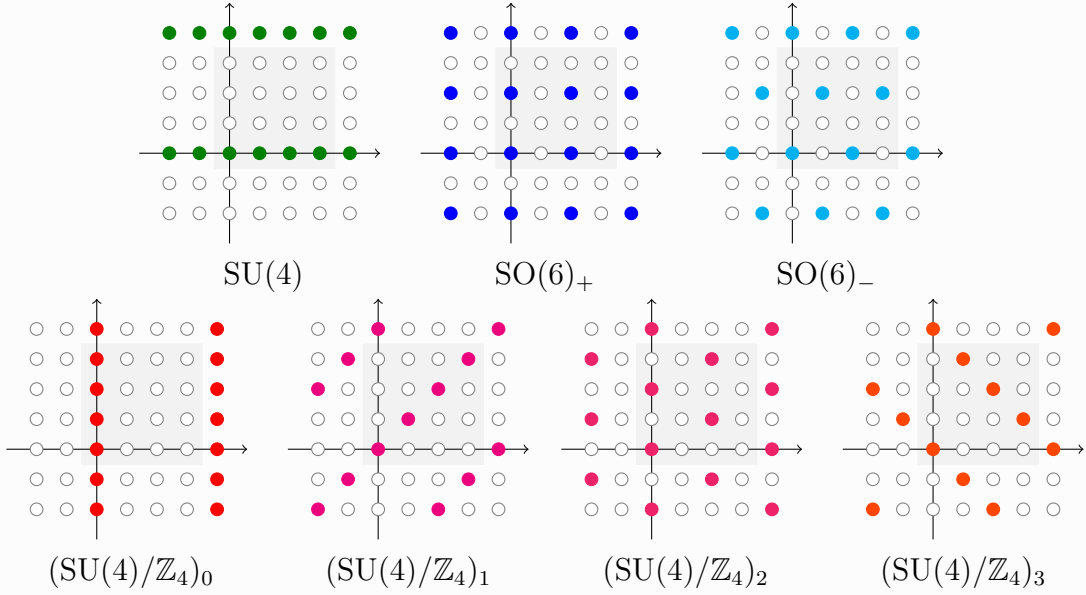

\centering
\begin{tabular}{ccc}
\lattice01{Green} &  \slattice{blue} & \lattice21{ProcessBlue} \\
$\SU(4)$ & $\SO(6)_+$ & $\SO(6)_-$
\end{tabular}

\begin{tabular}{cccc}
\lattice10{red} & \lattice11{RubineRed}& \lattice12{WildStrawberry} & \lattice13{OrangeRed}\\
$(\SU(4)/\bZ_4)_0$ & $(\SU(4)/\bZ_4)_1$ & $(\SU(4)/\bZ_4)_2$ &$(\SU(4)/\bZ_4)_3$
\end{tabular}

\caption{The weights of line operators of gauge theories with the Lie algebra $\mathfrak{su}(4)$.
The horizontal axis is for $\lambda_e$ and the vertical axis is for $\lambda_m$. 
\label{fig:su4lines}
}
\end{figure}

For an $\SU(4)$ theory, we demand the existence of electric line operators with charge $(\lambda_e,\lambda_m)=(\text{arbitrary},0)$;
for an $\SO(6)=\SU(4)/\bZ_2$ theory, we similarly require the operators with charge $(\lambda_e,\lambda_m)=(\text{even},0)$;
finally for an $\SU(4)/\bZ_4$ theory, there are operators with charge $(\lambda_e,\lambda_m)=(0,0)$ modulo 4.
We can then ask what are the maximally allowed set of charges in each case.
We find one for $\SU(4)$,
two for $\SO(6)=\SU(4)/\bZ_2$,
and four for $\SU(4)/\bZ_4$,
as shown in Fig.~\ref{fig:su4lines}.

Again, the distinction among four choices of $\SU(4)/\bZ_4$ can be understood by shifting the theta angle.
Indeed, shifting $\theta$ to $\theta+2\pi$,
we shift $(\lambda_e,\lambda_m)$ to $(\lambda_e+\lambda_m,\lambda_m)$, 
mapping \begin{equation}
(\SU(4)/\bZ_4)^{\theta+2\pi}_i =(\SU(4)/\bZ_4)^{\theta}_{i+1}
\end{equation} where the integer label $i$ is taken modulo 4.
This corresponds to the fact that the instanton number of an $\SU(N)/\bZ_N$ configuration on a spin manifold is in general in $(1/N) \bZ$, and that the true periodicity of the theta angle is $2N\pi$.

We note however that the two choices for $\SO(6)$ are \emph{not} mapped by shifting the theta angle by $2\pi$.
Indeed, one easily sees that $\SO(6)_+$ is mapped to $\SO(6)_+$ and that the same is true for $\SO(6)_-$.
Still, it is known that the distinction between $\SO(6)_\pm$ can be expressed as a topological term in the Lagrangian
\begin{equation}
\pi i \nu\frac12 \int_{M_4} \cP(w_2)
\label{pont}
\end{equation}
where $w_2$ is the second Stiefel-Whitney class of the $\SO(6)$ bundle,
$\cP:H^2(-,\bZ_2)\to H^4(-,\bZ_4)$ is the cohomology operation known as the Pontryagin square.
On a spin manifold, $\cP(w_2)$ integrates to an even number modulo 4.
Then, the choice $\nu=0$ or $1$ specifies whether we have $\SO(6)_+$ or $\SO(6)_-$.

\subsubsection{Full spectrum of line operators: $\Spin(N)$ vs $\SO(N)$}
\label{sec:spin-vs-so}

\begin{figure}
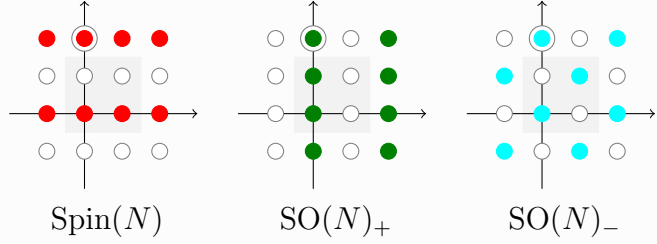

\centering
\begin{tabular}{ccc}
\xlattice01{Red} & \xlattice10{Green} & \xlattice11{Cyan} \\
$\Spin(N)$ & $\SO(N)_+$ & $\SO(N)_-$
\end{tabular}
\caption{The weights of line operators of gauge theories with the Lie algebra $\mathfrak{so}(N)$ for $N$ odd.}
\label{fig:soNodd}
\end{figure}

We can generalize the discussion to $\Spin(N)$ vs $\SO(N)$.
Here we only discuss the case $N$ is odd.
For the more complicated case of $N$ even, see \cite{Aharony:2013hda}.

When $N$ is odd, the spectra of the line operators look similar to $\SO(3)$,
as shown in Fig.~\ref{fig:soNodd}.
The main difference is that the smallest instanton in $\SO(5)$ is in fact smaller by a factor of 2 compared to the smallest instanton of $\SO(3)$ embedded to $\SO(5)$ via a natural embedding.
In other words, the instanton number in $\SO(3)$ is \begin{equation}
k:=\frac{1}{64\pi^2}\int_{\bR^4} \tr_{\mathbf{3}} F_{\mu\nu} \tilde F^{\mu\nu}  \in \bZ
\end{equation} while it is \begin{equation}
k:=\frac{1}{32\pi^2}\int_{\bR^4} \tr_{\mathbf{N}} F_{\mu\nu} \tilde F^{\mu\nu}  \in \bZ
\end{equation} for $\SO(N)$ with $N\ge 4$,
where $F_{\mu\nu}$ is an $N\times N$ anti-symmetric matrix.
This makes the $2\pi$ shift of $\theta$ preserves $\SO(N)_\pm$, instead of exchanging $\pm$.

In terms of the Lagrangian, this is explained as follows.
In general, for an $\SO(N)$ bundle, the topological classes satisfy 
the relation \begin{equation}
4\int p_1 = \int \cP(w_2)+2\int w_4 \mod 4,
\end{equation}  
where $\int \cP(w_2)$ is even on a spin manifold.
The difference between small $N$ and large $N$ is that  $\int p_1=4k$  for $N=3$ while $\int p_1=2k$  for $N\ge 4$, where $k$ is the instanton number in the standard normalization.
This means that $k$ can be half-integral for $N=3$ but not for $N\ge 4$, etc.
Therefore, the coefficient $\nu$ in \eqref{pont} can be traded for $\theta=2\pi$ in $\SO(3)$,
but this cannot be done for $\SO(N)$ with larger $N$.

\subsubsection{Line operators and the confinement of pure Yang-Mills theory}
\label{sec:nonsusypure}
Let us now study, at least heuristically, how the choice of $\SU(2)$ vs $\SO(3)_+$ vs $\SO(3)_-$ affects the dynamics of the pure Yang-Mills theory.
It is believed that on a flat $\bR^4$, the system confines and there are no finite-energy excitations charged under $\su(2)$.
The standard order parameter for the confinement is the area law of the Wilson loop operator in the doublet representation of $\su(2)$, which has the charge $(\lambda_e,\lambda_m)=(1,0)\in \bZ\times \bZ$ in our normalization.

One way to interpret the dynamics \cite{Mandelstam:1974vf,Mandelstam:1974pi,tHooft:1975krp} is that the confining vacuum has a condensate of magnetic monopoles.\footnote{%
In the non-supersymmetric pure Yang-Mills theory, it is difficult to make sense of an actual monopole excitation in the confining vacuum.
If we consider instead the \Nequals2 supersymmetric pure Yang-Mills theory with a small supersymmetry breaking term, which is continuously connected to the non-supersymmetric pure Yang-Mills theory, we can more directly identify the monopole excitation, following the seminal paper of \cite{Seiberg:1994rs}.
We discuss the \Nequals2 supersymmetric pure Yang-Mills with $\SO(3)$ gauge group in Sec.~\ref{sec:so3nf1}.
In the non-supersymmetric setting, the concept of the charge of the condensing monopole can be made more precise using the idea of 1-form symmetry and its spontaneous breaking.
We will not have time and space to develop this interesting idea; 
the readers will be referred to \cite{Gaiotto:2014kfa}.
}
We can ask what is the magnetic charge of the condensing monopole, in our normalization.
The quantization condition of a 't Hooft operator applies also to the charge of dynamical monopoles.
In addition, the property of dynamical monopoles should be the same across the choice of $\SU(2)$ or $\SO(3)_\pm$.
This means that the charge of condensing dynamical monopoles should be of the form $(\lambda_e,\lambda_m)=(0,\text{even})\in \bZ\times \bZ$,
since dynamical monopoles need to exist also in $\SU(2)$.
A minimal assumption is that the condensing monopole has charge $(\lambda_e,\lambda_m)=(0,2)\in \bZ\times \bZ$.

This value of the charge  is marked by $\odot$ in Fig.~\ref{fig:su2lines},
and the condensate renders line operators with odd $\lambda_e$ to follow the area law.
Other line operators follow the perimeter law.
One immediate difference between the $\SU(2)$, $\SO(3)_+$ and $\SO(3)_-$ cases which can be seen from Fig.~\ref{fig:su2lines} is that
the non-trivial line operators in $\SU(2)$ and $\SO(3)_-$ follow the area law,
but that the non-trivial line operator in $\SO(3)_+$ follows the perimeter law.
Therefore, in a certain sense,  we cannot detect the confinement of the $\SO(3)_+$ theory using the area law of the line operator.

Another manifestation of the difference appears when we consider the confined theory on a large spatial $T^3$.
For the $\SO(3)_\pm$ theory, one can introduce the 't Hooft magnetic flux, which are $\bZ_2$-valued, on each of the three faces of $T^3$.
In total, there are $2^3=8$ states.
Since magnetic fluxes are screened, they will give rise to eight degenerate vacua in the limit of large $T^3$ in the $\SO(3)_+$ theory. 
In the $\SO(3)_-$ theory, the states with non-zero 't Hooft magnetic flux also has non-zero electric charge, and therefore are projected out.
Therefore, on a large $T^3$, the $\SU(2)$ theory and the $\SO(3)_-$ theory have a single vacuum,
whereas the $\SO(3)_+$ theory has $2^3=8$ states.

One can understand these eight states from the charge of the magnetic monopole.
In the $\SO(3)_+$ theory, the charge of the condensing magnetic monopole is twice the minimal charge allowed by the periodicity of the ``magnetic $\U(1)$'' symmetry.
A condensate of charge 2 will break the $\U(1)$ gauge symmetry to $\bZ_2$.\footnote{%
This also happens in the BCS superconductivity, where the condensing Cooper pair has electric charge 2, and therefore the electric $\U(1)$ symmetry of the Maxwell field is broken to $\bZ_2$. 
}
Now, given a low-energy unbroken $\bZ_2$ gauge field on $T^3$,
one can choose the $\bZ_2$ holonomy around three directions of $T^3$.
This will again give rise to $2^3=8$ states.

What we discussed so far in this section was topological, and therefore it should apply to any theory continuously connected to the pure non-supersymmetric Yang-Mills theory.
Indeed, we will see in the next section that the softly-broken \Nequals1 Yang-Mills theory will show the expected behavior.
We will also study it from the point of view of the softly-broken \Nequals2 Yang-Mills theory in Sec.~\ref{sec:so3nf1}.

\subsection{4d \Nequals1 superfield formalism}
\label{sec:4dsuperfields}
The coefficients here in this subsection should not be trusted;
the discussion here is intended to remind the readers of the general idea behind the construction.
More details can be found in any of the reviews cited in Sec.~\ref{sec:introduction}.

\subsubsection{Two-component spinors}
Let us first recall the two-component spinors. 
They arise from the isomorphism \begin{equation}
\SO(3,1)_+\simeq \SL(2,\bC)/\{\pm1\},
\end{equation}
where $\SO(3,1)_+$ is the subgroup of $\SO(3,1)$ connected to the identity.
Let us remind ourselves how this isomorphism arises.
We consider \begin{equation}
g=\begin{pmatrix}
a & b\\
c & d
\end{pmatrix} \in \SL(2,\bC)
\end{equation} which naturally acts on a two-dimensional complex vector space $S\simeq \bC^2$.
There is a complex conjugate representation $\bar S$ on which $\SL(2,\bC)$ acts by $\bar g$.
We next consider the space $V$ of Hermitean $2\times 2$ matrices, which we parameterize as\begin{equation}
X=\begin{pmatrix}
t+z & x+iy\\
x-iy & t-z
\end{pmatrix} \in V.
\label{X}
\end{equation}
We note that \begin{equation}
\det X=t^2-x^2-y^2-z^2
\end{equation} is the standard norm of the Minkowski space $\bR^{3,1}$.
We have a natural action of $\SL(2,\bC)$ on $V$, given by \begin{equation}
X\mapsto g X g^\dagger.
\end{equation} 
This shows $V_\bC\simeq S\otimes \bar S$ as a representation.
Since $\det (gXg^\dagger)=(\det g)( \det X)( \det g^\dagger)= \det X$,
this determines a homomorphism $\SL(2,\bC)\to \SO(3,1)$.
A more careful study shows that the kernel is $\{\pm1\}$ and the image is $\SO(3,1)_+$.

We use the undotted index $\alpha$ for $S$,
the dotted index $\dot \beta$ for $\bar S$.
Then a Hermitean $X\in V\simeq S\otimes \bar S$ has the index structure $X_{\alpha\dot\beta}$.
Given $X_\mu=(t,x,y,z)$, we then have \begin{equation}
X_{\alpha\dot\beta} = \sigma^\mu_{\alpha\dot \beta} X_\mu,
\end{equation}where the group theory constants $\sigma^\mu_{\alpha\dot\beta}$ can be read off from \eqref{X}.

The supersymmetry algebra is obtained by adding $Q_\alpha$, $\bar Q_{\dot\alpha}$ to the Poincar\'e algebra $P_\mu$, $M_{\mu\nu}$.
The nontrivial anticommutation relation is \begin{equation}
\{Q_\alpha,\bar Q_{\dot\alpha}\}=2\sigma^\mu_{\alpha\dot\alpha} P_\mu.
\end{equation}
We would like to realize a theory which has this symmetry algebra.

\subsubsection{Superspace and superfields}
Let us review first how we write down a theory which has the symmetry  $P_\mu$.
We take a 4d space $\bR^{4}$ parameterized by $x_\mu$, and let $P_\mu$ act on it via $\delta x_\mu= \epsilon_\mu$.
We consider a function $\phi$ on $\bR^{3,1}$ as the basic dynamical variable.
As $P_\mu$ acts on $\bR^{4}$, it also acts on $\phi$, and we obtain $\delta\phi=-\epsilon^\mu \partial_\mu\phi$.
From this, we easily conclude \begin{equation}
\delta L  = -\epsilon^\mu\partial_\mu L \label{L}
\end{equation} for $L=\partial_\mu\phi \partial^\mu\phi$, say.
We now consider an action of the form $S=\int d^4x L$,
where $L$ has the transformation law \eqref{L}.
The important fact is that $\int d^{4}x \partial_\mu (\cdots) =0$.
Using this, we see that \begin{equation}
\delta S = \int d^4x \delta L=-\epsilon^\mu \int d^{4}x \partial_\mu L = 0,
\end{equation}
and we have an action invariant under $P_\mu$.

We imitate this construction with $Q_\alpha$ and $\bar Q_{\dot\alpha}$.
We consider the space $\bR^{4|4}$ parameterized by $x_\mu$, $\theta^\alpha$ and $\bar\theta^{\dot\alpha}$, and let the supertranslations act by \begin{equation}
\delta\theta^\alpha=\epsilon^\alpha,\qquad
\delta\bar\theta^{\dot\alpha}=\bar\epsilon^{\dot\alpha},\qquad
\delta x^\mu=-i\epsilon^\alpha\bar\theta^{\dot\alpha}\sigma_{\alpha\dot\alpha}^\mu -i\epsilon^{\dot\alpha}\theta^\alpha\sigma_{\alpha\dot\alpha}^\mu.
\end{equation} They correspond to differential operators \begin{equation}
Q_\alpha=\frac{\partial}{\partial \theta^\alpha}-i\sigma^\mu_{\alpha\dot\alpha}\bar\theta^{\dot\alpha}\partial_\mu,\qquad
\bar Q_{\dot\alpha}=\frac{\partial}{\partial \bar\theta^{\dot\alpha}}-i\sigma^\mu_{\alpha\dot\alpha}\theta^{\alpha}\partial_\mu.
\end{equation}
They satisfy \begin{equation}
\{ Q_\alpha,Q_{\dot\alpha} \} = -2\sigma^\mu_{\alpha\dot\alpha} i\partial_\mu.
\end{equation}

It is useful to introduce \begin{equation}
D_\alpha=\frac{\partial}{\partial \theta^\alpha}+i\sigma^\mu_{\alpha\dot\alpha}\bar\theta^{\dot\alpha}\partial_\mu,\qquad
\bar D_{\dot\alpha}=\frac{\partial}{\partial \bar\theta^{\dot\alpha}}+i\sigma^\mu_{\alpha\dot\alpha}\theta^{\alpha}\partial_\mu.
\end{equation}
They satisfy \begin{equation}
\{ D_\alpha,D_{\dot\alpha} \} = +2\sigma^\mu_{\alpha\dot\alpha} i\partial_\mu.
\end{equation}
and anticommute with $Q_\alpha$, $\bar Q_{\dot\alpha}$.
The point is that $\bR^{4|4}$ is the group manifold of the group formed by supertranslations and translations.
Then there are both the left action and the right action of the group on the group manifold, which naturally (anti)commute.
There is an $\SU(2)\times \SU(2)$ action on $S^3\simeq \SU(2)$ from the same reason.
Here, we pick the left action as the supersymmetry  $Q_\alpha$, $\bar Q_{\dot\alpha}$ inherited by the superfields,
and the right action as the superderivatives $D_\alpha$, $\bar D_{\dot\alpha}$ used in the construction of the supersymmetric terms in the Lagrangian.

A general superfield $X$ is a function on $\bR^{4|4}$.
The integral \begin{equation}
\int d^4x d^2\theta d^2\bar\theta K(X_i,\bar X_i, D_\alpha X_i, \bar D_{\dot\alpha}X_i, \cdots)
\label{kahler-integral}
\end{equation} is  supersymmetric, where $K$ is a function of  superfields $X_i$
and their superderivatives,
since \begin{equation}
\delta K = -(\epsilon^\alpha Q_\alpha+\bar\epsilon^{\dot\alpha}\bar Q_\alpha) K
\end{equation} and \begin{equation}
\int d^4x d^2\theta d^2\bar\theta Q_\alpha(\cdots)=0.
\end{equation}
Note that $K$ cannot depend on $Q_\alpha X$ or $\bar Q_{\dot\alpha}$ in \eqref{kahler-integral},
since the supertranslations $Q_\alpha$, $\bar Q_{\dot\alpha}$ do not (anti)commute among themselves.
$D_\alpha X$ and $\bar D_{\dot\alpha}X$ are allowed to appear, since they (anti)commute  with the supertranslations.

A chiral superfield $\Phi$ is a function on $\bR^{4|4}$ satisfying \begin{equation}
\bar D_{\dot\alpha} \Phi=0.
\end{equation}
 $\bar D_{\dot\alpha} $ annihilates 
$\theta^\alpha$ and $y^\mu := x^\mu +i\theta^\alpha \bar\theta^{\dot\alpha}\sigma^\mu_{\alpha\dot\alpha}$.
In fact, a chiral superfield on $\bR^{4|4}$ is equivalent to a function on $\bR^{4|2}$ parameterized by $y^\mu$ and $\theta^\alpha$ without $\bar \theta^{\dot\alpha}$.
The complex conjugate of $\Phi$ satisfies \begin{equation}
D_\alpha \bar\Phi=0
\end{equation} instead, and is called anti-chiral.
The term \begin{equation}
\int d^4x d^2\theta W(\Phi_i,\ldots,)
\end{equation} is also supersymmetric, where $W$ is an arbitrary chiral superfield.
This is called the superpotential term.
For example, an arbitrary polynomial of chiral superfields $\Phi_i$ will do as the superpotential.

\subsubsection{Gauge symmetry in superspace}
Let us next consider how to construct a supersymmetric gauge theory.
Again, we start by reviewing the non-supersymmetric case.
Let us say we have an $N$-component  scalar field $\phi$.
We would like to have a Lagrangian which is invariant under the gauge transformation \begin{equation}
\phi(x) \mapsto U(x)\phi(x) 
\end{equation} where $U(x)$ is a unitary matrix defined at each point $x$ in the spacetime.
The kinetic term of the ungauged theory $\partial_\mu \phi \partial^\mu\phi$ is not invariant under this transformation.
We need to add a matrix field $A_\mu$ taking values in $N\times N$ anti-Hermitean matrices, with the transformation property \begin{equation}
A_\mu(x)\mapsto U A_\mu(x) U^{-1} + U \partial_\mu U^{-1}
\end{equation}and modify the kinetic term to $
D_\mu\phi D^\mu\phi
$ where \begin{equation}
D_\mu \phi = \partial_\mu \phi -  A_\mu \phi.
\end{equation}
The kinetic term of $A_\mu$ is found by first finding a gauge-covariant combination constructed from $A_\mu$, which is $F_{\mu\nu} = [D_\mu,D_\nu]. $
We note that $F_{\mu\nu}$ is gauge-invariant when the gauge group is Abelian.

Let us see how it goes in a supersymmetric theory. 
We take an $N$-component chiral multiplet $\Phi(y^\mu,\theta)$.
We would like to find a Lagrangian invariant under the gauge transformation \begin{equation}
\Phi(y,\theta) \mapsto U(y,\theta) \Phi(y,\theta),
\end{equation}  where $U$ takes values in the space of invertible $N\times N$ matrices.
Note that, due to the chirality, we cannot impose the reality condition on $U$.
The gauge group is effectively complexified: $\U(N)$ is replaced by $\GL(N)$.
In general, this means that the chiral superfields are in a complex representation of the symmetry group.\footnote{%
In comparison, in a non-supersymmetric theory, scalars naturally transform in a strictly real representation, and in an \Nequals2 theory, hypermultiplets are naturally in a pseudoreal representation.
No non-gauge multiplets are allowed in an \Nequals4 theory.
}

The kinetic term $\bar\Phi\Phi$ is not invariant under this transformation.
We then need to introduce a superfield $e^V$, taking values in the space of real invertible $N\times N$ matrices, with the transformation law \begin{equation}
e^V \mapsto \bar U e^V U^{-1}.
\end{equation}
The kinetic term of the chiral superfield is now modified to $\bar\Phi e^V \Phi$.
The introduction of the superfield $e^V$ has the added bonus that it spontaneously breaks the complexified $\GL(N)$ symmetry down to $\U(N)$,
which is the unbroken subgroup of $\GL(N)$ at any value of $V$.

Using part of the gauge invariance, we can fix $e^V$ to be of the form \begin{equation}
e^V=1+A_{\mu} \bar\theta\sigma^\mu \theta + \cdots
\end{equation} so that the bottom component is $1$ and the $\theta$, $\bar\theta$,  $\theta\theta$ and $\bar\theta\bar\theta$ components vanish.
This is called the Wess-Zumino gauge.\footnote{%
Other gauge choices are often useful, see e.g. Sec~11.3 of \cite{ArgyresReview1996}, or Sec.~4.3 of \cite{Elvang:2009gk}.}

The kinetic term for the vector superfield $V$ is found by first finding a gauge-covariant combination constructed from $V$, which becomes gauge-invariant for Abelian gauge theories.
The required combination is a chiral superfield \begin{equation}
W_\alpha \propto \bar D^{\dot\alpha}  \bar D_{\dot\alpha} e^{-V} D_\alpha e^{V},
\end{equation} and the kinetic term is given by \begin{equation}
\propto \int d^2\theta  i \tr W_\alpha W^\alpha + \cc .
\end{equation}

For an Abelian vector multiplet $V$, the following term \begin{equation}
\int d^4\theta \xi V
\end{equation} is also gauge invariant, since the $d^4\theta$ integral kills the gauge variations $\Lambda$ and $\bar\Lambda$ since they are (anti)chiral.
This term is called the Fayet-Iliopoulos term \cite{Fayet:1974jb}.

\section{Pure super Yang-Mills}
\label{sec:pure}

Here we assume that the reader is familiar with the \Nequals1 superfield formalism.
Although we have given a brief review of the construction in Sec.~\ref{sec:4dsuperfields},
it needs to be properly learned somewhere else, e.g.~\cite{WessBagger}.

\subsection{Lagrangian for the vector multiplets}
\label{sec:purelag}
An $\cN{=}1$ vector multiplet consists of a Weyl fermion $\lambda_\alpha$ and a vector field $A_\mu$,
both in the adjoint representation of the gauge group $G$. We combine them into the superfield $W_\alpha$ with the expansion \begin{equation}
W_\alpha=\lambda_\alpha+F_\alpha^\beta\theta_\beta + D\theta_\alpha+\cdots
\end{equation} where $D$ is an auxiliary field, again in the adjoint of the gauge group. 
$F_{\alpha}^{\beta}=\frac \ii2 \sigma^\mu{}^\beta_{\dot\gamma} \bar\sigma^{\nu}{}^{\dot \gamma}_\alpha F_{\mu\nu}$ is the anti-self-dual part of the field strength $F_{\mu\nu}$.

The kinetic term for a vector multiplet is given by \begin{equation}
\int d^2\theta \frac{-\ii}{8\pi} \tau \tr W_\alpha W^\alpha + \cc  
\label{gaugekin}
\end{equation} where \begin{equation}
\tau=\frac{4\pi \ii}{g^2}+\frac{\theta}{2\pi}
\label{tau}
\end{equation} is a complex number combining the inverse of the coupling constant and the theta angle. We call it the complexified coupling of the gauge multiplet.
Expanding in components, we have \begin{equation}
\frac{1}{2g^2} \tr F_{\mu\nu}F^{\mu\nu} + \frac{\theta}{16\pi^2} \tr F_{\mu\nu} \tilde F^{\mu\nu} + \frac{1}{g^2}\tr D^2 -\frac{2\ii}{g^2}\tr \bar\lambda\slash D \lambda.
\end{equation} 
We use the convention that $\tr T^u T^v = \frac12 \delta^{uv}$
where $u,v=1,\ldots,\dim G$
 for the standard generators of gauge algebras, which explain why we have the factors $1/(2g^2)$ in front of the gauge kinetic term. 
The $\theta$ term is a total derivative of a gauge-dependent term. Therefore, it does not affect the perturbative computations. although it does affect non-perturbative computations.

The Lagrangian \eqref{gaugekin} as it stands defines a gauge theory consisting simply of a gauge field and a chiral massless fermion in the adjoint representation, minimally coupled to the gauge field.
This is known as the supersymmetric pure Yang-Mills theory.
Here we constructed the Lagrangian using superfields, 
but note that a theory  consisting of a gauge field and a massless adjoint fermion is automatically supersymmetric, independent of how one constructs it.

\exercise{Check the supersymmetry of this Lagrangian from various viewpoints.}

\answer{} We already know the supersymmetry from the superfield formulation.
In this answer we check the on-shell supersymmetry of the action \begin{equation}
\tr F_{\mu\nu}F^{\mu\nu}+ \ii \tr \bar\lambda\slash D \lambda.\label{onshell}
\end{equation}
From the consideration of the dimensions of the fields and the Lorentz transformation properties,
the only supersymmetry transformation one can write is \begin{equation}
\delta A_\mu \propto \bar \epsilon \sigma_\mu \lambda + \cc ,\qquad
\delta\lambda_\alpha \propto F_{\alpha\beta}\epsilon^\beta \label{onshell-trans}
\end{equation} where we expressed $F_{\mu\nu}$ using the spinor indices as $F_{\alpha\beta}$.
Then the supersymmetry transformation of the on-shell action \eqref{onshell} has the following structure:
\begin{equation}
\vcenter{\hbox{\begin{tikzpicture}[xscale=2.5,yscale=1.5]
\node(A) at (0,0) {$\tr FF$};
\node(B) at (2,0) {$\tr \bar\lambda \slash D \lambda$};
\node(P) at (1,-1) {$\epsilon^\alpha F_{\alpha\beta}\lambda^\beta + \cc $};
\node(Q) at (3,-1){$f_{abc} (\bar{\epsilon} \sigma_\mu \lambda^a ) (\bar\lambda^b \sigma^\mu \lambda^c) +\cc $ };
\draw[->] (A) to node[midway,left]{$\delta A$} (P);
\draw[->] (B) to node[midway,left]{$\delta \lambda$} (P);
\draw[->] (B) to node[midway,right]{$\delta A$} (Q);
\end{tikzpicture}}}
\end{equation}
The variation of the form $\epsilon^\alpha F_{\alpha\beta}\lambda^\beta$ can be canceled by appropriately choosing the proportionality coefficients in \eqref{onshell} and \eqref{onshell-trans}.
The variation of the form $f_{abc} (\bar{\epsilon} \sigma_\mu \lambda^a ) (\bar\lambda^b \sigma^\mu \lambda^c) $ needs to vanish by itself.
This follows if \begin{equation}
\sigma_\mu{}^{(\alpha}_{(\dot\alpha}
\sigma^\mu{}^{\beta)}_{\dot\beta)}=0\label{crucial}
\end{equation} where the parentheses in the subscripts and the superscripts mean separate symmetrization.

The crucial identity \eqref{crucial} can be checked by a brute-force computation;
there is only a finite number of components to check.\footnote{%
One should not stay away from such a brute-force direct computation, if it is really needed.
The very first paper  on four-dimensional supergravity \cite{Freedman:1976xh} by Freedman, van Nieuwenhuizen and Ferrara, has a \emph{note added} in its abstract, saying that the vanishing of the supersymmetry variation of the four-Fermi term was checked by a computer while the manuscript was being refereed. 
The author was  curious how this computation involving gamma matrices was done in a computer in 1976 when the paper was written.
The author had a chance to ask D. Z. Freedman and P. van Nieuwenhuizen directly.
They  kindly told the author that it was done by writing a Fortran program which computed every component of the product of the gamma matrices.
They also told the author that they soon started using \emph{Schoonschip} \cite{Veltman:1991xb}
developed by M. Veltman around that time.
\emph{Schoonschip} is still available  at \cite{SchoonschipArchive}.
}
A more conceptual derivation can be given as follows.
Note that this identity follows if $p^\mu p_\mu=0$ for $p^\mu= \bar v \sigma^\mu w$ for arbitrary \emph{bosonic} spinors $v$ and $w$.
Now,  for a four-vector $x^\mu=(t,x,y,z)$, we use the $2\times2$ matrix \begin{equation}
\underline{x} := x_\mu \sigma^\mu =\begin{pmatrix}
t+z & x+iy\\
x-iy & t-z
\end{pmatrix}
\end{equation}
we already introduced in \eqref{X}.
We already noted there that  \begin{equation}
x^\mu x_\mu = \det \underline{x} = t^2-x^2-y^2-z^2.
\end{equation}
Note that $\underline{p}$ for $p^\mu=\bar v\sigma^\mu w$  has the form \begin{equation}
\underline{p}\propto \begin{pmatrix}
\bar v_1 w_1 & \bar v_2 w_1 \\
\bar v_1 w_2 & \bar v_2 w_2 
\end{pmatrix} .
\end{equation}
Then we easily see \begin{equation}
p^\mu p_\mu = \det \underline{p} = 0.
\end{equation}
This is what we wanted to demonstrate.

Before proceeding, we note that this crucial identity \eqref{crucial} and our derivation of it underlies the spinor helicity formalism. For more on this interesting topic, see e.g.~\cite{Elvang:2013cua}.
We also note that this check of the on-shell supersymmetry of the pure Yang-Mills theory in $4=2+2$ dimensions can be naturally and uniformly extended to spacetime dimensions $3=2+2^0$, $4=2+2^1$, $6=2+2^2$ and $10=2+2^3$.
An interested reader can consult Sec.~\ref{sec:minimalSYM}.

\subsection{The dynamical scale $\Lambda$}
\label{sec:pureLambda}
\emph{There is a renormalization scheme where the superpotential remains a holomorphic function of the chiral superfields, including background fields whose vevs are the gauge and superpotential couplings.}
This is the core of Seiberg's holomorphy argument.\footnote{Recently an obstruction to this philosophy was found for 2d \Nequals{(4,4)} theories \cite{Gomis:2016sab}; but the analysis there confirms that there is no similar problem in 4d \Nequals1.
}

The gauge kinetic term of a supersymmetric theory is given in \eqref{gaugekin}
in terms of the complexified coupling $\tau$ given in \eqref{tau}.
The one-loop running coupling at the energy scale $\LambdaRG$ can be expressed as \begin{equation}
\tau(\LambdaRG )=\tau_\text{UV} - \frac{b}{2\pi i} \log\frac{\LambdaRG }{\Lambda_\text{UV}} + \cdots
\label{1looprunning}
\end{equation} where $b$ is the rational number appearing on the right hand side of  \eqref{b}.

Note that the coupling $\tau$ starts from  $1/g^2$, and therefore an $n$-loop diagram would have the dependence $g^{2(n-1)}$.  The leading term in \eqref{1looprunning} is then a one-loop effect.
In general, perturbation theory is independent of the $\theta$ angle, since $F_{\mu\nu}\tilde F_{\mu\nu}$ is a total derivative, although of a gauge-dependent quantity.
Therefore an $n$-loop effect is a function of $(\Im \tau)^{1-n}$.

Let us regard $\tau$ to be an expectation value of a background chiral superfield. 
In this scheme, the renormalized $\tau$ should also be a chiral superfield,
and therefore should be given by a holomorphic expression of the original $\tau$.
We now note that  $(\Im \tau)^{1-n}$ is not holomorphic unless $n=1$.
We conclude that the running \eqref{1looprunning} is one-loop exact in the holomorphic scheme. 
We find that the combination \begin{equation}
\eta := \Lambda^b : = \LambdaRG ^b e^{2\pi i  \tau(\LambdaRG )}
\end{equation} is invariant to all orders in perturbation theory.

We call this $\Lambda$ the complexified  dynamical scale of the theory.\footnote{A redefinition of the form $\Lambda\to c\Lambda$ by a real constant $c$  corresponds to a redefinition of the coupling of the form $1/g^2 \to 1/g^2 - c'$ where $c'$ is another constant, or equivalently $g^2 \to g^2 + c' g^4 + \cdots$. Therefore this is a redefinition starting at the one-loop order, keeping the leading order definition of $g^2$ fixed. In this lecture note, we do not track such finite renormalization of the coupling very carefully. } 
Note that $\Lambda$ is a complex quantity, and can be considered as a vev of a background chiral superfield. 
We also note that the single-valued quantity is $\eta = \Lambda^b$,
and $\Lambda$ has an ambiguity by $e^{2\pi i/b}$.

In the case of $\SU(N)$ pure Yang-Mills,  the one-loop running of the coupling is given by \begin{equation}
E\frac{\partial}{\partial E}\tau(E) = \frac{i}{2\pi} 3N,
\end{equation}
 and therefore we define the dynamical scale $\Lambda$ by the relation \begin{equation}
\eta=\Lambda^{3N}= e^{2\pi i\tau(E) }E^{3N}.
\label{suNtau}
\end{equation}
For general $G$, $N$ in the expression is replaced by the dual Coxeter number $C(\text{adj})=h^\vee(G)$.

One naturally wonders if there is any non-perturbative renormalization for this $\Lambda$ or $\eta$.
To study this, we need to consider the R-symmetry, to which we turn next.

\subsection{R-symmetry and the number of supersymmetric vacua}
\label{sec:numvac}
In a supersymmetric theory, a global symmetry is called an R-symmetry when it acts nontrivially on the supercharge.
In particular, the R-charge of the different component of a single supermultiplet will be different.
In contrast, a global symmetry which commutes with the supercharge is called a flavor symmetry.

Let us study the R-symmetry of the pure Yang-Mills theory with gauge group $\SU(N)$.
We assign R-charge zero to the gauge field, and R-charge 1 to the gaugino $\lambda_\alpha$. 
The phase rotation $\lambda_\alpha\to e^{i\varphi} \lambda_\alpha$ is anomalous, and needs to be compensated by $\theta\to \theta + 2N \varphi$.
The shift of $\theta$ by $2\pi$ is still a symmetry, therefore the discrete rotation \begin{equation}
\lambda_\alpha \to e^{\pi i/N} \lambda_\alpha, \qquad \theta\to \theta+2\pi
\end{equation} is a symmetry generating $\bZ_{2N}$. 
We already saw this at the end of Sec.~\ref{sec:chiral} in a non-supersymmetric context.

Another way to state this anomalous breaking of $\U(1)_R$ symmetry to $\bZ_{2N}$ is the following.
Note $\eta$ defined in \eqref{suNtau} contains $2\pi i\tau$ in the exponent.
Therefore, under the $\U(1)_R$ symmetry acting on the gaugino,
it has charge $2N$.
The subgroup of $\U(1)_R$ which keeps $\eta$ invariant is then the unbroken $\bZ_{2N}$ R-symmetry.
Note that under the unbroken $\bZ_{2N}$ R-symmetry, $\eta$ is invariant but $\Lambda$ has the transformation \begin{equation}
\Lambda \to e^{2\pi i/(3N)} \Lambda.\label{lambdatrans}
\end{equation}
Under the continuous $\U(1)_R$ symmetry, $\Lambda$ has charge $2/3$.

Now we can answer what will be a possible nonperturbative correction to $\eta$ or $\Lambda$.
We suppose that there is a regularization scheme which is holomorphic in Seiberg's sense and
furthermore preserves the $\U(1)_R$ action on the background fields $\eta$ and $\Lambda$.
There is only a single single-valued holomorphic expression which has charge $3N$, which is just $\eta$ itself and nothing else.
This means that there is no further nonperturbative correction either.\footnote{%
The perturbative non-renormalization of $\tau$ is a general result in supersymmetric theories,
but the non-perturbative non-renormalization of $\Lambda$ needs the fact that the R-charge of $\Lambda$ is nonzero.
For a theory for which $\Lambda$ is neutral under the $\U(1)$ R-symmetry,
which happens  when the gauge theory has zero beta function and is semiclassically superconformal,
there can be and indeed are non-perturbative corrections to $\Lambda$.
This fact was first noted in \cite{Dorey:1996ez} in the case of mass-deformed \Nequals4 $\SU(2)$ super Yang-Mills theory.
It is also known that the choice of the non-perturbative regularization schemes, such as whether one treats the $\SU(2)$ instanton using the $\SU(2)$ ADHM construction or $\Sp(1)$ ADHM construction, affects these non-perturbative corrections \cite{Hollands:2010xa}.
For a short summary, see Sec.~4.2 of the author's other review \cite{Tachikawa:2014dja}.
}

The pure super Yang-Mills theory is believed to confine, with nonzero gaugino condensate $\vev{\lambda_\alpha\lambda^\alpha}$.
What would be the value of this condensate?  This should be of mass dimension 3 and of R-charge 2. The only candidate is 
\begin{equation}
\vev{\lambda_\alpha\lambda^\alpha}=c \Lambda^3
\end{equation}  for some constant $c$.
The  symmetry \eqref{lambdatrans} acts in the same way on both sides by the multiplication by $e^{2\pi i/N}$.
Assuming that the numerical constant $c$ is non-zero, this $\bZ_{2N}$ is  spontaneously broken to $\bZ_2$,
 generating $N$ distinct solutions \begin{equation}
\vev{\lambda_\alpha\lambda^\alpha}=c e^{2\pi i \ell/N } \Lambda^3
\end{equation} where $\ell=0,1,\ldots, N-1$.  Unbroken $\bZ_2$ acts on the fermions by $\lambda_\alpha \to -\lambda_\alpha$, which is a $360^\circ$ rotation. This $\bZ_2$ symmetry is hard to break.

It is now generally believed that this theory has these $N$ supersymmetric vacua and not more.\footnote{%
In the mid 1990s, there was a proposal \cite{Kovner:1997im} by Kovner and Shifman suggesting that there is an additional vacuum with $\vev{\lambda\lambda}=0$.
This was suggested partly to reconcile  incompatible results coming from two microscopic methods to compute $\vev{\lambda\lambda}$, known as the strongly-coupled instanton computation and the weakly-coupled instanton computation,
and partly to reconcile the then-discrepancy in the number of vacua of pure $\SO(N)$ gauge theory, again computed in two different methods.
On the first point, later developments in the instanton computation such as \cite{Hollowood:1999qn} showed that the Kovner-Shifman vacuum failed to resolve this particular issue;
it is now believed that the strongly-coupled instanton computation cannot be trusted since it leads to results violating the cluster decomposition principle.
In general, an instanton computation is reliable only when the infrared limit of the theory is weakly coupled.
On the second point, the discrepancy was resolved in a paper \cite{Witten:1997bs}, which we outline in Sec.~\ref{sec:SpinSO}.
All things combined, the author does not think there is any more need to consider the proposal by Kovner and Shifman.
\label{foot:SCIvsWCF}
} 
For other gauge groups, the analysis proceeds in the same manner, by replacing $N$ by the dual Coxeter number $C(\text{adj})=h^\vee(G)$ of the gauge group under consideration.

\subsection{The theory in a box}\label{sec:box}

Let us perform a check of the number of supersymmetric vacua  we just obtained.
We put the \Nequals1 super Yang-Mills theory with gauge group $G$ in a spatial box of size $L\times L\times L$  with the periodic boundary condition in each direction. We keep the time direction as $\bR$. 
We then study the system both in the limit $L\gg \Lambda^{-1}$ and $L\ll \Lambda^{-1}$,
where $\Lambda$ is the dynamical scale of the theory.
As we will review soon, in the supersymmetric case,
we can argue that the number of vacua in both limits should agree.
Now, the system in the limit $L\ll \Lambda^{-1}$ is weakly-coupled,
and therefore an honest counting of the vacua should be possible.

For $G=\SU(N)$ and $=\Sp(N)$, this analysis was performed originally in \cite{Witten:1982df}.
In the same paper, a problem was noticed when $G=\Spin(N)$, which was later resolved in \cite{Witten:1997bs} by the same author.
Soon this analysis was generalized to arbitrary  connected gauge groups in \cite{Kac:1999gw,Borel:1999bx,Witten:2000nv,Tachikawa:2014mna}.
Here we will see some representative examples.

\subsubsection{Independence of the Witten index on the size of the box}

As we said, we consider the pure super Yang-Mills on $T^3$, with a periodic boundary condition.
This system preserves the translations $P^\mu$ and the supertranslations $Q_\alpha$ and $Q^\dagger_{\dot\alpha}$ unbroken. 
We only need to use, among the supertranslations,
a single  linear combination $\cQ$  of $Q_\alpha$ and $Q_\alpha^\dagger$, satisfying \begin{equation}
H=P^0=  \{\cQ,\cQ^\dagger\}.
\end{equation} 
We also make use of the fermion number operator $(-1)^F$ such that \begin{equation}
\{(-1)^F,\cQ\}=0.
\end{equation}
Consider eigenstates of the Hamiltonian $H$, given by \begin{equation}
H\ket{E}=E\ket{E}.
\end{equation} 
In general, the multiplet structure under the algebra of $\cQ$, $\cQ^\dagger$, $H$ and $(-1)^F$ is of the form 
\begin{equation}
\begin{array}{ccccc@{\qquad}c}
&\raisebox{-.8ex}{\rotatebox{30}{$\leftrightarrow$}}& \cQ^\dagger \ket{E} & \leftrightarrow & (\cQ^\dagger\cQ-\cQ\cQ^\dagger)\ket{E} \\
\ket{E} & \leftrightarrow & \phantom{^\dagger}\cQ\ket{E} &  \rotatebox{30}{$\leftrightarrow$} 
\end{array}
\end{equation}
involving four states.  When $\cQ\ket{E}=0$ or $\cQ^\dagger\ket{E}=0$, the multiplet only has two states.
If $\cQ\ket{E}=\cQ^\dagger\ket{E}=0$, the multiplet has only one state, and $E$ is automatically zero
due to the equality
\begin{equation}
E\vev{E|E}
=\vev{E|H|E}
=\vev{E|(\cQ \cQ^\dagger+\cQ^\dagger \cQ)|E}=
|\cQ\ket{E}|^2+
|\cQ^\dagger\ket{E}|^2 .
\end{equation} 
We see that a bosonic state is always paired with a fermionic state unless $E=0$.

This guarantees that  the Witten index \begin{equation}
Z:=\tr e^{-\beta H} (-1)^F =\tr\big|_{E=0} (-1)^F
\end{equation} is a robust quantity independent of the change in the size $L$ of the box: when a perturbation makes  a number of zero-energy states to non-zero energy $E\neq 0$, the states involved are necessarily  composed of  pairs of a fermionic state and a bosonic state. Thus it cannot change $\tr (-1)^F$. 

\subsubsection{$\SU(N)$}

Let us take $G=\SU(N)$.
In the large size limit $L \gg \Lambda^{-1}$,
we expect that there is a single vacuum for each value of $\vev{\lambda_\alpha\lambda^\alpha}=c \exp^{2\pi i\ell/N} \Lambda^3$,
giving $N$ zero energy states in total.
As they are all related by the $\bZ_N$ R-symmetry,
these $N$ states all have the same value of $(-1)^F$.
We therefore see that $|\text{Witten index}|=N$ in the large $L$ limit.

Next, let us consider the Witten index in the limit where the box size $L$ is far smaller than the scale $\Lambda^{-1}$ set by the dynamics.  
The system is weakly coupled, and we can use perturbative analysis. 
To have almost zero energy, we need to set  $F_{\mu\nu}=0$ for $m,n=1,2,3$, since nonzero   magnetic fields contribute to the energy.  
Then the only low-energy degrees of freedom in the system are the holonomies \begin{equation}
U_x, U_y, U_z \in \SU(N),
\end{equation} which commute with each other. 
A standard linear algebra says that they can be simultaneously diagonalized by a $\U(N)$ matrix.
In fact it is not difficult to see that we can use an $\SU(N)$ matrix to diagonalize them.
Therefore, any three commuting holonomies $U_{x,y,z}$ can be conjugated  to  the form \begin{align}
U_x&=\diag(e^{i\theta^x_1},\ldots,e^{i\theta^x_N}),\\
U_y&=\diag(e^{i\theta^y_1},\ldots,e^{i\theta^y_N}),\\
U_z&=\diag(e^{i\theta^z_1},\ldots,e^{i\theta^z_N}).\label{Uxyz}
\end{align} Gaugino zero modes are then \begin{equation}
\lambda^{\alpha=1}_1,\ldots,\lambda^{\alpha=1}_N,\quad
\lambda^{\alpha=2}_1,\ldots,\lambda^{\alpha=2}_N
\end{equation} with the condition that they are traceless, \begin{equation}
\sum_i \theta^x_i=\sum_i \theta^y_i=\sum_i \theta^z_i=0,\quad
\sum_i\lambda^{\alpha=1}_i=
\sum_i\lambda^{\alpha=2}_i=0.
\end{equation}
The wavefunction of this truncated quantum system is given by a linear combination of states of the form \begin{equation}
\lambda^{\alpha_1}_{i_1}\lambda^{\alpha_2}_{i_2}\cdots \lambda^{\alpha_\ell}_{i_\ell}\psi(\theta^x_i;\theta^y_i;\theta^z_i)
\end{equation} which is invariant under the permutation acting on the index $i=1,\ldots N$.
To have zero energy, the wavefunction cannot have dependence on $\theta^{x,y,z}_i$ anyway, since the derivatives with respect to them are the components of the electric field, and they contribute to the energy.
Thus the only possible zero energy states are just invariant polynomials of $\lambda$s.
We find $N$ states with the wavefunctions given by \begin{equation}
1,\ S,\ S^2,\ \ldots,\ S^{N-1}\label{tow}
\end{equation} where $S=\sum_i\lambda^{\alpha=1}_i\lambda^{\alpha=2}_i$. They all have the same Grassmann parity, and contribute to the Witten index with the same sign. 
Thus we found $|Z(L)|=N$  in the limit of small box, $L\ll \Lambda^{-1}$, too. 

\subsubsection{$\SO(3)$}
\label{sec:so3}
In the computation for $\SU(N)$ presented above,
we used the fact that three commuting holonomies $U_{x,y,z} \in G$
can be simultaneously conjugated into the Cartan subgroup when $G=\SU(N)$. 
This is \emph{not} true in general; $\SU(N)$ is very special.
The simplest counterexample is in fact $\SO(3)$, where we take \begin{equation}
U_x = \diag(+--),\quad
U_y = \diag(-+-),\quad
U_x = \diag(--+).
\end{equation}  
These three matrices are diagonalized in $\SO(3)$, but the diagonal matrices do not form the Cartan subgroup for $\SO(3)$. 

Now we refer the reader to the discussion in Sec.~\ref{sec:StiefelWhitney}:
this $\SO(3)$ configuration has a nontrivial Stiefel-Whitney class, or equivalently nontrivial 't Hooft magnetic fluxes.
Indeed, lifting from $\SO(3)$ to $\SU(2)$, we find that the holonomies $U_{1,2,3}$ lift to Pauli matrices $i\sigma_{1,2,3}$.  Note that $U_1U_2=U_2U_1$ but $\sigma_1\sigma_2=-\sigma_2\sigma_1$.
This extra minus sign  measures the second Stiefel-Whitney class $w_2$ of the $\SO(3)$ bundle: when evaluated on the face $C_{12}$ of the $T^3$, it gives $-1$.\footnote{%
This is the $w_2$ of the gauge bundle, and not to be confused with the $w_2$ of the spacetime.} 
Here and in the following,  $C_{ij}$ is the $T^2$ formed by the edges in the $i$-th and the $j$-th directions of $T^3$.  
We can similarly compute $w_2(C_{23})$ and $w_2(C_{31})$; we have  $(w_2(C_{23}),w_2(C_{31}),w_2(C_{12}))=(-1,-1,-1)$.

In general, the possible choices of $w_2$ are $(\pm1,\pm1,\pm1)$. The commuting triples in the class $(+1,+1,+1)$ are the ones that can be simultaneously conjugated to the Cartan torus $T\subset \SO(3)$ discussed above.
They behave basically the same as the $\SU(2)$ case, and therefore they give 2 states, by setting $N=2$ in the discussion of the last subsection.

For each of the other seven choices of $w_2$, there is one  isolated commuting triple, that gives one zero-energy state.
In total, we find \begin{equation}
|Z_{\SO(3)}(L)|=2+7=9 \qquad(L\Lambda \ll 1).\label{bar}
\end{equation} 

Therefore, we should find the same when $L$ is very, very big. 
To reproduce this, we need to study the vacua of $\SO(3)$ theory in more detail.
There are still two vacua, with $\vev{\tr\lambda\lambda}=\pm \Lambda^3$. 
As we know, they are exchanged by shifting $\theta$ by $2\pi$.
This does not charge the $\SU(2)$ theory, but it exchanges $\SO(3)_+$ theory and $\SO(3)_-$ theory, as we saw in \eqref{SO3+-}.
We also saw in Sec.~\ref{sec:nonsusypure} that
the $\SO(3)_+$ theory has magnetic $\bZ_2$  gauge symmetry and produces $2^3=8$ states on $T^3$,
while $\SO(3)_-$ theory only has one state.
In total, we find\footnote{%
Here and in \eqref{bar}, we need to argue that all these states have the same $(-1)^F$. For details, see \cite{Tachikawa:2014mna}.}
\begin{equation}
|Z_{\SO(3)}(L)|=2^3+1=9 \qquad(L\Lambda \gg 1).
\end{equation} 
This is again consistent with the computation in the opposite regime \eqref{bar}.

\subsubsection{$\Spin(N)$ vs $\SO(N)$}
\label{sec:SpinSO}

Here we consider the subtle distinction of pure $\Spin(N)$ gauge theory and $\SO(N)$ gauge theory.
We assume $N\ge 7$.

\paragraph{$\Spin(N)$:}
Let us first recall the situation when $G=\Spin(N)$, first studied in the Appendix I of \cite{Witten:1997bs}. 
The dual Coxeter number is $N-2$, and therefore, there are $N-2$ vacua in the far infrared, distinguished by the gaugino condensate \begin{equation}
\vev{\tr\lambda\lambda}=\Lambda^3, \quad \omega \Lambda^3,\quad \ldots,\quad \omega^{N-3} \Lambda^3
\end{equation} where $\omega=\exp(2\pi i/(N-2))$. Therefore when the size $L$ of $T^3$ is very big, we find \begin{equation}
|Z_{\Spin(N)}(L)|=N-2, \qquad (L\Lambda \gg 1).
\end{equation}

It is known that the commuting holonomies $(g_1,g_2,g_3)$ can be put into either of the two following standard forms.
\begin{itemize}
\item The first possibility is the usual one. Namely, we have \begin{equation}
g_a\in T\subset \Spin(N)
\end{equation} where $T$ is the Cartan torus of $\Spin(N)$.
\item The second possibility is the one missed until \cite{Witten:1997bs}:   \begin{equation}
g_a = g_a^{(7)} s_a
\end{equation} where $g_{1,2,3}^{(7)}$ is a lift to $\Spin(7)$ of the following $\SO(7)$ matrices \begin{equation}
\begin{array}{l}
\diag(+1,+1,+1,-1,-1,-1,-1), \\
\diag(+1,-1,-1,+1,+1,-1,-1), \\
\diag(-1,+1,-1,+1,-1,+1,-1),
\end{array} 
\end{equation}and $s_a \in T'$ where $T'$ is the Cartan torus of $\Spin(N-7)\subset \Spin(N)$ commuting with  $g_{1,2,3}^{(7)}$. 
\end{itemize} 

The former component gives $1+\rank T$  zero-energy states,
and the latter component gives $1+\rank T'$ zero-energy states. In total, we find \begin{equation}
|Z_{\Spin(N)}(L)|=(\lfloor \frac{N}2 \rfloor+1)+(\lfloor \frac{N-7}2 \rfloor+1)=N-2, \qquad (L\Lambda \ll 1).
\end{equation}

\paragraph{$\SO(N)$:}
Now, we move on to the case $G=\SO(N)$. In this case, there are two choices of the discrete theta angle, so there are two theories $\SO(N)_\pm$.
We explained this in the case of $N=6$ when $\SO(N)\simeq \SU(6)/\bZ_2$ in Sec.~\ref{sec:groupglobal}, and the general case is similar. 

As in the $\SO(3)$ theory, 
the vacua of $\SO(N)_+$ theory have unbroken $\bZ_2$ gauge symmetry,
while the vacua of the $\SO(N)_-$ theory do not.
The difference from the $\SO(3)$  case is that the shift $\theta\to \theta+2\pi$ maps
$\SO(N)_+$ to $\SO(N)_+$ and $\SO(N)_-$ to $\SO(N)_-$.
From this consideration, in the infrared, we simply find \begin{equation}
|Z_{\SO(N)_+} |= 8(N-2), \quad (L\Lambda \gg 1)
\end{equation} and 
\begin{equation}
|Z_{\SO(N)_-} |= (N-2), \quad (L\Lambda \gg 1).
\end{equation}

Let us confirm this result in a computation in the ultraviolet, $L\Lambda\ll 1$.  
The topological type of the bundle is given by the Stiefel-Whitney class 
evaluated on the faces, $(m_{23},m_{31},m_{12})\in\{\pm1\}^3$. 

When $(m_{23},m_{31},m_{12})=(+1,+1,+1)$, all the commuting holonomies are obtained by projecting the $\Spin(N)$ commuting holonomies down to $\SO(N)$. Then, these give $(1+\rank T)+(1+\rank T')=N-2$ zero-energy states as before. 

For seven other choices $(m_{23},m_{31},m_{12})\neq (+1,+1,+1)$, we can always apply $\SL(3,\bZ)$ to have $(m_{23},m_{31},m_{12})=(-1,+1,+1)$. 
In \cite{Borel:1999bx} it was proved that  the commuting holonomies are either of the following two forms:
\begin{itemize}
\item The first possibility is  \begin{equation}
g_a = g_a^{(3)} s_a
\end{equation} where $g_{1,2,3}^{(3)}$ is the following $\SO(3)$ matrices \begin{equation}
\begin{array}{l}
\diag(+1,+1,+1),\quad
\diag(-1,-1,+1), \quad
\diag(-1,+1,-1), 
\end{array} 
\end{equation}and $s_a \in T''$ where $T''$ is the Cartan torus of $\SO(N-3)\subset \SO(N)$ commuting with  $g_{1,2,3}^{(3)}$.
\item The second possibility is  \begin{equation}
g_a = g_a^{(4)} s_a
\end{equation} where $g_{1,2,3}^{(4)}$ is the following $\SO(4)$ matrices \begin{equation}
\begin{array}{l}
\diag(-1,-1,-1,-1),\quad
\diag(-1,-1,+1,+1), \quad
\diag(-1,+1,-1,+1),
\end{array} 
\end{equation}and $s_a \in T'''$ where $T'''$ is the Cartan torus of $\SO(N-4)\subset \SO(N)$ commuting with  $g_{1,2,3}^{(4)}$.
\end{itemize}

Quantization of the zero modes then give \begin{equation}
(1+\rank T'')+(1+\rank T''')=N-2
\end{equation} states for each of the seven choices $(m_{23},m_{31},m_{12})\neq (+1,+1,+1)$.
In the $\SO(N)_+$ theory they are all kept, but in the $\SO(N)_-$ theory, they have a nontrivial induced discrete electric charge $e=(m_{23},m_{31},m_{12})$ due to the non-zero theta angle. This causes these states to be projected out. 

In total, we find \begin{equation}
|Z_{\SO(N)_+} |= 8(N-2), \quad (L\Lambda \ll 1)
\end{equation} and 
\begin{equation}
|Z_{\SO(N)_-} |= (N-2), \quad (L\Lambda \ll 1)
\end{equation}  in the ultraviolet computation, agreeing with the infrared computations.

\subsection{Non-supersymmetric deformation}
We end this section on the pure super Yang-Mills by briefly discussing the non-supersymmetric deformation.
We add a gluino mass $m_g$ to the ${\cal N}=1$ supersymmetric theory discussed above.
For simplicity, we consider the case when $G=\SU(2)$ or $\SO(3)$.

\begin{figure}
\[
\includegraphics[width=.5\textwidth]{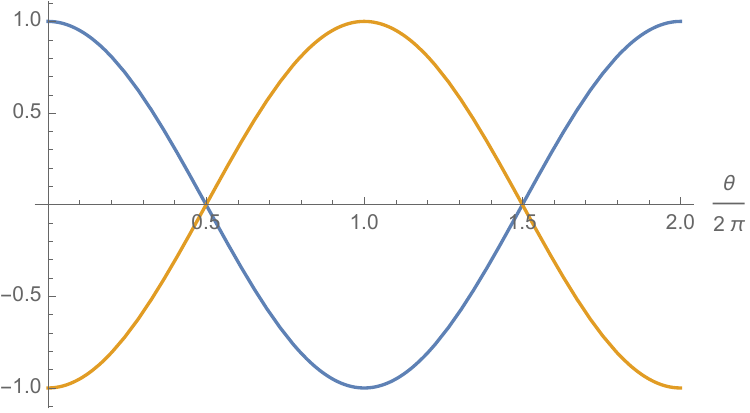}
\]
\caption{Energy of two vacua for softly broken \Nequals1 $\SU(2)$ SYM.\label{fig:sb}}
\end{figure}

Consider first the case $|m_g| \ll |\Lambda|$. Since we have a mass gap, the dynamics in each vacuum is essentially the same as above. 
Indeed,  the soft mass term for the gaugino is just \begin{equation}
\delta \cL = m_g \lambda \lambda + \cc 
\end{equation} and we have the condensate $\vev{\lambda\lambda}\simeq \pm\Lambda^3$.
So,  their vacuum energy is  \cite{Evans:1996hi,Konishi:1996iz}
\begin{equation}
\sim \pm\Re(m_g\Lambda^3).
\end{equation}
In particular, at $\theta=\pi$, two branches are exchanged and  the CP is spontaneously broken;
there is a first order phase transition there, realizing Dashen's idea \cite{Dashen:1970et} explicitly.
See Fig.~\ref{fig:sb} for an illustration.

We now consider the case of $G=\SO(3)$.
As we already discussed in Sec.~\ref{sec:so3}, 
before the non-supersymmetric deformation,
one of the two vacua has the perimeter law for its nontrivial line operator, with an unbroken $\bZ_2$ gauge symmetry,
while the other has the area law, without any unbroken gauge symmetry.
These properties are inherited by the two branches exchanged at $\theta=\pi$ after a small non-supersymmetric deformation $|m_g| \ll |\Lambda|$.
This change in the topological behavior of the vacuum at $\theta=\pi$ almost forces something nontrivial there, even when $|m_g|\gg |\Lambda|$. 
To study it in detail, we need to study the CP anomaly of the system more carefully \cite{Gaiotto:2017yup}.

\section{Preliminaries on theories with Matters}
\label{sec:matters}
\subsection{Lagrangian for the chiral multiplets}
An $\cN{=}1$ chiral multiplet $Q$ consists of a complex scalar $Q$ and a Weyl fermion $\psi_{\alpha}$, both in the same representation of the gauge group. 
It is represented by a chiral superfield satisfying $\bar D_{\dot\alpha} Q=0$, and schematically has the expansion \begin{equation}
Q(y)=Q\big|_{\theta=0} + \psi_\alpha(y) \theta^\alpha + F(y) \theta_\alpha \theta^\alpha
\end{equation} where $F$ is auxiliary. 
The complex conjugate of $Q$ is antichiral and satisfies $D_\alpha Q^\dagger=0$.

The (effective) action has the general form \begin{equation}
\int d^4\theta K(Q^\dagger,Q) + \int d^2\theta W(Q) + \cc 
\end{equation}  
where $K$ is the K\"ahler potential and $W$ is the superpotential.

Expanding in components, we find that the kinetic term contains $\propto F^i \bar F^{\bar j} g_{i\bar j}$ where $g_{i\bar j}=\partial_i \bar\partial_{\bar j} K$,
and the superpotential term contains $\propto F^i \partial_i W$.
Eliminating $F$, we have $\bar F^{\bar j} g_{i\bar j} \propto \partial_i W$, and the potential is \begin{equation}
V\sim g^{i\bar j} \partial_i W \bar\partial_{\bar j} \bar W.
\end{equation}
The potential is automatically positive.
The zeroes of the potential given by \begin{equation}
\partial_i W=0
\end{equation} are the supersymmetric vacua.

\subsection{Renormalization of the chiral multiplets}
\subsubsection{Wess-Zumino model}
As an interacting  UV Lagrangian, let us consider the Wess-Zumino model \cite{Wess:1974tw}:
\begin{equation}
\int d^4\theta \Phi^\dagger \Phi + \int d^2\theta g \Phi^3 + \cc 
\end{equation}
How does it renormalize? Promote $g$ to a background chiral superfield  $Y$: \begin{equation}
\int d^4\theta \Phi^\dagger \Phi + \int d^2\theta Y \Phi^3 + \cc 
\end{equation} Assign R-charge $+2$ to $Y$, zero to $\Phi$. 
Due to the R-charge conservation and the choice of holomorphic gauge, all-loop computations can only give \begin{equation}
\int d^4\theta K(\Phi^\dagger, \Phi) + \int d^2\theta Y f(\Phi) + \cc 
\end{equation} 
When $Y$ is small, the perturbation theory is applicable, and this means that there is only the tree term in the superpotential. So we conclude $f(\Phi)=\Phi^3$. This is called the non-renormalization theorem.

What happens to $K$? To see this, it is useful to note that $\Phi^\dagger \Phi$ is not only the kinetic term but also the superfield version of the $\U(1)_\Phi$ current associated to $\Phi\to e^{i\theta} \Phi$.
Indeed, the $\theta \sigma^\mu \bar\theta$ component of $\Phi^\dagger \Phi$ contains $ j^\mu=\phi^\dagger \partial_\mu \phi - (\partial_\mu \phi^\dagger) \phi$. 
In our Lagrangian, the term $\Phi^3$ breaks $\U(1)$ to $\bZ_3$. 
Accordingly, there is a source term in the (non)conservation equation: \begin{equation}
\bar D^2 (\Phi^\dagger \Phi) = 3g \Phi^3\label{cons}
\end{equation} 
This is obtained by taking the variation of $\Phi \to e^{X} \Phi$ where $X$ is an arbitrary chiral superfield;
we just have to repeat the standard procedure to obtain the Noether currents in a supersymmetric manner.

It turns out that the non-conservation equation \eqref{cons} is the basic ingredient which determines the running of the coupling $g$.
For this purpose, we recast this equation in terms of the OPE.
The coefficient $3$ on the right hand side of \eqref{cons} simply says that the operator $\Phi^3$ has charge $3$ under $\U(1)_\Phi$.
The same coefficient appears in the OPE of the current $J=\Phi^\dagger \Phi$ and the operator $\cO=\Phi^3$ charged under it:
\begin{equation}
(\Phi^\dagger \Phi)(x) \Phi^3(0) \sim 3\cdot \frac{1}{4\pi^2 |x|^2} \Phi^3(0) + \cdots
\end{equation}
where we remind the reader that the massless complex-scalar two-point function in $d$ dimension is $\vev{\phi(x)\phi(0)}=(d-2)^{-1} S_d^{-1} |x|^{-(d-2)} $ where $S_d$ is the surface area of the unit sphere in $\bR^d$. 
In our case $S_4=2\pi^2$; the denominator we see in the equation above is $(d-1)S_d=4\pi^2$.

Using $\vev{\Phi^3(x)\Phi^3(0)^\dagger} = 6\cdot (4\pi^2 |x|^2)^{-3}$, 
we rewrite the OPE above into the three-point function \begin{equation}
\vev{\Phi^3(x) \Phi^\dagger{}^3(y) (\Phi^\dagger \Phi)(z)} = \frac{3 \cdot 6}{(4\pi^2)^4 |x-y|^4 |x-z|^2 |y-z|^2}
\end{equation}
which then implies that the next-leading term in the OPE of $\Phi^3$ and $\bar\Phi^3$ has the form  \begin{equation}
\Phi^3(x) \Phi^\dagger{}^3(0) \sim \frac{6}{(4\pi^2)^3|x|^6} + \frac{3\cdot 6}{(4\pi^2)^2|x|^4} (\Phi^\dagger\Phi)(0) + \cdots.
\end{equation}

This is exactly what is needed to compute the leading perturbation in $gg^\dagger$:
\begin{equation}
\int d^4x d^2\theta g \Phi^3 (x) \int d^2\bar\theta g^\dagger \Phi^\dagger{}^3(0)
\sim \int d^4\theta gg^\dagger  \frac{3\cdot 6 \cdot (2\pi^2)}{(4\pi^2)^2} (\log \mu) \Phi^\dagger \Phi(0) 
\end{equation}
So, if we write the renormalized kinetic term as $\int d^4\theta Z \Phi^\dagger \Phi$, we see \begin{equation}
\mu\frac{\partial}{\partial\mu} Z=  \frac{3\cdot 6 \cdot (2\pi^2)}{(4\pi^2)^2}  gg^\dagger.
\end{equation}
Or equivalently, if you slightly lower the cutoff from $\mu'$ to $\mu$,  \begin{equation}
\delta K =   \frac{3\cdot 6 \cdot (2\pi^2)}{(4\pi^2)^2}  gg^\dagger \cdot \log(\mu'/\mu) \Phi^\dagger \Phi.
\end{equation}
This corresponds to the scaling dimension $1+\gamma$  of the operator $\Phi$ given by \begin{equation}
\gamma=\frac12   \frac{3\cdot 6 \cdot (2\pi^2)}{(4\pi^2)^2}  gg^\dagger = \frac{9}{8\pi^2} gg^\dagger.
\label{anomg}
\end{equation}

So far we used the holomorphic scheme. To keep $K$ fixed to have canonical kinetic term,
we use the (non)conservation equation \eqref{cons} again: \begin{align}
\int d^4\theta \delta K &= -\frac12\int d^2\theta \bar D^2 \delta K + \cc  \label{KDK}\\
&= - \frac12 3g [\frac{3\cdot 6 \cdot (2\pi^2)}{(4\pi^2)^2} gg^\dagger \dot \log(\mu'/\mu) ]  \int d^2\theta\Phi^3 +\cc 
\end{align}
Therefore we see that \begin{equation}
\mu\frac{\partial}{\partial\mu}  g= \frac12 3g[\frac{3\cdot 6 \cdot (2\pi^2)}{(4\pi^2)^2} gg^\dagger ] = \frac{27}{8\pi^2}g^3 
\label{poo}
\end{equation} to the leading order.
This means that $g$ renormalizes to zero in the infrared.

Of course the computation we just performed can also be done in the standard perturbation theory.\footnote{%
The two-, three- and four-loop computations were done in \cite{Townsend:1979ha,Abbott:1980jk,Sen:1981hk}, respectively.
}
\exercise{Carry out the perturbative computation in a standard manner.}

\subsubsection{A reformulation}

The renormalization of $g$ \eqref{poo} was computed in a somewhat unconventional manner here since the manipulation we used here can be readily generalized to perturbations around a strongly-coupled fixed point as follows \cite{Green:2010da}.
Let us consider a perturbation of the form \begin{equation}
\int d^2\theta \lambda \cO + \cc 
\label{lO}
\end{equation}
by a dimension-3 operator $\cO$ whose two-point function has the normalization \begin{equation}
\vev{\cO(x)\cO^\dagger(0)}= \frac{\cN_\cO}{ (4\pi^2)^3|x|^6}.
\end{equation}
Suppose that there is a broken $\U(1)$ current $J$ satisfying \begin{equation}
\bar D^2 J= q \lambda \cO.
\label{DDJ}
\end{equation} so that the charge of the operator $\cO$ under $J$ is $q$.
For the computation we also need to know the normalization of $J$, which we write as \begin{equation}
\vev{J(x)J(0)} = \frac{\cN_J} {(4\pi^2)^2 |x|^4}.
\end{equation}

The equation \eqref{DDJ} is translated to the OPE which says \begin{equation}
J(x) \cO(0) = \frac{q}{4\pi^2 |x|^2} \cO(0) + \cdots
\end{equation} which determines the three-point function to have the form \begin{equation}
\vev{\cO(x)\cO(y)^\dagger J(z)}=\frac{ q \cN_O }{(4\pi)^4 |x-y|^4 |x-z|^2 |y-z|^2}
\end{equation}  from which the OPE of $\cO$ with $\cO^\dagger$ can be read off as  \begin{equation}
\cO(x)\cO^\dagger(0)=\frac{\cN_O}{(4\pi^2)^3|x|^6} + \frac{q\cN_O}{\cN_J} \frac{1}{(4\pi^2)^2|x|^4} J(0)+\cdots
\end{equation} which then leads to 
 \begin{equation}
\delta K= \frac{ q \cN_\cO  } { 8\pi^2\cN_J } \lambda\lambda^\dagger \cdot \log(\mu'/\mu) J
\label{dK}
\end{equation}
which can be converted via \eqref{KDK} to \begin{equation}
\mu\frac{\partial}{\partial\mu}  \lambda=  \frac{q \cN_\cO}{16\pi^2 \cN_J} (\lambda\lambda^\dagger)\lambda .
\label{dl}
\end{equation}
Plugging in $\cN_{\cO}=6$, $\cN_J=1$ and $q=3$, we indeed reproduce \eqref{poo}.

As a strongly-coupled example, we take the $E_6$ symmetric theory of Minahan and Nemeschansky \cite{Minahan:1996fg}.
This theory has \Nequals2 supersymmetry. 
As an \Nequals1 theory, it has $\U(1)\times E_6$ as the flavor symmetry.
This theory has a chiral operator $u$ of dimension $3$, which is charged under the $\U(1)$ flavor symmetry.
We can try to deform the theory by adding a superpotential term \begin{equation}
W=g \int d^2\theta u + \cc 
\end{equation}
The computation as above shows that $g$ renormalizes to zero in the infrared.

\subsection{Renormalization of the gauge multiplets}
\label{sec:rengauge}
\subsubsection{Lagrangian of the SQCD}
Now we couple $N$ chiral fields to the $\SU(N)$ vector multiplet.
We recall that a vector multiplet $V$ has a component expansion schematically of the form \begin{equation}
V=A_\mu \bar \theta\sigma^\mu \theta + \lambda_\alpha \theta^\alpha \bar\theta^2 + D \theta^2\bar\theta^2.
\end{equation}
Here we want to gauge with $\SU(N)$, so $V$ is assumed to be traceless. 

When there are $N$ chiral multiplets $Q_a$ in the fundamental representation, 
$(Q_a)^\dagger Q_b$ is the $\U(N)$ current, as we saw. 
So, to the leading order, the term \begin{equation}
\int d^4\theta V^{\bar a b} (Q_a)^\dagger Q_b
\end{equation} is the coupling to the vector multiplet; the all-order version is \begin{equation}
\int d^4 \theta (Q_a)^\dagger  (e^{V})^{\bar ab} Q_b.
\end{equation}

With only $N$ chiral fields in the fundamental representation, the gauge symmetry is  anomalous due to the perturbative triangle anomaly for $N\ge 3$, and due to the global anomaly for $N=2$.
So we add $\tilde Q^b$ in the anti-fundamental representation to cancel the anomaly.

In fact we can add $N_f$ pairs $(Q_a^i,\tilde Q^a_i)$ for $a=1,\ldots,N$ and $i=1,\ldots, N_f$.
The kinetic term of the gauge field is given by \eqref{gaugekin}.
The resulting theory is known as the Supersymmetric Quantum ChromoDynamics  (SQCD).
We would like to know what happens to this theory in the infrared.

For the moment we keep the discussion general: we have the gauge group $G$ and $N_f$ pairs of chiral multiplets $Q$, $\tilde Q$ in the representation $R$ and $\bar R$.
The general formulas \eqref{nonsusyRG}, \eqref{b}  reduce in the \Nequals1 case to 
\begin{equation}
\LambdaRG\frac{d}{d\LambdaRG}g=-\frac{g^3}{(4\pi)^2} \left[
3C(\text{adj})-2N_fC(R).
\right] 
\end{equation} and to \begin{equation}
\LambdaRG\frac{d}{d\LambdaRG} \tau=
\frac{i}{2\pi}[3C(\text{adj})-2N_fC(R)]\label{n1running}
\end{equation}
where we remind the reader that $\tau=\frac{\theta}{2\pi}+\frac{4\pi i}{g^2}$.
Then \begin{equation}
\eta:= \Lambda^{3C(\text{adj})-2N_fC(R)} := E^{3C(\text{adj})-2N_fC(R)} e^{2\pi i \tau(E)}\label{one-instanton-factor}
\end{equation} is RG-invariant in the holomorphic scheme, and can be regarded as a background chiral superfield.
Note that $\eta$ contains $e^{i\theta}$ in the exponent, and represents the one-instanton contribution.
We can argue one-loop and non-perturbative exactness of $\eta$ and $\Lambda$ in the holomorphic scheme, just as in our discussion of the pure super Yang-Mills in Sec.~\ref{sec:pureLambda}.

\subsubsection{Anomalous dimensions}

When $G=\SU(N)$ and the matter fields are in the fundamental representation, $3C(\text{adj})-C(R)$ is $3N-N_f$.
So it is IR free when $N_f> 3N$ , it is  conformal to this order when $N_f=3N$, and $N_f <3N$ the coupling starts to grow.
We will soon see that $N_f=3N$ case is also IR free.

\def\clubsuit{\underline{c}}

Let us first analyze the region where $(3N-N_f)\ll N_f$. 
The one-loop renormalization of the chiral multiplets can be found by a standard one-loop computation and yields
\begin{equation}
\delta K = -\clubsuit \frac{g^2}{8\pi^2}\frac{2C(R) \dim G }{ \dim R}  \log(\mu'/\mu) \sum_i (Q_i^\dagger Q_i+\tilde Q^i{}^\dagger \tilde Q^i)
\label{renormK}
\end{equation} 
where \begin{itemize}
\item the factor $C'(R):=C(R)\dim G/\dim R$ arises when one computes $(T^u_R)^a_b (T^v_R)^b_c = C'(R)\delta^a_c$; compare with \eqref{CR}.
Here, $u,v=1,\ldots, \dim G$ are the indices for the adjoint representation
and $a,b=1,\ldots,\dim R$ are the indices for the representation $R$,
\item the factor $g^2/(8\pi^2)$ is the standard one-loop factor,
\item and the only factor which needs an explicit computation in this approach is the overall numerical factor $\clubsuit$ independent of $G$ and $R$.
\end{itemize}
The numerical factor $\clubsuit$ turns out to be just $\clubsuit=1$ from a direct computation.
We keep it as an unknown, and instead fixes it by a different means below.

The wave-function renormalization \eqref{renormK} corresponds to the anomalous dimension $1+\gamma$ for $Q_i$ which is \begin{equation}
\gamma_{Q_i}=-\clubsuit\frac{g^2}{8\pi^2} \frac{C(R)\dim G}{\dim R}.
\label{gammaQi}
\end{equation}
Note that this is negative, where as the anomalous dimension from the superpotential interaction \eqref{anomg} was positive.

\subsubsection{Konishi anomaly}

We can now demand that the kinetic terms of the chiral multiplets are kept in their canonical normalization.
This can be achieved by using the non-conservation law \begin{equation}
\bar D^2 (Q_i{}^\dagger Q_i) = 2C(R) \frac1{16\pi^2} \tr WW, 
\label{Konishi}
\end{equation}  
where we sum  over the gauge indices but not over the index $i=1,\ldots, N_f$.
This equation, known  as the Konishi anomaly \cite{Konishi:1983hf,Konishi:1985tu}\footnote{%
Due to this reason the operator $\Phi^\dagger \Phi$ in the \Nequals4 super Yang-Mills theory is often called the Konishi operator in the literature of the integrability of \Nequals4 super Yang-Mills.
}, is a supersymmetric version of the standard chiral anomaly \begin{equation}
\partial^\mu \bar\psi_Q \sigma_\mu \psi_Q = 2C(R) \frac{1}{16\pi^2} \tr F\tilde F.
\end{equation}

We can  now use the equation \eqref{KDK}, which was
$\int d^4\theta \delta K=-(1/2)\int d^2\theta \bar D^2 K+c.c$,
to rewrite the change \eqref{renormK} as a change in $\tau$:
 We  find \begin{multline}
\mu\frac{\partial}{\partial\mu} \frac{\tau}{8\pi i }
= \frac{1}{16\pi^2} (3C(\text{adj})-2C(R)N_f) - \\
2N_f  \cdot \frac12 \cdot  \clubsuit\cdot \frac{g^2}{8\pi^2} \frac{2C(R) \dim G }{  \dim R}   ( 2C(R)  \frac1{16\pi^2})  +\cdots\label{BZ}
\end{multline} to this order.\footnote{%
This is a beta function in the scheme where the chiral multiplets are fixed to have canonically-normalized kinetic terms
and the gauge multiplets are fixed to have a constant coupling in the covariant derivative.
In particular, the coefficient $\tau$ in front of the gauge kinetic term changes.
When we further go to the scheme where the coefficient of the gauge kinetic term is fixed and the running coupling is placed within the covariant derivative, we find the so-called Novikov-Shifman-Vainshtein-Zakharov (NSVZ) beta function \cite{Novikov:1983uc,Shifman:1986zi,Shifman:1991dz}. 
For an expository account on these matters, see the articles by Arkani-Hamed and Murayama \cite{ArkaniHamed:1997ut,ArkaniHamed:1997mj}.
}

\exercise{Confirm this in the standard perturbation theory. This is a two-loop effect. Where in the above computation was the two-loop computation carried out?}

\subsubsection{Banks-Zaks fixed point}
We find the zero of \eqref{BZ} at \begin{equation}
\frac{g^2}{8\pi^2}= \frac{3C(\text{adj})-2N_f C(R)}{N_f}  \frac{\dim R}{\clubsuit (2C(R))^2\dim G} + \cdots.
\end{equation} 
We find a conformal fixed point in a parametrically weak coupling region just below the threshold $3N \simeq N_f$ when $|3N-Nf|\ll N_f$.
This conformal fixed point is usually known as the Banks-Zaks fixed point \cite{Banks:1981nn}.\footnote{%
This fixed point was originally noticed by Caswell in \cite{Caswell:1974gg}.
 Banks and Zaks \cite{Banks:1981nn} pointed out that there is a controllable limit $|3N-N_f| \ll N_f$.
These works were done without supersymmetry.
}

At this conformal point, we find that the anomalous dimension $\gamma_Q$ \eqref{gammaQi} takes the value
 \begin{equation}
\gamma_Q= -\frac \clubsuit2 \frac{3C(\text{adj})-2N_f C(R)}{2N_f C(R) } +\cdots = -\clubsuit\frac{3N-N_f}{2N_f} + \cdots.
\label{deltaq}
\end{equation}
 We will soon see that the undetermined coefficient $\clubsuit=1$, and moreover the result is exact,
i.e.~the $\cdots$ simply vanishes.

\section{SQCD and Seiberg duality}
\label{sec:SUSQCD}
\subsection{Constraint of the superconformal symmetry}
At the zero of the beta function, the system is invariant under the scaling transformation $D$.
It is expected that the theory is not only invariant under scale symmetry, but also under conformal symmetry.%
\footnote{For the important distinction between scale invariance and conformal invariance, we refer the reader to the excellent review article \cite{Nakayama:2013is} by Nakayama.}
The conformal algebra contains the special conformal generators $K_\mu$ in addition to translation generators $P_\mu$.
In the supersymmetric case, we get an additional  supercharge $S_{\dot\alpha}$ from the commutator of  the ordinary supercharge  $Q_\alpha$ and $K_\mu$.

The anti-commutator of $Q$ and $S$ plays an important role: \begin{equation}
\{Q_\alpha, S^\dagger{}_\beta\} = \epsilon_{\alpha\beta}(2iD+3R)+M_{\alpha\beta}
\label{QS}
\end{equation} where $R$ is the generator of the superconformal R-symmetry.
The commutation relation \eqref{QS} implies that a chiral scalar operator, annihilated by $Q_\alpha$, satisfies $\Delta=(3/2)R$.
Here $\Delta$ is th scaling dimension and $R$ is the R-charge.
For example,  in a free theory, $\Phi$ has dimension $\Delta=1$ and R-charge $R=2/3$.

Recall that any $\U(1)$ symmetry that rotates $Q$ with charge $\pm1$ is called \emph{an} R-symmetry. 
Our convention is that the supercoordinate $\theta^\alpha$ has the R-charge $+1$.
The superconformal symmetry contains a particular R-symmetry, which needs to be conserved.
In favorable cases, this fact can be used to fix the scaling dimension of the operators.

So, let us consider $\SU(N)$ with $N_f$ flavors again. 
Let us determine the anomaly-free R-symmetry.
We are forced to assign R-charge 1 to the gaugino $\lambda$, since \begin{equation}
W_\alpha=\lambda_\alpha + F_{\alpha\beta} \theta^\beta + \cdots 
\end{equation}
and the gauge field strength $F_{\alpha\beta}$ has zero R-charge.
To make the R-gauge-gauge anomaly vanish, the R-charge $r=R(\psi)=R(\tilde\psi)$ of the fermion components $\psi$, $\bar \psi$ of $Q$, $\tilde Q$ should satisfy \begin{equation}
N +  r \cdot 2\cdot \frac12 \cdot N_f =0
\end{equation} meaning that $r=-N/N_f$.
Since the chiral superfield has the expansion \begin{equation}
Q(y,\theta)= Q(y) + \psi_\alpha(y) \theta^\alpha + \cdots,
\end{equation} we see that \begin{equation}
R(Q)=R(\tilde Q)=r+1 =1-\frac{N}{N_f}.
\end{equation}
We find therefore that 
\begin{equation}
\Delta(Q)=\frac32-\frac{3N}{2N_f}
\end{equation} and that 
\begin{equation}
\gamma=-\frac{3N-N_f}{2N_f}.\label{exactgamma}
\end{equation}
This reproduces the perturbative computation in \eqref{deltaq}.
Furthermore, we found that  the undetermined coefficient $\clubsuit$ there is in fact given by $\clubsuit=1$.
Not only that, we learned that the anomalous dimension \eqref{exactgamma} at the superconformal point is \emph{exact}. 
This fact was first noticed in \cite{Seiberg:1994bz}.

When $N_f$ is very close to $3N$, we have a weakly-coupled conformal fixed point.
What happens when we gradually lower $N_f$? 
Consider the gauge invariant operator $\tilde QQ$, which has dimension $3(1-N/N_f)$.
The unitarity bound which we recall very soon demands that the scaling  dimension of any scalar operator to be $\ge 1$.\footnote{%
The unitarity bound need not apply to gauge-dependent operators,
since the Hilbert space of gauge theory is only positive-definite after imposing the gauge (=BRST) invariance.
} 
Therefore, we need $N_f \ge 3N/2$ to have a conformal point.

\subsection{Aside: unitarity bound}
In general, the dimension $\Delta$ of a scalar operator needs to satisfy $\Delta\ge1$,
due to the unitarity of the theory.
This is the simplest example of unitarity bounds.
Let us review how this bound is derived, which is usually done using the (super)conformal algebra \cite{Mack:1975je,Dobrev:1985qv,Minwalla:1997ka}.
Here we explain a lesser-known version of the argument in \cite{Grinstein:2008qk},\footnote{The author thanks K.~Yonekura for the information.}
which is less powerful but does not require the full conformal algebra for the derivation.

Consider a scalar operator $\cO$ of dimension $\Delta$. 
Then its Euclidean two-point function behaves as \begin{equation}
\vev{\cO(x)\cO(0)} = \frac{C}{|x|^{2\Delta}}.
\end{equation} 
The constant $C$ is guaranteed to be positive: Consider smearing the operators around $x$ and $0$ with compact support. Then it should be the norm of a wave function, which should be positive.

A slightly more detailed use of unitarity leads to the condition $\Delta\ge 1$. To see this, we first Fourier-transform the two-point function and write \begin{equation}
\frac{C}{|x|^{2\Delta}}=C\frac{(2\pi)^2  \Gamma(2-\Delta)}{4^{\Delta-1}\Gamma(\Delta)} \int \frac{d^4k}{(2\pi)^4} e^{ikx} |k|^{2(\Delta-2)}.
\end{equation} 
Wick rotating back to Minkowski signature, we find the spectral density at 4-momentum $k$, which is given by \begin{align}
C\frac{(2\pi)^2  \Gamma(2-\Delta)}{4^{\Delta-1}\Gamma(\Delta)} \Im (-k^2-i\epsilon)^{\Delta-2}
&=C\frac{(2\pi)^2  \Gamma(2-\Delta)}{4^{\Delta-1}\Gamma(\Delta)} \sin (\pi(2-\Delta)) |k|^{2(\Delta-2)}\\
&=C\frac{(2\pi)^2 \pi  (\Delta-1)}{4^{\Delta-1}\Gamma(\Delta)^2} |k|^{2(\Delta-2)}.
\end{align}
This requires $\Delta \ge 1$.

\exercise{The derivation given above was somewhat imprecise, because of the divergence of $\Gamma(2-\Delta)$ at positive integer $\Delta> 2$ canceling against $\sin(\pi(2-\Delta))$ which appeared only later in the computation. Make this more precise. }

\subsection{Seiberg duality}
\label{sec:seiberg}
\subsubsection{Statement}
Long before $N_f$ hits the lower bound  $3N/2$ coming from the unitarity bound applied to the operator $Q\tilde Q$, the anomalous dimension is of order $1$ and we lose perturbative control. Is there any way out? Here comes Seiberg duality to the rescue \cite{Seiberg:1994pq}.
We simply make the statement here, and perform numerous consistency checks in the rest of the lecture.

\begin{claim}
The following two gauge theories are \emph{dual} when $N+N'=N_f$ and $N,N'\ge 2$:
\begin{itemize}
\item an $\SU(N)$ gauge theory with $N_f$ pairs of $Q^i$ in the fundamental and $\tilde Q_{\tilde \imath}$  in the anti-fundamental, and with zero superpotential $W=0$.
\item an $\SU(N')$ with $N_f$ pairs of $q_i$ in the fundamental and $\tilde q^{\tilde \imath}$ in the anti-fundamental, a set of gauge-singlet scalars $M^i_{\tilde \jmath}$, and with the superpotential $W=q_i\tilde q^{\tilde \jmath} M^i_{\tilde \jmath}$
\end{itemize}
\end{claim}
We call the first theory and the second theory as the original theory and the dual theory, respectively.

Two preliminary remarks are in order here.
First, when $N$ or $N'$ is $=1$ or $0$,  similar statements can be made with a small modification.

Second, the word \emph{dual} needs to be treated with some care. 
We assume $N_f< 3N$ so that the original theory is asymptotically free.
From $N+N'=N_f$, it is easy to see that the dual theory is asymptotically free when $N_f>(3/2)N$ 
and that it is infrared free when $N_f<(3/2)N$.
In the former case $N_f>(3/2)N$, the word \emph{dual} means that the two theories flow to the same superconformal theory in the infrared:
\begin{equation}
\begin{tikzpicture}[baseline=(X.east),every text node part/.style={align=center}]
\node(A) at (-4,1) {UV: the original $\SU(N)$ theory\\
with $Q$, $\tilde Q$ and $W=0$ };
\node(P) at (4,1) {UV: the dual $\SU(N')$ theory \\
with $q$, $\tilde q$, $M$ and $W=q\tilde q M$};
\node(X) at (0,0) {};
\node(B) at (0,-1) {IR: some superconformal theory.};
\draw[->] (A) to (B);
\draw[->] (P) to (B);
\end{tikzpicture}
\end{equation}
In the latter case $N_f<(3/2)N$, the word \emph{dual} needs to be interpreted  in the sense that the dual theory is the infrared limit of the original theory:
\begin{equation}
\begin{tikzpicture}[baseline=(X.east)]
\node(A) at (0,1) {UV: the original $\SU(N)$ theory with $Q$, $\tilde Q$ and $W=0$};
\node(X) at (0,0) {};
\node(B) at (0,-1) {IR: the dual $\SU(N')$ theory with $q$, $\tilde q$, $M$ and $W=q\tilde q M$.};
\draw[->] (A) to (B);
\end{tikzpicture}
\end{equation}
Note that this means that a non-Abelian $\SU(N')$ gauge field can emerge from a strong dynamics in the infrared,
and also that an asymptotically non-free $\SU(N')$ gauge theory can have a sensible ultraviolet completion.

\subsubsection{Check: structure of continuous symmetries}
Let us now start our checks of this duality.
First, we see the same type of conserved continuous symmetries on both sides: we can count, among them,
\begin{itemize}
\item a conserved $\U(1)$ R-symmetry,
\item an $\SU(N_f)_\text{untilded}$ symmetry acting on the indices $i$, $j$,
\item another  $\SU(N_f)_\text{tilded}$ symmetry acting on the indices $\tilde \imath$, $\tilde \jmath$,
\item and a non-R $\U(1)_B$ symmetry called the $\U(1)$ baryon symmetry,
which rotates $Q$ and $\tilde Q$  oppositely, and similarly $q$ and $\tilde q$ oppositely.
\end{itemize}

\subsubsection{Check: gauge-invariant chiral operators}
Next, we can list  gauge-invariant chiral operators with the same $\SU(N_f)\times \SU(N_f)$ charges: \begin{equation}
\begin{array}{ccc}
\text{original theory} & & \text{dual theory}\\[.1em]
\hline
M^i_{\tilde \jmath}:= Q^i \tilde Q_{\tilde \jmath} &\leftrightarrow& M^i_{\tilde \jmath}  \\
B^{i_1,\ldots,i_N} := \epsilon^{a_1,\ldots,a_N} Q_{a_1}^{i_1} \cdots Q_{a_N}^{i_N}
&\leftrightarrow &
b_{i_1,\ldots,i_{N'}} := \epsilon_{a_1,\ldots,a_{N'}} q^{a_1}_{i_1} \cdots q^{a_{N'}}_{i_{N'}}\\
\tilde B_{\tilde \imath_1,\ldots,\tilde \imath_N} := \epsilon_{a_1,\ldots,a_N} \tilde Q^{a_1}_{\tilde \imath_1} \cdots \tilde Q^{a_N}_{\tilde \imath_N}
&\leftrightarrow &
\tilde b^{\tilde \imath_1,\ldots,\tilde \imath_{N'}} := \epsilon^{a_1,\ldots,a_{N'}} \tilde q_{a_1}^{\tilde \imath_1} \cdots q_{a_{N'}}^{\tilde \imath_{N'}}
\end{array}
\label{chiralopmapping}
\end{equation}  where $B\leftrightarrow b$ and $\tilde B\leftrightarrow \tilde b$  are to be related by the $\epsilon$ symbol for $\SU(N_f)$, e.g.~\begin{equation}
B^{i_1,\cdots,i_N} \propto b_{j_1,\cdots,j_{N'}} \epsilon^{i_1,\ldots,i_N,j_1,\ldots,j_{N'}}.
\end{equation}
Note also that $\underline{M}_i^{\tilde \jmath}:=q_i \tilde q^{\tilde \jmath}$ on the dual theory side is killed by the superpotential:  \begin{equation}
\frac{\partial W}{\partial M^i_{\tilde \jmath}}\propto \underline{M}_i^{\tilde \jmath}.
\end{equation} 
The derivatives of the superpotential are zero on the supersymmetric vacua,
and therefore the gauge-invariant combinations $ \underline{M}_i^{\tilde \jmath}$ are zero on the supersymmetric vacua.

The R-charges of these operators do also match.
On the original theory side, the conserved R-charge assigns $R(Q)=R(\tilde Q)=1-N/N_f$, as we already saw.
The story is similar on the dual theory side:
the gauge-singlet fields $M$ do not contribute to the R-gauge-gauge anomaly anyway, and therefore  $R(q)=R(\tilde q)=1-N'/N_f=N/N_f$.
Since $R(W)=2$, we have $R(M)=2-2N/N_f$.
This establishes that $R(Q\tilde Q)=R(M)$.
We can similarly check that $R(b)=NN'/N_f=R(B)$.
We will perform a more extensive check of the agreement of operators in Sec.~\ref{sec:SCI}. 

\subsubsection{Check: anomaly polynomials}
Our next check is about the anomaly polynomial $\cA$
for the continuous conserved symmetry $\U(1)_R\times \U(1)_B \times \SU(N_f)_\text{untilded}\times \SU(N_f)_\text{tilded}$.
As we argued in Sec.~\ref{sec:nonrenanom}, the anomaly polynomial is not renormalized.
Therefore, two dual theories have to share exactly the same anomaly polynomial.

Let us first consider the $\SU(N_f)_\text{untilded}^3$. 
The anomaly polynomial is written in terms of the background gauge field $F^u_{\mu\nu}$ where $u=1,\ldots,\dim\SU(N_f)$ for the $\SU(N_f)$ flavor symmetry. 
On the original theory side, we simply have
\begin{equation}
\frac16 N \tr_{N_f} (\frac{F}{2\pi})^3
\end{equation}
where $F=\frac12 F^u_{\mu\nu} T^u dx^\mu dx^\nu$
and $T^u$ are the generators of $\SU(N_f)$ in the fundamental $N_f$-dimensional representation.
On the dual theory side, we have 
\begin{equation}
\frac16 (N'\tr_{\bar{N_f}} F^3 + N_f \tr_{N_f} F^3)
=\frac16 ( N_f-N') \tr_{N_f} F^3
\end{equation} 
where the first contribution comes from $q$ and the second comes from $M$;
$\tr_{\bar {N_f}}$ is the trace in the anti-fundamental representation.
They agree, thanks to the equality $N+N'=N_f$.

Let us also consider the R-$\SU(N_f)_\text{untilded}$-$\SU(N_f)_\text{untilded}$ part of the anomaly.
We introduce the background gauge field $F_R$ for the $\U(1)_R$ symmetry.
Since $R(Q)=1-N/N_f$, the fermion component $\psi_Q$ has the R-charge $R(\psi_Q)=R(Q)-1=-N/N_f$.
Therefore, on the original side, we have \begin{equation}
\frac12  N(-\frac{N}{N_f})  \frac{F_R}{2\pi} \tr_{N_f}(\frac{F}{2\pi})^2
\end{equation} 
where the factor $N$ comes from the fact that $Q$ has $N$ components under $\SU(N)$ gauge symmetry and the factor $-N/N_f$ is the R-charge of $\psi_Q$.
On the dual side, we similarly have \begin{equation}
\frac12 \left[ N'(-\frac{N'}{N_f}) \frac{F_R}{2\pi}\tr_{\bar N_f}(\frac{F}{2\pi})^2+
  N_f  (1-\frac{2N}{N_f}) \frac{F_R}{2\pi}\tr_{N_f}(\frac{F}{2\pi})^2\right]
\end{equation} where the first term is from $q$ and the second is from $M$.
Again the two expressions agree, thanks to $N+N'=N_f$.

\exercise{Check the agreement of other parts of the anomaly polynomial, such as R-R-R.}

\subsubsection{Check: decoupling a flavor}
Let us next check that the duality is compatible with decoupling a flavor.
Consider giving to the original theory a non-zero mass to the last flavor $i=\tilde \imath=N_f$, \begin{equation}
\delta W=m Q^{i=N_f} \tilde Q_{\tilde \imath=N_f}.
\label{lastmass}
\end{equation} 
In the scale far below $m$, we get $N_f{}^\text{new}=N_f-1$, keeping the gauge group $N$ fixed.

What happens on the dual theory side?  The  superpotential is now \begin{equation}
W= qM\tilde q+ mM^{i=N_f}_{\tilde \imath=N_f},
\end{equation} 
where we added to the original superpotential $qM\tilde q$ the part $mM^{i=N_f}_{\tilde \imath=N_f}$
obtained by translating \eqref{lastmass} using the correspondence of operators \eqref{chiralopmapping}.
Setting $\partial W/\partial M^{i=N_f}_{\tilde \imath=N_f}=0$, we have \begin{equation}
\tilde q_{\tilde \imath=N_f}^a q^{i=N_f}_a+m=0.
\end{equation}
This gives a vev to $\tilde q^{\tilde \imath=N_f}$ and $ q_{i=N_f}$, breaking $\SU(N')$ to $\SU(N'-1)$.
Seiberg duality is compatible with this, since $N'{}^\text{new}=N_f{}^\text{new}-N=N'-1$.

\subsubsection{Check: re-dualization}
Finally, let us consider re-dualizing the dual theory.
The dual theory has the following structure: \begin{enumerate}
\item An $\SU(N')$ theory with $N_f$ pairs $q_i$, $\tilde q^{\tilde \imath}$.
\item Then add $N_f^2$ singlets $M^i_{\tilde \jmath}$ and add the coupling $\delta W=M^i_{\tilde \jmath} q_i \tilde q^{\tilde \jmath}$.
\end{enumerate}
Let us dualize the first part. Then we have \begin{enumerate}
\item an $\SU(N)$ theory with $N_f$ pairs $\mathscr{Q}^i$, $\tilde{\mathscr{Q}}_{\tilde \imath}$ with $N_f^2$ singlets $\mathscr{M}_i^{\tilde \jmath}$ with the coupling $W=\mathscr{M}_i^{\tilde \jmath} \mathscr{Q}^i \mathscr{\tilde Q}_{\tilde \jmath}$,
\item then add $N_f^2$ singlets $M^i_{\tilde\jmath}$ and add the coupling $\delta W=M^i_{\tilde \jmath} \mathscr{M}_i^{\tilde \jmath}$.
\end{enumerate}
The total coupling is now \begin{equation}
W_\text{total}=\mathscr{M}_i^{\tilde \jmath} \mathscr{Q}^i \tilde{\mathscr{  Q}}_{\tilde \jmath}+M^i_{\tilde \jmath} \mathscr{M}_i^{\tilde \jmath}.
\end{equation} 
Taking the variation with $M^i_{\tilde \jmath}$, we see $\mathscr{M}_i^{\tilde \jmath}$ is now massive, and 
taking the variation with $\mathscr{M}_i^{\tilde \jmath}$, we have 
 \begin{equation}
M^i_{\tilde\jmath} = -\mathscr{Q}^i \tilde{\mathscr{Q}}_{\tilde \jmath}
\end{equation} and can eliminate $M^i_{\tilde\jmath}$. 
We end up eliminating both $M^i_{\tilde\jmath}$ and $\mathscr{M}_i^{\tilde\jmath}$, and we simply have 
\begin{enumerate}
\item an $\SU(N)$ theory with $N_f$ pairs $\mathscr{Q}^i$, $\tilde{\mathscr{Q}}_{\tilde\imath}$.
\end{enumerate}
We thus re-obtain the original theory.

\subsection{Behavior of SQCD}
Assuming now the validity of Seiberg duality, let us try to understand the behavior of SQCD for various values of $N_f$.
We give the summary at the end with a  Table~\ref{table:SU}.

\subsubsection{When $N<N_f<3N$: gauge theory duals}
\label{sec:sudualregion}
When $N/N_f$ is close to $1/3$, the original theory flows to a weakly-coupled superconformal theory, as we already saw.
The dual theory has $N'/N_f$ slightly above $2/3$. 
Therefore the dual theory as defined in the ultraviolet  has a large one-loop beta function for $\SU(N')$ gauge coupling.
The statement of the duality is that it ends up in a weakly-coupled conformal theory of $\SU(N)$ gauge bosons and quarks.

We raise $N/N_f$ gradually, making the original theory more and more strongly coupled.
The dual theory's $N'/N_f$ decreases accordingly, making it more and more weakly coupled.
When $N/N_f$ becomes $2/3$, we hit the unitarity bound of $M=Q\tilde Q$.
The dual theory's $N'/N_f$ hits $1/3$, which is very weakly coupled. 
This region where $ (3/2)N < N_f < 3N$ is known as the \emph{conformal window}.

Now, we can raise $N/N_f$ even further. 
Since the operator $M=Q\tilde Q$ violates the unitarity bound,
we can no longer expect superconformal symmetry in the infrared. 
But note that the check of Seiberg duality performed above only cared about having a conserved R-symmetry, not that this conserved R-symmetry is in the superconformal symmetry.
So we can continue: the dual theory has $N'/N_f$ below $1/3$.
The dual theory is infrared free from the start; recall that the one-loop beta function coefficient is $3N'-N_f<0$.
There is a logarithmic running of the coupling toward zero in the infrared, and indeed this is not superconformal. 

How far can we go? Of course we can only have $N'\ge 0$. 
What happens when $N'$ is very low can be understood by being more careful about the decoupling.

\subsubsection{When $N_f=N+1$: weakly interacting mesons and baryons}
\label{sec:n+1}
Let us start from $\SU(N)$ with $N_f$ flavors, and decouple $N'-k$ flavors by giving mass \begin{equation}
W=  m^{\tilde\jmath}_i M_{\tilde\jmath}^i = m^{\tilde\jmath}_i Q^i \tilde Q_{\tilde\jmath},
\label{decoupling-mass-E}
\end{equation} where the sum over indices go over the last $N'-k$ of $1,\dots, N_f$.
Integrating out the quarks, we have $\SU(N)$ with $N+k$ flavors.

On the dual side, we have $\SU(N')$ with $N_f$ flavors, and the superpotential is \begin{equation}
W=qM\tilde q+ m^{\tilde\jmath}_i M_{\tilde\jmath}^i,
\label{decoupling-mass-M}
\end{equation} 
where again the sum in the second term is over the last $N'-k$ of $1,\ldots, N_f$.
As before, this gives vevs to $q_i$ and $\tilde q^{\tilde\imath}$ for the last $N'-k$ flavors, breaking 
$\SU(N')$ with $N_f$ flavors to $\SU(k)$ with $N+k$ flavors. 
Assume that the resulting $\SU(k)$ theory after decoupling flavors is in the infrared free region.
Then everything is weakly coupled in the infrared.
This means that the instanton computation in the dual $\SU(N')$ gauge theory is reliable.

Recall from \eqref{one-instanton-factor} that one-instanton effects come with the factor $\Lambda_{\SU(N')}^{3N'-N_f}$, which contains the factor $e^{i\theta}$.
Here the subscript $\SU(N')$ emphasizes that this is the instanton factor in the dual $\SU(N')$ theory.
Under the $\U(1)_i$ rotation $q_i \to q_i e^{\ii\varphi}$ fixing other $q$ and $\tilde q$, this factor $\Lambda_{\SU(N')}^{3N'-N_f}$ is also of charge 1. 
We note that $m_i^{\tilde\jmath}$ has charge $+1$ under this $\U(1)_i$,
and $M^i_{\tilde\jmath}$ has charge $-1$ under this $\U(1)_i$.

The form of a one-instanton contribution to the superpotential 
can be found by demanding that it is invariant under $\U(1)_i$,
 under $\SU(N+k)\times \SU(N'-k)$ apparent in the superpotential \eqref{decoupling-mass-M},
 and their counterparts acting on fields with tildes.
We find that it is given by 
\begin{equation}
 { \Lambda_{\SU(N')}^{3N'-N_f} }\frac{\det M_{(N+k)\times (N+k)} }{\det m_{(N'-k)\times (N'-k)}}
\end{equation} 
 where $M_{(N+k)\times (N+k)}$ is the submatrix of $M_{\tilde\jmath}^i$ whose indices are restricted to the first $N+k$ from $1,\ldots,N_f$,
 and $m_{(N'-k)\times (N'-k)}$ is the mass matrix $m_i^{\tilde\jmath}$ we introduced in 
 \eqref{decoupling-mass-E}, \eqref{decoupling-mass-M}, which we took to be an $(N'-k)\times (N'-k)$ matrix.
This has R-charge \begin{equation}
\frac{2N'}{N_f} (N+k) - \frac{2N}{N_f} (N'-k) =2k.
\end{equation}
Therefore, if and only if $k=1$, one-instanton configurations can produce this superpotential,
since the superpotential has to have R-charge 2.

Let us take $k=1$, and we assume that the one-instanton computation does generate this superpotential with nonzero coefficient.
From the point of view of the low energy $\SU(1)$ theory with $N+1$ flavors, $\Lambda_{\SU(N')}^{3N'-N_f} / \det m$ is just a numerical factor.
This means that only in this edge case $k=1$, the superpotential on the dual  $\SU(1)$ side is modified to be \begin{equation}
W=q M \tilde q - \det M.
\end{equation}

Note also that in this case $q_i\propto b_i=\epsilon_{ii_1\ldots i_{N}} B^{i_1\ldots i_N}$ and similarly for the tilded variables.
Therefore we conclude:
\begin{claim}
The infrared limit of the $\SU(N)$ theory with $N_f=N+1$ flavors 
are described by an almost free theory of the mesons $M^i_{\tilde \jmath}$ and baryons $B_i$, $\tilde B^{\tilde \jmath}$ with the superpotential \begin{equation}
W=\frac{1}{\Lambda^{3N-(N+1)}}(B_i M^i_{\tilde \jmath}B^{\tilde \jmath} - \det M)\label{BMB}
\end{equation}
where everything is written in the variables of the original theory,
and  the powers of $\Lambda$ is introduced to match the mass dimension.
\end{claim}
Those who know supersymmetric instanton calculus might worry:  
this looks like a $(-1)$-instanton effect in the original theory, while in any instanton computation only a positive-instanton contribution generates the superpotential.
It is fine, since the instanton computation is only applicable in the weakly-coupled theories, whereas this is an extremely strongly coupled situation in the original variables.
See also the footnote \ref{foot:SCIvsWCF}.

The appearance of $\det M$ can be also checked by considering the dual of $\SU(2)$ with $N_f=3$. 
In this case, $Q^{i=1,2,3}_a$ and $\tilde Q_{\tilde\imath=1,2,3}^a$ both transform in the doublet representation of $\SU(2)$, so can be combined to $\sQ^a_{I=1,2,3,4,5,6}$ with an $\SU(6)$ flavor symmetry.
The baryon $B_i$ and $\tilde B^{\tilde\imath}$ are just quadratic, and can be combined with $M_i^{\tilde\jmath}$ to form \begin{equation}
\mathsf{M}_{[IJ]}= \sQ^a_I \sQ^b_J \epsilon_{ab}.
\end{equation}
Then the superpotential \eqref{BMB} can be written as \begin{equation}
W=\epsilon^{IJKLMN} \mathsf{M}_{IJ}\mathsf{M}_{KL}\mathsf{M}_{MN}.\label{Nf=N+1}
\end{equation} 
Without $\det M$, the superpotential would not be $\SU(6)$ invariant as it should be.

\subsubsection{When $N_f=N$: deformed moduli space}
\label{sec:deformed}
The behavior with less flavors can be understood by decoupling the flavors further. 
Before proceeding, it is useful to understand how the instanton factors are related,
when $N_f$ is lowered by one by the decoupling.
We compare $\SU(N)$ with $N_f$ flavors and $\SU(N)$ with $N_f^\text{new}=N_f-1$ flavors.
Let us add $mQ^{i=N_f} \tilde Q_{\tilde\imath=N_f}$ to decouple one flavor to get the latter from the former.

\begin{figure}
\[
\includegraphics[width=.5\textwidth]{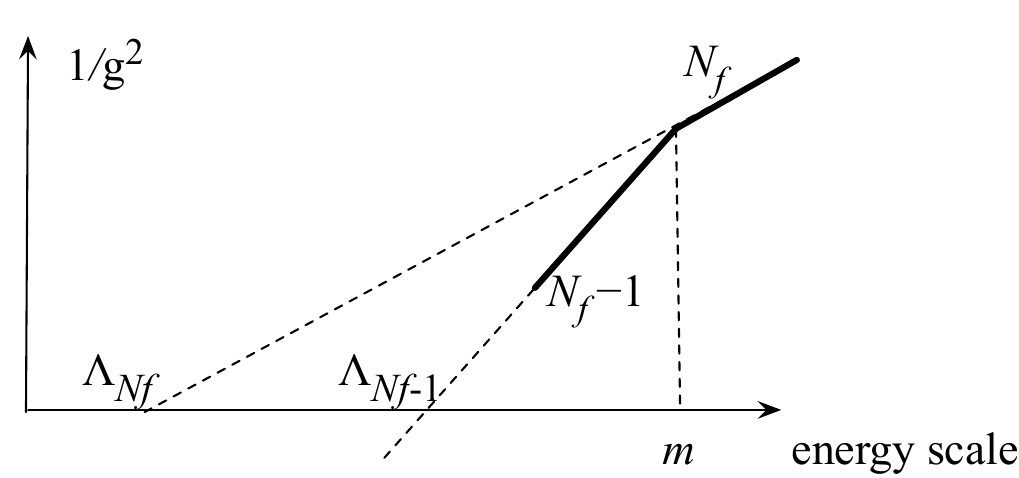}
\]
\caption{Running coupling under the decoupling.
The dynamical scales $\Lambda_{N_f}$, $\Lambda_{N_f'}$ are where a naive continuation of the graph gives a divergent coupling, and the slope of the running becomes steeper when the energy scale is lowered across the mass scale $m$.
\label{fig:running}}
\end{figure}

The one-instanton factors are respectively $\eta_{N_f}=\Lambda_{N_f}^{3N-N_f}$ and $\eta_{N_f-1}=\Lambda_{N_f-1}^{3N-(N_f-1)}$.  
The only relation between them we can write is, by considering their transformation properties under the symmetries, \begin{equation}
\eta_{N_f-1}= m \eta_{N_f},\quad\text{equivalently}\quad
\Lambda_{N_f}{} ^{3N-N_f} =m\Lambda_{N_f-1}{}^{3N-(N_f-1)},
\end{equation} 
up to a dimensionless proportionality coefficient.
This is also natural because \begin{align}
\Lambda_{N_f}{} ^{3N-N_f} &= E^{3N-N_f} e^{2\pi i \tau_{N_f} (E)},\\
\Lambda_{N_f-1}{} ^{3N-(N_f-1)} &= E^{3N-(N_f-1)} e^{2\pi i \tau_{N_f-1} (E)},
\end{align} and 
the running coupling $\tau(E)$ needs to match around $E=m$, see Fig.~\ref{fig:running}.

Now, let us add $m M^{i=N+1}_{\tilde\imath=N+1}$ to \eqref{Nf=N+1}. Taking the variation with respect to $M^{i=N+1}_{\tilde\imath=N+1}$, we get \begin{equation}
\det M_{N\times N} - B_{i=N+1}\tilde B^{\tilde\imath=N+1} = m\Lambda_{N_f=N+1}{}^{3N-N+1} .
\end{equation} Translating to the notation for $N_f=N$, we have the following statement:
\begin{claim}
The $\SU(N)$ theory with $N_f=N$ flavors has the moduli space of vacua given by 
 \begin{equation}
\det M- B\tilde B=\Lambda^{2N}.
\end{equation}
This is a constraint  rather than a superpotential, generated by an instanton effect.
\end{claim}

Note that classically, $B=\det Q^i_a$, $\tilde B=\det \tilde Q^a_{\tilde\imath}$ and $M^i_{\tilde \jmath}=Q^i_a \tilde Q^a_{\tilde\jmath}$. 
Therefore $\det M=\det Q\det \tilde Q = B\tilde B$.
We found that this relation is deformed by the one-instanton effect.
This deformation can be checked directly by an instanton computation \cite{Beasley:2004ys}.

\subsubsection{When $0<N_f<N$: ADS superpotential}
Let us decouple another flavor. 
To do this, implement the constraint above by a Lagrange multiplier $X$
and add $m Q^{i=N_f=N} \tilde Q_{\tilde \imath=N_f=N}$ :\begin{equation}
W= X(\det M-B\tilde B-\Lambda^{2N}) + m M^{i=N}_{\tilde\imath=N}.
\end{equation}
Eliminating $X$ and $M^{i=N}_{\tilde\imath=N}$, we get \begin{equation}
W= \frac{m \Lambda_{N_f=N}{}^{2N}}
{\det M_{(N-1)\times (N-1)}} 
\end{equation} where $M_{(N-1)\times (N-1)}$ is the submatrix of $M^i_{\tilde\jmath}$
where the indices run only over  $1,\ldots,N-1$.
In the variables appropriate for $N_f=N-1$, we have \begin{equation}
W= \frac{\Lambda_{N_f=N-1}{}^{3N-(N-1)}}{\det M}.
\end{equation} This has the one-instanton form, correctly invariant under the rotation $Q^i \to Q^i e^{\mathrm{i}\varphi}$, with the correct R-charge. 
This result was originally obtained  by Affleck, Dine and Seiberg \cite{Affleck:1983rr}, and therefore this is called the Affleck-Dine-Seiberg superpotential.
It is known that this can be reproduced from an honest instanton computation.
A generic vev to $Q$ and $\tilde Q$ breaks $\SU(N)$ to $\SU(1)$, and therefore it is reliable.

Note that the potential computed from this superpotential is nonzero as long as $M$ is nonzero and finite, and decreases toward infinity. This behavior is called the runaway.

We can equally decouple $k$ flavors from $N_f=N$. Then we  have the following statement:
\begin{claim}
The $\SU(N)$ theory with $N_f=N-k$ flavors generates the Affleck-Dine-Seiberg superpotential
 \begin{equation}
W=k \left[\frac{\Lambda_{N_f=N-k}{^{3N-(N-k)}}}{\det M}\right]^{1/k}
\label{ADSk}
\end{equation}  on its supersymmetric moduli space of vacua.
\end{claim}
 This might look more puzzling when $k>1$: it looks like a $1/k$-instanton effect.
Again this is fine. 
When we give a generic vev to $M$, the gauge group $\SU(N)$ is broken to $\SU(k)$ with zero massless flavors, which becomes strongly coupled and the instanton computation is unreliable.
Having $k$ branches in \eqref{ADSk} is also natural from this point of view, since we learned in Sec.~\ref{sec:numvac} that the pure $\SU(k)$ theory has $k$ distinct supersymmetric vacua.

If we decouple all $N$ flavors, we get \begin{equation}
W= N (\Lambda^{3N})^{1/N}.
\end{equation} This reproduces $N$ vacua of the pure $\SU(N)$ theory we saw earlier.
Indeed, as UV Lagrangian is $\int d^2\theta \tau_\text{UV} \tr WW$, 
we have \begin{equation}
\vev{\tr WW}= \frac{\partial}{\partial \tau_\text{UV}} W_\text{effective} \propto (\Lambda^{3N})^{1/N}.
\end{equation} with $N$ branches.

\subsubsection{Summary}

\begin{table}
\[
\begin{array}{c||c|c|l}
N_f & \text{unbroken} & \text{dual} &\text{behavior}\\
\hline
\hline
3N & - & \SU(2N) &  \\ 
\hline
3N-1 & - & \SU(2N-1) & \text{superconformal} \\
\vdots & \vdots & \vdots & \vdots \\
2N & - & \SU(N) & \text{superconformal, selfdual} \\
\vdots & \vdots & \vdots& \vdots\\
(3/2)N & - & \SU(N/2) & \text{superconformal} \\
\hline
(3/2)N-1 & - & \SU(N/2-1) &  \text{IR free with $W=q\tilde qM$}\\
\vdots & \vdots & \vdots& \vdots\\
N+2 & - & \SU(2) &  \text{IR free with $W=q\tilde qM$}\\
N+1 & - & \SU(1) & W=BM\tilde B+\det M \\
N &  - &  - & \det M - B\tilde B=\Lambda^{2N} \\
\hline
N-1 & \SU(1) & -& \text{ADS superpotential, computable} \\
N-2 & \SU(2) & -& \text{ADS superpotential, 2 branches} \\
\vdots & \vdots & \vdots& \vdots\\
1 & \SU(N-1) & -& \text{ADS superpotential, $N-1$ branches} \\
0 & \SU(N) & - & \text{$N$ vacua}
\end{array}
\]
\caption{Behavior of $\SU$ SQCD.\label{table:SU}}
\end{table}

We end our discussion of $\SU(N)$ SQCD with a big Table~\ref{table:SU}
summarizing the behavior of this class of theories with various values of $N_f$.
In the Table, the column `unbroken' gives the generic unbroken gauge group when a vev for $M=Q\tilde Q$ is given,
and the column `dual' lists the gauge group of the Seiberg dual theory.

\subsection{Kutasov duality}
\label{sec:kutasov}
As the last topic in this section on $\SU$ SQCD,
let us briefly consider a slightly different model originally considered by Kutasov and collaborators \cite{Kutasov:1995ve,Kutasov:1995np,Kutasov:1995ss}.
\subsubsection{Statement of the duality}
The model is an $\SU(N)$ gauge theory with $N_f$ pairs of fundamental chiral multiplets $Q^i$, $\tilde Q_i$ ($i=1,\ldots,N_f$) together with another chiral multiplet $\Phi$ in the adjoint, and the superpotential $W=\tr \Phi^{k+1}$.\footnote{%
When $k\ge 3$, the superpotential generates non-renormalizable interactions.
In those cases, this superpotential needs to be understood more precisely as follows.
We first consider a model with the same set of fields but without the superpotential.
For a suitable choice of $N$ and $N_f$,
this model with zero superpotential goes to an infrared superconformal field theory, in which the operator $\tr \Phi^{k+1}$ is relevant.
Then we deform this infrared superconformal theory by this relevant operator.
The model without the superpotential will be treated briefly in Sec.~\ref{sec:adjSQCD}.
}

When $k=1$, the field $\Phi$ is massive and can be integrated out. 
Assuming that the mass is vey large,  we come back to the standard $\SU$ SQCD below the mass scale, and in this sense this model is a natural generalization of the standard $\SU$ SQCD.
For general $k$, the conserved $\U(1)$ R-charge can be fixed as follows.
First, $\tr \Phi^{k+1}$ needs to have R-charge $2$. This fixes 
\begin{equation}
R(\Phi)=\frac{2}{k+1}.
\end{equation}
Then the condition that the $\U(1)_R\SU(N)^2$ anomaly vanishes fixes 
\begin{equation}
R(Q)=R(\tilde Q)=1-\frac{2}{k+1}\frac{N}{N_f}.
\label{rqk}
\end{equation}
As gauge-invariant operators, we can consider generalized mesons \begin{equation}
M^{(\ell)}{}^i_j = \tilde Q_j \Phi^\ell Q^i, \qquad (\ell=0,1,\ldots,k-1),
\end{equation}  baryons, and generalizations.

As the dual of this theory, the above cited papers proposed an $\SU(N')$ theory with $N_f$ pairs of fundamental chiral multiplets $q_i$ and $\tilde q^i$, an adjoint chiral multiplet $\phi$,
together with a number of gauge-singlet fields $M^{(\ell)}{}^i_j$ as above,
and the superpotential involving them: \begin{equation}
W=\tr \phi^{k+1} + \sum_{\ell=0}^{k-1} M^{(\ell)}{}^i_j \tilde q^j \phi^{k-1-\ell} q_i.
\end{equation}
By matching the anomaly-free $\U(1)$ R-charges of $M^{(\ell)}{}^i_j$ on the original theory and on the dual theory, we find \begin{equation}
N+N'=kN_f.
\end{equation}
It is a good exercise to check that the original theory and the dual theory have the same anomaly polynomial.

\exercise{Carry this out.}

\subsubsection{Unitarity bound and the decoupling}
\label{sec:kutasov-prescription}
Let us study the implication of the unitarity bound. 
The R-charge of the operator $M^{(0)}$ is twice the value of \eqref{rqk},
and falls below the unitarity bound when \begin{equation}
\frac{N}{N_f} > \frac{k+1}3.
\label{origb}
\end{equation}
This translates in terms of $N'$, the number of colors of the dual theory, to the expression \begin{equation}
\frac{N'}{N_f} < \frac{2k-1}{3}.
\label{dualb}
\end{equation}

This relation \eqref{dualb}, for the standard case of Seiberg duality $k=1$,
corresponds to the fact that the dual gauge theory is now infrared free.
as we saw in Sec.~\ref{sec:sudualregion}.
As an infrared-free gauge theory has a logarithmic running,
it is not superconformal. 
This removes the contradiction with unitarity.

However, starting with $k=2$, the condition \eqref{dualb} no longer corresponds to the dual gauge theory becoming infrared free;
the dual theory still looks interacting.
To see what is going on, let us take $k=2$ for definiteness. 

The dual theory has the superpotential \begin{equation}
W=\tr \phi^3 + M^{(0)} (\tilde q \phi q) + M^{(1)} (\tilde q q),
\end{equation}
where $M^{(0)}$ and $M^{(1)}$ are now gauge-singlet chiral superfields.
From the point of view of the dual theory, the R-charge of $M^{(0)}$ is fixed
only because we demand that the superpotential term $M^{(0)} (\tilde q \phi q)$ has R-charge two, or equivalently that this term is marginal.
This assumption leads to the violation of the unitarity bound,
when the condition \eqref{origb} or equivalently \eqref{dualb} is satisfied. 
A simple way out is to assume that $M^{(0)}$ has R-charge $2/3$ and scaling dimension $1$ and becomes free.
The superpotential term $M^{(0)} (\tilde q \phi q)$ now has R-charge more than $2$ and scaling dimension more than $3$, making it an irrelevant interaction.
Our interpretation when the operator $M^{(0)}$ appears to go below the unitarity bound
can then be schematically given as follows:
\begin{equation}
\begin{tikzpicture}[baseline=(X.east),every text node part/.style={align=center}]
\node(A) at (-4,1) {The original $\SU(N)$ theory\\
 with $Q,\tilde Q$ and $\Phi$ \\
 with $W=0$};
\node(P) at (4,1) {The dual $\SU(N')$ theory, \\
 with $q,\tilde q$, $\phi$, $M^{(0)}$ and $M^{(1)}$\\
 with $W=\tr \phi^3 + M^{(0)} (\tilde q \phi q) + M^{(1)} (\tilde q q)$
 };
\node(X) at (0,0) {};
\node(B) at (0,-1) {a superconformal theory $X$ + free $M^{(0)}$.};
\draw[->] (A) to (B);
\draw[->] (P) to (B);
\end{tikzpicture}
\end{equation}

It would be useful to have a Lagrangian description of the infrared theory $X$ itself.
This can be done \cite{Benvenuti:2017lle} by introducing an additional  set of gauge singlet field $\underline{M^{(0)}}^j_i$
and introduce a superpotential interaction $\delta W=\underline{M^{(0)}}^j_i {M^{(0)}}^i_j$ to the system, either to the original Lagrangian or to the dual Lagrangian.
On the original side, this interaction is $\delta W=\underline{M^{(0)}}^j_i Q^i \tilde Q_j$,
whereas on the dual side, this interaction introduces a mass term coupling  gauge singlet fields $\underline{M^{(0)}}^j_i$ and $M^{(0)}{}^i_j$ and effectively removing them.
We then have the following situation: \begin{equation}
\begin{tikzpicture}[baseline=(X.east),every text node part/.style={align=center}]
\node(A) at (-4,1) {The original $\SU(N)$ theory\\
 with $Q,\tilde Q$, $\Phi$ and $\underline{M^{(0)}}^j_i$ \\
with $W= \underline{M^{(0)}}^j_i Q^i \tilde Q_j$
 };
\node(P) at (4,1) {The dual $\SU(N')$ theory, \\
 with $q,\tilde q$, $\phi$, and  $M^{(1)}$ \\
 with $W=\tr \phi^3+ M^{(1)} (\tilde q q)$};
\node(X) at (0,0) {};
\node(B) at (0,-1) {a superconformal theory $X$,};
\draw[->] (A) to (B);
\draw[->] (P) to (B);
\end{tikzpicture}
\end{equation}
again assuming that we are in the region where the R-charge of $M^{(0)}=\tilde QQ$ drops below the unitarity bound.
There is no problem associated to the unitarity bound any more,
since the fact that \begin{equation}
\frac{\partial W}{\partial \underline{M^{(0)}}} = \tilde QQ
\end{equation} removes the problematic operator $\tilde QQ$ from the set of superconformal primary operators.

\subsubsection{Flipping of operators}
\label{sec:flipping}
This technique of 
\begin{itemize}
\item  picking a gauge-invariant operator $\cO$,
\item adding a corresponding gauge-singlet operator $\underline{\cO}$,
\item and introducing a superpotential interaction $\delta W=\underline{\cO}\cO$
\end{itemize}
is now called the \emph{flipping} of the operator $\cO$ of a theory.\footnote{%
As far as the author knows, this terminology was first introduced in \cite{Dimofte:2012pd}.
}
Note that flipping an operator $\cO$ by introducing $\underline{\cO}$,
and then re-flipping the introduced operator $\underline{\cO}$ 
by further introducing the operator $\underline{\underline{\cO}}$,
we get back the theory we started with, since the combined superpotential interaction\begin{equation}
	W=\underline{\underline{\cO}}\underline{\cO}+\underline{\cO}\cO+\cdots
\end{equation} allows us to identify $\underline{\underline{\cO}}=-\cO$
since \begin{equation}
	\frac{\partial W}{\partial \underline{\cO}}=\underline{\underline{\cO}}+\cO.
\end{equation} This operation also removes $\underline{\cO}$ since \begin{equation}
	\frac{\partial W}{\partial \underline{\underline{\cO}}}={\underline{\cO}}.
\end{equation}
In other words, the flipping is a reversible operation.

The technique of flipping looks innocuous, but has become one of the essential tools in modern study of \Nequals1 supersymmetric theories.
This might not be too surprising in the end: the Seiberg dual of the $\SU(N)$ SQCD is itself the $\SU(N')$ SQCD with the gauge-invariant operator $\tilde qq$ flipped by the gauge-singlet $M$.

\subsection{Other dualities}
\label{sec:other}
After Seiberg's original discovery \cite{Seiberg:1994pq},
many other dualities have been found.
We  just treated Kutasov's duality \cite{Kutasov:1995ve,Kutasov:1995np,Kutasov:1995ss}.
Here we simply mention some other notable dualities about which we do not have time to treat in detail.

We will discuss the $\Sp$ version \cite{Intriligator:1995ne} and $\SO$ version \cite{Intriligator:1995id} of the Seiberg duality in some detail
in Sec.~\ref{sec:SpSQCD} and in Sec.~\ref{sec:SOSQCD}, respectively.
There are also $\Sp$ and $\SO$ versions of Kutasov duality \cite{Intriligator:1995ff,Leigh:1995qp}.

The $G_2$ version of the SQCD was first considered in \cite{Giddings:1995ns},
and the analogue of Seiberg duality was found in \cite{Pouliot:1995zc};
the dual gauge group is$\SU(N_f-3)$ and it has  $S$ in the symmetric tensor representation and $\tilde Q$ in the anti-fundamental representation.
The analysis of SQCD with other exceptional gauge groups was carried out in \cite{Distler:1996ub,Karch:1997jp}.

The main content of \cite{Pouliot:1995zc} was the analysis of $\Spin(7)$ gauge theory with a matter in the spinor representation. 
This was then extended to $\Spin(8)$ in \cite{Pouliot:1995sk}, to $\Spin(10)$ in \cite{Pouliot:1996zh,Kawano:1996bd}, to $\Spin(11)$ in \cite{Cho:1997sa}, and to $\Spin(12)$ in \cite{Maru:1998hp}.
$\Spin(N)$ gauge theories with multiple chiral fields in the spinor representation was considered in \cite{Csaki:1997cu,Cho:1997kr,Berkooz:1997bb,Strassler:1997fe}.

In a slightly different direction, we considered the adjoint SQCD with $W=\Phi^k$ in Sec.~\ref{sec:kutasov}.
The infrared behavior of a generalization, namely the $\SU(N)$ theory with two adjoints and $N_f$ flavors, was studied in \cite{Brodie:1996vx,Intriligator:2003mi}.
The original superpotential $\Phi^k$ is considered as the $A_{k-1}$ singularity,
and the two-adjoint models correspond to simple singularities of type $D_n$ and $E_{6,7,8}$.\footnote{%
More classic usage of simple singularities in 2d \Nequals{(2,2)} models will be discussed in Sec.~\ref{sec:LG} below.
}
For further studies of these models, see \cite{Kutasov:2014yqa,Kutasov:2014wwa,Intriligator:2016sgx}.
It seems fair to say that we have a good understanding for $A_{k-1}$ and $D_\text{odd}$, but not for the other cases.

\section{Behavior of $\Sp$ SQCD with $N_f$ flavors}
\label{sec:SpSQCD}
You think you understood the behavior of $\SU(N)$ SQCD? 
Let us try to check if you really understand, by considering other groups and other matters.\footnote{%
In Japanese physics community there is a term called \begin{uCJK}銅鉄主義\end{uCJK}, which can be translated as copper-iron-ism. 
This means that any finding concerning a metal needs to be re-examined for any other metal, such as copper and iron.
Such a secondary research might not be totally original, but in the process one might have a new discovery. 
At least one learns the details and intricacies of the original research by one's own hand.
Here the author is trying to justify the $\Sp$-$\SO$-ism in the study of gauge theory.
}
The simplest generalization turns out to be to consider $\Sp(N)$.\footnote{%
The group is also often denoted as $\mathrm{USp}(2N)$.
Unfortunately there are sometimes also papers which denote by $\Sp(2N)$ what we denote by $\Sp(N)$.
} 
The analysis was originally done in \cite{Intriligator:1995ne}.

\subsection{What is the group $\Sp(N)$?}
\label{sec:sp}
$\bC$ has the absolute value function that satisfies $|z||w|=|zw|$.
Writing $z=a+bi$ and $w=s+ti$ for $a,b,s,t\in \bR$,
one finds the formula \begin{equation}
(a^2+b^2)(s^2+t^2)=(as-bt)^2 + (at+bs)^2.
\end{equation} 
This comes from regarding $\bC$ as a two-dimensional real vector space and considering its product as a function $(a,b)\circ (s,t) = (as-bt,at+bs)$.

It is natural to wonder if it is possible to introduce  a bilinear product on $\bR^n$ of the form \begin{equation}
\bR^n\ni a,s\quad
\Longrightarrow\quad
a\circ s =:x \in \bR^n
\end{equation} such that \begin{equation}
x_k = \sum_{i,j}c^{ij}_k a_i s_j
\end{equation}
and to have a formula \begin{equation}
\sum a_i{}^2 + \sum s_i{}^2 = \sum_k (\sum_{i,j} c^{ij}_k a_i s_j)^2 .
\end{equation}
A deep mathematical theorem says that 
it is possible only for $n=1,2,4,8$.
For a readable account, see e.g.~Part B of \cite{Numbers}.

The cases $n=1$ and $n=2$ are the familiar $\bR$ and $\bC$.
The $n=4$ case is known as the quaternion $\bH$ and the $n=8$ case is known as the octonion $\bO$.
The product \begin{equation}
\circ: \bR^n \times \bR^n \to \bR^n
\end{equation} defined by $c^{ij}_k$ loses commutativity for $n=4$, and associativity for $n=8$.

It is standard to use the basis $1$, $i$, $j$, $k$ over $\bR$ for $\bH$. 
The multiplications are \begin{equation}
i^2=j^2=k^2=-1,\qquad ij=-ji=k
\end{equation} and cyclic permutations.

A general element is \begin{equation}
q=a+bi+cj+dk,\qquad a,b,c,d\in \bR.
\end{equation}  The conjugate is defined as \begin{equation}
\bar q=a-bi-cj-dk
\end{equation} and we define \begin{equation}
|q|^2= q\bar q=a^2+b^2+c^2+d^2.
\end{equation} 
We can check $|qq'|=|q||q'|$.

Consider $\bH^n$, consisting of column vectors with $n$ elements of $\bH$.
This is a $\bH$-linear space, where the scalar multiplication is \emph{from the right}.
A $\bH$-linear transformation is then the matrix multiplication \emph{from the left} \begin{equation}
q_i \mapsto m_i^j q_j.
\end{equation} This commutes with the scalar multiplication thanks to the associativity \begin{equation}
m_i^j (q_j c) = (m_i^j q_j)c.
\end{equation} 
Even this type of linearity fails over $\bO$ since it is non-associative.
This makes it hard to perform linear algebra over $\bO$.
Also, already for $\bH$, it is difficult to define the determinant of a matrix, due to noncommutativity.

Now, $\bR^n$, $\bC^n$ and $\bH^n$ have a natural norm \begin{equation}
|v|^2 = \sum_{i=1}^n |x_i|^2.
\end{equation}
$\bR$-, $\bC$-, $\bH$- linear transformations which preserve the norm are respectively called $\mathrm{O}(n)$, $\mathrm{U}(n)$, $\Sp(n)$.
Note that \begin{itemize}
\item For the first two, we can demand that the determinant is 1, which defines the subgroups $\SO(n)$ and $\SU(n)$.
\item When $n=1$, they respectively become $\mathrm{O}(1)=\bZ_2$, $\UU(1)$, $\Sp(1)=\SU(2)$.
\end{itemize}

Another way to represent $\Sp(n)$ is as follows. $\bH^{n}=\bC^{2n}$, and therefore  $\Sp(n)\subset \mathrm{U}(2n)$.
An element $g\in \mathrm{U}(2n)$ is in $\Sp(n)$ when $g$ commutes with
the left multiplication by $j$ on $\bH^n=\bC^{2n}$.
This translates to the condition that $g$ preserves $\sJ_{ij}$, $i,j=1,\ldots, 2n$ given by \begin{equation}
\sJ=\underbrace{\begin{pmatrix}
0 & -1 \\
1 & 0
\end{pmatrix}\oplus
\begin{pmatrix}
0 & -1 \\
1 & 0
\end{pmatrix}\oplus
\cdots\oplus
\begin{pmatrix}
0 & -1 \\
1 & 0
\end{pmatrix}}_\text{$n$ copies}.
\label{J}
\end{equation}
In particular, $\Sp(1)=\SU(2)$.

\subsection{$\Sp(N)$ with $N_f$ flavors}
Consider \Nequals1 supersymmetric $\Sp(N)$ gauge theory with chiral multiplets $\sQ^a_I$, where $a=1,\ldots,2N$.
To avoid Witten's global anomaly, we need to have $I=1,\ldots,2N_f$. 
The one-instanton factor is \begin{equation}
\eta=\Lambda^{3(N+1)-N_f}.
\end{equation}
Under the anomaly-free R-symmetry, $R(\sQ)=1-(N+1)/N_f$.

The continuous conserved symmetries are the conserved R-symmetry $\U(1)_R$ 
and the flavor symmetry $\SU(2N_f)$.
The basic gauge invariant chiral scalar operators are mesons \begin{equation}
\sM_{IJ}= \sQ_I^a \sQ_J^b \sJ_{ab}
\end{equation} which is automatically antisymmetric under $I\leftrightarrow J$.
There are no independent baryon operator, since the epsilon tensor can be written as a polynomial in $J$: \begin{equation}
\epsilon_{i_1 i_2 \cdots i_{2N}}\propto\sJ_{[i_1 i_2} \sJ_{i_3 i_4} \cdots \sJ_{i_{2N-1}i_{2N}]},
\end{equation}
where $J$ is the invariant antisymmetric tensor of the $\Sp$ group given in \eqref{J}.
This property makes the structure of gauge invariant operators of an $\Sp$ gauge theory particularly simple, since baryons can be decomposed as a polynomial of mesons.

\paragraph{When $N_f>N+2$:}
When $N_f$ is close to the upper bound $3(N+1)$, it is in the weakly-coupled conformal phase.
We analyze what happens when we lower $N_f$, we need to invoke the duality:
\begin{claim}
The following two gauge theories are \emph{dual} when $(N+1)+(N'+1)=N_f$:
\begin{itemize}
\item an $\Sp(N)$ gauge theory with $2N_f$ chiral multiplets $\sQ_I$ in the fundamental representation, and with zero superpotential $W=0$.
\item an $\Sp(N')$ with $2N_f$ chiral multiplets $\sq^I$  in the fundamental representation, a set of gauge-singlet scalars $\sM_{[IJ]}$, and with the superpotential $W=\sq^I\sq^{J} \sM_{[IJ]}$. 
\end{itemize}
\end{claim}
The operators match rather obviously.
The matching of the R-charge of the operator $\sM_{IJ}= \sQ_I^a \sQ_J^b$
 fixes the relation $(N+1)+(N'+1)=N_f$.
The  anomaly polynomials then also match.
Deforming both sides by $m\sM_{i=2N_f-1,j=2N_f}$, we can check the consistency under the decoupling.

\exercise{Confirm these statements.}

\paragraph{When $N_f=N+2$:}
Decoupling $N_f-(N+2)$ flavors, as before, we see that the superpotential $W\propto \Pf \sM$ can be generated on the dual side by an instanton effect, where $\Pf$ denotes the Pfaffian:
\begin{equation}
\Pf \sM = \epsilon^{I_1I_2\cdots I_{2N_f}} \sM_{I_1I_2}\cdots \sM_{I_{2N_f-1} I_{2N_f}}.
\end{equation}
In terms of the variables of the original theory, this means that $\Sp(N)$ with $N+2$ flavors in the infrared becomes almost free theories of mesons $\sM_{[IJ]}$ with the superpotential \begin{equation}
W=\frac{\Pf \sM}{\Lambda^{3(N+1)-(N+2)}},
\end{equation} which has the correct R-charge and mass dimension.

\paragraph{When $N_f=N+1$:}
Decoupling another, one finds the constraint \begin{equation}
\Pf \sM=\Lambda^{3(N+1)-(N+1)}.
\end{equation} 

\paragraph{When $N_f \le N$:}
When $N_f=N$, we find \begin{equation}
W=\frac{\Lambda^{3(N+1)-N}}{\Pf \sM}
\end{equation} which is produced by one instanton.
Decoupling further, one finds \begin{equation}
W=(N+1-N_f)\left[\frac{\Lambda^{3(N+1)-N_f}}{\Pf \sM}\right]^{1/(N+1-N_f)}
\end{equation}
and in the extreme $N_f=0$, one just finds \begin{equation}
W=(N+1)(\Lambda^{3(N+1)})^{1/(N+1)}.
\end{equation}

\begin{table}
\[
\begin{array}{c||c|c|l}
N_f & \text{unbroken} & \text{dual} &\text{behavior}\\
\hline
\hline
3(N+1) & - & \Sp(2N+1) &  \\ 
\hline
3N+2 & - & \Sp(2N) & \text{superconformal} \\
\vdots & \vdots & \vdots & \vdots \\
2N+2 & - & \Sp(N) & \text{superconformal, selfdual} \\
\vdots & \vdots & \vdots& \vdots\\
(3/2)(N+1) & - & \Sp((N+1)/2) & \text{superconformal} \\
\hline
(3/2)(N+1)-1 & - & \Sp((N+1)/2-1) &  \text{IR free with $W=qqM$}\\
\vdots & \vdots & \vdots& \vdots\\
N+3 & - & \Sp(1) &  \text{IR free with $W=qqM$}\\
N+2 & - & - & W=\Pf M \\
N+1 &  - &  - & \Pf M=\Lambda^{2(N+1)} \\
\hline
N & - & -& \text{ADS superpotential, computable} \\
N-1 & \Sp(1) & -& \text{ADS superpotential, 2 branches} \\
\vdots & \vdots & \vdots& \vdots\\
1 & \Sp(N-1) & -& \text{ADS superpotential, $N$ branches} \\
0 & \Sp(N) & - & \text{$N+1$ vacua}
\end{array}
\]
\caption{Behavior of $\Sp$ SQCD.\label{table:Sp}}
\end{table}

\paragraph{Summary:}
Let us summarize the discussions so far in a big Table~\ref{table:Sp}.
As before, the column `unbroken' shows the generic unbroken subgroup when $\sM$ is given a vev, and the column `dual' shows the dual gauge group, if available.

\subsection{$\SU(2)\simeq \Sp(1)$ revisited}
\label{sec:su2sp1}
This has been totally un-surprising so far, but there is a small surprise when one recalls $\SU(2)\simeq \Sp(1)$.
Then, the same theory can be analyzed both as an $\SU(2)$ theory with $N_f$ pairs $Q^i_a$, $\tilde Q_i^a$ or an $\Sp(1)$ theory with $2N_f$ chiral multiplets $\sQ^a_I$.
The moral of this short section will be the following:
\begin{itemize}
\item  Seiberg duality tells us that there can be  \emph{two} different Lagrangians which describe the same theory in the infrared.
\item In fact there can be \emph{more than two} different Lagrangians which describe the same theory in the infrared.
\item This also suggests that there might be \emph{no} Lagrangian theory that describe a given theory in the infrared.
\end{itemize}

\subsubsection{$N_f=0,1,2,3$}
When $N_f=0$, it is just the pure theory.
When $N_f=1$, there is only one meson $M$ and \begin{equation}
W=\frac{\Lambda^{6-1}}{M}.
\end{equation} 
In these cases $\SU(2)$ and $\Sp(1)$ behave completely in the same way.

When $N_f=2$, as an $\SU(2)$ theory, the gauge invariant variables are $B$, $M^i_j$ and $\tilde B$. We have $1+2^2+1=6$ operators here.
As an $\Sp(1)$ theory, they are combined into $\mathsf{M}_{[IJ]}$, having $(4\cdot 3)/2=6$ operators in total. 
The constraint can be written in two ways as\begin{equation}
\det M- B\tilde B=\Pf \mathsf{M}=\Lambda^{6-2}.
\end{equation} 

The case $N_f=3$ we get \begin{equation}
W=\frac{\det M -B_i M^i_j \tilde B^j}{\Lambda^{6-3}}
=\frac{\Pf \mathsf{M}}{\Lambda^{6-3}}.
\end{equation}
We already mentioned this in Sec.~\ref{sec:n+1}.

In the cases$N_f=0,1,2,3$ we just studied, the results obtained either as $\SU(2)$ or as $\Sp(1)$ agreed on the nose, without further ado.
This is as it should be, since we were supposed to be describing the system in the infrared using the variables adapted for the infrared.
As we will see, the situation will be different for $N_f=4$ and $5$,
where we have two dual theories with different variables, which flow to a single theory in the infrared.

\subsubsection{$N_f=4$}
\label{SU2Sp1Nf4}

Now consider the case $N_f=4$. 
As $\SU(2)$ with 4 flavors, the dual $\SU(4-2)=\SU(2)$ theory has the fundamentals $q_i$  , $\tilde q^i$ and the $4\times 4$ singlets $M^i_j$ with the superpotential \begin{equation}
W_1=q_i  \tilde q^i M^i_j.
\end{equation} As $\Sp(1)$ with 4 flavors, the dual theory is again $\Sp(1)$, has fundamentals $\mathsf{q}^I$ and a gauge-singlet antisymmetric $\mathsf{M}_{[IJ]}$ which has six $(8\cdot 7)/2=28$ components, with the superpotential \begin{equation}
W_2=\mathsf{q}^I \mathsf{q}^J \mathsf{M}_{[IJ]}.
\end{equation}
They are clearly different as Lagrangian theories. 
In particular, only $\SU(4)^2\times\UU(1)_B$ flavor symmetry is manifest in the former, while the full $\SU(8)$ flavor symmetry is manifest in the latter.

There are two important lessens to be learned here.
One is that the flavor symmetry in the infrared limit can be enhanced compared with the symmetry manifest in the ultraviolet definition. 
In the example above, the Seiberg dual of $\SU(2)$ with $N_f=4$ only had $\SU(4)^2\times \UU(1)_B$ in the ultraviolet Lagrangian, but this should enhance to $\SO(8)$ in the infrared, for the consistency of the whole framework.
The importance of this enhancement was first emphasized in \cite{Leigh:1996ds},
and this type of symmetry enhancement is ubiquitous in the modern study of supersymmetric theories.

Another important point here is that we have found two distinct duals of a single theory. 
We can produce more, in fact.
In the $\SU(2)$ variables, we can rewrite the superpotential $W_2$ for the $\Sp(1)$ dual as \begin{equation}
W_2 = q_i \tilde q^i M^i_j + (q_i q_j) B^{[ij]} + (\tilde q^i \tilde q^j) \tilde B_{[ij]}.
\end{equation}
Now, this theory has the structure \begin{enumerate}
\item We have the $\SU(2)$ with four flavors $q_i$, $\tilde q^i$.
\item Add $M^i_j$, $B^{[ij]}$, $\tilde B_{[ij]}$ and add the coupling \begin{equation}
\delta W=q_i \tilde q^i M^i_j + (q_i q_j) B^{[ij]} + (\tilde q^i \tilde q^j) \tilde B_{[ij]}.
\end{equation}
\end{enumerate}
Now, let us dualize the first entry as $\SU(2)$ with four flavors. We get \begin{enumerate}
\item We have the $\SU(2)$ with four flavors $\mathscr{Q}^i$, $\mathscr{\tilde Q}_i$,
and $4\times4$ singlets $\mathscr{M}_i^j$, and the coupling 
\begin{equation}
W=\mathscr{Q}^i\mathscr{\tilde Q}_j \mathscr{M}_i^j.
\end{equation}
\item Add $M^i_j$, $B^{[ij]}$, $\tilde B_{[ij]}$ and the coupling \begin{equation}
\delta W=\mathscr{M}_i^j M^i_j + \epsilon_{ijkl}(\mathscr{Q}^k \mathscr{Q}^l) B^{[ij]} + \epsilon^{ijkl}(\mathscr{\tilde Q}_k \mathscr{\tilde Q}_l) \tilde B_{[ij]}.
\end{equation}
\end{enumerate}
Adding the superpotential, we can eliminate $M^i_j$ and $\mathscr{M}_i^j$, we obtain the third dual \cite{Csaki:1997cu}: it is again $\SU(2)$ with four flavors $\mathscr{Q}^i$, $\mathscr{\tilde Q}_i$,
and gauge singlets $B^{[ij]}$, $\tilde B_{[ij]}$, with the superpotential \begin{equation}
W=\epsilon_{ijkl}(\mathscr{Q}^k \mathscr{Q}^l) B^{[ij]} + \epsilon^{ijkl}(\mathscr{\tilde Q}_k \mathscr{\tilde Q}_l) \tilde B_{[ij]}.
\end{equation}
In the language of Sec.~\ref{sec:flipping}, we can say that this theory is obtained by flipping the baryon operators $B^{[ij]}$ and $\tilde B_{[ij]}$ of the $\SU(2)$ theory with $N_f=4$ flavors.

In fact there are many ways to split $\sQ_{I=1,\ldots,8}$ to $(Q^{i=1,\ldots,4},\tilde Q_{i=1,\ldots,4})$ and there are many more duals one can consider, the entire web of which is controlled by the Weyl group of $E_7$, as shown in \cite{Dimofte:2012pd}.
We will come back to this story in Sec.~\ref{sec:E7surprise}.

\subsubsection{$N_f=5$}
We will end this section by briefly mentioning the case $N_f=5$.
The dual as $\SU(2)$ is an $\SU(3)$ gauge theory with five flavors in the fundamental $\mathbf{3}$ and the anti-fundamental $\mathbf{3}$, whereas the dual as $\Sp(1)$ theory is an $\Sp(2)\simeq \Spin(5)$ gauge theory with five flavors in the fundamental $\mathbf{4}$. 
So the two duals clearly have different Lagrangians.

\section{Behavior of $\SO$ SQCD with $N_f$ flavors}
\label{sec:SOSQCD}
We have so far studied the behaviors of $\SU$ and $\Sp$ SQCD.
Let us now consider $\SO(N)$ SQCD with $N_f$ flavors $Q_i$, $i=1,\ldots,N_f$ in the vector representation.
There are a lot of surprises in this case. 
The analysis was first carried out in \cite{Intriligator:1995id}.

\subsection{$\SO(3)$ with one flavor}
\label{sec:so3nf1}
The simplest nontrivial example is $\SO(3)$ with $N_f=1$ flavor $Q$.  
The Lagrangian of this system in fact automatically has an enhanced \Nequals2 supersymmetry.
The study from this point of view was first done in \cite{Seiberg:1994rs}.
For a leisurely and somewhat modern review, readers are referred to Chapter 4 of the author's \Nequals2 review \cite{Tachikawa:2013kta}.
Here we analyze the system from the \Nequals1 point of view.
This analysis alone will require a few pages.

\subsubsection{Setup}

We normalize the gauge kinetic term as follows: \begin{equation}
\int d^2\theta \frac{-i\tau_\text{UV}}{8\pi} \tr_{\mathbf{2}} W_\alpha W^\alpha 
=
\int d^2\theta \frac{-i\tau_\text{UV}}{32\pi} \tr_{\mathbf{3}} W_\alpha W^\alpha 
\end{equation}
where $W_\alpha$ on the left hand side are considered as  $2\times2$ matrices for $\su(2)$
and $W_\alpha$ on the right hand side are $3\times 3$ matrices for $\so(3)$.
This makes $\tau_\text{UV}$ to have the periodicity $\tau_\text{UV}\sim \tau_\text{UV}+1$ on the flat $\bR^4$.

Let us give a vev to $Q$. This breaks $\SO(3)$  to $\SO(2)$,
and therefore the low energy theory contains a massless Abelian gauge field $W_\alpha$.
It also has massive W-bosons and the 't Hooft-Polyakov monopoles.

The coupling of the low-energy Abelian gauge field depends on the vev of $Q$.
The gauge-independent combination of the vev is $u:=Q\cdot Q$.
The low-energy Lagrangian would have the form \begin{equation}
\propto \int d^2\theta \tau_{\U(1)}(u) W_\alpha W^\alpha +\cc 
\end{equation} 
We  would like to determine $\tau_{\U(1)}(u)$ as a locally holomorphic function of $u$.

\subsubsection{Normalization of electric and magnetic $\U(1)$  charges}
For this end, we need to discuss the normalization of the low-energy $\U(1)$ field.
Recall that the Maxwell theory has an $\SL(2,\bZ)$ duality symmetry
acting on the electric charge and the magnetic charge $(\lambda_e,\lambda_m)\in \bZ\times \bZ$ of particles.\footnote{%
For a brief and leisurely introduction for this important duality symmetry, see e.g.~Chapter 1 of \cite{Tachikawa:2013kta}.
}
Here we normalize the charges so that massive W-bosons have $(\lambda_e,\lambda_m)=(\pm 2,0)$
and 't Hooft-Polyakov monopoles have $(\lambda_e,\lambda_m)=(0,\pm 2)$.
This is the convention we used in Sec.~\ref{sec:su2so3} of this note.
A nice feature of this choice for our purpose is that the charges of the dynamical particles are always even.

If the gauge group is $\SU(2)$, we can introduce an external electric source in the fundamental representation of $\SU(2)$, which has $(\lambda_e,\lambda_m)=(1,0)$.
If the gauge group is $\SO(3)$, we can instead introduce an external magnetic source whose charge is half that of a 't Hooft-Polyakov monopole, again as discussed in Sec.~\ref{sec:su2so3}.
This magnetic source has $(\lambda_e,\lambda_m)=(0,1)$.
Note that one cannot introduce a static electric source of charge 1 and a static magnetic source of charge 1 at the same time, since the pair violates the Dirac quantization condition.

We normalize $\tau_{\U(1)}$ so that the change $\tau_{\U(1)} \to \tau_{\U(1)}+1$ induces the Witten effect $\lambda_e\to \lambda_e+\lambda_m$.
It is known that a 't Hooft-Polyakov monopole of charge $(\lambda_e,\lambda_m)=(0,2)$ becomes
a dyon of charge $(\lambda_e,\lambda_m)=(2,2)$ under $\tau_\text{UV}\to \tau_\text{UV}+1$.
Therefore we should have $\tau_{\U(1)}=\tau_\text{UV}$ classically.\footnote{%
This is different from the normalization used in Chapter 4 of the author's \Nequals2 review \cite{Tachikawa:2013kta}, where the convention $\tau_{\U(1)}^\text{there}=2\tau_\text{UV}$  was used instead.
Our choice here leads to the form of the Seiberg-Witten curve originally found in \cite{Seiberg:1994rs}, which the author stupidly wrote to be `not very well motivated' in \cite{Tachikawa:2013kta}. 
But as we will see below it is well motivated and useful.
}

Now that we fixed the normalization, we can compute $\tau_{\U(1)}(u)$ when $|u|$ is large.
The gauge coupling runs logarithmically from the ultraviolet to the scale set by the vev $u$,
at which $\SO(3)$ is Higgsed to $\U(1)$.
At that scale, the charged fields all become massive, and the coupling stays constant below that.
With this consideration, we  find
\begin{equation}
\tau_{\U(1)}(u) = -\frac{1}{2\pi i}\log \frac{u^2}{\Lambda^4} + \cdots
\label{tauu}
\end{equation}   where $\Lambda^4$ is the one-instanton factor.

\exercise{Check this logarithmic running.}

When $|u|$ is small enough, the system is strongly coupled.
The leading term of \eqref{tauu} would have negative imaginary part.
If this were the sole term determining $\tau_{\U(1)}(u)$,
this would mean that $g^2_{\U(1)}$ is negative. 
This is strange, and something needs to happen there. 
We are going to invoke the Abelian duality.

\subsubsection{The duality group $\Gamma(2)\subset \SL(2,\bZ)$}

Our infrared Abelian theory is embedded in a bigger $\SO(3)$ theory. 
In our normalization, odd charges can only come from external sources,
and all dynamical excitations have even electric and magnetic charges $(\lambda_e,\lambda_m)$.
Therefore, the action of the dynamical duality transformation fixes $(\lambda_e,\lambda_m)$ modulo 2.
In other words, we can restrict the duality transformation of the Abelian theory from the full $\SL(2,\bZ)$ to the subgroup  \begin{equation}
\Gamma(2):= \left\{
\begin{pmatrix}
a & b\\
c& d
\end{pmatrix} \in \SL(2,\bZ)
\mid
\begin{pmatrix}
a & b\\
c& d
\end{pmatrix} \equiv
\begin{pmatrix}
1 & 0\\
0 & 1
\end{pmatrix} \pmod 2
\right\}.
\end{equation}
$\Gamma(2)$ is the kernel of the natural projection from $\SL(2,\bZ)$ to $\SL(2,\bZ_2)=S_3$.

The coupling $\tau$ modulo the action of $\Gamma(2)$ has a nice geometrical representation.
Take a torus $T^2$ obtained by identifying $z\sim z+1\sim z+ \tau$ of the complex plane.
The group $\SL(2,\bZ)$ is the group of the change of the basis of the torus.
The subgroup  $\Gamma(2)$ is the group that fixes the four points  $P$, $Q$, $R$ and $S$ on the torus invariant under $z\mapsto -z$.
Now, take the quotient of the torus $T^2$ by $z\mapsto -z$. 
There are four points fixed by this action of $z\mapsto -z$,
namely at $z=(n+m\tau)/2$ with $n,m=0,1$.
We call them $P$, $Q$, $R$ and $S$; we choose to put $P$ at $z=0$.

\begin{figure}
\centering
\includegraphics[width=.5\textwidth]{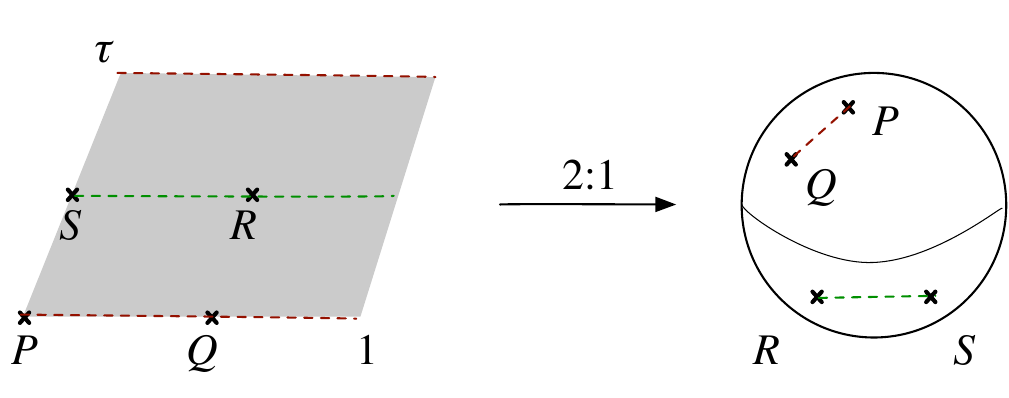}
\caption{The quotient of a torus $z\sim z+1\sim z+\tau$ by $z\mapsto -z$ is a sphere with four branch points. \label{fig:doublecover}}
\end{figure}

This makes the torus a double cover of a sphere, whose coordinate we call $x$, together with four branch points $P$, $Q$, $R$ and $S$. 
Without loss of generality we can put $P$ at $x=0$, $Q$ at $x=1$, $R$ at $x=\infty$ on the sphere.
Then the position $x=\lambda$ of $S$ is the only remaining freedom.
The original torus is given by the double cover \begin{equation}
y^2=x(x-1)(x-\lambda).
\end{equation}
See Fig.~\ref{fig:doublecover} for an illustration.

The action of $\SL(2,\bZ)$ keeps $P$ fixed, and permutes $Q$, $R$ and $S$.
This realizes the homomorphism $\SL(2,\bZ)\to \SL(2,\bZ_2)=S_3$.
Therefore $\Gamma(2)$ is the subgroup of $\SL(2,\bZ)$ which fixes $P$, $Q$, $R$ and $S$ pointwise.
From the construction we see that \begin{equation}
\lambda \stackrel{1:1}{\longleftrightarrow} \text{$\tau$ up to the action of $\Gamma(2)$}.
\end{equation}
In other words, the information of $\tau$ which is invariant under the duality group $\Gamma(2)$ is contained in the parameter $\lambda$.

When the coupling is very weak, an explicit computation shows that \begin{equation}
\lambda\sim e^{\pi i\tau}.
\end{equation}
When $\lambda$ is very small, one can redefine the coordinate on the sphere as $x'=1/x$. Then the torus becomes \begin{equation}
y'{}^2=x'(x'-1)(x'-1/\lambda).
\end{equation} 
Note that this change of variables is \emph{not} in $\Gamma(2)$
and in fact corresponds to the S transformation \begin{equation}
S=\begin{pmatrix}
0 & 1 \\
 -1 & 0
\end{pmatrix} \in \SL(2,\bZ).
\end{equation}
Correspondingly, we have \begin{equation}
\lambda \sim e^{-\pi i\tau_S}.\label{magnetic}
\end{equation}
Similarly, when $\lambda\to 1$, we can change the coordinate system, and we have another dual coupling $\tau_{ST}$.

We see that our aim is to fix $\tau(u)$ up to $\Gamma(2)$, or equivalently to fix $\lambda(u)$, as a meromorphic function.
The leading behavior is \begin{equation}
\lambda(u) \sim e^{\pi i\tau(u)} \sim c \frac{u}{\Lambda^2}
\label{lambdafoo}
\end{equation} which is just a funny way of expressing the one-loop running.
The full solution would be given by adding correction terms.

\subsubsection{Mass deformation}
\label{sec:masdef}
We need another step before we complete the analysis.
Consider the anomalous $\U(1)$ symmetry rotating $Q\to e^{i\varphi} Q$.
This shifts $\theta$ by $4\varphi$. Equivalently the instanton factor $\eta=\Lambda^4$ has charge 4.
Then the $\bZ_4$ subgroup remains a conserved symmetry.
There is also a corresponding anomalous conservation law, the Konishi anomaly,
which we already studied in Sec.~\ref{sec:rengauge}:
\begin{equation}
\bar D^2 Q^\dagger Q=4 \frac{1}{16\pi^2} \tr_{\mathbf2} WW.
\end{equation} 

Now, let us add the mass term $mQ^2$ to the original chiral multiplet $Q$. 
The Konishi anomaly equation is modified by the presence of the explicit $\U(1)_Q$ breaking term, and becomes \begin{equation}
\bar D^2 Q^\dagger Q=4 \frac{1}{16\pi^2} \tr WW + 2mQ^2.
\end{equation}
Taking the vev on a supersymmetric vacuum, we find \begin{equation}
\vev{\frac{1}{16\pi^2} \tr \lambda\lambda}=  -\frac12 m\vev{Q^2}.\label{gkonishi}
\end{equation}

We can analyze the effect of this mass term for $Q$ in another way.
When we integrate out $Q$ first, we find the pure \Nequals1 $\SO(3)$ theory.
We learned in Sec.~\ref{sec:numvac} that this theory has two possible values $\pm\Lambda^3_\text{pure}$ of the gaugino condensate $\vev{\lambda\lambda}$.
We also have the matching $\Lambda_\text{pure}^3=m \Lambda^2$.
This matching condition can be derived as in Sec.~\ref{sec:deformed}, where we studied the relation between the dynamical scales before and after decoupling a fundamental flavor.

Combining with \eqref{gkonishi}, we find an important constraint.
Namely, only two points $\vev{u}\propto \pm\Lambda^2$  survives the deformation of the theory by the mass term $mQQ$.
We have not been careful about the normalizations of $u$ and $\Lambda^2$;
we here demand that the normalizations are such that $\vev{u}=\pm\Lambda^2$.

\subsubsection{The Seiberg-Witten curve}

After these preparations, we can finally determine the full form of $\lambda(u)$.
As long as $\lambda(u)$ is neither 0, 1 nor $\infty$, we just have an Abelian gauge multiplet and that is it.
Let us now add the mass term $mQ^2$ to the system.
This is a term $mu$ in the infrared description. 
This is linear in $u$, and therefore, generic choices of $u$ do not survive as  supersymmetric vacua.

The only values of $u$ which survive are where $\lambda(u)$ is either 0, 1, or $\infty$.
Let us say $\lambda(u)=0$ at $u=u_*$.  Write $\lambda(u)=c(u-u_*)+\cdots$.
From \eqref{magnetic}, we see that \begin{equation}
\tau_S(u) = +\frac{1}{\pi i}\log c(u-u_*).
\end{equation}
This is an infrared free logarithmic running of the Abelian coupling,
which can be reproduced assuming that there is a charged particle of mass proportional to $|u-u_*|$.
Since this is the dual coupling, this particle is a monopole.
The charge can be fixed by carefully following the conventions; we find that this is due to a charge-2 monopole.
Let us denote by $q_\pm$ these monopole chiral multiplets. Then we have the superpotential \begin{equation}
W\sim (u-u_*) q_+ q_-.
\end{equation}

Let us add $mQQ=mu$ again. The total superpotential is \begin{equation}
W\sim mu+(u-u_*) q_+ q_-.
\end{equation}
Taking the variations, one find that $q_\pm$ gets a vev and $u=u_*$.
Since the magnetic particles condense, this is the dual Higgs mechanism and the original electric charge is confined.
From the Konishi equation \eqref{gkonishi} we see \begin{equation}
\vev{\lambda\lambda} \simeq m u_*.\label{ll}
\end{equation}

The same analysis can be carried out when $\lambda(u)=1$ or $\lambda(u)=\infty$ at $u=u_*$.
In each case, we find that a dyon of charge 2 or a electric particle of charge 2 has a mass proportional to $|u-u_*|$.
Adding $mQ^2$ to the superpotential, there is either the oblique confinement or the Higgs mechanism, and the relation \eqref{ll} holds.

Therefore, only points where $\lambda(u)=0$, $1$ or $\infty$ survive the mass deformation.
We already know from a different analysis in Sec.~\ref{sec:masdef} that only two points $u\propto \pm\Lambda^2$ survive the mass deformation.
This implies that  there are exactly two finite values  $u=\pm\Lambda^2$ such that $\lambda(u)=0$, $1$ or $\infty$.

We also know from \eqref{lambdafoo} that $\lambda(u)$ is linear in $u/\Lambda^2$ when $|u|$ is large.
Now we have  sufficient information to fix $\lambda(u)$: it has the form \begin{equation}
\lambda(u)=\frac{u}{2\Lambda^2}+\frac12
\end{equation} with the torus given by \begin{equation}
y^2=x(x-1)(x-\frac{u}{2\Lambda^2}-\frac12).
\end{equation} 
By a slight change of the variables we can also write \begin{equation}
\tilde y^2=(\tilde x^2-\Lambda^4)(\tilde x-u).
\end{equation}
This is called the Seiberg-Witten curve of the system.

Note that, during the derivation, we did not use \Nequals2 supersymmetry which this system secretly has. 
In general, for any \Nequals1 theory which has Abelian gauge fields on the generic points of the supersymmetric moduli space, finding the Seiberg-Witten curve would be a major step in the analysis of the theory.
We will see below in Sec.~\ref{sec:sec} that the $\SO(N)$ theory with $N_f=N-2$ flavors has a similar Seiberg-Witten curve.

The \Nequals1 Seiberg-Witten curves have not been studied very extensively. 
Here we provide an almost complete list of references up to 2010 \cite{Intriligator:1994sm,Kapustin:1996nb,Kitao:1996mb,Giveon:1997gr,Csaki:1997zg,Gremm:1997sz,Lykken:1997gy,Giveon:1997sn,deBoer:1997zy,Burgess:1998jh,Csaki:1998dp,Hailu:2002bg,Hailu:2002bh}.

\subsection{Seiberg duality for $\SO$ SQCD}
Consider an $\SO(N)$ gauge theory with $N_f$ chiral fields $Q^i_a$ in the vector representation, where $a=1,\ldots,N$ and $i=1,\ldots,N_f$.
The theory is asymptotically free when $N_f\le 3(N-2)$.
The one-instanton factor is \begin{equation}
\eta=\Lambda^{3(N-2)-N_f}.
\end{equation}
We have an anomaly-free $\U(1)$ R-symmetry,
together with the  flavor symmetry $\SU(N_f)$ acting on the index $i$ of the chiral multiplets $Q^i$.

We have the $\SO$ version of the Seiberg duality:
\begin{claim}
The following two gauge theories are \emph{dual} when $(N-2)+(N'-2)=N_f$:
\begin{itemize}
\item an $\SO(N)$ gauge theory with $N_f$ chiral multiplets $Q^i$ in the vector representation, and with zero superpotential $W=0$.
\item an $\SO(N)$ with $N_f$ chiral multiplets $q_i$  in the vector representation, a set of gauge-singlet scalars $M^{(ij)}$, and with the superpotential $W=q_i q_j M^{(ij)}$. 
\end{itemize}
\end{claim}

Let us perform some checks of the duality.
We identify the gauge-invariant operator $M^{ij}=Q^iQ^j$ of the original theory
and the gauge-singlet chiral field $M^{ij}$ of the dual theory.
Both transform in the two-index symmetric tensor of the $\SU(N_f)$ flavor symmetry.

The R-charge of $Q$ and $q$ can be easily found to be $R(Q)=1-(N-2)/N_f$
and $R(q)=1-(N'-2)/N_f$, respectively, from the anomaly-free condition.
We require that $2R(q)+R(M)=2$,
which leads to the important condition $(N-2)+(N'-2)=N_f$.
We can then check the agreement of the anomaly polynomials of the original theory and of the dual theory.

\exercise{Carry this out.}

Let us next try to match the baryonic operators on both sides.
The standard baryonic operators on both sides are: \begin{align}
\epsilon_N Q^N &:=\epsilon^{a_1 a_2 \cdots a_N} Q^{i_1}_{a_1} \cdots Q^{i_N}_{a_N},\\
\epsilon_{N'} q^{N'} &:=\epsilon_{a_1 a_2 \cdots a_{N'}} q_{i_1}^{a_1} \cdots q_{i_{N'}}^{a_{N'}}.
\end{align}
Their quantum numbers  do not match.
What saves the day is the existence of another pair of operators \begin{align}
\epsilon_N W_\alpha W^\alpha Q^{N-4} &:= \epsilon^{a_1 a_2 \cdots a_N} W_{a_1a_2,\alpha}W_{a_3a_4}^\alpha Q^{i_1}_{a_5} \cdots Q^{i_{N-4}}_{a_N},\\
\epsilon_{N'} W_\alpha W^\alpha q^{N'-4} &:= \epsilon_{a_1 a_2 \cdots a_{N'}} W^{a_1a_2}_{\alpha}W^{a_3a_4,\alpha} q_{i_1}^{a_5} \cdots Q_{i_{N'-4}}^{a_{N'}}.
\end{align}
where the reader should recall that the gauge multiplets $W_\alpha$ are in the adjoint, i.e.~the two-index antisymmetric representation of $\SO$.
Now we see that the quantum numbers \emph{do} match under the mapping \begin{align}
\epsilon_N Q^N & \longleftrightarrow \epsilon_{N'} W_\alpha W^\alpha q^{N'-4},\\
\epsilon_N W_\alpha W^\alpha Q^{N-4}  & \longleftrightarrow \epsilon_{N'} q^{N'},
\end{align}
once one uses the epsilon symbols for the $\SU(N_f)$ flavor symmetry.

\subsection{Aside: generalized Konishi anomaly}
\label{sec:yagi}
But this begs the question: why did not we have to consider such chiral scalar operators involving $W_\alpha W^\alpha$ in $\SU(N)$ and $\Sp(N)$ SQCD, to match the spectrum of the chiral operators on both sides?
After all, we should have listed all gauge-invariant scalar chiral operators, without restricting that  they are composed of scalar chiral operators alone.\footnote{The content of this section is based on an unpublished discussion with Futoshi Yagi in 2007, following the joint paper \cite{Kawano:2005nc} with Ookouchi, Kawano and the lecturer and \cite{Kawano:2007rz} by Kawano and Yagi.}
For example, what happens to 
\begin{equation}
\epsilon_N (W_\alpha W^\alpha Q) Q^{N-1}:=
\epsilon^{a_1\cdots a_N} (W_\alpha{}_{a_1}^b W^\alpha{}^c_b Q_c^{i_1})
Q^{i_2}_{a_2}\cdots Q^{i_N}_{a_N}
\end{equation}
 in the $\SU(N)$ SQCD?

The simplest case to consider is $\tr W_\alpha W^\alpha$ itself. For this, we already know the answer. 
As we saw in Sec.~\ref{sec:rengauge}, the Konishi anomaly, the anomalous transformation law under $Q\to e^{i\varphi} Q$, says \begin{equation}
\bar D^2 (Q^\dagger Q)=  Q \frac{\partial W}{\partial Q} + 2C(R) \frac{1}{16\pi^2} \tr W_\alpha W^\alpha .
\end{equation} Therefore, $\tr W_\alpha W^\alpha$ is essentially $Q\frac{\partial W}{\partial Q}$ in supersymmetric vacua, and does not have to be treated independently.

This relation can be generalized even further \cite{Cachazo:2002ry}, by considering the infinitesimal variation $\delta Q^a=\epsilon f(Q)^a$ for an arbitrary holomorphic function $f(Q)$.
Let us say that $Q$ is in the representation $R$ of the gauge group $G$,
and we regard the partial derivative $\partial f(Q)/\partial Q$ also as a matrix acting on the representation space of $R$: \begin{equation}
\delta f(Q)^a = (\frac{\partial f(Q)}{\partial Q} )^a_b \delta Q^b.
\end{equation}
Then the corresponding equation is \begin{equation}
\bar D^2 Q^\dagger{}_a f(Q)^a = f(Q)^a(\frac{\partial W}{\partial Q})_a +\frac{1}{16\pi^2} \tr_R (W_\alpha W^\alpha \frac{\partial f(Q)}{\partial Q} ).
\end{equation}

This relation allows us to eliminate $\epsilon_N (W_\alpha W^\alpha Q) Q^{N-1}$ in $\SU(N)$ SQCD by considering the variation for $\tilde Q$ given by \begin{equation}
\delta \tilde Q^a_{i} = \tilde Q^b_i Q_c^{i_1} \epsilon^{a a_2 a_3\cdots a_N} Q_{a_2}^{i_2}\cdots Q_{a_N}^{i_N}.
\end{equation}
Note that the transformation on $\tilde Q$ is used to eliminate an operator which does not contain $\tilde Q$.

Similarly, we can eliminate the operator $\epsilon_N W_\alpha W^\alpha Q^{N-4}$ in an $\Spin(N)$ theory with $Q$ in the vector representation, if there is a chiral field $\Psi$ in the chiral spinor representation.
Indeed, we can consider the variation \begin{equation}
\delta\Psi =\slash{Q}^{N-4} \Psi,
\end{equation} where $\slash Q=Q_a \gamma^a$.
The corresponding generalized Konishi identity has the form \begin{equation}
\bar D^2 \Psi^\dagger \Psi \sim \tr_\text{chiral spinor} \slash{Q}^{N-4} \Gamma^{ab} W_{ab}^\alpha
\Gamma^{cd} W_{cd,\alpha}  \sim \epsilon_N W_\alpha W^\alpha Q^{N-4}.
\end{equation}
This resolves a puzzle mentioned in the footnote 1 of \cite{Kawano:2010hd},
which is about the case $N=10$.
This fact for $N=8$ was already noticed in \cite{Pouliot:1995sk} using more indirect means, see the discussions around (1.5) there.

\subsection{$\SO(N)$ with $N_f$ flavors}

\subsubsection{When $N_f\ge N$}
Let us now study  the infrared behavior of the $\SO$ SQCD with various number of flavors $N_f$.
Close to the maximum $N_f\sim 3(N-2)$, the theory is in the weakly-coupled conformal phase.
As before, let us consider lowering $N_f$ gradually.
The system becomes more and more strongly coupled, and eventually the gauge-invariant operator $Q^iQ^j$ hits the unitarity bound,
and the dual description becomes infrared free. 
Nothing of note happens up to and including $N_f=N$, for which the dual is $\SO(N'=4)$, with the superpotential \begin{equation}
W=q_i q_j M^{ij}
\end{equation} as always.

\subsubsection{When $N_f=N-1$}
Decoupling one flavor, we get to $N_f=N-1$ for which we have $N'=3$. 
It is straightforward to see that a new term can be generated by the one-instanton configuration in the broken gauge group on the dual side much as in Sec.~\ref{sec:n+1}, and we get \begin{equation}
W=q_i q_j M^{ij} +\tilde\Lambda^{6-2(N-1)} \det M,
\end{equation}
where $\tilde \Lambda$ is the instanton factor of the dual theory.

\subsubsection{When $N_f=N-2$}
\label{sec:sec}
We now add the term $\delta W=m M^{N-1,N-1}$ to decouple one more flavor. 
Close to the origin of $M$, we see that it gives a vev to $q_{N-1}$, breaking $\SO(3)$ to $\SO(2)$.
The remaining superpotential is $W=q_i q_j M^{ij}$ where the indices $i,j$ now run over $1,\ldots,N-2$.
Note that $q_i$ here are in the two-dimensional representation of $\SO(2)$, in addition to being in the fundamental of the $\SU(N_f)$ flavor symmetry.
As field charged under $\UU(1)$, it can be written as\footnote{%
There can be a complicated instanton correction $f(\det M/\Lambda^{2(N-2)})$ multiplying the whole expression, with $f(0)\neq 0$. We neglect these issues below.
} \begin{equation}
W=q^+_i q^-_j M^{ij}.
\end{equation} 

There is another region where something happens. 
To see this, we imitate our analysis of $\SO(3)$ with one flavor, which was already given in detail in Sec.~\ref{sec:so3nf1}.
Note that for the rest of the analysis of $N_f=N-2$, we directly work in the original description.

The one-instanton factor is\footnote{%
Note that this is valid only for $N\ge 4$.
For $N=3$, the one-instanton factor is $\Lambda^4$.
This is due to the fact that the one-instanton configuration of $\SO(3)$ embedded into $\SO(4)$ in the standard manner has instanton number 2.
We encountered the same issue before in Sec.~\ref{sec:spin-vs-so}.
} \begin{equation}
\eta=\Lambda^{3(N-2)-(N-2)}=\Lambda^{2(N-2)}.
\end{equation} 
The field $Q$ is neutral under the anomaly-free R-charge.
We can instead consider the $\U(1)$ flavor symmetry acting on $Q$.
This is anomalous due to the $\U(1)$-$\SO(N)^2$ anomaly,
and the unbroken subgroup is  $\bZ_{2(N-2)}$.

Giving generic vevs to $M$, we get an $\SO(2)$ gauge field in the infrared.
We need to determine its coupling as a function of $M^{ij}$.
The flavor symmetry says that the coupling can only depend on $U=\det M$,
on which the unbroken $\bZ_{2(N-2)}$ acts trivially.

Just as in the case of $\SO(3)$ with $N_f=1$,
we know that the dynamical duality group is $\Gamma(2)\subset \SL(2,\bZ)$, and the coupling can be usefully represented in terms of $\lambda(U)$, or equivalently in terms of the equation of the torus.
We can in fact give big vevs to $Q_{2,\ldots,N-1}$, then the Higgsed theory is $\SO(3)$ with $N_f=1$.
Using this, we can find the large $U$ behavior of $\lambda(U)$ to be \begin{equation}
\lambda(U) \sim \frac{U}{\Lambda^{2(N-2)}}.
\label{lambdaU}
\end{equation}
We already know from the decoupling argument from $N_f=N-1$ that at $U=0$ we have $N-2$ pairs of fields $q_i^\pm$ of charge $\pm1$, 
making the coupling at $U=0$ to go to zero. 
Therefore, $\lambda(U=0)=0$.

We have not been careful about the normalization of $U$;
here we choose it so that we exactly have \begin{equation}
\lambda(U)=\frac{U}{\Lambda^{2(N-2)}}.
\end{equation} In other words, the Seiberg-Witten curve is given by
\begin{equation}
y^2=x(x-1)(x-U/\Lambda^{2(N-2)}).
\end{equation}

We already discussed the physics when $U\to \infty$ and $U\to 0$.
There is another singular locus where $U=\Lambda^{2(N-2)}$. 
The fact that $\lambda\to 1$ there means that there is a light dyon $E_\pm$, with the superpotential term \begin{equation}
W\sim (\det M-\Lambda^{2(N-2)})  E_+ E_-.
\end{equation}

\subsubsection{When $N_f=N-3$}
Now we add $\delta W=m Q^{N-2}Q^{N-2} = m M^{N-2,N-2}$
and decouple one flavor.
This term forces the vacua to be either on $\det M=0$ or $\det M=\Lambda^{2(N-2)}$.
In the former branch, $q_{N-2}^\pm$ condenses, while $q_{i}$ and $M^{ij}$ with $i,j=1,\ldots,N-3$ remain massless, with the superpotential \begin{equation}
W\sim q_i q_j M^{ij}.
\end{equation} 
 This $q_i$ is naturally identified with the operator on the dual side as follows:  \begin{equation}
q_i = \epsilon_{N_f=N-3}\epsilon_N W_\alpha W^\alpha Q^{N-4}.
\end{equation}
In the latter branch, $E^\pm$ condenses. We find that we have the Affleck-Dine-Seiberg superpotential \begin{equation}
W \sim \frac{\Lambda^{3(N-2)-(N-3)}}{\det M}.
\end{equation}

\subsubsection{When $N_f=N-4$}
Let us add $\delta W=m Q^{N-3}Q^{N-3} = m M^{N-3,N-3}$
and decouple another flavor.
From the former branch, we just condense $q_{N-3}=\pm \sqrt{m}$, eliminating $M^{i,N-3}$ and the rest of $q_i$. We  obtain two branches of vacua, parameterized by $M^{ij}$ with $i,j=1,\ldots,N-4$, with zero superpotential $W=0$.

From the latter branch, we get a behavior familiar from the analysis of $\SU(N)$ and $\Sp(N)$: we just get the ADS superpotential \begin{equation}
W = \pm \left[\frac{\Lambda^{3(N-2)-(N-4)}}{\det M}\right]^{1/2},
\end{equation} which again has two branches.

In the end we found \emph{four} branches. They can also be understood as follows: giving a generic vev to $Q$, $\SO(N)$ is broken to pure $\SO(4)\simeq \SU(2)_1\times \SU(2)_2$ gauge theory, and the one-instanton factor of both is $\Lambda'{}^6= \Lambda^{3(N-2)-(N-4)}/{\det M}$.
Each of $\SU(2)$ can have $W=\epsilon_{1,2}\Lambda'{}^3$ where $\epsilon_{1,2}=\pm1$, with the total superpotential \begin{equation}
W=(\epsilon_1+\epsilon_2) \Lambda'{}^3 = (\epsilon_1+\epsilon_2) \left[\frac{\Lambda^{3(N-2)-(N-4)}}{\det M}\right]^{1/2}.
\end{equation}
Therefore, we find two branches with zero superpotential,
and two branches with non-zero ADS superpotential.

\subsubsection{When $N_f<N-4$}

Let us  add $\delta W=m Q^{N-4}Q^{N-4} = m M^{N-4,N-4}$
and decouple another flavor.
The branch with $W=0$ is eliminated, while from the branch with $W\neq 0$ we get the standard  ADS superpotential \begin{equation}
W=3\left[\frac{\Lambda^{3(N-2)-(N-5)}}{\det M}\right]^{1/3},
\end{equation} with three branches.

From this point on, the structure regularize, and we just have the ADS superpotential\begin{equation}
W=(N-N_f-2)\left[\frac{\Lambda^{3(N-2)-N_f}}{\det M}\right]^{1/(N-N_f-2)}
\end{equation} up to and including $N_f=1$. 
When $N_f=0$ we have the pure $\SO(N)$ Yang-Mills.

\subsubsection{Summary}

\begin{table}
\[
\begin{array}{c||c|c|l}
N_f & \text{unbroken} & \text{dual} &\text{behavior}\\
\hline
\hline
3(N-2) & - & \SO(2N-2) &  \\ 
\hline
3N-7 & - & \SO(2N-3) & \text{superconformal} \\
\vdots & \vdots & \vdots & \vdots \\
2N-4 & - & \SO(N) & \text{superconformal, selfdual} \\
\vdots & \vdots & \vdots& \vdots\\
\frac32(N-5) & - & \SO(\frac{N-7}2) & \text{superconformal} \\
\hline
\frac32(N-5)-1 & - & \SO(\frac{N-7}2-1) &  \text{IR free with $W=qqM$}\\
\vdots & \vdots & \vdots& \vdots\\
N & - & \SO(4) &  \text{IR free with $W=qqM$}\\
N-1 & \SO(1) & \SO(3) &  \text{IR free with $W=qqM+\det M$}\\
\hline
N-2 & \SO(2) & \SO(2)  & \text{Coulomb phase, $y^2=x(x-\Lambda^{2(N-2)})(x-U)$}\\
\hline
N-3 & \SO(3) & \SO(1)  & \text{one branch with $W=qqM$, another with ADS}\\
N-4 & \SO(4) & -  & \text{two branches with $W=0$, two more with ADS}\\
N-5 & \SO(5) & -  & \text{ADS with three branches}\\
N-6 & \SO(6) & -  & \text{ADS with four branches}\\
\vdots & \vdots & \vdots& \vdots\\
1 & \SO(N-1) & -& \text{ADS superpotential, $N-3$ branches} \\
0 & \SO(N) & - & \text{$N-2$ vacua}
\end{array}
\]
\caption{Behavior of $\SO$ SQCD.\label{table:SO}}
\end{table}

Let us summarize the behavior of $\SO$ SQCD with $N_f$ flavors
in a big Table~\ref{table:SO}.
As always, the column `unbroken' shows the generic unbroken subgroup when $M$ is given a vev,
and the column `dual' gives the dual gauge group, when available.
A comparison of the $\SO$ table with the $\SU$ version, Table~\ref{table:SU},
and the $\Sp$ version, Table~\ref{table:Sp},
shows that the behavior here is significantly more complicated.
Furthermore, we see here that the unbroken gauge group and the dual gauge group can both be nontrivial when $N_f=N-2$, in contrast to the $\SU$ and $\Sp$ cases.
In this particular case of $N_f=N-2$, the unbroken $\SO(2)$ and the dual $\SO(2)$ are electromagnetic dual to each other.
In this sense, the Seiberg duality is a generalization of the standard electromagnetic duality of Maxwell theory.

\subsection{$\Spin(N)$, $\SO(N)_+$ and $\SO(N)_-$}

So far in this section we did not distinguish the three possible choices of the gauge group, $\Spin(N)$, $\SO(N)_+$ and $\SO(N)_-$, which we reviewed in Sec.~\ref{sec:spin-vs-so}.
In this last part we see how they are mapped under the Seiberg duality \cite{Aharony:2013hda}. 

For this, it is useful to start by considering pure $\so(4)$ theory with the common coupling $\Lambda_1^6=\Lambda_2^6=:\Lambda^6$.
The representation of $\so(4)=\su(2)\times \su(2)$ can be classified by $\bZ_2\times \bZ_2$.
To compare with the $\SO(N)$ theory with matters in the vector representation,
 we consider a rougher classification of $\so(4)$ representations, by considering $\bZ_2\times \bZ_2$ modulo the vector representation of $\so(4)$, which corresponds to $(1,1)\in \bZ_2\times \bZ_2$.
 We also use a similar classification scheme for magnetic charges.

The $\Spin(4)$ theory has the spinor Wilson line operator,
$\SO(4)_+$ theory has the 't Hooft line operator,
and the $\SO(4)_-$ theory has the dyonic line operator.
We have four vacua, distinguished by the superpotential  \begin{equation}
W=(\epsilon_1+\epsilon_2) \Lambda^3 
\end{equation} where $\epsilon_{1,2}=\pm1$.
For each $\su(2)_i$ factor, there is a monopole condensation if $\epsilon_i=+1$
and a dyon condensation if $\epsilon_i=-1$.
In terms of our rougher classification of charges as described above,
monopoles of $\so(N)$ condense in the vacua where $\epsilon_1=\epsilon_2$,
whereas dyons of $\so(N)$ condense in the vacua where $W=0$.
Therefore, in the vacua where $W\neq 0$, 
the 't Hooft line has the perimeter law but the dyonic line is confined.
Meanwhile, in the vacua where $W=0$, the 't Hooft line has the area law while the dyonic line is unconfined. 
The spinor Wilson line is always confined.

Now, consider $\so(N)$ with $N_f$ flavors $Q$, whose dual is $\so(N')$ with $N_f$ flavors $q$  where $N'=N_f-N-4$, together with mesons $M$.
Give a large vev to $M$. 
On the original side, we have completely Higgsed vacua.
Therefore, the three types of line operators have the following behavior:
\begin{equation}
\begin{array}{c|c|c}
\text{Wilson} & \text{'t Hooft} & \text{dyonic} \\
\hline
\text{perimeter} & \text{area} & \text{area}
\end{array}.
\end{equation}

On the dual side, the vev to $M$ gives masses to $N$ out of $N_f$ of $q$,  breaking $\so(N')$ to $\so(4)$.
The branch with $\epsilon_1+\epsilon_2\neq 0$ has a runaway superpotential, 
and the supersymmetric vacua come from the branch with $\epsilon_1+\epsilon_2=0$.
Therefore, 
\begin{equation}
\begin{array}{c|c|c}
\text{Wilson} & \text{'t Hooft} & \text{dyonic} \\
\hline
\text{area} & \text{area} & \text{perimeter}
\end{array}.
\end{equation}
Comparing the behavior of the line operators on the original side and on the dual side,
we see that the spinor Wilson line operator of the original theory
is mapped to the dyonic line operator of the dual theory.
This means that we have \begin{equation}
\Spin(N)\leftrightarrow \SO(N')_-
\end{equation} under the Seiberg duality, and then by exhaustion, we see that
\begin{equation}
\SO(N)_+\leftrightarrow \SO(N')_+.
\end{equation}

\section{Supersymmetric index on $S^3\times S^1$}
\label{sec:SCI}
Let us perform more detailed checks of Seiberg duality, by considering the supersymmetric index of the system on $S^3$.\footnote{%
This quantity is often called the superconformal index and abbreviated as SCI in the literature.
However, only the existence of a conserved $\U(1)_R$ symmetry is necessary in the construction, and therefore we prefer not to use this terminology. 
Under the assumption of the superconformal symmetry, the states on $S^3$ are mapped to point operators of the theory via the state-operator correspondence,
and the derivations given below can be and often are phrased in that language.
}
The idea is analogous to the analysis performed for the pure super Yang-Mills in Sec.~\ref{sec:box} on $T^3$,
but the analysis of SQCD on $T^3$ is afflicted with many technical pitfalls due to various zero modes, and has not been successfully carried out.
In contrast, the system on $S^3$ is better behaved and is easier to analyze.
This analysis was originally done in two papers by R\"omelsberger \cite{Romelsberger:2005eg,Romelsberger:2007ec}
in the context of \Nequals1 superconformal symmetry, and by Kinney-Maldacena-Minwalla-Raju in \cite{Kinney:2005ej} in the context of $\cN{>}1$ superconformal symmetry.

\subsection{Supersymmetry on $S^3\times \bR$ }
We first need to study the supersymmetry on $S^3\times \bR_t$ and $S^3\times S^1$.
Since this spacetime is curved, we cannot just use the flat-space result.
The full systematic treatment was performed in \cite{Festuccia:2011ws}. 
We will just sketch the argument.

The bosonic part of the symmetry algebra on $S^3\times \bR_t$ consists of $J^i_\ell$, $J^i_r$ for the $\SO(4)\simeq \SU(2)_\ell\times \SU(2)_r$ acting on $S^3$, together with a time translation, which we temporarily denote by $P$.
Let us further assume that we have a supercharge $Q_\alpha$ and its complex conjugate $\bar Q^{\beta}$.
We take the convention that they are doublets under $\SU(2)_\ell$ but are neutral under $\SU(2)_r$.
A consistent set of commutation relations is \begin{align}
\{Q_\alpha,\bar Q^{ \beta}\} &= 2\sigma^0_{\alpha}{}^{\beta} P+\frac 2\rho \sigma^i_{\alpha}{}^{\beta} J^i_\ell,&
\{Q_\alpha,Q_\beta\}&=0,&
\{\bar Q^{ \alpha},\bar Q^{ \beta}\}&=0,\label{QQ}\\
[P,Q_\alpha]&=\frac 1\rho Q_\alpha, & 
[P,\bar Q^{  \beta}]&=-\frac 1\rho \bar Q^{ \beta}.\label{PQ}
\end{align}
Here we took $J$ to be dimensionless, $P$ to have dimension 1, $Q$ to have dimension $1/2$,
and introduced a  parameter $\rho$ with dimension $-1$ proportional to the radius of $S^3$.
The symmetry algebra is now $\SU(2|1)_\ell\times \SU(2)_r$.

The relations \eqref{PQ} mean that the supercharge rotates under the time translation.
Therefore, if we compactify the time direction as $t\simeq t+\beta$ with generic $\beta$ to consider the partition function,
no supersymmetry will be preserved. 
To have a supersymmetric partition function, we assume that we have an additional $\U(1)_R$-symmetry generator $R$ such that \begin{equation}
[R,Q_\alpha]=-Q_\alpha,\qquad
[R,\bar Q^{ \beta}]= +\bar Q^{ \beta}
\end{equation} and use \begin{equation}
H=P+\frac1\rho R
\end{equation} as the Hamiltonian of the system,
so that the supercharges are preserved by $H$.

We can now consider the supersymmetric partition function \begin{equation}
Z_{S^3\times S^1}(\beta,\mu_r,\mu_i)=\tr (-1)^F e^{-\beta H + \mu_r J_r^3 + \sum \mu_i J_i}
\end{equation}
where $\mu_r$ is the chemical potential for the $\SU(2)_r$ spatial rotation
and $\mu_i$ are the chemical potentials for the other global symmetry generators $J_i$.
These chemical potentials are often called the fugacities in the literature.
By the standard argument using the commutation relation \eqref{QQ},
the only states contributing to this partition function are those with $P=(2/\rho) J^3_\ell$.
In the following we set $\rho=1$ for brevity.

We can write Lagrangians with the symmetries discussed above, and use them to perform supersymmetric localization. 
The \Nequals4 supersymmetric case was carried out in detail in \cite{Nawata:2011un}.
In this lecture note we just have to assume that there are such Lagrangians.
The computation itself can be done without actually using the Lagrangian,
since the partition function is a generalized version of the Witten index and
does not depend on the details of the theory.

\subsection{Indices of chiral multiplets}
\label{sec:SCIchiral}
Let us compute the $S^3$ index of a chiral multiplet $\Phi$ of R-charge $R$.
This multiplet contains a complex scalar $\phi$ and a Weyl fermion $\psi$ of R-charge $R$ and $R-1$, respectively.
We need to find their Hilbert space on $S^3$.
This can be done entirely analogously to the analysis of free fields on $\bR^3$,
which everybody should have learned in their first textbook on quantum field theory.
There, one expands the fields by the Fourier modes, convert them to quantum operators, and then build the Fock space.

On $S^3$, the starting point is to find the eigenmodes of $\phi$ and $\psi$ on $S^3$.
This can be done by studying the theory of scalar and spinor spherical harmonics on general-dimensional spheres.
Essentially equivalently, the result can be found by using the theory of induced representations.\footnote{%
Briefly, it goes as follows. Given a group $G$, its subgroup $H$, and a representation $\rho$ of $H$, we can consider the space $G/H$ and a vector bundle on $G/H$ such that each fiber at a point on $p\in G/H$ transforms as $\rho$ under the symmetry $H$ preserving $p$.
The space of sections of this vector bundle forms a representation of $G$ denoted by $\Ind^G_H\rho$,  called the representation of $G$ induced from $\rho$ of $H$.
The reverse procedure would be more familiar: take a representation $\rho'$ of $G$, and regard it as a representation of $H$. This is known as the restriction of a representation and is denoted by $\Res^G_H \rho'$.

These two constructions are adjoint of each other, in the sense that $(\rho',\Ind^G_H \rho)=(\Res^G_H\rho',\rho)$ where the parenthesis denotes the standard inner product of the characters of two representations.
This in particular implies that the number of times that an irreducible representation $\rho'$ of $G$ appears in $\Ind^G_H \rho$ for an irreducible representation $\rho$ of $H$ 
is equal to the number of times the irreducible representation $\rho$ appears in the decomposition of the irreducible representation $\rho'$ of $G$ under $H$.

The case we need is $G=\SU(2)\times \SU(2)$, $H=\SU(2)_\text{diagonal}$ so that $G/H=S^3$.
The scalar wavefunction is when $\rho$ is trivial and the spinor wavefunction is when $\rho$ is the doublet of $\SU(2)$.
}
In the end, we find that, under $\SU(2)_\ell \times \SU(2)_r$, the one-particle Hilbert spaces of $\phi$ and $\psi$ and their complex conjugates have the following decompositions under $\SU(2)_\ell\times \SU(2)_r$:
\begin{equation}
\phi,\phi^\dagger: V_j\otimes V_j,\qquad
\psi: V_j\otimes V_{j+1/2},\qquad
\psi^\dagger: V_{j+1/2}\otimes V_j,
\end{equation}
where $V_j$ is the spin-$j$ representation of $\SU(2)$ and $j$ runs over $0,1/2,1,\ldots$. 

We now want  to organize them into representations of the supersymmetry $\SU(2|1)_\ell\times \SU(2)_r$.
This can be easily done by noticing that two $\SU(2)_\ell$ representations $V_j$ and $V_{j-1/2}$ form an irreducible representation of $\SU(2|1)_\ell$.
Here  the $P$ eigenvalues on two subspaces $V_j$ and $V_{j-1/2}$ are $2j$ and $2j+1$, respectively. 
The quantum numbers are given as follows:
\def\gr{\cellcolor{black!10}}
\begin{equation}
\begin{array}{|c|cccc|c|cccc|}
\hline
& P & \SU(2)_\ell & \SU(2)_r &  \U(1)_R & & P&  \SU(2)_\ell & \SU(2)_r  &  \U(1)_R \\
\hline
\phi:&  2j &\gr V_j & V_j  & R & \phi^\dagger: &2j+2 & \gr V_j& V_j &  -R \\
\psi:& 2j+1 &  \gr V_{j-1/2} & V_j  & R-1 & \psi^\dagger: & 2j+1& \gr V_{j+1/2}& V_j  & -R +1\\
\hline
\end{array}
\end{equation}
where two representations of $\SU(2)_\ell$ forming a representation of $\SU(2|1)_\ell$ are shaded.
Before proceeding, we note that from the table above we see $P=D_\text{UV}-(3/2)R_\text{UV}$, where $D_\text{UV}$ and $R_\text{UV}$ are the scaling dimension and the R-charge in the ultraviolet.

Inside the supersymmetric index, within a supermultiplet $V_j \oplus V_{j-1/2}$, only the top component of $V_j$ survives. 
This means that the only one-particle states contributing to the index are those depicted in the following schematic diagram: \begin{align}
\phi: & 
\begin{tikzpicture}[baseline=(A.west),yscale=.5,xscale=2]
\node(A) at (0,0) {$\circ$};
\node at (1,1) {$\circ$};
\node at (1,-1) {$\circ$};
\node at (2,2) {$\circ$};
\node at (2,0) {$\circ$};
\node at (2,-2) {$\circ$};
\node at (2.5,0) {$\cdots$};
\node at (0,-4) {$V_{j_r{=}0}$};
\node at (1,-4) {$V_{j_r{=}1/2}$};
\node at (2,-4) {$V_{j_r{=}1}$};
\draw[->] (0,.5) to node[near end,above] {$m$} (2,2.5);
\draw[->] (0,-.5) to node[near end,below] {$n$} (2,-2.5);
\end{tikzpicture}
&\psi^\dagger: & 
\begin{tikzpicture}[baseline=(A.west),yscale=.5,xscale=2]
\node(A) at (0,0) {$\circ$};
\node at (1,1) {$\circ$};
\node at (1,-1) {$\circ$};
\node at (2,2) {$\circ$};
\node at (2,0) {$\circ$};
\node at (2,-2) {$\circ$};
\node at (2.5,0) {$\cdots$};
\node at (0,-4) {$V_{j_r{=}0}$};
\node at (1,-4) {$V_{j_r{=}1/2}$};
\node at (2,-4) {$V_{j_r{=}1}$};
\draw[->] (0,.5) to node[near end,above] {$m$} (2,2.5);
\draw[->] (0,-.5) to node[near end,below] {$n$} (2,-2.5);
\end{tikzpicture}
\end{align}
tensored with the top component of the supermultiplet for $\SU(2|1)_\ell$.
Here,  the horizontal axis labels the subscript of $V_j$ for the $\SU(2)_r$ representation
and the vertical axis is for the weights under $\SU(2)_r$;
It is customary to label the states contained by two integers $m,n=0,1,\ldots$.

We then need to build up the Fock space from these one-particle states. 
Assigning the $\U(1)$ charge $+1$ to the entire multiplet, the supersymmetric partition function is then\footnote{%
Here and in the following we neglect a possibly-nontrivial vacuum energy, called the supersymmetric Casimir energy.
It needs to be taken in account in a more proper analysis.
} \begin{equation}
Z_\text{chiral}=\prod_{m,n\ge 0} \frac{1-t^{m+n-R+2}y^{m-n}/z} {1-t^{m+n+R} y^{m-n}z},
\end{equation}
where we introduced $t:=e^{-\beta/\rho}$, $y:=e^{\mu_r}$, $z:=e^{\mu_{\U(1)}}$.

Somewhat surprisingly, this is  exactly what the mathematicians call the elliptic gamma function, defined as an infinite product of the form \begin{equation}
\Gamma_{p,q}(z):= \prod_{m,n} \frac{1-z^{-1} p^{m+1}q^{n+1}}{1-z p^m q^n}.
\end{equation}
Using this, we simply have \begin{equation}
Z_\text{chiral}=\Gamma_{ty,t/y}(t^R z).
\end{equation}
It is instructive to expand this to the first few orders, by taking the standard value $R=2/3$: \begin{equation}
Z_\text{chiral}=1+tz^{2/3} + (z^2-z^{-1})t^{4/3}+(y+y^{-1})z t^{5/3}+\cdots.
\end{equation}
The first few terms come from the vacuum, the constant mode of $\phi$, the square of the constant mode of $\phi$,
the lowest mode of $\psi^\dagger$, and then the second mode of $\phi$, and so on.

\subsection{Indices of gauge theories}

Let us now move on to the index of gauge theories.
For brevity we only consider the gauge group $\U(N)$ and $\SU(N)$.
The index contribution from the vector multiplets can be worked out as in the case of the chiral multiplets.
A vector multiplet contains a vector field $A$ and the gaugino $\lambda$.
We then also need to include the contributions from the ghosts used in gauge fixing. 
Now, the constant mode of the gauge field $A$ needs to be treated with care.
This mode corresponds to the holonomy of $A$ around $S^1$ of $S^3\times S^1$.
We call the coordinate around $S^1$ as the Euclidean time and denote by $t$.
We can take a gauge in which $A^t=\diag(A^t_1,A^t_2,\ldots, A^t_N)$ and keep it time-independent. 
In the end we need to integrate over $A^t$, projecting the spectrum to the gauge-invariant states,  but for the moment we keep them to a fixed value.

Then the  contribution from the $\U(N)$ vector multiplet to the index can be computed to be \begin{align}
Z_\text{vector}&:=\tr (-1)^F e^{-\beta H - \beta \sum_a  A^t_a  J_a  + \mu_r J_r^3  } \\
&=\left[\prod_{a,b}^N \prod_{m,n=0}^{\infty}\right]' \frac{1-t^{m+n} y^{m-n} z_a/z_b}{ 1-t^{m+n+2} y^{m-n} z_a/z_b}
\label{vander}
\end{align}
where $J_a$ is the gauge charge at the $a$-th diagonal entry, and we  defined $z_a:=e^{-\beta A^t_a}$.
The prime $'$ on the product symbol $\prod$ is to remind ourselves that 
the factors in the numerator of the $a=b$, $m=n=0$ terms need to be omitted 
in the last expression.
The omitted modes come from the constant modes and need to be integrated later.

Using the elliptic gamma function,  we write the expression above  as \begin{equation}
Z^{\U(N)}_\text{vector}= \frac{1}{\Gamma'_{ty,t/y}(1)^{N}} \prod_{a\neq b}\frac{1}{\Gamma_{ty,t/y}(z_a/z_b)}.
\end{equation}
We clearly see the contributions from the diagonal components and the off-diagonal components of gauge fields.
For $\SU(N)$, we have \begin{equation}
Z^{\SU(N)}_\text{vector}= \frac{1}{\Gamma'_{ty,t/y}(1)^{N-1}} \prod_{a\neq b}\frac{1}{\Gamma_{ty,t/y}(z_a/z_b)}.
\end{equation}
where we removed one contribution from the diagonal component 
and $z_N$ needs to be eliminated via the relation $z_1z_2\cdots z_N=1$.

To compute the index of a gauge theory, we need to extract the gauge-invariant part of the combined contributions from the vector multiplet and the chiral multiplets. 
For this purpose, we use the relation \begin{equation}
\oint \frac{dz}{2\pi iz} z^n= \delta_{n0}.
\end{equation}
The final result is that \begin{equation}
Z_\text{$\SU(N)$ gauge theory} = \frac{1}{N!}( \prod_{a}\oint \frac{dz_a}{2\pi i z_a} ) Z^{\SU(N)}_\text{vector} \prod_{\Phi} Z_\text{chiral}^\Phi.
\end{equation}
Note that the Vandermonde factor is already included in the $m=n=0$ term of the numerator of \eqref{vander},
and that the factor $1/N!$ in front takes into account the remaining Weyl symmetry.

\subsection{Case study: $\SU(N)$ with $N_f$ flavors}
Let us now write down the index of $\SU(N)$ theory with $N_f$ flavors $Q$, $\tilde Q$. 
Recall that the R-charge of $Q$ and $\tilde Q$ is $1-N/N_f$.
There is an $\SU(N_f)\times \SU(N_f)$ symmetry acting on $Q$ and $\tilde Q$.
We introduce the corresponding fugacities \begin{equation}
(\mu_1,\ldots,\mu_{N_f}),\qquad
(\tilde\mu_1,\ldots,\tilde\mu_{N_f}),\qquad
\end{equation} with $\prod \mu_i=\prod\tilde \mu_i=1$.
We also have $\U(1)$ baryonic symmetry under which $Q$ has charge $1$ and $\tilde Q$ has charge $-1$.
We introduce the fugacity $\nu$ for it.
Finally, we use $(z_1,\ldots, z_N)$ with $z_1\cdots z_N=1$ for the fugacity of $\SU(N)$ gauge group.
Then the index is \begin{multline}
Z_{\SU(N),N_f} =
 \frac1{N!}\prod(\oint \frac{dz_a}{2\pi iz_a}) \frac{1}{\Gamma'(1)^{N-1} \prod_{\pm,a\neq b} \Gamma(z_a/z_b)} \\
\times\prod_{\pm,i,a} \Gamma(t^{1-2/N_f} z_a \mu_i\nu)
\prod_{\pm,i,a} \Gamma(t^{1-2/N_f} z_a^{-1} \tilde\mu_i\nu^{-1})
\end{multline}
where we abbreviated the elliptic gamma function $\Gamma_{ty,t/y}(x)$ simply as $\Gamma(x)$.
This is a horrific expression, but with a help of symbolic computer algebra system
it is easy to expand it in series of $t$, to any given order.

Take in particular $\SU(2)$ with $N_f=3$. 
We discussed before that the low-energy limit is a theory of $M_{ij}=Q_i \tilde Q_j$, $B^k=Q_i Q_j \epsilon^{ijk}$, $\tilde B^k=\tilde Q_i\tilde Q_j \epsilon^{ijk}$:
\begin{equation}
\begin{tikzpicture}[baseline=(X.east)]
\node(A) at (0,1) {UV: $\SU(2)$ with 3 flavors};
\node(X) at (0,0) {};
\node(B) at (0,-1) {IR: Theory of $M$, $B$ and $\tilde B$};
\draw[->] (A) to (B);
\end{tikzpicture}
\end{equation}
Then we should have the equality
\begin{multline}
\frac12\oint \frac{dz}{2\pi iz} 
\underbrace{\frac{1}{\Gamma'(1) \prod_{\pm} \Gamma(z^{\pm2})}}_\text{from $\SU(2)$}
\underbrace{\prod_{\pm,i} \Gamma(t^{1/3} z^{\pm1} \mu_i\nu)}_\text{from $Q$}
\underbrace{\prod_{\pm,i} \Gamma(t^{1/3} z^{\pm1} \tilde\mu_i/\nu)}_\text{from $\tilde Q$}\\
= \underbrace{\prod_i \Gamma(t^{2/3}\mu_i^{-1}\nu^{2} )}_\text{from $B$}
\underbrace{\prod_i \Gamma(t^{2/3}\tilde\mu_i^{-1}\nu^{-2} )}_\text{from $\tilde B$}
\underbrace{\prod_{i,j} \Gamma(t^{2/3}\mu_i\tilde\mu_j )}_\text{from $M$}.
\end{multline}
It is a fun exercise to confirm the equality in an explicit expansion in $t$.

\exercise{Carry out this computation.}

Next, consider the $\SU(2)$ theory with $N_f=4$ flavors $Q$ and $\tilde Q$.
We saw above that the infrared limit is a nontrivial superconformal theory,
and there is a Seiberg dual description, which is again an $\SU(2)$ theory with $N_f=4$ flavors $q$, $\tilde q$, together with meson fields $M$, with a superpotential $W=Mq\tilde q$:
\begin{equation}
\begin{tikzpicture}[baseline=(X.east),every text node part/.style={align=center}]
\node(A) at (-4,1) {UV$_1$: $\SU(2)$ with \\ 4 flavors $Q$, $\tilde Q$};
\node(P) at (4,1) {UV$_2$: $\SU(2)$ with \\ 4 flavors  $q$, $\tilde q$, mesons $M$ and $W=Mq\tilde q$};
\node(X) at (0,0) {};
\node(B) at (0,-1) {IR: a superconformal theory};
\draw[->] (A) to (B);
\draw[->] (P) to (B);
\end{tikzpicture}
\end{equation}

We can write down the indices of two $\SU(2)$ gauge theories; they should be equal. 
Explicitly, the $\SU(2)$ theory with four flavors has the index \begin{align}
Z_{\SU(2),N_f=4}&=\frac12 \oint \frac{dz}{2\pi iz} \frac{1}{\Gamma'(1)\prod_{\pm}\Gamma(z^{\pm2})}
\prod_{\pm,i} \Gamma(t^{1/2} z^{\pm1} \mu_i \nu )
\prod_{\pm,i} \Gamma(t^{1/2} z^{\pm1} \tilde \mu_i /\nu )\\
&=:
I(\mu_i\nu; \tilde \mu_i /\nu).
\end{align}
Then the dual theory has the index \begin{equation}
Z_{\SU(2),N_f=4,\text{dual}} = 
I(\mu_i{}^{-1}\nu;\tilde\mu_i{}^{-1} /\nu) \prod_{i,j} \Gamma(t \mu_i \tilde \mu_j).
\end{equation}
Therefore, we should have the equality \begin{equation}
I(\mu_i\nu; \tilde \mu_i /\nu)
=I(\mu_i{}^{-1}\nu;\tilde\mu_i{}^{-1} /\nu) \prod_{i,j} \Gamma(t \mu_i \tilde \mu_j).\label{24}
\end{equation}

\exercise{Check this to your satisfaction.}

Exactly in the same way, we can write down the $S^3$ indices of a Seiberg-dual pair of gauge theories.
They should be equal, if the Seiberg duality is true.
We will obtain in this way a complicated equality of multiple integrals of products of elliptic gamma functions.
Surprisingly, the equality expressing the Seiberg duality of $\SU(N)$ with $N_f$ flavors
was proved by mathematicians 
completely independently of our physics context.
On the mathematics side, 
this type of integral equalities was first introduced by Spiridonov in \cite{Spiridonov1}, who also proved the identity for  $N=2$, $N_f=3$ in the same paper. 
Various generalizations were discussed in 2003 in  \cite{Spiridonov2}, and then the cases with general $N$, $N_f$ were settled by Rains in 2003 \cite{Rains}.
On the physics side, as already mentioned, the first study of the supersymmetric index on $S^3\times S^1$ was done in 2005 by R\"omelsberger \cite{Romelsberger:2005eg}.\footnote{%
It is very mysterious that the closely related facts were found independently in mathematics and in physics around the same time.\label{foot:mystery}}
The relation of these two works were first noticed in \cite{Dolan:2008qi} in 2008.

We note that other \Nequals1 dualities, some of which we mentioned in Sec.~\ref{sec:other},
lead to many other identities among the integrals of products of elliptic gamma functions.
Some of them were proved, but many of them were still not proved. 
For an extensive discussion on this point, see \cite{Spiridonov:2009za}.
It is also conceivable that some of the claimed duality would be wrong, which can be shown by the mismatch of the supersymmetric indices.
We will see one example soon.

\subsection{Case study: the fate of $\SU(2)$ with matter in the spin 3/2 representation}
The $S^3$ index can be used to decide the behavior of a confusing supersymmetric gauge theory.
Consider the $\SU(2)$ theory with one chiral superfield $Q$ in $\mathrm{4}$, i.e.~the 3-index symmetric traceless tensor, or equivalently the spin $3/2$ representation.
The global anomaly is absent, so it is OK to consider this theory.
The one-loop beta function is the same as $N_f=5$ flavors of doublets. So this is asymptotically free.
Under the anomaly-free R-symmetry, $R(Q)=3/5$.

The basic gauge-invariant chiral superfield is $U:=Q^4$ with all indices contracted; there is essentially one way to do so.
The R-charge is $R(U)=12/5$.
One can entertain two possibilities for the IR behavior of the theory at this point:
\begin{enumerate}
\item It is given just by $U$ as a chiral scalar.
\item It is a nontrivial superconformal theory.
\end{enumerate}
We now know the answer is the second. 
But historically,  this conclusion was reached in a complicated process \cite{Intriligator:1994rx,Brodie:1998vv,Intriligator:2005if,Vartanov:2010xj}.

As a support for the first possibility, the authors of \cite{Intriligator:1994rx} computed the 't Hooft anomaly for the anomaly-free $\U(1)_R$ symmetry, and compared the values in the IR and in the UV. 
They magically agreed. 
That did not prove the description 1, but at least was a piece of support.

Gradually, it was noticed that the description 2 is more plausible \cite{Brodie:1998vv,Intriligator:2005if}. 
The definitive argument in favor of the latter came after the supersymmetric index was introduced. 
In \cite{Vartanov:2010xj} the supersymmetric index in the UV gauge description and that in the proposed IR free description were computed.
They were clearly different, therefore the choice 1 was ruled out.

\subsection{Case study: the $E_7$ surprise}
\label{sec:E7surprise}
Let us describe another physics we can glean from the $S^3$ index, following Dimofte and Gaiotto \cite{Dimofte:2012pd}.
We start from the equality \eqref{24} expressing the Seiberg duality of $\SU(2)$ with $N_f=4$.
The contribution of the mesons can be rewritten as follows \begin{equation}
\prod_{i,j} \Gamma_{ty,t/y}(t \mu_i \tilde \mu_j)
= \prod_{i,j} \frac{(t \mu_i{}^{-1} \tilde\mu_i{}^{-1})_{ty,t/y}}{  (t \mu_i \tilde \mu_j)_{ty,t/y}}
\end{equation}
where \begin{equation}
(a)_{p,q}=\prod_{m,n}(1-ap^m q^n)
\end{equation} is the $(p,q)$-Pochhammer symbol.
This allows us to express \eqref{24} in a slightly more symmetric manner: \begin{equation}
J(\mu_i\nu; \tilde \mu_i /\nu)=J(\mu_i{}^{-1}\nu;\tilde\mu_i{}^{-1} /\nu)
\end{equation} where we defined\begin{equation}
J(m_1,\cdots,m_8) :=   I(m_1,\ldots,m_4; m_5,\cdots, m_8) \prod_{i<j} \frac{1}{(t m_i{}^{-1} m_j{}^{-1})_{ty,t/y}}. \label{scJ}
\end{equation}
Here and in the following we assume $\prod m_i=1$.

We now recall the discussion of Sec.~\ref{SU2Sp1Nf4}: the Seiberg duality gives a different dual when we regard $\SU(2)$ as $\Sp(1)$. The corresponding equality of the $S^3$ index is \begin{equation}
I(m_1,\ldots, m_8) = I(m_1{}^{-1},\ldots, m_8{}^{-1}) \prod_{i<j} \Gamma(t m_i m_j).
\end{equation}
This can again be written in terms of $J$: \begin{equation}
J(m_1,\cdots,m_8) = J(m_1{}^{-1},\ldots, m_8{}^{-1}).
\end{equation}

Thus we found that the function $J$ has the following properties: 
\begin{enumerate}
\item $J(m_1,\ldots,m_8)$ is invariant under the permutation of $m_i$. 
\item $J(\mu_i\nu; \tilde \mu_i /\nu)=J(\mu_i{}^{-1}\nu;\tilde\mu_i{}^{-1} /\nu)$,  from the duality as an $\SU(2)$ gauge theory.
\item $J(m_1,\cdots,m_8) = J(m_1{}^{-1},\ldots, m_8{}^{-1})$, from the duality as an $\Sp(1)$ gauge theory.
\end{enumerate}
The first transformation is simply the Weyl group of the $\SU(8)$ flavor symmetry of the chiral multiplets $\sQ^I_a$ for $I=1,\ldots,8$, coming from the fact that the doublet and the anti-doublet of $\SU(2)$ gauge group are the same.
The second and the third transformations extend this to the Weyl group of the $E_7$ symmetry,
as already noticed in \cite{Spiridonov2,Rains} and mentioned in \cite{Spiridonov:2008zr}.

What is the physical significance of this $E_7$ enhancement? 
The function $J$ is not quite the gauge theory index, since we introduced an additional factor 
\begin{equation}
\prod_{i<j} \frac1{(t m_i{}^{-1} m_j{}^{-1})_{ty,t/y}}\label{add}
\end{equation}
 in its definition \eqref{scJ}.
Dimofte and Gaiotto noticed the following \cite{Dimofte:2012pd}.

We have been considering a four-dimensional gauge theory on $S^3\times S^1$.
We can add to it a five-dimensional bulk of the form $D^4\times S^1$, where $D^4$ is a four-dimensional hemisphere, whose boundary is $S^3$.
As the boundary has 4d \Nequals1 supersymmetry,
it is natural to consider 5d \Nequals1 supersymmetry in the bulk.\footnote{%
We refer the reader to Sec.~\ref{sec:various} for an exposition of supersymmetry in various dimensions.
}
In the five-dimensional bulk we consider a theory of free hypermultiplets.
Consider then a five-dimensional hypermultiplet in a representation $R$ of a symmetry $G$.
If we restrict it to the boundary, it consists of a 4d chiral multiplet $X$ in the representation $R$
and another 4d chiral multiplet $Y$ in the conjugate representation $\bar R$.

There is a half-supersymmetric boundary condition of the form \begin{equation}
W_\text{boundary}= X \cO, \qquad Y|_\text{boundary} = \cO
\end{equation} where $\cO$ is an operator on the boundary in the representation $\bar R$
and $W$ is the boundary superpotential coupling between the bulk field $X$ and the boundary operator $\cO$.
When $\cO=0$, this puts the Neumann condition on $X$ and the Dirichlet condition on $Y$.
A nontrivial $\cO$ deforms this boundary condition.

The contribution from the bulk hypermultiplet to the index can be computed. When $G=\U(1)$ and there is a single hypermultiplet of charge $1$, it turns out to be given by \begin{equation}
\frac{1}{(tz)_{ty,t/y}}
\end{equation} where $z$ is the fugacity of the $\U(1)$ flavor symmetry.

We can then identify the factor \eqref{add} to be the contribution to the index of a hypermultiplet $(X^{IJ},Y_{IJ})$ in the bulk five-dimensional space, in the two-index antisymmetric tensor of $\SU(8)$ flavor symmetry, with the boundary condition \begin{equation}
W_\text{boundary}= X^{IJ} \sQ_I \sQ_J ,\qquad Y_{IJ}|_\text{boundary} = \sQ_I \sQ_J.
\end{equation}
Then the function $J$ can be identified with the index of the total system of this bulk hypermultiplet, coupled to the boundary $\SU(2)$ theory with four flavors,
and the invariance of $J$ under the $E_7$ Weyl group strongly suggests that the $\SU(8)$ symmetry of this coupled system enhances to $E_7$, at least at some special value of the coefficient of the boundary superpotential.

There are various other pieces of evidence and consistency checks of this enhancement to $E_7$ given in the original paper \cite{Dimofte:2012pd}. 
A simple check is that the bulk hypermultiplet transforms nicely under $E_7$ as a half-hypermultiplet of $\mathbf{56}$, which is possible since $\mathbf{56}$ is a pseudoreal representation.

Another check\footnote{%
This is based on an unpublished work with Ken Kikuchi.
} is about the anomaly polynomial.
A four-dimensional theory cannot have an $(E_7)^3$ term in the anomaly polynomial,
because the gauge group $E_7$ simply does not have such a degree-3 invariant.
Therefore, by restricting to the $\SU(8)\subset E_7$ symmetry,
we see that an $\SU(8)$ symmetry which can enhance to $E_7$ should not have the $\SU(8)^3$ term in the anomaly polynomial.

Let us confirm this in our setup.
The fields $\sQ_I^a$ contains two fundamentals in $\SU(8)$, and therefore contributes to the anomaly polynomial by \begin{equation}
2\cdot \frac16\tr_{\mathbf{8}} (\frac{F}{2\pi})^3.
\end{equation}
This is clearly nonzero, and we need something which cancel this.

The cancelling contribution in fact comes from the boundary condition of the five-dimensional bulk theory.
Consider a 5d massless fermion in the representation $R$ of $G$, put on $\bR^4\times [0,L]$.
We place chiral boundary conditions on two ends of $[0,L]$,
so that there is a single zero mode of 4d Weyl fermion in the representation $R$ of $G$.
By taking $L\to 0$, we simply isolate this 4d massless fermion, decoupling all the Kalzua-Klein modes.
This system has the anomaly polynomial $(1/6)\tr_R (F/2\pi)^3$.
Interpreted from the 5d point of view, they should come from two equal contributions from the two boundaries.
We conclude that a single boundary has the anomaly\footnote{%
This fact that the chiral boundary condition of a massless fermion in an odd-dimensional spacetime has half the anomaly of a Weyl fermion in the one-dimension-lower even-dimensional spacetime has been known for some time.
For example, the fact that the half-supersymmetric boundary of the 11d supergravity has half the anomaly of a 10d  \Nequals1 supergravity multiplet was an essential part of the deduction of the $E_8$ gauge symmetry on this boundary in the seminal paper by Ho\v rava and Witten \cite{Horava:1995qa,Horava:1996ma}.
} \begin{equation}
\frac12\cdot \frac16\tr_R (\frac{F}{2\pi})^3.
\end{equation}

In our case, we have $R=\overline{\mathbf{28}}$ of $\SU(8)$, which is the antisymmetric two-index tensor constructed from the anti-fundamental representation $\overline{\mathbf{8}}$.
For $\SU(N)$, such an antisymmetric two-index tensor has the contribution of $4-N$ times the anomaly of the fundamental.
We therefore have \begin{equation}
2\cdot \frac16\tr_{\mathbf{8}} (\frac{F}{2\pi})^3 +\frac12\cdot \frac16\tr_{\overline{\mathbf{28}}} (\frac{F}{2\pi})^3
= \left[2 + \frac12(4-8)\right]\tr_{\mathbf{8}} (\frac{F}{2\pi})^3 = 0.
\end{equation}

Before proceeding, we note that the $E_7$ surprise can be formulated purely in 4d without introducing the 5d bulk, at the price of doubling the number of fields.
Namely, in 4d, we consider one $\SU(2)$ theory with eight doublets $\sQ_I$, $I=1,\ldots, 8$
transforming in the fundamental of $\SU(8)$,
and another $\SU(2)$ theory with eight doublets $\sq^I$ in the anti-fundamental of $\SU(8)$.
We then introduce the superpotential  \begin{equation}
W=c (\sq^I \sq^J)(\sQ_I \sQ_J)
\end{equation} coupling two sectors.
The $S^3$ index shows the enhancement from $\SU(8)$ to $E_7$ in 4d,
and in \cite{Dimofte:2012pd} it was argued rather convincingly that the symmetry does enhance to $E_7$ at a special value of the coefficient $c$ of the superpotential.
This 4d version of the $E_7$ surprise has been generalized, see e.g.~\cite{Razamat:2017hda,Razamat:2018gbu}.

\section{The $a$-maximization}
\label{sec:amax}
In our analysis of the infrared dynamics of supersymmetric gauge theories so far,
conserved $\U(1)$ R-symmetries played many important roles.
In particular, when the low-energy limit is superconformal,
the R-symmetry in the superconformal algebra is tautologically a conserved $\U(1)$ R-symmetry.

In the gauge theories  considered so far, there is always a unique such conserved $\U(1)$ R-symmetry.
However, more complicated gauge theories often have a continuous family of such conserved $\U(1)$ R-symmetries. 
Then, it is of fundamental importance to decide exactly which $\U(1)$ R-symmetry is \emph{the} superconformal R symmetry in the superconformal algebra.
The tool to determine it is the $a$-maximization introduced in \cite{Intriligator:2003jj}.\footnote{%
One of the authors of \cite{Intriligator:2003jj} is now better known as Ninja Brian of the comedy songwriting pair, Ninja Sex Party: \url{https://www.youtube.com/user/NinjaSexParty}.
The author is proud that he wrote many papers with Ninja Brian.
}
The aim of this section is to explain this technique.
But we first need to introduce the quantity $a$ itself.

\subsection{The central charges $a$ and $c$}
\label{sec:ca}
For two-dimensional conformal field theories, the central charge $c$
is defined to be the leading coefficient for the operator product expansion (OPE)
of the energy momentum tensor with itself \begin{equation}
T(z)T(0)\sim \frac{c}{2z^4}+\frac2{z^2}T(0)+\frac{1}{z} \partial T(0)+\cdots.
\end{equation} $c$ is positive for unitary theories, adds up if we combine
two decoupled CFTs, and is $1$ for the free  CFT with one bosonic scalar field.
Thus $c$ can be  said to `count' the number of degrees of freedom in CFT.
It can also be measured by the trace anomaly caused by the coupling to
the external gravitational field, i.e.  \begin{equation}
\vev{T^\mu_\mu}=-\frac{c}{12}R
\end{equation}where $R$ is the scalar curvature of the metric. 

The central charge $c$ is known to decrease under the renormalization group flow.
This statement is known as Zamolodchikov's $c$ theorem.\cite{Zamolodchikov:1986gt}.
This is in accord with the intuition that the number of degrees of freedom should decrease by coarse-graining.

In four-dimensional conformal field theory,
there are two central charges $c$ and $a$ defined as the coefficients
appearing in the equation \begin{equation}
\vev{T^\mu_\mu}=\frac{c}{16\pi^2} (W_{\mu\nu\rho\sigma})^2
-\frac{a}{16\pi^2}  E_{(4)}
\end{equation}where $W_{\mu\nu\rho\sigma}$ is the Weyl tensor and \begin{equation}
E_{(4)}=\frac{1}{4}\epsilon^{\mu_1\nu_1\nu_2\nu_2}
\epsilon^{\rho_1\rho_1\rho_2\sigma_2} 
R_{\mu_1\nu_1\rho_1\sigma_1}
R_{\mu_2\nu_2\rho_2\sigma_2} 
\end{equation} is the Euler density in four dimensions.
For details, see e.g.~\cite{Deser:1993yx,Deser:1996na}.
Cardy conjectured that the central charge $a$  decreases along the renormalization group flow in 1988 \cite{Cardy:1988cwa}, which was convincingly demonstrated only in 2011 by Komargodski and Schwimmer \cite{Komargodski:2011vj}.

Here we are only concerned with \Nequals1 supersymmetric cases.
Now the energy-momentum tensor
is combined with the superconformal $R$-symmetry current and the supersymmetry current
to form the supercurrent $R_{\alpha\dot \alpha}$, whose lowest component
is the $R$-current itself.
The anomaly  for the R-currents by the external fields
can be summarized by the equation \cite{Anselmi:1997am,Anselmi:1997ys} \begin{equation}
\bar D^{\dot\alpha} R_{\alpha\dot \alpha} =\frac{1}{24\pi^2}
(c\cW ^2 - a \Xi)\label{Ranomaly}
\end{equation} where $\cW$, $\Xi$  are the superfields which contain
the Weyl tensor and the Euler density in the appropriate places.
The same problem was studied from the component formalism in
\cite{Osborn:1998qu} in which the three-point correlator of 
the supercurrent \begin{equation}
\vev{R_\mu (x_1,\theta_1,\bar \theta_1)
R_\nu (x_2,\theta_2,\bar \theta_2)
R_\rho (x_3,\theta_3,\bar \theta_3)
}
\end{equation} was found to be expressible in a linear combination
of two superconformal invariants.

The discussion above implies that the central charges $a$ and $c$
for a superconformal theory is a linear combination 
of $\U(1)_R^3$ and $\U(1)_R$-gravity-gravity anomalies.
The coefficients can be fixed by considering the CFT
consisting of free chiral and vector multiplets, with the result \begin{equation}
a=\frac 3{32}(3\tr R^3-\tr R),\qquad
c=\frac 1{32}(9\tr R^3-5\tr R).\label{fieldtheoryresults}
\end{equation}
Here $\tr R^3$ and $\tr R$ are certain coefficients in the anomaly polynomial of the theory so that \begin{equation}
\cA=\frac{\tr R^3}6 (\frac{F_{\U(1)R}}{2\pi})^2
+\frac{\tr R}{24} (\frac{F_{\U(1)R}}{2\pi}) p_1.
\end{equation}
When the theory under consideration has a Lagrangian description,
then $\tr$ is the summation over the labels of the left-handed Weyl fermions.

For example, a free chiral multiplet $\Phi$ has $\Delta(\Phi)=1$ and therefore $R(\Phi)=2/3$.
Then the fermion component has $R(\psi_\Phi)=-1/3$. 
Plugging in to the formula above, one finds \begin{equation}
(a,c)_\text{free chiral}=(\frac1{48},\frac1{24}).
\end{equation}
Similarly, for a free vector multiplet, $R(\lambda)=+1$, and we have \begin{equation}
(a,c)_\text{free vector}=(\frac3{16},\frac18).
\end{equation}

We can also compute the central charges of the $\SU(N)$ SQCD with $N_f$ flavors, assuming that the theory is in the superconformal window.
The superconformal R-symmetry is the unique anomaly-free $\U(1)$ R-symmetry,
and therefore we have $R(Q)=R(\tilde Q)=1-N/N_f$ and therefore $R(\psi_Q)=R(\psi_{\tilde Q})=-N/N_f$. 
Together with the gauginos with $R(\lambda)=1$, we see that \begin{equation}
(a,c)_\text{$\SU$ SQCD}=(
-\frac{9N^4}{16N_f^2}+\frac{3N^2}{8}-\frac3{16},
-\frac{9N^4}{16N_f^2}+\frac{7N^2}{16}-\frac1{8}
).
\end{equation}
We obtain the same central charges when we use the Seiberg dual description.
This is guaranteed from the matching of the anomaly polynomials across the duality,
since the central charges are simply determined in terms of the anomaly polynomial.

Before proceeding, we mention some known universal bounds on $a$ and $c$ from the conformal symmetry and unitarity.
Hofman and Maldacena showed the following upper and lower bounds for the ratio $a/c$:
\begin{equation}
\begin{array}{c|cccc}
&\cN{=}0 & \cN{=}1 & \cN{=}2 & \cN{=}4\\
\hline
\text{upper bound} & 31/18 & 3/2 & 5/4 & 1 \\
\text{lower bound} & 1/3 & 1/2 & 1/2 & 1 
\end{array}
\end{equation}
where the upper bound corresponds to free vector fields or vector multiplets,
and the lower bound corresponds to a free scalar; a free chiral multiplet; a free hypermultiplet, respectively.
The original derivation was innovative but indirect; a much more direct derivation was given later in \cite{Hofman:2016awc}.
With \Nequals2 supersymmetry, more bounds are known, such as 
\begin{equation}
c\ge \frac{11}{30}
\label{LiendoBound}
\end{equation}
derived by \cite{Liendo:2015ofa}.

We already mentioned above that $a/c=1$ for \Nequals4 theories.
This holds in particular in the large $N$ limit of \Nequals4 $\SU(N)$ super Yang-Mills,
which has the $AdS_5\times S^5$ dual.
In general, for a 4d conformal theory with a weakly-curved holographic dual on $AdS_5$,
we can show that $a/c\sim 1$.
In this regime, one can further show \cite{Kats:2007mq,Buchel:2008vz} that \begin{equation}
\frac{\eta}{s} = \frac{1}{4\pi} \frac{a}{c} + O(\frac1{N^2})
\end{equation} where $\eta/s$  is the ratio of the shear viscosity to the entropy density.
The value $1/(4\pi)$ was originally found in \cite{Kovtun:2003wp}, which was the lowest value of shear viscosity in any known medium at that time.
This led to a conjecture that this might actually be a universal lower bound,
which aroused the interest in the possible value of $a/c$ in 4d (super)conformal field theories.

It turns out that most of the known large $N$ \Nequals1 superconformal theories at that time had $a/c$ slightly \emph{below} 1, thus disproving the conjecture. 
This was the main point of \cite{Kats:2007mq,Buchel:2008vz}.
A few years later, a large class of large $N$ \Nequals2 superconformal theories with $a/c$ slightly \emph{above} 1 was constructed by Gaiotto and Maldacena \cite{Gaiotto:2009gz}.
For more details on the relation of $\eta/s$ and $a/c$, see \cite{Cremonini:2011iq}.

\subsection{The $a$-maximization}
In more general cases, it often happens that there is a whole family of anomaly-free $\U(1)$ R-symmetry.
As a main example, we take the $\SU(N)$  adjoint SQCD, i.e.~the $\SU(N)$ gauge theory
with a chiral multiplet $\Phi$ in the adjoint representation
together with $N_f$ pairs of fundamental chiral multiplets $Q^i$, $\tilde Q_i$, $(i=1,\ldots,N_f)$.
The anomaly-free condition says that \begin{equation}
N_f R(\psi_Q) + N R(\psi_\Phi) + N R(\lambda) =0
\end{equation} where $R(\lambda)=1$.
This allows us to express $R(Q)=1-s N/N_f$ where $s=R(\Phi)$,
but we cannot eliminate $s$.
Exactly which $s$ gives the superconformal R-symmetry in the infrared?

The important insight of Intriligator and Wecht \cite{Intriligator:2003jj} is that,
for any flavor symmetry $G$,
the $G\cdot\U(1)_\text{SC}^2$ anomaly
is proportional, by a universal constant,
to the $G\cdot \text{gravity}^2$ anomaly,
where $\U(1)_\text{SC}$ is the superconformal R symmetry.
This is because that the background gauge field for the superconformal R symmetry
and the background metric are in a single  supermultiplet $\cW$ we already saw in \eqref{Ranomaly}, 
and both anomalies are encoded in a single supersymmetric equation
\begin{equation}
\bar D^2 J= \frac {k}{384\pi^2}\cW^2
\end{equation}
where  $J$ is the Konishi current for the flavor symmetry $G$
and $k$ is a numerical coefficient.

Assuming these general properties,
the proportionality coefficient between the $G\cdot\U(1)_\text{SC}^2$ anomaly
and the $G\cdot \text{gravity}^2$ anomaly can be fixed using a free SCFT.
The result is 
\begin{equation}
	9\tr Q R_\text{SC} R_\text{SC} = \tr Q.\label{amax1}
\end{equation} 
where $Q$ is the charge  of the flavor symmetry
and $R_\text{SC}$ is the superconformal R-charge,
and the trace is as always over the labels of the left-handed Weyl fermions.
Another requirement is the negative definiteness
\begin{equation}
	\tr Q Q R_\text{SC}<0\label{amax2}
\end{equation}which comes from the positivity of the two point
function of currents $\vev{J(x_1)J(x_2)}$.

Now suppose we have a continuous family $R(s)$ of $\U(1)_R$ symmetries,
and introduce the trial $a$-function by the formula \begin{equation}
a(s)=\frac3{32}(3\tr R(s)^3 -\tr R(s) ),
\end{equation} generalizing the case of the superconformal R-symmetry \eqref{fieldtheoryresults}.
The two conditions \eqref{amax1}, \eqref{amax2} mean that the superconformal R-symmetry $R(s_0)$ corresponds to the value of the parameter $s_0$ where $a(s)$ has a local maximum.
This is why the method is called the $a$-maximization.\footnote{%
The $a$-maximization  in \Nequals1 superconformal theory involves a solution of a coupled set of quadratic equations whose coefficients are determined in terms of the anomaly polynomial,
whose coefficients are in turn quantized from topological reasons.
Therefore, the resulting superconformal R-charge is always an algebraic number, i.e.~a solution to a polynomial equation with rational coefficients.
Then, the scaling dimensions of chiral operators are also algebraic numbers.

We note that one can derive, under more or less plausible assumptions, that the scaling dimensions of chiral operators of  \Nequals2 theories are rational \cite{Argyres:2018urp,Caorsi:2018zsq}.
Finally,  all known \Nequals4 superconformal theories  are super Yang-Mills theories for some gauge group $G$, and the scaling dimensions of chiral operators are integers.
We can summarize the situation in the following table:\[
\begin{array}{c|c|c|c}
$\Nequals0$ & $\Nequals1$ & $\Nequals2$  & $\Nequals4$ \\
\hline
\text{???} & \text{algebraic} & \text{rational} & \text{integral}
\end{array}.
\]

This leads to the following natural question: can we place any restriction on the scaling dimensions of operators of isolated non-supersymmetric conformal field theories in four dimensions (or in other dimensions).
The author only expects countably many such theories, and therefore such scaling dimensions will be in a countable subset of real numbers.
Are they computable numbers, in the sense of computing science, i.e.~is there a program which computes the $N$-th digit given $N$ as the input? 
Are they periods in the sense of Kontsevich and Zagier \cite{KZ}, which are known to include all coefficients of perturbative quantum field theory computations using Feynman diagrams?
}

\subsection{Case study: the adjoint SQCD}
\label{sec:adjSQCD}
Let us apply this method to the $\SU$ adjoint SQCD, briefly introduced at the beginning of the last subsection. 
This analysis was first carried out in \cite{Kutasov:2003iy}.

Calling the unknown $s=R(\Phi)$, the trial $a$ function is \begin{equation}
a(s)=\underbrace{(N^2-1)\mathsf{a}(1)}_\text{from $\lambda$} 
+ \underbrace{(N^2-1)\mathsf{a}(s-1)}_\text{from $\psi_\Phi$}
+ \underbrace{2 N_f N \mathsf{a}(-s {N}/{N_f})}_\text{from $\psi_Q$, $\psi_{\tilde Q}$}
\end{equation}
where \begin{equation}
\mathsf{a}(r)=(3/32)(3r^3-r)
\end{equation}
is the contribution from a single fermion of charge $r$.
There is one local maximum and one local minimum.
For simplicity, we analyze the theory in the limit $N,N_f\to \infty$ with the ratio $x=N/N_f$ fixed.
Let us also introduce $\epsilon=N/N_f-1/2$.
The theory is asymptotically free when $\epsilon>0$.
Furthermore, the theory is in a very weakly coupled, Banks-Zaks regime when $|\epsilon|\ll 1$.
Explicitly, we find  \begin{equation}
s=\frac{10}3 \cdot \frac{1}{3+\sqrt{20x^2-1}}=
\frac23-\frac23\epsilon+O(\epsilon^2)
\label{s}
\end{equation}  at the local maximum of $a(s)$.

Let us confirm this value of $s$ to the first order by the perturbation theory.
We know from \eqref{gammaQi} that 
\begin{align}
\gamma(Q)&=-\frac{g^2}{8\pi^2}\frac{N^2-1}{2N} +O(g^4), &
\gamma(\Phi)&=-\frac{g^2}{8\pi^2}N+O(g^4).
\end{align}
Converting them into the running of $\tau$ as in \eqref{BZ} and we see that the beta function is zero when \begin{equation}
2N-N_f + 2N_f\gamma(Q) +2N\gamma(\Phi)=0,
\end{equation}which fixes \begin{equation}
\frac{g^2}{8\pi^2}=\frac{\epsilon}{N}+O(\epsilon^2).
\end{equation}
This means 
\begin{equation}
D(\Phi)=1+\gamma(\Phi)= 1-\epsilon+O(\epsilon)^2,
\end{equation} 
which reproduces \eqref{s} since $D(\Phi)=(3/2)R(\Phi)=(3/2)s$.

As we increase $x$, the scaling dimensions of various operators decrease.
Eventually, at $x=3+\sqrt{7}$, the gauge-invariant operator $M:=Q\tilde Q$ hits the unitarity bound.
Above $x>3+\sqrt{7}$, we interpret that this operator becomes free and decouple from the system, leaving an interacting theory, as we discussed in the case of the adjoint SQCD with the superpotential $W=\tr \Phi^3$ in Sec.~\ref{sec:kutasov-prescription}.
The interacting theory without the free decoupled field can be obtained by flipping the operator $M$, as we discussed in Sec.~\ref{sec:flipping}.
This was the operation to introduce an additional gauge-singlet field $\underline{M}^j_i$ 
and the accompanying superpotential $W=\underline{M}^j_i M^i_j$ to the system.

The trial $a$-function is now \begin{equation}
a(s)=\underbrace{(N^2-1)\mathsf{a}(1)}_\text{from $\lambda$} 
+ \underbrace{(N^2-1)\mathsf{a}(s-1)}_\text{from $\psi_\Phi$}
+ \underbrace{2 N_f N \mathsf{a}(-s {N}/{N_f})}_\text{from $\psi_Q$, $\psi_{\tilde Q}$}
+ \underbrace{N_f^2 \mathsf{a}(2s N/N_f-1)  }_\text{from $\underline{M}$}.
\end{equation} whose local maximum is now at \begin{equation}
s=\frac{\sqrt{20x^4-48x^3+87x^2-16x}-15x}{3(2x^3-8x^2-x)}.
\end{equation}
As we further increase $x$, the operator $M^{(1)}:= Q \Phi \tilde Q$ hits the unitarity bound next, and decouples.
To continue the analysis, we add a flipping operator $\underline{M^{(1)}}$ and the accompanying superpotential $W=\underline{M^{(1)}} M^{(1)}$ to isolate the interacting part in the infrared.
This process can be continued indefinitely.
For details, see the original article \cite{Kutasov:2003iy}.\footnote{%
The original article \cite{Kutasov:2003iy} did not introduce the flipping fields;
the flipping fields in the analysis of this class of models were introduced much later in \cite{Benvenuti:2017lle}.
In the original article \cite{Kutasov:2003iy}, the would-be contribution from the decoupled gauge-invariant composite operators was subtracted by hand.
Either way, one ends up with exactly the same computation.
}

\subsection{Case study: Lagrangians for  Argyres-Douglas theories}
Let us consider next the simplest of the adjoint SQCD theory,
by taking $N=2$ and $N_f=1$.
In this case it is convenient to combine $Q$ and $\tilde Q$ into a single object $\sQ^I_a$ with $I=1,2$ for the flavor $\SU(2)$ and $a=1,2$ for the gauge $\SU(2)$, as we did in Sec.~\ref{sec:su2sp1}.
As basic gauge-invariant operators we can find $\sM:=\epsilon_{IJ}\epsilon^{ab}\sQ_a^I \sQ_b^J$,
$X^{(IJ)}:=\sQ_a^I \Phi^{(ab)} \sQ_b^J$,
and $U:=\tr \Phi^2$, among others.

By performing the $a$-maximization, we find that $s=(\sqrt{1009}-9)/87\sim 0.26$.
At this value the operator $U$ would violate the unitarity bound.
This requires us to introduce a flipper $\underline{U}$ and the superpotential $W=\underline{U}U$.
Redoing the $a$-maximization, we find that $s=(\sqrt{601}+3)/111\sim 0.25$.
Assuming that the low-energy theory is superconformal, we can use the state-operator correspondence to read off the spectrum of supersymmetric operators from the supersymmetric index on $S^3$, 
and their scaling dimensions should all satisfy the unitarity bound. 
Indeed, by computing  the supersymmetric index  $S^3$ of this theory using the methods explained in Sec.~\ref{sec:SCI}, we find no unitarity violating operators.
This gives us confidence that  the system flows to a superconformal field theory in the infrared.

This does not stop us to include a flipper $\underline{\sM}$ to the operator $\sM$, although this is not necessary from the point of view of the unitarity bound violation.
This still leads to an interesting result as noticed first by \cite{Maruyoshi:2016tqk} and as we will see soon.

The model is now \cite{SongTalk,Benvenuti:2017lle} an $\SU(2)$ gauge theory with one flavor $\sQ^{I=1,2}$ and an adjoint $\Phi$, together with gauge-singlets $\underline{U}$ and $\underline{\sM}$, and the superpotential\begin{equation}
W=\underline{U}\tr\Phi^2 + \underline{\sM}(\sQ \sQ).
\end{equation}
The $a$-maximization gives $s=2/9$, a rational number.
The resulting central charges are \begin{equation}
(a,c)=(\frac{11}{24},\frac12).
\end{equation}
Various gauge-invariant operators have the scaling dimensions \begin{equation}
\Delta(\underline{\sM})=\frac43,\quad
\Delta(\underline{U})=\frac73,\quad
\Delta(X^{(IJ)})=2.
\end{equation}
These are exactly the central charges and the operator spectrum
of a particular \Nequals2 superconformal theory in the class of theories known as the Argyres-Douglas theories.
This particular theory is known under various names: the $H_1$ Argyres-Douglas theory, or the $(A_1,A_3)$ Argyres-Douglas theory, or the Argyres-Douglas point of the \Nequals2 $\SU(2)$ with two flavors,
or the Argyres-Douglas point of the \Nequals2 pure $\SU(4)$ theory,
all of which refers to the same theory.

A slight variant \cite{SongTalk,Maruyoshi:2018nod} is to consider the $\SU(2)$ gauge theory with $\sQ^{I=1,2}$ and an adjoint $\Phi$, together with singlets $\underline{U}$ and $\underline{X_{11}}$, so that the superpotential is \begin{equation}
W=\underline{U}\tr\Phi^2 + \underline{X_{11}} (\sQ^1\Phi  \sQ^1) + \sQ^2 \Phi \sQ^2.
\end{equation}
By $a$-maximization, we find the superconformal theory with \begin{equation}
(a,c)=(\frac{43}{120},\frac{11}{30})
\end{equation} and \begin{equation}
\Delta(\underline{X_{11}})=\frac65,\qquad
\Delta(\sQ^1 \sQ^2)=\frac{11}5.
\end{equation}
These data match with those of another Argyres-Douglas theory, again known under various names: 
the $H_0$ Argyres-Douglas theory, or the $(A_1,A_2)$ Argyres-Douglas theory, or the Argyres-Douglas point of the \Nequals2 $\SU(2)$ with one flavor,
or the Argyres-Douglas point of the \Nequals2 pure $\SU(3)$ theory.
This theory saturates the lower bound of the $c$ central charge of \Nequals2 theories, which we already mentioned above in \eqref{LiendoBound}, originally derived in \cite{Liendo:2015ofa}.

Originally, the Argyres-Douglas theories were found by tuning Coulomb branch vacuum expectation values of  \Nequals2 gauge theories so that electric and magnetic particles become simultaneously massless \cite{Argyres:1995jj,Argyres:1995xn}.
The $\U(1)$ R-symmetry of these  theories can be identified directly in the infrared
thanks to the \Nequals2 supersymmetry.
For a gentle introduction to Argyres-Douglas theories from the \Nequals2 point of view,
the readers are referred to e.g.~Chapter 10 of the author's review \cite{Tachikawa:2013kta}.

The superconformal $\U(1)$ R-symmetry of the Argyres-Douglas theories emerges only in the infrared in the manifestly \Nequals2 supersymmetric formulation, and cannot be directly identified in the ultraviolet Lagrangian description of the theory.
This made it hard to compute the central charges, which was only later found in \cite{Aharony:2007dj,Shapere:2008zf} using non-Lagrangian techniques.
The Lagrangian description we presented here, originally found in \cite{Maruyoshi:2016tqk} and greatly simplified in \cite{SongTalk,Benvenuti:2017lle}, does not manifest the full \Nequals2 supersymmetry in the ultraviolet.
Instead it makes the superconformal $\U(1)$ R-symmetry  available at the ultraviolet,
and allows us to compute  the supersymmetric index of $S^3\times S^1$,
which again required a non-Lagrangian technique if one uses the \Nequals2 description of the theory.

\Nequals1 supersymmetric Lagrangian descriptions of \Nequals2 Argyres-Douglas theories are being studied further in e.g.~\cite{Maruyoshi:2016aim,Agarwal:2016pjo,Benvenuti:2017kud,Agarwal:2017roi,Benvenuti:2017bpg,Giacomelli:2017ckh,Maruyoshi:2018nod,Giacomelli:2018ziv,Agarwal:2018oxb},
where the readers can find e.g.~generalizations to various other Argyres-Douglas theories,
discussions on string theory embeddings, and analyses of $S^1$ compactifications to 3d.
The talks \cite{SongTalk,MaruyoshiTalk} are also quite helpful.
The field is still young, and we do not even clearly understand the mechanism behind the enhancement of supersymmetry to \Nequals2. 
A lot remain to be uncovered.

\subsection{Case study: an AdS/CFT correspondence}

Our last example is about the AdS/CFT correspondence.
The prototypical example \cite{Maldacena:1997re} comes from considering $N$ D3-branes in the flat space $\bR^{1,3}\times \bR^6$ in the large $N$ limit.
This  gives rise to the \Nequals4 $\SU(N)$ super Yang-Mills theory on the field theory side
and the spacetime $\mathrm{AdS}_5\times S^5$ on the gravity side.

To generalize, we note that $\bR^6$ has the metric of the form \begin{equation}
ds^2_{\bR^6} = dr^2 +r^2 ds^2_{S^5}
\end{equation} where $ds^2_{S^5}$ is the metric of the unit sphere, which is an Einstein manifold with $R_{\mu\nu}=4g_{\mu\nu}$.
We then replace $ds^2_{S^5}$ by another Einstein manifold $X$ with the same normalization $R_{\mu\nu}=4g_{\mu\nu}$: \begin{equation}
ds^2_{C(X)} = dr^2 + r^2 ds^2_{X}.
\end{equation}
This space is called the cone over $X$.
We consider $N$ D3-branes on $\bR^{1,3}\times C(X)$.
On the gravity side we have the AdS spacetime of the form $\mathrm{AdS}_5\times X$,
and we have a field theory determined by the geometry of $X$ on the field theory side.
One standard consequence of the AdS/CFT correspondence is that \begin{equation}
a\sim c \sim \frac{N^2\pi^3}{4\Vol X}.
\label{vol}
\end{equation}
It is easy to check that the \Nequals4 $\SU(N)$ super Yang-Mills theory 
and $X=S^5$ satisfy this relation.

\begin{figure}
\centering
\includegraphics[width=.3\textwidth,valign=t]{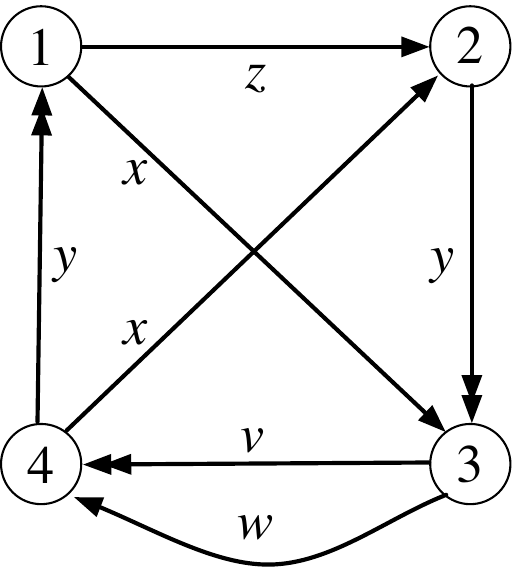}\qquad\qquad\includegraphics[width=.3\textwidth,valign=t]{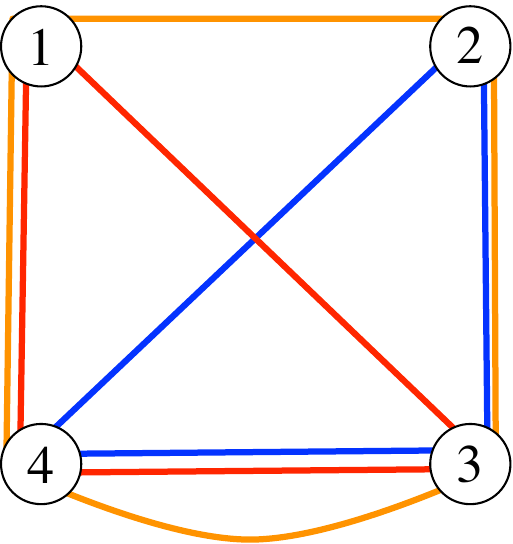}
\caption{Left: the matter content of the gauge theory with the symbols we use for their R-charges; Right: the structure of the superpotential \label{fig:quiver}}
\end{figure}

The setup preserves an \Nequals1 supersymmetry if $X$ is a Sasaki-Einstein manifold.\footnote{%
A  manifold $X$ is Sasaki-Einstein if the cone $C(X)$ is Calabi-Yau.
A manifold $X$ is Sasaki if the cone $C(X)$ is K\"ahler.
}
There is a particular Sasaki manifold called the $Y^{2,1}$ space.
The cone over it, $C(Y^{2,1})$, is also known as the complex cone over $dP_1$,
and the corresponding field theory was constructed in \cite{Feng:2000mi} in 2000,
whose structure is summarized in Fig.~\ref{fig:quiver}.

The figure is read as follows. Each node $i$ corresponds to an $\SU(N)_i$ gauge group.
An arrow from $\SU(N)_i$ to $\SU(N)_j$ corresponds to a chiral multiplet $X_{ij}$ in the fundamental of $\SU(N)_i$ times the anti-fundamental of $\SU(N)_j$,
and if the arrow is double-headed, the chiral multiplet comes in a doublet under an $\SU(2)$ flavor symmetry, which we denote by $X^{\alpha}_{ij}$ where $\alpha=1,2$.
For $(i,j)=(3,4)$, we have both $X_{34}$ and $X^\alpha_{34}$,
corresponding to the single-headed arrow and the double-headed arrow.
We then introduce a superpotential term for each closed path given in the figure, with a carefully chosen coefficient: \begin{equation}
W=\epsilon_{\alpha\beta} \tr\left[ 
X^\alpha_{34}X^\beta_{41}X_{13}
-X^\alpha_{34}X_{42}X^\beta_{23}
+X_{34} X^\alpha_{41}X_{12}X^\beta_{23}
\right].
\end{equation}

Can we test the AdS/CFT prediction \eqref{vol}?
This was not possible when the gauge theory was first constructed in 2000,
since the volume of $Y^{2,1}$ was not known on the gravity side,
and the method to compute the central charge $a$ for such a complicated gauge theory was not known on the field theory side either.
Things changed a few years later, since in 2003 the $a$-maximization was devised in \cite{Intriligator:2003jj},
and the Einstein metric on $Y^{2,1}$ was constructed in \cite{Gauntlett:2004yd} in 2004,
where the authors constructed a first-ever example of an irregular Sasaki-Einstein manifold,
which was sometimes conjectured not to exist in the literature before that.\footnote{%
It is very mysterious that the closely related facts were found independently in mathematics and in physics around the same time. See also footnote \ref{foot:mystery}.}
With these developments, the AdS/CFT prediction \eqref{vol} can now be tested \cite{Bertolini:2004xf}.

On the gravity side, we just quote the value of the volume: \begin{equation}
\Vol(Y^{2,1})=\frac14\cdot\frac{1}{13\sqrt{13}-46}.
\label{volY21}
\end{equation}
Let us perform the $a$-maximization on the field theory side.
We first need to determine the conserved $\U(1)$ R-symmetries.
The symmetry of Fig.~\ref{fig:quiver} implies that it can be parameterized by 
at most five variables  \begin{gather}
x:=R(X_{13})=R(X_{42}), \quad y:=R(X_{41}^\alpha)=R(X_{23}^\alpha), \\
z:=R(X_{12}), \quad
v:=R(X^\alpha_{34}), \quad w:=R(X_{34}),
\end{gather}
see Fig.~\ref{fig:quiver} for the assignment of symbols.

We demand that all superpotential terms have R-charge two: \begin{equation}
2y+z+w=x+y+v=2.
\end{equation}
We also demand that the $\SU(N)_i\U(1)_R^2$ anomaly vanishes: \begin{equation}
z+x+2y=2, \qquad x+2y+2v+w=4
\end{equation}
where the first condition comes from $\SU(N)_{1,2}$ and the second from $\SU(N)_{3,4}$.
We can use them to eliminate $z,v,w$ in favor of $x$ and $y$: \begin{equation}
z=2-x-2y,\quad v=2-x-y,\quad w=x.
\label{zvw}
\end{equation}

The trial $a$ function is given by, in the large $N$ limit, \begin{equation}
a(x,y)=N^2\left[
2\mathsf{a}(x-1)+4\mathsf{a}(y-1)+\mathsf{a}(z-1)+2\mathsf{a}(v-1)+\mathsf{a}(w-1)+4\mathsf(1)
\right],
\end{equation}
where $z$, $v$, $w$ are eliminated using \eqref{zvw}.
We maximize it with respect to $x$ and $y$, and find \begin{equation}
x=\sqrt{13}-3,\qquad y=\frac43(4-\sqrt{13})
\end{equation} at which \begin{equation}
a(x,y)=(13\sqrt{13}-46)N^2.
\label{aY21}
\end{equation}
We see that the geometric result \eqref{volY21}
and the field theory result \eqref{aY21} 
satisfy the AdS/CFT prediction \eqref{vol}.

Here we treated only one example, $Y^{2,1}$.
There are natural generalizations called the $Y^{p,q}$ metrics 
and corresponding quiver gauge theories, for which the same analysis can be repeated
and the $a$-maximization reproduces the inverse of the volume of the manifold \cite{Benvenuti:2004dy,Franco:2005sm}.
The geometric dual of the $a$-maximization process have also been identified \cite{Martelli:2005tp,Martelli:2006yb}.
Soon, the check of the agreement was generalized to all toric Sasaki-Einstein manifolds and corresponding quiver gauge theories \cite{Butti:2005vn,Benvenuti:2006xg,Lee:2006ru,Eager:2010yu}.
For more on this fascinating subject, the readers are referred e.g. to Chapter 5 of Yamazaki's review on brane tilings \cite{Yamazaki:2008bt}, or Chapters 6 and 7 of the author's PhD thesis \cite{TachikawaThesis}.

\part{A glimpse into 2d \Nequals{(2,2)} and \Nequals{(0,2)} dynamics}

In this Part 2, which is significantly shorter than the Part 1, we will have a brief look at 2d supersymmetric dynamics, with \Nequals{(2,2)} and \Nequals{(0,2)} supersymmetry.
In 2d, we can have strong dynamics without gauge fields, whereas in 4d, it was necessary to have non-Abelian gauge fields for  strong dynamics.
This makes the study in 2d somewhat simpler.
At the same time, in 2d, the low energy conformal theory is controlled by the Virasoro symmetry, which is infinite dimensional, while in 4d, the superconformal algebra is finite dimensional.
This allows us to analyze the infrared dynamics in 2d in much more detail.

Furthermore, many of the techniques we learned to analyze 4d \Nequals1 theories have direct analogues, as we will see.
We note that our presentation is somewhat ahistorical, since these techniques were often first developed in 2d, which were later extended to 4d.
Our discussion in this part will hopefully exemplify that it is fruitful to study supersymmetric dynamics in various dimensions at the same time.

\section{Landau-Ginzburg and minimal models}
\label{sec:LGmin}
\subsection{2d \Nequals{(2,2)} superfields}
\label{sec:2dlag}
We will start by quickly reviewing the structure of the superfields of 2d \Nequals{(2,2)} theories.
As in Sec.~\ref{sec:4dsuperfields} where we summarized 4d \Nequals1 superfields, our discussion will be extremely brief and the coefficients here  should not be trusted.
For more details, we refer the reader e.g.~to the influential article \cite{Witten:1993yc}, or a short review by Hori \cite{Hori:2002fa}, or to the extensive review \cite{MirrorBook}.

A 2d \Nequals{(2,2)} theory has four supercharges, which is the same with a 4d \Nequals1 theory.
In particular, the dimensional reduction from 4d \Nequals1 on $T^2$ will give a 2d \Nequals{(2,2)}  theory.
Therefore, many aspects of 2d \Nequals{(2,2)} theories, such as the superspace, the superfields and the supermultiplets, can be studied by the dimensional reduction.

One starts from the 4d \Nequals1 superspace, with coordinates \begin{equation}
x^{\mu=0,1,2,3},\quad 
\theta^{\alpha=1,2},\quad
\bar\theta^{\dot\alpha=\dot 1,\dot 2}.
\end{equation}
To obtain the coordinates of 2d \Nequals{(2,2)} superspace, one simply discards $x^2$ and $x^3$:
\begin{equation}
x^{\mu=0,1},\quad
\theta^+,\quad \theta^-,\quad
\bar\theta^+,\quad \bar\theta^-.
\end{equation}
Here we used the lightcone combinations $\pm$ for the spinor indices,
so that $\theta^+$ and $\theta^-$ are complex Weyl spinors of positive and negative chirality, respectively.
Note that we have $\overline{\theta^+}=\bar\theta^+$ and $\overline{\theta^-}=\bar\theta^-$.

A chiral superfield $\bar D_\pm \Phi=0$ has the expansion schematically of the form \begin{equation}
\Phi\sim\phi+\theta^+\psi_+ + \theta^- \psi_- + \theta^+\theta^- F.
\end{equation}
We refer to $\theta^+$ and $\psi_+$ as right-moving
and $\theta^-$ and $\psi_-$ as left-moving.\footnote{%
In other words, we always put the superscripts to the supercoordinates
and the subscripts to the fermions.
Then the plus sign is declared to be right-moving 
and the minus sign to be left-moving.
}
We assign right-moving  R-charge $+1$ to $\theta^+$
and left-moving R-charge $+1$ to $\theta^-$.

Since the smallest spinor in 2d is a real object,
a complex right-moving Weyl spinor $\theta^+$ counts as \Nequals{(0,2)}
and the left-moving $\theta^-$ counts as \Nequals{(2,0)}.
In total we have \Nequals{(2,2)} supersymmetry.

There are three basic supersymmetric integrals. 
The first two are familiar from 4d \Nequals1, namely the K\"ahler potential term
\begin{equation}
\int d^2\theta d^2\bar\theta (\text{arbitrary superfield}) 
\end{equation}
and the superpotential term \begin{equation}
\int d^2\theta  (\text{arbitrary chiral superfield})  + \cc 
\end{equation} where a chiral superfield $\Phi$ satisfies \begin{equation}
\bar D_\pm \Phi=0.
\end{equation}
A supersymmetric Lagrangian can then be formed from a single chiral superfield $\Phi$ just as in 4d: \begin{equation}
\int d^2\theta d^2\bar \theta K(\bar\Phi,\Phi) + \int d^2\theta W(\Phi)+ \cc  
\end{equation}
The superpotential $W(\Phi)$ has left-moving R-charge $1$ and right-moving R-charge $1$.

In two dimensions one can also have  the twisted superpotential term, which is given by \begin{equation}
\int d\theta^+ d\bar\theta^- (\text{arbitrary twisted chiral superfield}) + \cc 
\end{equation} where a twisted chiral superfield $\Sigma$ satisfies \begin{equation}
\bar D_+ \Sigma = D_- \Sigma=0.
\end{equation}

A gauge field Lagrangian looks differently from the one in 4d, due to the following reason.
For simplicity we only discuss the Abelian case.
We start from the vector superfield $V$ which transforms under the gauge transformation as \begin{equation}
V \to V+\Lambda+\bar\Lambda.
\end{equation} where $\Lambda$ is a chiral superfield.
The basic gauge-covariant combination in 4d is the chiral superfield containing three superderivatives:\begin{equation}
W_\alpha = \bar D_{\dot\beta} \bar D^{\dot\beta}  D_\alpha V.
\end{equation}
In 2d, the combination with two superderivatives is already a gauge-invariant scalar,  \begin{equation}
\Sigma= \bar D_+  D_- V.
\end{equation}
This is twisted chiral and has the expansion \begin{equation}
\Sigma \sim \sigma - \theta^+ \bar\lambda_+ -  \bar\theta^- \lambda_- + \theta^+\bar\theta^- (D-iF_{01})+ \cdots.
\end{equation}
Here, $D$ is the $D$ auxiliary field, $F_{\mu\nu}$ is the field strength,
and $\sigma$ is a complex scalar, all in the adjoint of the gauge group.

The appearance of the complex scalar $\sigma$ can be understood from 4d as follows:
we start from a 4d vector multiplet which contains the vector potential $A_{0,1,2,3}$.
We reduce it down to 2d. Then the components $A_{0,1}$ remain as a 2d gauge field,
and $A_{2,3}$  is re-interpreted as two real 2d scalars,
which combine to form a complex adjoint scalar.

To treat the non-Abelian case properly, we need to introduce gauge-covariant superderivatives $\cD_\pm:=e^{-V} D_\pm e^V$ and $\bar\cD_\pm:=e^V \bar D_\pm e^{-V}$.
and (twisted) chiral superfields constrained by them.
Then the combination $\Sigma = \{\bar \cD_+,\cD_-\}$ becomes twisted chiral in the gauge-covariant sense.
For details of the non-Abelian case, see e.g.~Sec.~4.1 of \cite{Witten:1993xi}.

The kinetic term of a gauge field is given in terms of $\Sigma$ by \begin{equation}
\sim \frac1{g^2}\int d^2\theta d^2\bar\theta \tr\Sigma\bar\Sigma.
\end{equation}
For $\U(1)$ fields, $\Sigma$ itself is gauge invariant and one can introduce \begin{equation}
\int d\theta^+ d\bar\theta^- i t\Sigma + \cc 
\end{equation} where \begin{equation}
t=i\xi +\frac{\theta}{2\pi}
\end{equation} is a complex number. Expanding into components, one finds that \begin{equation}
\int d\theta^+ d\bar\theta^- i t\Sigma + \cc  \sim \xi D + \frac{\theta}{2\pi} F_{01},
\end{equation}
meaning that $\xi$ is the Fayet-Iliopoulos term and $\theta$ is the 2d theta angle.

Consider now a $\U(1)$ gauge theory with chiral multiplets $\Phi_i$ of $\U(1)$ charge $Q_i$. Let $W$ be the superpotential and $\xi$ be the coefficient of the Fayet-Iliopoulos term.
The classical potential  is then roughly of the form \begin{equation}
V \sim \sum_i |F_i|^2 + D^2 + |\sigma|^2(\sum_i Q_i^2 |\phi_i|^2)
\end{equation} where \begin{equation}
F_i =\partial W/\partial \Phi_i, \qquad  D=(\sum_i Q_i |\phi_i|^2)-\xi.\label{2dFD}
\end{equation}
One can determine the one-loop running of the Fayet-Iliopoulos term $\xi$  from this potential.
We first note that  the loop of $\phi_i$ contributes
to the operator $|\phi_i|^2$ of the form \begin{equation}
\vev{|\phi_i|^2} \sim \int\frac{d^2k}{(2\pi)^2} (\frac{1}{k^2+Q_i^2\sigma^2}-\frac{1}{k^2+Q_i^2\mu^2})
\sim +\frac1{2\pi} \log \frac{\mu}{|\sigma|} 
\end{equation} where $\mu$ is the cutoff.
To make the D-term  \eqref{2dFD} independent of the cutoff $\mu$,
one needs the renormalization of the Fayet-Iliopoulos term
\begin{equation}
\frac{\partial}{\partial\log\mu} \xi = \frac1{2\pi} \sum_i Q_i.
\label{xi-rg}
\end{equation}

\subsection{\Nequals2 superconformal algebra}
The conformal algebra in general $d$ spacetime dimensions is $\so(d,2)$,
and for $d=2$ this splits as $\so(2,2)\simeq \so(2,1)\oplus\so(2,1)$,
one factor associated to the left movers and the other associated to the right movers.
Each factor is enhanced to the Virasoro algebra which is infinite dimensional.

We consider \Nequals{(2,2)} supersymmetric systems.
This means that we will have  a copy of \Nequals2 super Virasoro symmetry for the left movers
and another copy for the right movers.
For definiteness, we map the left movers to the holomorphic side
and the right movers to the anti-holomorphic side.
Here we only discuss the holomorphic side.

The \Nequals2 super Virasoro symmetry contains the energy-momentum tensor $T(z)$, 
an $\so(2)\simeq \mathfrak{u}(1)$ current $J(z)$,
and the complex supercharges $G^\pm(z)$.
Our convention is that the supercoordinates have $\U(1)$ R-charge $\pm1$.
Therefore the supercharges $G^\pm(z)$ also have $\U(1)$ R-charge $\pm1$.
Note that, contrary to  the notation in Sec.~\ref{sec:2dlag},
we use the indices $\pm$ to denote the charge under the $\U(1)$ R-symmetry.

They have the following operator-product expansions. 
First, the energy-momentum tensor $T(z)$ has  the standard OPE:
\begin{align}
T(z)T(w) &= \frac{c/2}{(z-w)^4} + \frac{2T(w)}{(z-w)^2}+\frac{\partial T(w)}{z-w}+\cdots
\end{align}
where $c$ is the central charge.
The supercurrent $G^\pm(z)$ and the $\U(1)$ current $J(z)$ are primaries of dimension $3/2$ and $1$, respectively:
\begin{align}
T(z) G^\pm (w) &= \frac{3/2}{(z-w)^2} G^\pm(w)+\frac{\partial G(w)}{z-w}+\cdots,\\
T(z)J(w)&= \frac{J(w)}{(z-w)^2}+\frac{\partial J(w)}{z-w}+\cdots.
\end{align}
We normalized the $\U(1)$ current  $J(z)$ so that 
the supercurrent $G^\pm(z)$ has charge $\pm1$ under it: \begin{equation}
J(z) G^\pm(w)=\pm \frac{G^\pm(w)}{z-w}+\cdots.
\end{equation}
Then the Jacobi identities demand that the $\U(1)$ current $J(z)$ has the following self-OPE: \begin{equation}
J(z) J(w) = \frac{c/3}{(z-w)^2}+\cdots\label{JJ}.
\end{equation}
The final nontrivial OPE is \begin{equation}
G^+(z)G^-(w)= \frac{2c/3}{(z-w)^3}+\frac{2J(w)}{(z-w)^2}+\frac{2T(w)+\partial J(w)}{z-w}+ \cdots.
\end{equation}
In the \Nequals{(2,2)} literature one also encounters $\hat c=c/3$.

Independent of \Nequals2 Virasoro symmetry, consider a $\U(1)$ current with the OPE \begin{equation}
J(z)J(w) = \frac{\kappa}{(z-w)^2}+\cdots.
\end{equation}
This corresponds to the anomaly polynomial
 \begin{equation}
\frac12 \kappa (\frac{F}{2\pi})^2.
\label{u1r-anom}
\end{equation} 

To derive it, for the moment let us denote the anomaly polynomial as $(1/2) k (F/2\pi)^2$.
From general principles, it is clear that $k\propto \kappa$ with a theory-independent proportionality coefficient.
What we need to do is to show that this coefficient is one.
For this purpose it suffices to consider a single example, which we take to be
a complex Weyl fermion $\psi^\pm(z)$ with charge $\pm q$.
This has the anomaly polynomial  $(1/2) (qF/(2\pi))^2$ in our convention in Sec.~\ref{sec:anomalypoly}.
Now, the Weyl fermions have the OPE $\psi^+(z)\psi^-(w)\simeq 1/(z-w)$. 
The $\U(1)$ current is then $J(z)=q\psi^+\psi^-(z)$, whose OPE with $\psi^\pm$ is indeed $J(z)\psi^\pm(w)= \pm q \psi^\pm(w)/(z-w)$.
Then we find $J(z)J(w) = q^2/(z-w)^2+\cdots$.
This argument also shows that   \begin{equation}
\kappa = \tr_\text{left movers} R^2 -\tr_\text{right movers} R^2
\end{equation}
where the trace is over the labels of complex Weyl fermions in the theory and $R$ is the matrix of R-charges.

Comparing with \eqref{JJ}, this means that if the anomaly polynomial of the left-moving R-symmetry is given by \eqref{u1r-anom},
the left-moving central charge of the system is given by \begin{equation}
c=3 \kappa.
\label{ckappa}
\end{equation}
This relation between the central charge and the R-charge anomaly is the 2d analogue of the 4d relation \eqref{fieldtheoryresults} we studied in Sec.~\ref{sec:ca}.

\subsection{Landau-Ginzburg models}
\label{sec:LG}
Let us now consider the following Lagrangian of a single chiral superfield $\Phi$ \cite{Lerche:1989uy}:
\begin{equation}
\int d^2\theta d^2\bar\theta \bar \Phi \Phi + \int d^2\theta \Phi^d + \cc \label{LG}
\end{equation}
The bosonic potential is then given by  $V\propto |\Phi^{d-1}|^2$,
and one can expect to have an interesting low-energy conformal field theory.

Let us determine the central charge of the low-energy theory.
The left-moving R-charge of the chiral multiplet $\Phi$ is fixed to be $1/d$.
The expansion of $\Phi$ into components is of the form \begin{equation}
\Phi=\phi+\theta^+ \psi_+ + \theta^-\psi_- + \theta^+\theta^- F,
\end{equation} and therefore 
the left-moving fermion $\psi_-$ has left-moving R-charge $1/d-1$,
while 
the right-moving fermion $\psi_+$ has left-moving R-charge $1/d$.
It means that the anomaly polynomial of the left-moving R-charge is \begin{equation}
\frac12\left[ (\frac1d-1)^2-(\frac1d)^2  \right](\frac{F}{2\pi})^2 = \frac12 (1-\frac2d) (\frac{F}{2\pi})^2.
\end{equation}
From \eqref{ckappa}, we conclude that the central charge of the low-energy Virasoro symmetry must be \begin{equation}
c=3(1-\frac 2d).\label{minimal}
\end{equation}

For 4d theories, we studied the supersymmetric index on $S^3\times S^1$ in Sec.~\ref{sec:SCI}.
A 2d analog is the supersymmetric index on $S^1\times S^1$.
This quantity is known under the name of the elliptic genus.\footnote{%
This is partly because $S^1\times S^1\simeq T^2$ is also known as an elliptic curve in mathematics.
Also note that the plural of the term `elliptic genus' is `elliptic genera'.
}
For  a general \Nequals{(2,2)} superconformal field theory, it is defined by
\begin{equation}
Z_\text{ell}(y,q):=\tr_\text{RR} (-1)^{F} y^{J_L} q^{L_0-c/24}\bar q^{\overline{L_0}-\bar c/24},
\end{equation}
where 
$F$ is the  fermion number,
and $J_{R,L}$ are the right-moving and left-moving $\U(1)_R$ charges.
The trace is taken in the R-R sector, i.e.~the sector where the fermions are periodic around the spatial $S^1$ when there is no $\U(1)_R$ background.
We also used the standard definition\begin{equation}
q:=e^{2\pi i \tau},\quad\text{and}\quad y:=e^{2\pi iz}.
\end{equation}

On the right-moving side, this is the Witten index as in Sec.~\ref{sec:box},
since this is the trace of $(-1)^{F}$ regularized by the right-moving energy $\bar L_0$.
As such the elliptic genus is automatically independent of $\bar q$.
This makes this quantity rigid against continuous changes, and make it computable using the ultraviolet Lagrangian description.

On the left-moving side, with the insertion of $y^{J_L}$,
the elliptic genus depends in general on $q$ and $y$.
When one sets $y=1$ or equivalently $z=0$, the left-moving side also becomes the Witten index, and the $q$ dependence also drops out.

Let us now compute the elliptic genus of the model \eqref{LG}, following \cite{Witten:1993jg}.
We simply use the Hilbert space of the model where the potential is neglected.
A chiral multiplet $\Phi$ of left-moving R-charge $r$ contains the following fields:
\begin{equation}
\begin{array}{c|cc}
&\text{left-moving}&\text{right-moving}\\
\hline
\text{boson} & r & r \\
\text{fermion} & r-1 & r
\end{array}
\end{equation} where the entries denote the left-moving R-charge.
Due to the common left-moving R-charge, the contributions from the right-moving fermions and the right-moving bosons cancel out.
Therefore the contribution to the elliptic genus from $\Phi$ of left-moving R-charge $r$ is simply \begin{equation}
Z_\text{ell,$\Phi$} (y,q)=\frac{\theta_1(y^{r-1},q)}{\theta_1(y^{r},q)}
\label{Phi-ell}
\end{equation} where \begin{equation}
\theta_1(y,q):=-\ii q^{1/8} y^{1/2} \prod_{n=1}^\infty (1-q^n)(1-y q^n)(1-y^{-1} q^{n-1}).
\end{equation}
The elliptic genus of the model \eqref{LG} is then given by setting $r=1/d$.
When $d=2$ and $r=1/2$, the elliptic genus is trivial and identically $-1$.\footnote{%
The sign factor comes from our convention of the fermion number of the vacuum.
We could have included a minus sign on the right hand side of \eqref{Phi-ell} to remove this sign.
}
This is as it should be, since with $W=\Phi^2$, the field $\Phi$ is massive,
and does not survive in infrared.

Note that the logic leading to this formula is the same as the method we used to compute
the $S^3\times S^1$ partition function for a 4d chiral multiplet in Sec.~\ref{sec:SCIchiral}.
The only difference is whether we use Fourier modes on $S^3$ or  Fourier modes on $S^1$,
and the computation on $S^1$ is definitely simpler.

We will study below how the central charge \eqref{minimal} and the elliptic genus \eqref{Phi-ell} stand up against a direct analysis using the superconformal algebra.
For this we need to review basic facts concerning \Nequals2 minimal models.

Before getting there, let us perform a generalization to multi-field models,
where we have chiral fields $\Phi_i$ for $i=1,2,\ldots$ 
and the superpotential $W(\Phi_i)$.
We assume that we can assign left-moving R-charges $r_i$ to $\Phi_i$, so that the superpotential has the transformation law \begin{equation}
\lambda W(\Phi_1,\Phi_2,\ldots ) = W(\lambda^{r_1} \Phi_1,\lambda^{r_2}\Phi_2,\ldots,).
\end{equation}
The computation of the central charge and the elliptic genus carries over easily to this general case. 
We simply have \begin{equation}
c=3\sum_i (1-2r_i)
\end{equation} and \begin{equation}
Z_\text{ell}=\prod_i \frac{\theta_1(y^{r_i-1},q)}{\theta_1(y^{r_i},q)} .
\end{equation}

A nice subset of such multi-field models consists of those with $c<3$.
First consider the case when \begin{equation}
W=\sum \Phi_i{}^{d_i}, \label{Phik}
\end{equation}
where we assume $d_i \ge 3$ because those $\Phi_i$ with $d_i=2$ are massive and do not survive in the infrared limit.
We need to solve $\sum_i(1-2/d_i) <1$. 
The full set of solutions is given by $(d_i)=(d)$ for $d\ge 3$, $(d_i)=(3,3)$, $(d_i)=(3,4)$ and $(d_i)=(3,5)$.

\begin{table}
\[
\begin{array}{c|r@{}l|cc
c|l}
\text{type} & \multicolumn{2}{c|}{\text{polynomial}} & h & a & b 
  & \multicolumn{1}{c}{\text{exponents}}\\
\hline
A_{n-1} & X^n& +Y^2
 & n & 1 & n/2 
  &  1,2,3,\ldots, n-1 \\
D_{n+1} & X^n &+ XY^2
 & 2n & 2& n-1
& 1,3,5,\ldots, 2n-1; n \\
E_6 & X^4 &+ Y^3
 & 12 & 3 & 4 
& 1 , 4, 5, 7, 8, 11  \\
E_7 & X^3Y &+ Y^3 
& 18 & 4& 6
 &  1, 5, 7, 9, 11, 13, 17 \\
E_8 & X^5&+Y^3
 & 30 & 6 & 10
& 1, 7, 11, 13, 17, 19, 23, 29
\end{array}
\]
\caption{Data of the singularities of type $A$, $D$, $E$.
See the main text for  the significance of the  numbers $h$, $a$, $b$.
\label{table:ADEsingularities}}
\end{table}

Mathematicians have shown that we only get a few more solutions,
even allowing for more general superpotentials than \eqref{Phik}.
All such polynomials for which $c<3$ are tabulated in Table~\ref{table:ADEsingularities}.
In the table,  the fields are renamed from $\Phi_{1,2}$ to $X$ and $Y$, 
and we set the number of variables to two. 
The central charge is parametrized by a number $h$ so that \begin{equation}
c=3(1-\frac{2}{h}),
\end{equation}
and the left-moving R-charges of $X$ and $Y$
are $a/h$ and $b/h$, 
respectively.
The cases \eqref{Phik} we studied above correspond to the following types:
$(d_i)=(d)$ is of type $A_{d-1}$,
$(d_i)=(3,3)$ is of type $D_4$,\footnote{%
The type $D_4$ polynomial in the Table is $W=X^3+XY^2$,
whereas $(d_i)=(3,3)$ gives  $W=X^3+Y^3$.
They can be transformed to each other by a change of variables: one simply needs to replace
$X\to X+Y$, $Y\to X-Y$ in the latter equation, and then to perform some rescalings.
}
$(d_i)=(3,4)$ is of type $E_6$, and 
$(d_i)=(3,5)$ is of type $E_8$.

The theory behind the classification is the following.
The quantity $\sum_i (1-2r_i)$ is also known in mathematics as the singularity index,
and the singularities whose index is less than one is known as simple singularities.
The point is that simple singularities have been classified, 
and they follow the ADE classification.
For more details, see e.g.~the textbook \cite{SingularityTheory}.

As is well-known, simply-laced Lie algebras $\fg$, namely those Lie algebras whose Dynkin diagram does not contain double or triple lines, admit the same ADE classification:
$\fg=A_{n-1}=\su(n)$, or $D_n=\so(2n)$, or $E_{6,7,8}$.
The number $h$ is the (dual) Coxeter number of the ADE type, for example,
and appears prominently in any gauge theory computation.

In the Table \ref{table:ADEsingularities}, the exponents are also listed.
Take $E_6$ as an example. We enumerate all monomials under the relation $\partial W/\partial X=\partial W/\partial Y
=0$.
The independent monomials are then $1$, $X$, $X^2$, $Y$, $XY$, $X^2Y$,
whose left-moving R-charges are $0/12$, $3/12$, $6/12$,
$4/12$, $7/12$, $10/12$.
The subscript of the type, here $6$ of $E_6$, is the number of independent such monomials.
The exponents $\{e_i\}$ are such that the left-moving R-charge of these monomials are $(e_i-1)/h$.

The exponents also have significance in gauge theories.
Let $\fg$ be one of those ADE algebras.
We consider a gauge theory $\fg$ with a scalar field $\phi$ in the adjoint representation.
We then enumerate gauge-invariant operators constructed from $\phi$.
Then  the exponents plus one, $\{e_i+1\}$, are the dimensions of generators of such gauge-invariant operators.
For example, take $A_{n-1}$. Then $\phi$ is simply a traceless $n\times n$ matrix,
and gauge-invariant operators are generated by $\tr \phi^2$, $\tr \phi^3$, \ldots, $\tr\phi^n$.
As another example, take $D_n$. Then $\phi$ is a $2n\times 2n$ antisymmetric matrix,
and gauge-invariant operators are generated by $\tr \phi^2$, $\tr\phi^4$, \ldots, $\tr\phi^{2n-2}$, and $\Pf \phi$.

\subsection{\Nequals2 Virasoro minimal models}
\label{sec:minimal}
Let us now quote basic results in the representation theory of \Nequals2 Virasoro symmetry.
The readers are referred to Greene's review \cite{Greene:1996cy}, or the textbook by Eguchi and Sugawara \cite{EguchiSugawara} if they read Japanese.

First, any unitary representation with $c<3$ should necessarily has the form $c=3(1-2/h)$ with an integer $h\ge 2$.\footnote{%
It is also often parameterized by $k$, where $h=k+2$.
}
Note that the central charge \eqref{minimal} we found above is exactly of this particular form.

The unitary irreducible representation in the R-sector for a particular $h$ is classified as follows \cite{Gepner:1987qi}.
They are labeled by $\ell=1,2,\ldots, h-1$ and $m=-\ell, 2-\ell,\cdots, +\ell-2,+\ell$.\footnote{%
Instead of $\ell$ it is often parameterized by $l=\ell-1$.
}
Let us denote this representation by $V^\ell_m$.
When $|m|\neq \ell$, the representation contains two lowest-dimension operators,
whose $L_0$ eigenvalue  and  the $\U(1)$ charge $Q$ are given by the formula \begin{equation}
L_0=\frac{\ell^2-1}{4h}-\frac{m^2}{4h}+\frac18,\qquad
Q=\frac{m}h\pm \frac12.
\end{equation}
These two operators are  paired by the action of the zero modes $G_0^\pm$ of the supercharges.
When $m=\ell$, the primary state with $Q=m/h+1/2$ drops out,
and the primary state with $Q=m/h-1/2$ is killed by the action of $G_0^+$, making it a chiral primary.
Similarly, when $m=-\ell$, it becomes an anti-chiral primary.
In both cases we simply have $L_0=c/24$, meaning that the energy on a circle is zero.

Let us denote the character of this representation, with an insertion of $(-1)^F$, by $I^\ell_m(y,q)$: 
\begin{equation}
I^\ell_m(y,q)=\tr_{V^\ell_m} (-1)^F y^Q q^{L_0-c/24} .
\end{equation}
It is common to extend the range of $m$ from $-\ell\le m \le +\ell$ to $m \in \bZ_{2h}$
by the rule $I^\ell_{m+h}:=-I^{h-\ell}_m$.
The explicit infinite-product formula for $I^\ell_m$ can be found in \cite{Matsuo:1986cj}.
We only need the $q\to 0$ limit:
\begin{equation}
I^\ell_{m}(y,q=0) = \begin{cases}
+y^{\ell/h-1/2}, & (m=\ell) \\
-y^{-\ell/h+1/2}, & (m=-\ell) \\
0 & (\text{otherwise}).
\end{cases}
\label{minimal-chiy}
\end{equation}

To have a full 2d superconformal theory with both the left movers and the right movers,
one needs to combine representations of holomorphic and anti-holomorphic copies of
the \Nequals2 Virasoro symmetry.
All possible modular-invariant combinations with $c<3$ was classified \cite{Cappelli:1986ed,Gepner:1986hr,Gepner:1987qi}, 
and the partition function in the RR-sector, with an insertion of $(-1)^{F_L+F_R}$, has the following form: \begin{equation}
Z_\text{\Nequals2 Virasoro}(y,q; \bar y, \bar q)=\frac12 \sum_{\ell,{\ell'}}\sum_{m\in \bZ_{2h}} N_{\ell,{\ell'}} I^\ell_m(y,q)\overline{I^{{\ell'}}_m ( y, q)}
\label{um}
\end{equation} where $N_{\ell,{\ell'}}$ are certain non-negative integers 
and $I^\ell_m$ with odd $\ell+m$ is considered to be zero.

The $N_{\ell,{\ell'}}$ follows the ADE pattern:
\begin{itemize}
\item The so-called diagonal invariant of type $A_{h-1}$,  
for which we simply have $N_{\ell,{\ell'}}=\delta_{\ell,{\ell'}}$.
\item the $D_{n+1}$-type invariant, with $h=2n$, and 
\item the $E_{6,7,8}$-type invariant, with $h=12, 18, 30$ respectively. 
\end{itemize}
Note that $h$ in each case is the dual Coxeter number of the ADE type.

\begin{table}
\[
\begin{array}{c|c|l}
h& \text{type} & \multicolumn{1}{c}{Z} \\
\hline 
h & A_{h-1} &  |\chi_1|^2 + |\chi_2|^2+ \cdots + |\chi_{h-1}|^2  \\
2n & D_{n+1}^\text{($n$ odd)} & |\chi_1+\chi_{2n-1}|^2+|\chi_3+\chi_{2n-3}|^2+\cdots + |\chi_{n-2}+\chi_{n+2}|^2+2|\chi_n|^2 
\\
2n & D_{n+1}^\text{($n$ even)} &  |\chi_1|^2 +|\chi_3|^2 + \cdots +|\chi_{2n-1}|^2+
 \chi_2\overline{\chi_{2n-2}}+\chi_4\overline{\chi_{2n-4}}+\cdots+\chi_{2n-2}\overline{\chi_2}
 \\
12&  E_6 &|\chi_1+\chi_7|^2+|\chi_4+\chi_8|^2+|\chi_5+\chi_{11}|^2 \\
18 & E_7 & |\chi_1+\chi_{17}|^2+|\chi_5+\chi_{13}|^2+|\chi_7+\chi_{11}|^2+|\chi_9|^2+(\chi_3+\chi_{15})\overline{\chi_9}+\cc \\
30 & E_8 & |\chi_1+\chi_{11}+\chi_{19}+\chi_{29}|^2
+|\chi_7+\chi_{13}+\chi_{17}+\chi_{23}|^2
\end{array}
\]
\caption{$N_{\ell,{\ell'}}$ for $\SU(2)_{h-2}$ modular invariants. 
This also determines $N_{\ell,{\ell'}}$ for unitary minimal \Nequals{(2,2)} models.
\label{table:su2modularinv}}
\end{table}

To describe the $D$-type and $E$-type invariants, it is useful to note that  the same sets of integers   appear in the modular invariant partition function of $\SU(2)_{h-2}$, \begin{equation}
Z_{\SU(2)_{h-2}} = \sum_{\ell,{\ell'}}N_{\ell,{\ell'}} \chi_\ell(y,q) \overline{\chi_{{\ell'}}(y,q)}
\end{equation}
where $\ell=1,\ldots, h-1$ now labels the irreducible representations of $\SU(2)_{h-2}$.
The $\SU(2)$ modular invariants were first classified in \cite{Cappelli:1986hf,Cappelli:1987xt,Kato:1987td}, and have the values given in Table~\ref{table:su2modularinv}.

We see that, in the Table~\ref{table:su2modularinv} of modular invariants,
$h$ is the dual Coxeter number of the type,
and the term $\chi_\ell \overline{\chi_\ell}$ appears exactly once
for each exponent $\ell$ of the type.
These results follow from an analysis using \Nequals2 Virasoro algebra and the modular invariance,
without using any Lagrangian description.

Before proceeding, let us compute the elliptic genus of a minimal model.
This is obtained by restricting the arguments of the partition function \eqref{um}
by setting $\bar y=1 $ or equivalently $\bar z=0$. 
From \eqref{minimal-chiy}, we see that $I^\ell_m(y{=}1,q{=}0)$ is $+1$ if $\ell=m$, $-1$ if $\ell=-m$, and zero otherwise.
Therefore we have \begin{equation}
Z_\text{ell}=\sum_{\ell,{\ell'}} N_{\ell, {\ell'}} I^\ell_{{\ell'}}(\tau,z).
\label{minimal-ell}
\end{equation}

\subsection{Minimal models as Landau-Ginzburg models}
\label{sec:mlg}
We have seen that the Landau-Ginzburg models with $c<3$ have an ADE classification in Sec.~\ref{sec:LG}.
We have also reviewed in Sec.~\ref{sec:minimal} that the minimal models, again with $c<3$ have an ADE classification.
It is natural to expect that they correspond to each other. 
Let us check this correspondence.

Let us first consider the single-field model, where the superpotential is simply $W=\Phi^h$.
We saw in Sec.~\ref{sec:LG} that the central charge is $c=3(1-2/h)$, and the elliptic genus is \begin{equation}
\frac{\theta_1(y^{1/h-1},q)}{\theta_1(y^{1/h},q)}.
\end{equation}
We identify the low energy limit to be the diagonal modular invariant at this central charge,
or equivalently the modular invariant of type $A_{h-1}$,
for which $N_{\ell,{\ell'}}=\delta_{\ell,{\ell'}}$.
For this model, the elliptic genus is, from \eqref{minimal-ell}, 
\begin{equation}
\sum_\ell I^\ell_\ell(q,y).
\end{equation}
For this identification to hold, we should have
 \begin{equation}
\frac{\theta_1(y^{1/h-1},q)}{\theta_1(y^{1/h},q)}=\sum_{\ell=1}^{h-1} I_\ell^\ell(y,q). 
\label{a-eq}
\end{equation}

The equality \eqref{a-eq} can be checked by expanding both sides in terms of $q$.
For example, using $\theta_1(y,q=0) = y^{1/2}-y^{-1/2}$ and \eqref{minimal-chiy},
the equation \eqref{a-eq} reduces to the well-known equality \begin{equation}
\frac{t^{h-1}-t^{1-h}}{t-t^{-1}} = t^{h-2} + t^{h-4} + \cdots + t^{4-h}+t^{2-h},
\label{th}
\end{equation}where $t=y^{1/(2h)}$.
In fact this agreement at $q\to 0$ is enough to prove the equality \eqref{a-eq} for arbitrary $q$, roughly due to the following.
Both sides of \eqref{a-eq} transform in the same manner under the shift of $z$ by $n+m\tau$ and also under the $\SL(2,\bZ)$ modular transformation on $\tau$.
Due to the low value of $c$, the possible form of such functions is quite limited,
and the agreement at $q\to 0$ forces the equality for all $q$.
For details, see \cite{DiFrancesco:1993dg,Kawai:1993jk,Kawai:2009ci}.

We can generalize this analysis to  the multi-field Landau-Ginzburg polynomials of type $D$ and  $E$, given in Table~\ref{table:ADEsingularities}.
We expect them to flow in the infrared to the unitary minimal model of the corresponding type  \cite{Vafa:1988uu}.
The agreement of the central charge is easy to check.
The agreement of the elliptic genus is more interesting to confirm.
The equality to check is the following: \begin{equation}
\frac{\theta_1(t^{h-a},q)}{\theta_1(t^{a},q)}
\frac{\theta_1(t^{h-b},q)}{\theta_1(t^{b},q)}
\stackrel{?}{=}
\sum_{\ell,{\ell'}} N_{\ell,{\ell'}} I^\ell_{{\ell'}}(t^h,q)
\end{equation}
where $t=y^{1/(2h)}$ as before;
the left hand side comes from the Landau-Ginzburg description
and the right hand side comes from the minimal model expression.
In the limit $q\to 0$, this reduces to the equation \begin{equation}
\frac{t^{h-a} - t^{a-h}}{t^a-t^{-a}}
\cdot
\frac{t^{h-b} - t^{b-h}}{t^b-t^{-b}}
=
\sum_{\ell:\text{exponents}} t^{h-2\ell}
\end{equation} where $\ell$ runs over the exponents of the corresponding type.
This generalizes the standard formula \eqref{th} to type $D$ and $E$,
and  can be checked by a direct computation.
This equation goes back at least to \cite{saito1983}.

For more details of the content in this section, we refer the readers again to \cite{DiFrancesco:1993dg,Kawai:1993jk,Kawai:2009ci}.
Before closing this section, we note that these central charges have directly been measured in numerical lattice simulations, see  \cite{Kawai:2010yj,Kamata:2011fr,Morikawa:2018ops,Morikawa:2018zys}.

\section{Calabi-Yau models}
\label{sec:CY}
Another important class of 2d \Nequals{(2,2)} theories are Calabi-Yau models,
which describe type II strings propagating on Calabi-Yau manifolds.
Here we will have a brief look, mainly concentrating on their elliptic genera.
For more detailed information, the readers are advised to consult Greene's  review \cite{Greene:1996cy}.

\subsection{Calabi-Yau sigma models}
A large class of 2d \Nequals{(2,2)} supersymmetric models can be constructed by taking a
K\"ahler manifold $X$ and performing the path-integral over the space of maps from the 2d spacetime to $X$. 
Let $X$ be a complex $d$-dimensional K\"ahler manifold.
Then we introduce $d$ chiral superfields $\Phi^{1,2,\ldots, d}$, and consider the model described by the Lagrangian \begin{equation}
\int d^2\theta d^2\bar\theta K(\bar\Phi^{\bar i},\Phi^i)
\end{equation} where $K$ is the K\"ahler potential of the K\"ahler manifold $X$.

The metric of the target space $X$  gets renormalized. 
To the one-loop order, one finds\begin{equation}
\frac{d}{d \log\Lambda} g_{i\bar j} \propto  R_{i\bar j}.
\label{run}
\end{equation}
The sign of the coefficient is such that a positively-curved manifold such as $S^2$ shrinks as one lowers the energy scale $\Lambda$;
the $S^2$ model is known to show the dimensional transmutation and develops a mass gap. 
This renormalization group equation was first studied by Friedan in \cite{Friedan:1980jf,Friedan:1980jm}  in non-supersymmetric setting.

The renormalization group equation \eqref{run} implies that if the K\"ahler manifold $X$ is Ricci flat, i.e.~when $R_{i\bar j}=0$, 
the model is free of renormalization and therefore conformal, at least to the one-loop order.
A Ricci-flat K\"ahler manifold is also known as a Calabi-Yau manifold, due to the following historical reason.
Since $R_{i\bar j}$ is the differential form which represents the first Chern class $c_1(T_\bC X)$ of the tangent bundle,
$R_{i\bar j}=0$ implies $c_1(T_\bC X)=0$.
Calabi conjectured the converse that any K\"ahler manifold with $c_1(T_\bC X)=0$ has a Ricci flat metric.
This difficult conjecture was later proved by Yau.

Let us give some examples of Calabi-Yau manifolds.
We start from the projective space $\CP^n$, parameterized by ratios $[x_1:x_2:\cdots:x_{n+1}]$.
There is a natural line bundle $L=\mathcal{O}(1)$ on it, such that $x_i$ can be thought of as holomorphic sections.
We now consider $L^{\otimes k}=\cO(k)$, whose section $f$ can be thought of as a degree-$k$ homogeneous polynomial of $x_i$.
The equation $f=0$ within $\CP^n$ determines a complex $(n-1)$-dimensional space $X$.
The restriction of the K\"ahler form of $\CP^n$ on $X$ makes $X$ a K\"ahler space.

A standard computation in mathematics tells us that this K\"ahler space $X$ is Calabi-Yau if and only if $k=n+1$.\footnote{%
The computation goes as follows.
We write the total Chern class of $L$ as $c(L)=1+H$, where $H$ is proportional to the K\"ahler class.
Now, note that $T\CP^n \oplus \bC = L^{\oplus(n+1)}$.
Therefore $c(T\CP)=(1+H)^{n+1}$.
Furthermore, denoting by $N$ the normal bundle to $X$  in $\CP^n$ 
we have $TX|_X \oplus N = T\CP^n|_X$ 
and $N=L^{\otimes k}|_X$.
We conclude that $
c(TX)= \frac{c(T\CP^n)}{c(L^{\otimes k})}=\frac{(1+H)^{n+1}}{1+kH} = 1+ (n+1-k) H + \cdots,
$
i.e. $c_1(TX)=(n+1-k) H$.
As $H$ is nonzero, the vanishing of $c_1(TX)$ is equivalent to $k=n+1$.
}
The simplest example is given by a cubic equation $(k=3)$ in $\CP^2$.
This is a complex one-dimensional space, and is in fact a torus in disguise.
It clearly has a flat metric, and in particular $R_{i\bar j}=0$.

The next example is given by a quartic equation $(k=4)$ in $\CP^3$.
This is a complex two-dimensional surface, and is known as a K3 surface.\footnote{%
It was named by A. Weil after three mathematicians Kummer, K\"ahler and Kodaira involved in the early study of the K3 surfaces, also alluding to the mountain K2 in the Himalayas.
}
The \Nequals{(2,2)} sigma-model on a K3 surface in fact has \Nequals{(4,4)} supersymmetry.
For physics applications of K3 surfaces, see the review article by Aspinwall \cite{Aspinwall:1996mn}.

The third example is given by a quintic equation $(k=5)$ in $\CP^4$.
This is the famous quintic Calabi-Yau manifold,
and is the prototypical example everyone always comes back to when one studies a Calabi-Yau manifold.
Arguably, the most famous one is given by the equation \begin{equation}
x_1^5+x_2^5+x_3^5+x_4^5+x_5^5=0
\end{equation} within $\CP^4$ parameterized by $[x_1:x_2:x_3:x_4:x_5]$.
This manifold is called the Fermat quintic.

We already mentioned that the one-loop running of the metric is given by the equation \eqref{run},
and therefore that the theory is conformal to this order when $R_{i\bar j}=0$,
i.e.~when the manifold is Calabi-Yau.
In fact, it has been shown that  a Calabi-Yau manifold $X$
defines a conformal field theory well-defined to all order in perturbation theory \cite{Nemeschansky:1986yx} and also non-perturbatively \cite{Dine:1986zy}.

Let us determine some of the properties of the low-energy conformal field theory.
Take a K\"ahler manifold of complex dimension $d$.
Classically, there is an R-symmetry which assigns charge zero to all $\Phi^{i=1,\ldots, d}$.
The left-moving fermions $\psi_-^{i}$ are 
charged under the target-space curvature $F_X$ valued in the adjoint of $\mathfrak{u}(d)$,
has charge $-1$ under the left-moving R-symmetry, and is neutral under the right-moving R-symmetry.
Similarly, the right-moving fermions $\psi_+^i$ are charged under $F_X$,
is neutral under the left-moving R-symmetry, and has charge $-1$ under the right-moving R-symmetry.
The anomaly polynomial is then  given by \begin{equation}
\frac d2 (\frac{F_L}{2\pi})^2 - \frac d2 (\frac{F_R}{2\pi})^2  + (\frac{F_L}{2\pi}-\frac{F_R}{2\pi}) \tr \frac{F_X}{2\pi} + \cdots,
\end{equation}
where $F_{L}$ and $F_R$ are the background gauge field for the left-moving and right-moving $\U(1)$ R-symmetries.

Note that $F_L$ and $F_R$ are background fields, but $F_X$ is the curvature of the target space $X$ and is part of the dynamical operators of the theory;
this is an analogue of the chiral anomaly of a 4d gauge theory where a chiral symmetry is broken due to the $\SU(N)_\text{gauge}^2\U(1)_\text{chiral}$ anomaly.
This means that  the right-moving and the left-moving R-symmetries  survive as the conserved symmetries
only when $\tr F_X\propto R_{i\bar j}$ vanishes,
i.e.~when  $X$ is Calabi-Yau.
Assuming this, one can safely identify the left-moving R-symmetry in the Lagrangian
with the left-moving R-symmetry of the low energy superconformal theory.
We find that  
\begin{equation}
c=3d, 
\label{CY-c}
\end{equation}
which is an exact answer without any further correction.

\subsection{Geometric computation of the elliptic genera}
Let us compute the elliptic genus of the Calabi-Yau sigma model, in the large volume limit.
Since the elliptic genus is rigid against continuous deformations,
the result we obtain here applies also to the Calabi-Yau models in deep quantum regimes.
A reader can skip this subsection if s/he is not familiar with basic algebraic geometry and algebraic topology;
only the final results will be needed later.

In the large volume limit, the curvature of the target $X$ is very small, and therefore we can analyze the system perturbatively.
The wavefunction of the zero modes can be identified with 
the differential forms on $X$,
and the wavefunctions of non-zero modes are sections of more complicated bundles on $X$.
This consideration leads to the  formula\footnote{%
The computation was originally outlined in Sec.~10 of \cite{Witten:1982df} and was given a detailed treatment e.g.~in \cite{AlvarezGaume:1983at},
in the case of  the 1d supersymmetric model whose target space is $X$,  or equivalently in the case of the supersymmetric quantum mechanics of a particle moving on $X$.
The formula \eqref{geom-ell} can be obtained by  carrying out the same analysis in the context of the 2d field theory.
} \begin{equation}
Z(y,q) :=\underbrace{\int_X \prod_{a=1}^d x_a}_\text{from zero modes} 
 \underbrace{\prod_{a=1}\frac{\theta_1(y^{-1} e^{x_a},q)}{\theta_1(e^{x_a},q)}}_\text{from non-zero modes}.
\label{geom-ell}
\end{equation}
Here,  $x_{a=1,\ldots, d}$ are the Chern roots of $T_\bC X$,
which means that the target space curvature is formally thought of as diagonalized in the form \begin{equation}
\frac{F_X}{2\pi}=\diag(x_1,x_2,\ldots,x_d).\label{FX}
\end{equation}
Note that the contributions from the non-zero mode   is  the contribution \eqref{Phi-ell} from  the R-charge-neutral chiral fields $\Phi_{1,\ldots,d}$ coupled to the target space curvature \eqref{FX}.
We then perform the zero-mode integral, which is the  integral  over $X$ after multiplying it with  the Euler class $\prod x_a$.

In the $y\to 1$ limit, one has \begin{equation}
Z(y,q) \to \int_X \prod x_a =: \chi(X),
\end{equation} which is the Euler number of the manifold $X$.
In the $q\to 0$ limit, one has \begin{equation}
Z(y,q) \to \int_X \prod x_a \frac{y^{-1/2} e^{x_a/2}-y^{1/2} e^{-x_a/2}}{ e^{x_a/2}-e^{-x_a/2}} =: \chi_y(X).
\end{equation}
This quantity is known as the $\chi_y$ genus.
In terms of the Betti and the Hodge numbers, we have the formulas \begin{equation}
\chi(X)=\sum_n (-1)^n \dim H^n(X),\qquad
\chi_y(X)=y^{-d/2} \sum_{p,q} (-y)^p (-1)^q \dim H^{p,q}(X).
\end{equation}

Let us consider in particular the quintic Calabi-Yau $X$,
which is the zero locus of a quintic polynomial in $M=\CP^4$.
Note that the equation \eqref{geom-ell} has the form $\int_X \varphi(TX)$
where $\varphi$ is a multiplicative characteristic class satisfying $\varphi(V\oplus V')=\varphi(V)\varphi(V')$ for two vector bundles $V$ and $V'$.
We also assume that for a line bundle $L$, we have $\varphi(L)=f(c_1(L))$ for some function $f(x)$.
A geometric quantity of this form is called a genus,
and the elliptic genus \eqref{geom-ell} is a special case when $f(x)=x \theta_1(e^x y^{-1},q)/\theta_1(e^x,q)$.
The following computation works for any $f(x)$.
For a general introduction to genera, see e.g.~\cite{Hirzebruch}.

We perform the following manipulation: \begin{multline}
Z=\int_X \varphi(TX)
\stackrel{(1)}{=}\int_X \frac{\varphi(T\CP^4)}{\varphi(L^{\otimes 5})}\\
\stackrel{(2)}{=}\int_{\CP^4} e(L^{\otimes 5}) \frac{\varphi(T\CP^4)}{\varphi(L^{\otimes 5})}
\stackrel{(3)}{=}\frac{1}{f(0)}\int_{\CP^4} e(L^{\otimes 5}) \frac{\varphi(L^{\oplus 5})}{\varphi(L^{\otimes 5})}
\stackrel{(4)}{=}\frac{1}{f(0)}\int_{\CP^4} 5H \frac{f(H)^5}{f(5H)}.
\end{multline}
Here, the equalities (1) and (2) use the fact that $X$ is the zero locus of a section of $L^{\otimes 5}$,
 the  equality (3) uses the fact $T\CP^4\oplus \bC=L^{\oplus 5}$,
 and the equality (4) uses $\varphi(L)=f(H)$.
Now, note that \begin{equation}
\int_{\CP^4} H^n=\begin{cases}
1 & (n=4),\\
0 & \text{(otherwise)}
\end{cases}
\end{equation} which means \begin{equation}
\int_{\CP^4} H^5 g(H)=\oint_{v=0} \frac{dv}{2\pi i} g(v).
\end{equation}
We therefore find \begin{equation}
Z=\frac{1}{f(0)}\oint_{v=0} \frac{dv}{2\pi i} \frac{5}{v^4}\frac{f(v)^5}{f(5v)}.
\end{equation}
For the particular case of the elliptic genus, one finds \begin{align}
Z
&=\frac{\theta_1'(y=1,q)}{\theta_1(y^{-1},q)} \oint_{w=0} dw \frac{\theta_1(e^{-5\cdot 2\pi i w},q)}{\theta_1(e^{-5\cdot 2\pi iw}y^{+1},q)} 
\left[
\frac{\theta_1(e^{2\pi i w} y^{-1},q)}{\theta_1(e^{2\pi i w},q)}
\right]^5
\label{geom-quintic}
\end{align}
where we used the variable $w=2\pi i v$ and the relation $\theta(y^{-1},q)=-\theta(y,q)$ for the first factor inside the integral.
We note that the integrand is periodic under $w\to w+1$ and $w\to w+\tau$,
namely it is an elliptic function.
The integrand has an order-5 pole at $w=0$,
whose residue we took in \eqref{geom-quintic}.

\begin{figure}
\centering
\includegraphics[width=.3\textwidth]{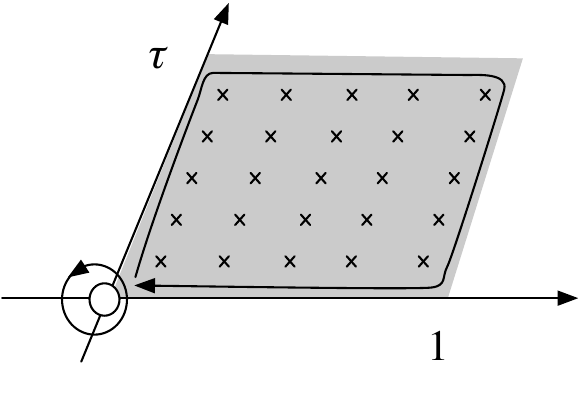}
\caption{The relation of \eqref{geom-quintic} and \eqref{orb-quintic}.
\label{fig:ell}}
\end{figure}

There are other poles in the fundamental region of $w\to w+1$ and $w\to w+\tau$.
Indeed, 
there are $5^2=25$ simple poles when $5 w = z$  mod $\bZ + \tau \bZ$, where $y=e^{2\pi i z}$.
As shown in Fig.~\ref{fig:ell}, from Cauchy's theorem, one finds an alternative expression of the elliptic genus, which is \begin{equation}
Z=\frac15\sum_{a,b=0}^4 
\left[
y^{-b/5} \frac{\theta_1(y^{-4/5} e^{2\pi i (a+b \tau)/5},q)}{\theta_1(y^{+1/5} e^{2\pi i (a+b \tau)/5},q)}
\right]^5.
\label{orb-quintic}
\end{equation}

From either expression, one finds  \begin{equation}
Z\to -100(y^{-1/2}+y^{1/2})
\end{equation} in the $q\to 0$ limit.
Further taking $y\to 1$, one finds that the Euler number of the quintic Calabi-Yau is $-200$.

We will find in the next section a more physical understanding of both formulas \eqref{geom-quintic}
and \eqref{orb-quintic}.
Before proceeding, we note that the geometric computation of the elliptic genus of complete intersections within projective spaces leading to \eqref{geom-quintic} goes back to \cite{Kawai:1994np,MaZhou}.
Similarly, the finite-sum form \eqref{orb-quintic} goes back to  \cite{Berglund:1993fj}.

\subsection{As Landau-Ginzburg orbifolds}
Let us proceed semi-historically.
Is there an alternative, more concrete way to realize the quintic Calabi-Yau sigma model?
A naive guess would be to consider the Landau-Ginzburg model with five chiral superfields 
$X_{1,\ldots,5}$ with the superpotential of degree 5. 
In the Fermat case, we simply have
\begin{equation}
W=X_1{}^5+X_2{}^5+\cdots + X_5{}^5.\label{5c5}
\end{equation}
In the low energy limit, this is just five copies  of the unitary minimal model with $h=5$.

One finds that it is not completely off the mark. 
The total central charge is \begin{equation}
c=5\times 3(1-\frac 25)=3\cdot 3,
\end{equation}
which agrees with \eqref{CY-c}.
However, the elliptic genus of this model is simply given by taking five copies of \eqref{Phi-ell} and is \begin{equation}
Z=\left[
\frac{\theta_1(y^{-4/5},q)}{\theta_1(y^{1/5},q)}
\right]^5.
\end{equation}
This does not reproduce \eqref{orb-quintic},
but it at least reproduces the $a=b=0$ term without a prefactor of $1/5$.

Another related problem is that the Landau-Ginzburg model \eqref{5c5} contains many operators with fractional R-charge, such as $X_i$'s themselves which have left-moving R-charge $1/5$.
In the geometric Calabi-Yau sigma model, there are only operators with integral R-charges.
We need to somehow kill the operators with fractional R-charge.

For this purpose, we consider a non-R $\bZ_5$ symmetry which acts on $X_i$'s by \begin{equation}
(X_1,\ldots,X_5) \mapsto e^{2\pi i/5} (X_1,\ldots, X_5).
\label{z5action}
\end{equation}
We then gauge this $\bZ_5$ symmetry.
In the spectrum, only the $\bZ_5$-invariant states survive. 
As the fractional part of the R-charge is equal to the $\bZ_5$-charge, this guarantees that
only the states with integral R-charge survive. 

Let us study  the gauging process slightly more carefully.
Suppose we have a theory with a $\bZ_k$ symmetry.
On a $T^2$ with the spatial $S^1$ and the temporal $S^1$,
we can introduce a background $\bZ_k$ gauge field,
which can be characterized by the holonomy $a\in \bZ_k$ around the temporal circle
and the holonomy $b\in \bZ_k$ around the spatial circle.
Let us denote the partition function with this background by $Z[a,b]$.
In the Hamiltonian formalism, we can write \begin{equation}
Z[a,b]=\tr_{\cH_b} g^a e^{-\beta H} 
\end{equation}
where 
$\cH_b$ is the Hilbert space of the $b$-twisted sector, i.e. the Hilbert space on $S^1$ with the holonomy $b\in \bZ_k$ around it,
$g$ is the generator of $\bZ_k$ acting on $\cH_b$ satisfying $g^k=1$,
and $H$ is the Hamiltonian.
Now, the sum over the temporal holonomy $a$ with fixed $b$,\begin{equation}
\frac1k \sum_{a=1}^k Z[a,b] =\tr_{\cH_b} \frac1k(g+g^2+\cdots+g^k) e^{-\beta H},
\end{equation} 
gives the trace over the $\bZ_k$ invariant states of $\cH_b$, 
since $(g+g^2+\cdots + g^k)/k$ removes all $\bZ_k$ charged states and keeps only the $\bZ_k$ invariant states.
To get a modular invariant combination, we are forced to sum over the spatial holonomy $b$ in addition.
This leads us to consider\begin{equation}
Z_\text{$\bZ_k$ gauged} = \frac1k \sum_{a=1}^k\sum_{b=1}^k Z[a,b] .
\end{equation}
Note that we ended up summing over all possible $\bZ_k$ gauge fields over $T^2$.
Stated differently, this is a path integral over the space of $\bZ_k$ gauge fields.
Equivalently, this is a $\bZ_k$ gauge theory.
This operation has traditionally been called as the $\bZ_k$ orbifold procedure in the literature,
and the resulting model is  often called the Landau-Ginzburg orbifold.

Let us come back to our $\bZ_5$ gauge theory. 
We consider the $\bZ_5$ gauge theory of five chiral multiplets $X_{1,\ldots,5}$
under the superpotential \eqref{5c5} and the $\bZ_5$ transformation \eqref{z5action}.
Let us compute its elliptic genus.

The contribution from a single $X$ with the $\bZ_5$ holonomy specified by $a,b\in \bZ_5$ can be found from the following consideration.
The modes of $X$ are now shifted to $X_{n+b/5}$,
and each mode of $X$ gets the phase $e^{2\pi i a/5}$ under the $\bZ_5$ transformation $g^a$.
The left-moving R-charge of $X$  is $1/5$.
Then the contribution from the Fock space is 
$\theta_1(y^{-4/5} e^{2\pi i(a+b\tau)/5 },q)/\theta_1(y^{+1/5} e^{2\pi i(a+b\tau)/5 },q)$.
In addition, we need to compute the the R-charge  of the vacuum, which depends on the twist $b$.
After a standard computation, say using the $\zeta$ function regularization, one finds that the R-charge of the vacuum is $-b/5$.
Raising to the power $5$ and summing over $a,b\in \bZ_5$, we reproduces the geometric result  
\eqref{orb-quintic}.

\subsection{As gauged linear sigma models}

Let us now consider a gauge theoretic realization of the geometric construction we are using.
Firstly, the $\CP^n$ sigma model can be realized using a $\U(1)$ vector multiplet $V$ and 
$n+1$ charge $+1$ chiral superfield $X_1$, \ldots, $X_{n+1}$.
We introduce the Fayet-Iliopoulos term $\xi$, which we assume to be positive.
Then the D-term equation is \begin{equation}
|X_1|^2+|X_2|^2+\cdots + |X_{n+1}|^2 = \xi.
\label{CP-D}
\end{equation}
We further need to identify under the $\U(1)$ action, i.e.~we make the identification \begin{equation}
(X_1,\cdots X_{n+1}) \sim e^{i\varphi} (X_1,\cdots,X_{n+1}).
\label{CP-U1}
\end{equation}

The resulting space is $\CP^n$: this is defined as the space of ratios $[x_1:\cdots : x_{n+1}]$,
namely, one identifies  \begin{equation}
(X_1,\cdots X_{n+1}) \sim \alpha (X_1,\cdots,X_{n+1}) \label{CP-X}
\end{equation} under an arbitrary nonzero complex number $\alpha\in\bC$.
To see that the identification \eqref{CP-X} leads to the same space as the condition \eqref{CP-U1} imposed on the subspace \eqref{CP-D},
we first use  $\alpha\in\bR$ so that the condition \eqref{CP-D} is satisfied.
Then the remaining identification by  $|\alpha|=1$ is exactly what the relation \eqref{CP-U1} does.

Now, we pick a degree-$k$ polynomial $f_k(X_1,\ldots,X_{n+1})$.
We also introduce another chiral superfield $P$ of $\U(1)$ charge $-k$.
We then add the superpotential \begin{equation}
W=Pf_k(X_1,\ldots,X_{n+1}).
\end{equation}
Let us analyze this model, following \cite{Witten:1993yc}.
Before proceeding, we note that this model is known in the literature under the name of the gauged linear sigma model.
This is simply a 2d \Nequals{(2,2)} $\U(1)$ gauge theory with a specific matter content and a specific superpotential.

We first consider the system classically.
The D-term equation is modified to be \begin{equation}
|X_1|^2+|X_2|^2+\cdots + |X_{n+1}|^2 - k|P|^2= \xi.
\label{CY-D}
\end{equation}
The F-term equations are \begin{equation}
\frac{\partial W}{\partial P} = f_k =0,\qquad
\frac{\partial W}{\partial X_i}= P\frac{\partial f_k}{\partial X_i} =0.
\end{equation}
If $P\neq 0$, these are $n+2$ equations for $n+1$ variables $X_{1,\ldots,n+1}$,
and not satisfied for a generic choice of the polynomial $f_k$,
except at the very special point $X_1=\cdots = X_{n+1}=0$.

When $\xi>0$, the D-term condition \eqref{CY-D} forces at least one of $X_i$ to be nonzero.
We then have $P=0$, and have successfully realized the hypersurface $f_k(X_1,\ldots, X_{n+1})=0$
within $\CP^n$.

When $\xi<0$, the D-term condition forces $P\neq 0$.
This breaks $\U(1)$ to its subgroup $\bZ_k$.
Therefore we have the Landau-Ginzburg model for the superfield $X_1,\ldots,X_{n+1}$ with the superpotential $W=f_k(X_1,\ldots, X_{n+1})$ which is now coupled with dynamical $\bZ_k$ gauge field.
We have thus obtained the Landau-Ginzburg orbifold.

Therefore, this single model not only realizes the geometric Calabi-Yau sigma model when $\xi \gg 0$,
but also it reduces to the Landau-Ginzburg orbifold when $\xi \ll 0$.
It is an ultraviolet Lagrangian description which reduces to either, depending on the sign of $\xi$.

A one-loop computation shows that \begin{equation}
\frac{d\xi }{d\log \Lambda} = \frac{1}{2\pi}(\underbrace{n+1}_\text{from $X$}-\underbrace{k}_\text{from $P$}) + \cdots
\end{equation}  as we reviewed in \eqref{xi-rg}.
Here  the sign of the proportionality coefficient is such that $\xi$ decreases in the low energy when $n+1>k$, is constant for $n+1=k$, and increases when $n+1<k$.
When $n+1=k$, the Lagrangian is believed to define a superconformal field theory.

The Lagrangian has an obvious left-moving R-symmetry which assigns charge $1$ to $P$ and charge $0$ to $X_{1,\ldots,n+1}$.
The vector multiplet $V$ is neutral under the R-symmetry, and therefore it contains one right-moving gaugino of left-moving R-charge $+1$.
The terms in the anomaly polynomial involving the left-moving R-symmetry are then \begin{equation}
\frac12 (\underbrace{n+1}_{X}-\underbrace{1}_P-\underbrace{1}_\text{gaugino}) (\frac{F_L}{2\pi})^2 - (\underbrace{n+1}_X-\underbrace{k}_P) \frac{F_L}{2\pi} \frac{F_\text{gauge}}{2\pi},
\end{equation}
where $F_L$ is the background for the left-moving R-symmetry
and $F_\text{gauge}$ is the $\U(1)$ gauge field.
We see that the left-moving R-symmetry is free of mixed gauge anomaly only when $n+1=k$,
Assuming this, we can identify the left-moving R-symmetry we see in the Lagrangian description as the low-energy left-moving R-symmetry in the superconformal algebra.
We conclude that the central charge is given by $c=3(n-1)$.

Classically, one cannot continuously connect the region $\xi>0$ and $\xi<0$,
since there is a singularity at $\xi=0$.
In the quantum theory, the parameter $\xi$ is combined with the $\theta$ angle to be the parameter $t=\xi +i\theta/2\pi$ which appears in the twisted superpotential $\int d\theta_- d\bar\theta_+ t\Sigma$.
It is known that the singularity is complex codimension-1 in the space of $t$,
and therefore one can continuously connect the region $\xi>0$ and $\xi<0$, avoiding hitting any singularity, by turning on a non-zero $\theta$ angle.

The elliptic genus can also be computed from the gauge-theoretic description.
The derivation is intricate and the reader is referred to the original paper  \cite{Benini:2013nda}.
It would be sufficient here to note the following.
The elliptic genus of a chiral multiplet $\Phi$ of $\U(1)$ charge $n$ and the R-charge $r$ can be found by slightly generalizing \eqref{Phi-ell} and is given by  \begin{equation}
Z_{\Phi;n,r}(y,q;w)=\frac{\theta_1(y^{r-1} e^{2\pi i nw}, q)}{\theta_1(y^{r} e^{2\pi i nw}, q)}
\end{equation} where $w$ parameterizes a constant $\U(1)$ background field on $T^2$,
so that $w\simeq w+1$ and $w\simeq w+\tau$.
We note that $Z_{\Phi;n,r}$ has $n^2$ poles as a function of $w$ in the fundamental region,
at the solutions of $r z + n w = 0 $ mod $\bZ + \tau \bZ$,
where $y=e^{2\pi i z}$.

Let us consider now a $\U(1)$ gauge theory with chiral fields $\Phi_i$ of $\U(1)$ charge $n_i$ and the R-charge $r_i$.
The contribution to the elliptic genus from the chiral fields is then \begin{equation}
\prod Z_{\Phi_i;n_i,r_i}(y,q;w).
\end{equation}
From the transformation law of the theta function, one finds that this is periodic under $w\sim w+1$  and $w\sim w+\tau$ only when $\sum n_i(r_i-1)$ is zero.
This condition also follows when one requires that the left-moving R-symmetry is free of the mixed gauge anomaly.
It is reasonable to guess that the path integral over the $\U(1)$ gauge multiplet leads to some form of an integral over $z$.
A somewhat complicated argument leads to the result that it is given by \begin{align}
Z&=\sum_{z_a} \frac{\theta_1'(y=1,q)}{\theta_1(y^{-1},q)} \oint_{w=u_a}dw
\prod Z_{\Phi_i;n_i,r_i}(y,q;w) \label{fooA}\\
&= -\sum_{w_b} \frac{\theta_1'(y=1,q)}{\theta_1(y^{-1},q)} \oint_{w=v_b} dw
\prod Z_{\Phi_i;n_i,r_i}(y,q;w),\label{fooB}
\end{align}
where $w\in \{u_a,v_b\}$ is the set of the poles of the integrand $\prod Z_{\Phi_i;n_i,r_i}(y,q;w)$,
such that $\{u_a\}$ is from chiral fields with positive charge $n_i>0$
and $\{v_a\}$ is from those with negative charge $n_i<0$.
The equality of the two expressions \eqref{fooA} and \eqref{fooB} is a consequence of Cauchy's theorem.

The explicit example we have been studying in this section was a $\U(1)$ gauge theory with
a chiral field $P$ of $\U(1)$ charge $-5$ and R-charge $1$,
and five chiral fields $X_{1,\ldots,5}$ of $\U(1)$ charge $+1$ and R-charge $0$.
It is easy to check that the general formula above reproduces \eqref{geom-quintic} and \eqref{orb-quintic}
depending on whether one chooses the pole at $z=0$ from the fields $X_i$
or the $5^2=25$ poles from the field $P$.

\section{\Nequals{(0,2)} triality}
\label{sec:triality}
Our last topic is the \Nequals{(0,2)} triality of Gadde, Gukov and Putrov \cite{Gadde:2014ppa}.
We will start by briefly reviewing the \Nequals{(0,2)} multiplets.
Our convention is that the right-moving side is supersymmetric.

\subsection{\Nequals{(0,2)} superfields}
To introduce \Nequals{(0,2)} multiplets,
it is helpful to start from the multiplets of \Nequals{(2,2)} supersymmetry
and decompose them.
An \Nequals{(2,2)} chiral multiplet $\Phi$ decomposes as follows:
\begin{equation}
\begin{array}{ccccc@{\qquad}c}
&\raisebox{-.8ex}{\rotatebox{30}{$\leftrightarrow$}}& \psi_-& \leftrightarrow & F & \text{\Nequals{(0,2)} Fermi multiplet},\\
\Phi & \leftrightarrow & \psi_+ &  \rotatebox{30}{$\leftrightarrow$} &  & \text{\Nequals{(0,2)} chiral multiplet},
\end{array}
\end{equation}
where the horizontal arrows show the action of the right-moving supersymmetry corresponding to the $\theta^+$ direction,
while the diagonal arrows correspond to the left-moving $\theta^-$ direction.

Similarly, an \Nequals{(2,2)} twisted chiral multiplet $\Sigma$ constructed from a vector multiplet decomposes as \begin{equation}
\begin{array}{ccccc@{\qquad}c}
&\raisebox{-.8ex}{\rotatebox{30}{$\leftrightarrow$}}& \bar\lambda_-& \leftrightarrow & D+iF_{01} & \text{\Nequals{(0,2)} Fermi multiplet},\\
\sigma & \leftrightarrow & \lambda_+ &  \rotatebox{30}{$\leftrightarrow$} &  & \text{\Nequals{(0,2)} chiral multiplet},
\end{array}
\end{equation} where the horizontal and the diagonal arrows correspond to the supersymmetries for the $\theta^+$ and $\bar\theta^-$ directions, respectively.

The \Nequals{(0,2)} superspace has bosonic coordinates $x^{0,1}$ and fermionic coordinates $\theta^+$ and $\bar\theta^+$.
An \Nequals{(0,2)} chiral multiplet $\Phi$ satisfies $\bar D_+\Phi=0$
and contains a complex scalar $\phi$ and a chiral fermion $\psi_+$, with the expansion \begin{equation}
\Phi=\phi+\theta^+\psi_+ + \cdots.
\end{equation}
An \Nequals{(0,2)} Fermi multiplet $\Psi_-$ satisfies $\bar D_+ \Psi_- = E(\Phi)$ where 
$E$ is a holomorphic function of all chiral multiplets in the theory, and has the expansion \begin{equation}
\Psi_- = \psi_- + \bar\theta^+ E(\Phi).
\end{equation}
The kinetic term of chiral and fermi multiplets is then of the form \begin{equation}
\int d\theta^+ d\bar\theta^+ \left[
\bar\Phi_a \partial_{--} \Phi^a
+ \bar\Psi_-{}_i \Psi_-{}^i
\right].
\end{equation}

The interaction term can be written in the form \begin{equation}
\int d\bar\theta^+ \Psi_-^i J_i(\Phi^a)+ \cc  ;
\end{equation} its supersymmetry variation is controlled by \begin{equation}
\bar D_+(\Psi_-^i J_i(\Phi^a))=E^i(\Phi^a)J_i(\Phi^a).
\end{equation} Therefore we need a consistency condition 
\begin{equation}
E^i(\Phi^a)J_i(\Phi^a)=0.
\end{equation}
The scalar potential is schematically of the form \begin{equation}
V\propto |E^i|^2 + |J_i|^2,
\end{equation} where the $E$-term and the $J$-term appear symmetrically.
In fact, we can regard $\Psi_-'{}_i := \bar \Psi_-{}_i$ instead of $\Psi_-{}^i$ as the fundamental ingredient;
then the $E$-term and the $J$-term for $\Psi_-'{}_i$ is given by $E'_i(\Phi)=J_i(\Phi)$ and $ J'{}^i(\Phi)=E^i(\Phi)$.
When an \Nequals{(2,2)} chiral multiplet $\Phi^{(2,2)}$ is decomposed into
an \Nequals{(0,2)} chiral $\Phi$ and an \Nequals{(0,2)} Fermi $\Psi_-$,
$E(\Phi)=0$ and $J(\Phi)=\partial W/\partial \Phi$.

As for the Fermi  multiplet $\Upsilon_-$ constructed from a vector multiplet, 
it has the expansion \begin{equation}
\Upsilon_- = \lambda_- + \theta^+(D+iF_{01}) +\cdots,
\end{equation} the kinetic term is given by \begin{equation}
\frac1{g^2}\int d\theta^+ d\bar\theta^+ \tr\bar\Upsilon_-\Upsilon_-.
\end{equation}  
For a $\U(1)$ multiplet one can introduce the term \begin{equation}
\int d\theta^+ t \Upsilon_-,
\end{equation} where $t=\xi + i\theta/2\pi$ is the complexified combination of the Fayet-Iliopoulos term $\xi$ and the theta angle $\theta$.

\subsection{\Nequals{(0,2)} triality}
Gadde, Gukov and Putrov considered the following \Nequals{(0,2)} gauge theory \cite{Gadde:2013lxa,Gadde:2014ppa}.
We use $\U(N_c)$ as the gauge group.
Then, as matter fields, we introduce:\footnote{%
S. Nawata notified the author that the names of the fields $\Phi$ and $P$ used here are swapped with respect to the names used in the original papers. 
The author thanks him and apologizes the readers for the inconveniences this might cause.
}
\begin{itemize}
\item  $N_1$ chiral multiplets $\Phi$ in the fundamental,
\item  $N_2$ chiral multiplets $P$ in the antifundamental, 
\item $N_3$ Fermi multiplets $\Psi$ in the antifundamental,
\item  $N_1N_2$ gauge-singlet Fermi multiplets $M$ with the superpotential interaction $\int d\theta_+ \tr M \Phi P$.
\item and two Fermi multiplets $\Omega_{1,2}$ in the determinant representation.
\end{itemize}
The cancellation of the $\SU(N_c)$ anomaly requires $N_c=(N_1+N_2-N_3)/2$.
 The $\U(1)$ part of the gauge anomaly is canceled thanks to the Fermi multiplets $\Omega_{1,2}$.
 
 There are flavor symmetries $\SU(N_{1,2,3})$; in addition, we have three  anomaly-free $\U(1)$ symmetry under which various fields are charged as follows: \begin{equation}
\begin{array}{c|ccccccccccccccccc}
& \Phi & P & \Psi  & M & \Omega_1 & \Omega_2 \\
\hline
\U(1)_1 & 1 & 0 & 0 & -1 & -N_1 & 0 \\
\U(1)_2 & 0 & -1 & 0 & 1 & -N_2 & 0 \\
\U(1)_3 & 0 & 0 & 1  & 0 & 0 & N_3
\end{array}.
\label{threeU1s}
\end{equation}
 
 Let us denote this theory by the following diagram: \begin{equation}
A_{2d}= \begin{tikzpicture}[my]
\draw (0,0) node[circle,draw] (A) {$n_3$};
\draw (0,1) node[rectangle,draw] (B) {$N_2$};
\draw (-1,-1) node[rectangle,draw] (C) {$N_1$};
\draw (1,-1) node[rectangle,draw] (D) {$N_3$};
\draw[->] (C)--(A);
\draw[->] (A)--(B);
\draw[->,dotted] (A)--(D);
\draw[->,dotted,curve] (B) to (C);
\draw (1.2,0) node (E) {$\Omega_{1,2}$};
\draw[-,snake it] (A)--(E);
\end{tikzpicture},
\label{triality-original}
\end{equation}
 where we introduced $n_i=N-N_i$ and $N=(N_1+N_2+N_3)/2$. 
 Note that $N_c=n_3$, and also that $n_1+n_2+n_3=N$.
Here, the central circle denotes the gauge symmetry, the squares flavor symmetries, the solid lines chiral multiplets, the dotted lines Fermi multiplets; the presence or the absence of the arrow tips at an end of an edge shows that the corresponding matter fields are fundamentals or anti-fundamentals, respectively; and curvy lines show that the fields are Fermi multiplets in the determinant representation.

Gadde, Gukov and Putrov noticed that this theory has a triality,
namely that this theory  in the infrared is equivalent to the theories given by the following diagrams, obtained by cyclically permuting the objects: 
\begin{equation}
B_{2d}=\begin{tikzpicture}[my]
\draw (0,0) node[circle,draw,] (A) {$n_2$};
\draw (0,1) node[rectangle,draw] (B) {$N_2$};
\draw (-1,-1) node[rectangle,draw] (C) {$N_1$};
\draw (1,-1) node[rectangle,draw] (D) {$N_3$};
\draw[<-] (C)--(A);
\draw[->,dotted] (A)--(B);
\draw[<-,] (A)--(D);
\draw[->,dotted,curve] (C) to (D);
\draw (1.2,0) node (E) {$\Omega_{1,2}$};
\draw[-,snake it] (A)--(E);
\end{tikzpicture}, 
\qquad
C_{2d}=\begin{tikzpicture}[my]
\draw (0,0) node[circle,draw,] (A) {$n_1$};
\draw (0,1) node[rectangle,draw] (B) {$N_2$};
\draw (-1,-1) node[rectangle,draw] (C) {$N_1$};
\draw (1,-1) node[rectangle,draw] (D) {$N_3$};
\draw[<-,dotted] (C)--(A);
\draw[<-] (A)--(B);
\draw[->] (A)--(D);
\draw[->,dotted,curve] (D) to (B);
\draw (-1.2,0) node (E) {$\Omega_{1,2}$};
\draw[-,snake it] (A)--(E);
\end{tikzpicture}.
\label{triality-after}
\end{equation}

Let us check that all three theories have the same 't Hooft anomalies.
In the original description, we have \begin{equation}
\begin{array}{llr}
\SU(N_1)^2 : & n_3-N_2 &=-n_1,\\
\SU(N_2)^2 : & n_3-N_1 &=-n_2,\\
\SU(N_3)^2 : & -n_3 &=-n_3.
\end{array}
\label{SUN_i-anom}
\end{equation}
As the results are cyclically symmetric, the anomaly is invariant under the triality.

Similarly, the anomaly polynomial for $\U(1)_{1,2,3}$ is given by a cyclically-invariant expression \begin{equation}
- \frac12NN_1 (\frac{F_1}{2\pi})^2
- \frac12NN_2 (\frac{F_2}{2\pi})^2
- \frac12NN_3 (\frac{F_3}{2\pi})^2
\end{equation}
where $F_i$ is the background  gauge field  for $\U(1)_i$.

Let us determine next $c_L$ and $c_R$ of the low-energy theory.
$c_R$ can be determined once one finds the low-energy  R-symmetry.
How should we identify it in the Lagrangian?
First, the superpotential should have the correct charge: \begin{equation}
R_P+R_\Phi+R_M=1
\label{eq1}
\end{equation} and then it needs to be free of mixed gauge anomaly: \begin{equation}
(R_\Phi-1)N_1-(R_P-1)N_2-R_\Psi N_3 -2R_\Omega =0.
\label{eq2}
\end{equation}
Since we have five unknowns $R_{P,\Phi,M,\Psi,\Omega}$
and two equations \eqref{eq1} and \eqref{eq2},
we see that there is a three-parameter family of conserved R-symmetries.

To identify exactly which conserved R-symmetry is in the superconformal algebra in the infrared,
we need to use the $c$-extremization,
which is an analogue of the $a$-maximization in four dimensions we reviewed in Sec.~\ref{sec:amax}.
In \eqref{ckappa}, we saw that the central charge $c$ is three times the coefficient of the $\U(1)_R^2$ anomaly, where $\U(1)_R$ is the R-symmetry in the superconformal algebra.
We can extend this formula to define a trial central charge $c$
for an arbitrary conserved R-symmetry.
Then the $c$-extremization simply says that the low-energy R-charge extremizes the trial central charge $c$ \cite{Benini:2012cz}.

The $c$-extremization is in fact rather easy to derive.
Suppose we have a $\U(1)$ current $J_F$ which does not act on the supercharge.
In the infrared, this means that the operators $J_F(z)$ and $G^\pm(z)$ have zero OPE.
From supersymmetry, this means that $J_F(z)$ and $J_R(z)$ have zero OPE.
Since the OPE coefficient between $J_F$ and $J_R$ is the derivative of the trial central charge $c$ 
in the direction of $J_F$, the $c$-extremization follows.

In our case, we obtain \begin{equation}
R_P=\frac{n_1}N,\quad
R_\Phi=\frac{n_2}N,\quad
R_M=\frac{n_3}N,\quad
R_\Psi=R_\Omega=0.
\end{equation}
Then we have \begin{equation}
\frac{c_R}3=n_3((R_\Phi-1)^2N_1+(R_P-1)^2N_2-R_\psi^2 N_3) -R_M N_1N_2 -2 R_\Omega^2 - n_3^2
=\frac{n_1n_2n_3}N,
\label{cR}
\end{equation}
which is cyclically symmetric.
$c_R-c_L$ can also be computed, since it is proportional to the pure gravitational anomaly, which can be computed in the Lagrangian description by simply counting the difference of the number of right-moving and left-moving fermions: \begin{align}
c_R-c_L&=n_3(N_1+N_2-N_3)-N_1N_2-2-n_3^2 \\
&= \frac14(N_1^2+N_2^2+N_3^2-2N_1N_2-2N_2N_3-2N_3N_1)-2.
\label{cL}
\end{align}

The elliptic genus can also be computed as before\cite{Benini:2013xpa};
the contributions from a fermi multiplet, from a chiral multiplet, and from a vector multiplet are
\begin{align}
\text{a Fermi}:&\frac{\theta_1(e^z,q)}{\eta(q)}, &
\text{a chiral}: & \frac{\eta(q)}{\theta_1(e^z,q)}, &
\text{$\U(N)$ vector}: &(\eta(q))^{2N} \prod_{i,j} \frac{\theta_1(e^{z_i-z_j},q)}{\eta(q)}.
\end{align} 
Then the gauge theory formula for the elliptic genus is \begin{multline}
Z=\frac{1}{n_3!} (\eta(q)^2)^{n_3} \oint 
\underbrace{\frac{dz_i}{2\pi \ii}\prod\frac{\theta_1(e^{z_i-z_j},q)}{\eta(q)}}_\text{vector}
\underbrace{\frac{\eta(q)}{\theta_1(e^{a_i-z_j},q)}}_P \times \\
\underbrace{\frac{\eta(q)}{\theta_1(e^{z_i-b_j},q)}}_\Phi
\underbrace{\frac{\theta_1(e^{-z_i-c_j},q)}{\eta(q)}}_\Psi
\underbrace{\frac{\theta_1(e^{-a_i+b_j},q)}{\eta(q)}}_M 
\left[
\underbrace{\frac{\theta_1(e^{z_1+\cdots+z_{n_3}},q)}{\eta(q)}}_\Omega
\right]^2 
\label{tri-ell}
\end{multline}
where we are supposed to take only the poles coming from the field $P$.
We end up summing over the subset $\{z_1,\ldots,z_{n_3}\}\subset \{a_1,\ldots, a_{N_2}\}$,
cancelling the $1/n_3!$ prefactor by the permutation.
The final formula is a gigantic sum over products of theta functions,
which happens to be manifestly invariant under the triality
without using any complicated identity of theta functions.

In fact, it is possible to propose a well-motivated complete description of  the low-energy limit.
First, from \eqref{SUN_i-anom},
one knows that there should be an $\SU(N_i)$ level $n_i$ current algebra on the left movers.
The Sugawara construction says that an $\SU(N)$ level $k$ current algebra contributes to the central charge by
\begin{equation}
c_\text{$\SU(N)$ level $k$} = \frac{k\dim \SU(N)}{N+k}
\end{equation}
We also have three $\U(1)_i$ currents, each of which contribute $1$ to the central charge.
It so happens that their sum saturates $c_L=(c_L-c_R)+c_R$ computed from \eqref{cR} and \eqref{cL}:
\begin{equation}
c_L=3+\sum_i \frac{n_1(N_i^2-1)}{N_i+n_i}.
\end{equation}
This means that the left movers can be described by the current algebras $\prod_i \SU(N_i)\times\U(1)_i$ alone.

Second, there is a nice \Nequals{2}-supersymmetric right-moving theory which can be combined with this left-moving current algebra.
This is given by the Kazama-Suzuki supercoset \cite{Kazama:1988qp,Kazama:1988uz} based on the coset \begin{equation}
\frac{\U(N)_N}{\prod_i \U(n_i)_N}
\end{equation} which also made an appearance in \cite{Gaiotto:2013gwa}.
Here, a super $\U(n)_N$ model consists of bosonic $\U(n)_{N-n}$ together with $n^2$ free fermions,
with the central charge \begin{equation}
c=\frac{(N-n)(n^2-1)}{N}+1+\frac{n^2}2.
\end{equation}
Then the super Kazama-Suzuki coset shown above has the central charge \begin{equation}
c_R=(\frac{N^2}2+1)-\sum_i(
\frac{(N-n_i)(n_i^2-1)}{N}+1+\frac{n_i^2}2
)
=3\frac{n_1n_2n_3}{N}
\end{equation} which magically agrees with \eqref{cR}.
As mentioned above,  the super $\U(n_i)_N$ contains the bosonic $\U(n_i)_{N-n_i}=\U(n_i)_{N_i}$,
and its level-rank dual $\U(N_i)_{n_i}$ appears on the left-moving side.
This fact allows us to glue the left-moving side and the right-moving side
in a modular-invariant way,
which gives a complete spectrum of the low-energy \Nequals{(0,2)} superconformal theory,
which is manifestly invariant under the triality.
One can compute the elliptic genus from this spectrum, which reproduces \eqref{tri-ell}.

This is a much more elaborate version of the agreement of the elliptic genus computed from the Landau-Ginzburg model $X^d$ and the elliptic genus computed from the unitary minimal model of type $A_{d-1}$ we saw in Sec.~\ref{sec:mlg}.
It would be fantastic if one can understand the low-energy superconformal theory of even a single 4d \Nequals1 Lagrangian theory, in a way similar to this case of 2d triality. 

\subsection{2d dualities from 4d dualities on $S^2$: generalities}
\label{sec:caution}
Next, we try to connect the triality of 2d \Nequals{(0,2)} theory we have been discussing so far to the Seiberg duality of 4d \Nequals1 theory we studied in detail in Sec.~\ref{sec:seiberg}.
We consider putting a 4d \Nequals1 theory on $S^2$ with a unit $\U(1)$ R-charge flux, and making the radius very small.
In Sec.~\ref{sec:expansion},
we study the 2d Lagrangian $X_\text{2d}$ obtained from putting a 4d \Nequals1 theory $X_\text{4d}$ on $S^2$ and keeping only the zero modes.
Then in Sec.~\ref{sec:trialityfromduality}, we see that, by putting Seiberg dual pairs of 4d $\U(N)$ gauge theories on $S^2$ and keeping only the zero modes, we obtain three 2d Lagrangians participating in the triality of Gadde, Gukov and Putrov.

We hasten to add that this \emph{does not} fully explain the 2d triality from the 4d duality.
The 4d duality is the statement that the low-energy limit of two 4d theories describes the same physics.
As long as we keep the radius of $S^2$ finite, we expect that the two dual 4d theories give rise to the same infrared physics. 
But keeping only the zero modes along $S^2$ is too crude.
For example, there are known cases where two dual 4d theories  $A_\text{4d}$, $B_\text{4d}$ on $S^2$, if we only keep the zero modes, give rise to two \emph{non}-dual 2d theories $A_\text{2d}$, $B_\text{2d}$.
For example, if we take a Seiberg  dual pair of $\SU$ theories as $A_\text{4d}$, $B_\text{4d}$,
the resulting 2d theories $A_\text{2d}$, $B_\text{2d}$ have different elliptic genera.

In \cite{Gadde:2015wta}, the supersymmetric index of 4d \Nequals1 theory on $S^2\times T^2$ was studied\footnote{%
There was an earlier but not quite complete analysis of the same setup in \cite{Honda:2015yha}.
}. 
It was observed there that 
\begin{itemize}
\item the supersymmetric index on $S^2\times T^2$ is independent of the radius of $S^2$,
\item two dual \Nequals1 theories have the same supersymmetric index on $S^2\times T^2$,
\item and there is a certain large class of 4d theories $X_\text{4d}$ where the supersymmetric index of $X_\text{4d}$ on $S^2\times T^2$ equals the elliptic genus of $X_\text{2d}$ on $T^2$,
where $X_\text{2d}$ is obtained by keeping only the zero modes along $S^2$.
\end{itemize}
The 4d dual pair leading to the 2d triality happens to be in this large class; the \Nequals{(0,4)} dualities of Putrov, Song and Yan \cite{Putrov:2015jpa} are another example.
It is fair to say, however, that the physical mechanism selecting this large class where the naive dimensional reduction works is not yet understood.

There is a similar issue in deriving 3d dualities by putting 4d dual theories on $S^1$,
or deriving 2d dualities from 3d dual theories on $S^1$.
In those cases, there is a satisfactory physical understanding given in \cite{Aharony:2013dha,Aharony:2013kma} for the reduction from 4d  to 3d  and in \cite{Aharony:2017adm} for the reduction from 3d to 2d.
It would be highly desirable to have a similar analysis for the reduction from 4d to 2d on $S^2$.

In this lecture note, we will be content on presenting 4d dual pairs on $S^2$
such that the 2d Lagrangians obtained by keeping only the zero modes along $S^2$ give rise to the 2d \Nequals{(0,2)} gauge theories related by triality.
The reader is welcome to come up with a physical explanation why this works.

\subsection{2d spectrum of 4d \Nequals1 theory on $S^2$}
\label{sec:expansion}
A generic 4d \Nequals1 theory on $S^2$ will not preserve any supersymmetry.
We require that the 4d \Nequals1 theory in question has a conserved $\U(1)$ R-symmetry.
The details of the Lagrangian can be found in e.g.~\cite{Closset:2013sxa,Closset:2014uda,Nishioka:2014zpa} 
and we only need its existence.
The essential point is that by introducing a unit flux of $\U(1)_R$ on $S^2$,
the curvature of the spinor bundle coming from the round metric of $S^2$ is canceled against the curvature coming from the R-charge, preserving 2d \Nequals{(0,2)} supersymmetry.
Note that we do \emph{not} have to use the $\U(1)_R$ in the superconformal group; rather, we just need  to pick a conserved and integrally-quantized $\U(1)$ R-symmetry, such that  the supercharges have charge $\pm1$.

We are interested in the massless modes in 2d when the radius of $S^2$ is very small. They can be read off 
from the spinor zero modes on $S^2$.  Let us put a 4d Weyl fermion with charge $q$ on $S^2$ with a unit monopole flux. Then there are $|q|$ 2d left-moving (right-moving) complex Weyl fermions when $q>0$ ($q<0$), while there are no zero modes when $q=0$. In either case, they transform in an irreducible representation of $\SU(2)$ rotating $S^2$.  

Recall that the chiral supercharge has R-charge $-1$. From this, we see that there is one complex right-moving supercharge in 2d, i.e.~it is \Nequals{(0,2)} supersymmetric.
We also find that when a chiral multiplet of R-charge $r$ is put on $S^2$, we have the following rule: \begin{itemize}
\item when $r>1$, there are $|r-1|$ Fermi multiplets, and
\item when $r<1$, there are $|r-1|$ chiral multiplets.
\end{itemize} 
Similarly, when a vector multiplet is put on $S^2$, the R-charge of the gaugino is fixed to 1, and therefore we find one \Nequals{(0,2)} vector multiplet in 2d.

\subsection{Triality from duality}
\label{sec:trialityfromduality}
The triality of Gadde, Gukov and Putrov is about $\U$ gauge theories, rather than $\SU$ theories.
This motivates us to consider 4d $\U(N_c)$ duality first.
An example can be obtained by starting from the 4d $\SU$ Seiberg duality
and then gauging the $\U(1)$ part.

More precisely,  we start from 4d $\SU(N_c)$ theory with $N_1$ flavors $Q^i$ and $\tilde Q_i$, $i=1,\ldots, N_1$.
We assign $\U(1)$ charge $\pm1$ to $Q^i$ and $\tilde Q_i$, respectively.
With this matter content, there is a nonzero $\U(1)^2\U(1)_R$ anomaly of strength \begin{equation}
2 (R(Q) -1 ) N_c N_f\cdot \frac13(\frac{F_{\U(1)}}{2\pi})^2\frac{F_{\U(1)R}}{2\pi} = -2N_c^2
\cdot  \frac13(\frac{F_{\U(1)}}{2\pi})^2\frac{F_{\U(1)R}}{2\pi}.
\end{equation} 
To cancel it, we introduce a pair of chiral superfields $\Omega$ and $\tilde\Omega$,
such that $\Omega$ transforms as the determinant of the fundamental of $\U(N_c)$,
and $\tilde\Omega$ as the determinant of the anti-fundamental,
both of which has R-charge 2.
This charge assignment also guarantees that the $\U(1)^3$ gauge anomaly cancels.
This procedure can be done simultaneously on the Seiberg dual pair of $\SU$ theories;
we obtain a dual pair of $\U$ theories in 4d.

We now pick  an R-charge assignment that is integral and anomaly-free, so that we can put the theory on $S^2$ preserving the 2d \Nequals{(0,2)} supersymmetry.
Let us say the R-charges of individual components of $Q_i$ and $\tilde Q_i$ are $r_i$ and $\tilde r_i$. Then the anomaly-free condition is \begin{equation}
\sum (r_i + \tilde r_i )= 2N_1-2N_c. \label{anomalyfree}
\end{equation}
A nice class of solutions is to take \begin{equation}
r_i=0; \qquad \tilde r_1=1-N_2,\ \tilde r_2=1+N_3,\ \tilde r_{i>2}=1.
\label{goodchoice}
\end{equation} From \eqref{anomalyfree},  we need $N_c=(N_1+N_2-N_3)/2$.
Let us add gauge-singlet mesons $M^{ij}$ where $i=1,\ldots, N_1$ but $j=1$ only, with the coupling $W=M^{ij}Q_{i}\tilde Q_j$. 
We assign the superpotential $W$ the R-charge 2. Then we see that $M$ has R-charge $1+N_2$. 
Call this 4d theory $A_\text{4d}$. 

In this 4d setup, we find the following non-R conserved symmetries: $\SU(N_1)$ acting on $Q$ and $\SU(2)$ rotating $S^2$, together with three $\U(1)$s: \begin{itemize}
\item $\U(1)_1$ acts on $Q_i$ with charge $+1$ and on $\tilde Q_3$ with charge $-N_f$,
\item $\U(1)_2$ acts on $\tilde Q_1$ with charge $+1$ and on $\tilde Q_3$ with charge $-1$, and 
\item $\U(1)_3$ acts on $\tilde Q_2$ with charge $+1$ and on $\tilde Q_3$ with charge $-1$.
\end{itemize}

By applying the rule above, from the theory $A_\text{4d}$ we obtain a 2d \Nequals{(0,2)} theory with the following spectrum: \begin{itemize}
\item $\U(N_c)$ vector multiplet,
\item $N_1$ chirals $\Phi$ in the fundamental representation,
\item $N_2$ chirals $P$ in the  anti-fundamental representation,
\item $N_3$ Fermis $\Psi$ in the anti-fundamental representation, 
\item $N_1N_2$ Fermis $M$ that are gauge neutral, and finally
\item two Fermis $\Omega$, $\tilde\Omega$ in the determinant representation,
\end{itemize} with a superpotential coupling $\int d\theta_+  \tr M \Phi P$ coming from the 4d superpotential $W=QM\tilde Q$.  

The 2d theory we obtained can be summarized in the following quiver diagram 
\begin{equation}
A_{2d}=
\begin{tikzpicture}[my]
\draw (0,0) node[circle,draw] (A) {$n_3$};
\draw (0,1) node[rectangle,draw] (B) {$N_2$};
\draw (-1,-1) node[rectangle,draw] (C) {$N_1$};
\draw (1,-1) node[rectangle,draw] (D) {$N_3$};
\draw[->] (C)--(A);
\draw[->] (A)--(B);
\draw[->,dotted] (A)--(D);
\draw[->,dotted,curve] (B) to (C);
\draw (1.2,0) node (E) {$\Omega_{1,2}$};
\draw[-,snake it] (A)--(E);
\end{tikzpicture}.
\end{equation}
Here, we used $N=(N_1+N_2+N_3)/2$ and $n_i=N-N_i$ as before; we have $N_c=n_3$.
This is exactly the original theory \eqref{triality-original}.

The 4d Seiberg dual of theory $A_\text{4d}$ has the gauge group $\U(n_2)$ with $N_1$ flavors $q^i$ and $\tilde q^i$,  transforming in the conjugate representations of the original $\SU(N_1)\times\SU(N_1)$. 
The dual R-charge assignment can be worked out by comparing the charges of baryons and mesons:
\begin{equation}
r_i= 0; \qquad
\tilde r_1=1+N_2,\ \tilde r_2=1-N_3, \ \tilde r_{i>2}=1.
\end{equation} 
We also have gauge-singlet mesons $M_{ij}'$ where $i=1,\ldots,N_1$  but now $j>1$,
and two fields $\Omega$, $\tilde\Omega$ cancelling the $\U(1)_\text{gauge}$ anomaly.
Call this theory $B'_\text{4d}$.

We put the theory $B'_\text{4d}$ on $S^2$, and the following 2d theory is obtained:  
\begin{equation}
B_{2d}=\begin{tikzpicture}[my]
\draw (0,0) node[circle,draw,] (A) {$n_2$};
\draw (0,1) node[rectangle,draw] (B) {$N_2$};
\draw (-1,-1) node[rectangle,draw] (C) {$N_1$};
\draw (1,-1) node[rectangle,draw] (D) {$N_3$};
\draw[<-] (C)--(A);
\draw[<-,dotted] (A)--(B);
\draw[<-,] (A)--(D);
\draw[->,dotted,curve] (C) to (D);
\draw (1.2,0) node (E) {$\Omega_{1,2}$};
\draw[-,snake it] (A)--(E);
\end{tikzpicture}
=\begin{tikzpicture}[my]
\draw (0,0) node[circle,draw,] (A) {$n_2$};
\draw (0,1) node[rectangle,draw] (B) {$N_2$};
\draw (-1,-1) node[rectangle,draw] (C) {$N_1$};
\draw (1,-1) node[rectangle,draw] (D) {$N_3$};
\draw[<-] (C)--(A);
\draw[->,dotted] (A)--(B);
\draw[<-,] (A)--(D);
\draw[->,dotted,curve] (C) to (D);
\draw (1.2,0) node (E) {$\Omega_{1,2}$};
\draw[-,snake it] (A)--(E);
\end{tikzpicture}.
\end{equation}
Here we used the fact that in 2d \Nequals{(0,2)} theories, having a Fermi multiplet in a representation $R$ is the same as having a Fermi multiplet in a representation $\bar R$, after exchanging its $J$ and $E$ interaction.  

This 2d theory $B_{2d}$ is exactly what Gadde, Gukov and Putrov discussed, see \eqref{triality-after}.
A natural question is how the triality can be obtained from the duality. The answer is that the 2d theory $B_{2d}$ can be obtained from a 4d theory $B_\text{4d}$ different from the theory $B'_\text{4d}$ used above. 
Namely, consider a 4d theory $B_\text{4d}$ with gauge group $\U(n_2)$ with $N_3$ flavors $q_i$, $\tilde q_i$ 
with the R-charge assignment \begin{equation}
r_i=0;\qquad
\tilde r_1=1-N_1,\ 
\tilde r_2=1+N_2,\ 
\tilde r_{i>2}=1,
\end{equation} together with gauge-singlet mesons $M_{ij}$ for $j=1$. 
and $\Omega$, $\tilde\Omega$.

By putting it on $S^2$, you get the 2d theory $B_{2d}$ exactly as before, by following the same procedure. 
Now, we apply the Seiberg duality to the 4d theory $B_\text{4d}$.  What you get is the 4d theory $C'_\text{4d}$ which is $\U(n_1)$ with $N_3$ flavors, with the R-charge assignment \begin{equation}
r_i=0;\qquad
\tilde r_1=1+N_1,\ 
\tilde r_2=1-N_2,\ 
\tilde r_{i>2}=1,
\end{equation} which gives the 2d theory $C_{2d}$ given by 
\begin{equation}
C_{2d}=\begin{tikzpicture}[my]
\draw (0,0) node[circle,draw,] (A) {$n_1$};
\draw (0,1) node[rectangle,draw] (B) {$N_2$};
\draw (-1,-1) node[rectangle,draw] (C) {$N_1$};
\draw (1,-1) node[rectangle,draw] (D) {$N_3$};
\draw[<-,dotted] (C)--(A);
\draw[<-] (A)--(B);
\draw[->] (A)--(D);
\draw[->,dotted,curve] (D) to (B);
\draw (-1.2,0) node (E) {$\Omega_{1,2}$};
\draw[-,snake it] (A)--(E);
\end{tikzpicture},
\end{equation}
which reproduces $C_{2d}$ of \eqref{triality-after}.

Again, this 2d theory $C_{2d}$ comes from another 4d theory $C_\text{4d}$ which is $\U(n_1)$ with $N_2$ flavors, with the R-charge assignment \begin{equation}
r_i=0;\qquad
\tilde r_1=1+N_1,\ 
\tilde r_2=1-N_3,\ 
\tilde r_{i>2}=1.
\end{equation}
Its Seiberg dual is the theory $A'_\text{4d}$, which is $\U(n_3)$ with $N_2$ flavors, with the R-charge assignment \begin{equation}
r_i=0;\qquad
\tilde r_1=1-N_1,\ 
\tilde r_2=1+N_3,\ 
\tilde r_{i>2}=1.
\end{equation} 
When we put this theory $A'_\text{4d}$ on $S^2$, we find the original 2d theory $A_{2d}$.

Summarizing, the 2d theory $A_{2d}$ comes either from the 4d theory $A_\text{4d}$ or $A'_\text{4d}$,
$B_{2d}$ from $B_\text{4d}$ or $B'_\text{4d}$, and $C_{2d}$ from $C_\text{4d}$ or $C'_\text{4d}$. 
The 4d Seiberg dual pairs are $A_\text{4d}\leftrightarrow B'_\text{4d}$, $B_\text{4d}\leftrightarrow C'_\text{4d}$, $C_\text{4d}\leftrightarrow A'_\text{4d}$. 
Therefore, we have the triality $A_{2d}\to B_{2d}\to C_{2d}\to A_{2d}$ in 2d. 

As already emphasized in Sec.~\ref{sec:caution},
this \emph{does not} fully explain the 2d triality as a consequence of the 4d duality on $S^2$.
Hopefully, one of the readers will get interested and identify the physical mechanism
behind the selection rule presented in Sec.~\ref{sec:caution}.

\part{Appendix}

\section{Supersymmetry in diverse dimensions}
\label{sec:various}
In this appendix, we will give  a brief summary of the structure of supersymmetry and superconformal symmetry in various dimensions.\footnote{%
Another good reference is Appendix B of Polchinski's volume 2 \cite{PolchinskiII}.
The ultimate reference in the case of supergravity in various dimensions is \cite{SugraDiverse}.
}
Recall that the basic ingredient of supersymmetry is the anti-commutator $ 
\{ Q_\alpha, Q_{\dot \beta}\}\sim P_\mu \gamma^\mu_{\alpha\dot \beta}.
$
Therefore it is imperative for us to learn the property of spinors in various dimensions.

\subsection{Spinors in various dimensions}
\subsubsection{Construction of gamma matrices}
We need to study spinors of both $\so(d)$ and $\so(d-1,1)$.
Let us start with $\so(d)$.
We first need the gamma matrices satisfying the relation\begin{equation}
\{\Gamma^i,\Gamma^j\} = +2\delta^{ij}
\end{equation}
for $i=1,\ldots, d$.
This anticommutation relation guarantees that the generators \begin{equation}
M^{ij}=\frac12[\Gamma^i,\Gamma^j]\label{spinor-generators}
\end{equation} satisfy the commutation relation of the $\so(d)$ Lie algebra.

Let us show inductively that we can choose such a set of gamma matrices as $2^n\times 2^n$ matrices when $d=2n$ and $d=2n+1$.
We start from $\so(2)$, for which we can take \begin{equation}
\Gamma^1=\sigma^x=\begin{pmatrix}
0 & 1 \\
1 & 0
\end{pmatrix},\quad 
\Gamma^2=\sigma^y=\begin{pmatrix}
0 &- i \\
i & 0.
\end{pmatrix}
\end{equation}

Now, suppose we already have a set of gamma matrices $\Gamma^1,\ldots,\Gamma^d$ for $d=2n$. We can define \begin{equation}
\Gamma^{2n+1} := \mathrm{i}^n \Gamma^1 \Gamma^2 \cdots \Gamma^{2n}
\end{equation} to upgrade this set to gamma matrices for $d={2n+1}$.
Note that for $d=2$ we have \begin{equation}
\Gamma^3= \sigma^z=\begin{pmatrix}
1 & 0 \\
0 & -1
\end{pmatrix}.
\end{equation}

We can combine gamma matrices $\Gamma^1,\ldots,\Gamma^{2n}$ for $d=2n$
and $\tilde\Gamma^1,\ldots,\tilde\Gamma^{2\tilde n}$ for $d=2\tilde n$
to obtain $\hat\Gamma^1,\ldots,\hat\Gamma^{2n+2\tilde n}$ for $d=2n+2\tilde n$.
The explicit formula is the following: 
\begin{equation}
\begin{array}{llllr@{\,}l}
\hat\Gamma^i &= \Gamma^i &\otimes& \mathbf{1}, & (i&=1,\ldots, 2n) \\
\hat\Gamma^{2n+j}&=\Gamma^{2n+1} &\otimes& \tilde\Gamma^j & (j&=1,\ldots, 2\tilde n)
\end{array}
\label{combine-gamma}
\end{equation}
where we remind the reader that for two matrices $A_{st}$ and $B_{uv}$,
their tensor product $A\otimes B$ is the matrix whose elements are given by $(A\otimes B)_{su,tv}:=A_{st} B_{uv}$.
This completes the proof that for $d=2n$ we can take $\Gamma^{i}$ to be $2^n\times 2^n$ matrices.
At the same time, we have given explicit forms of gamma matrices.

The $2^n$-dimensional space on which these gamma matrices act is called the Dirac spinor,
 for $\so(2n)$ and $\so(2n+1)$.
For $\so(2n)$, the generators \eqref{spinor-generators} commute with $\Gamma^{2n+1}$.
Therefore the $2^{n-1}$-dimensional eigenspaces satisfying $\Gamma^{2n+1}=\pm1$ are themselves representations of $\so(2n)$.
These two representations are called positive/negative or left-handed/right-handed Weyl spinors.
Indices for the positive Weyl spinor and for the negative Weyl spinors are often distinguished by using undotted $\alpha$ for the former and dotted $\dot\beta$ for the latter.

\subsubsection{Reality properties}
Let us study the reality properties of spinors of $\so(d)$.
This can be done in many ways. Here we use accidental isomorphisms of $\so(d)$ with other Lie algebras for low $d$:
\begin{itemize}
\item For $\so(1)$, the Dirac spinor is clearly strictly real, since we can simply take $\Gamma^1=1$.
\item For $\so(2)$, we have $\so(2)\simeq \mathfrak{u}(1)$. 
Under the identification that the standard 2-dimensional vector representation of $\so(2)$ splits as charge-$(\pm1)$ representations of $\mathfrak{u}(1)$,
the Weyl spinors of $\so(2)$ are charge-$(\pm\frac12)$ representations.
In particular, the complex conjugate of one Weyl spinor is the other Weyl spinor.
\item For $\so(3)$, we have $\so(3)\simeq\su(2)$, 
and the spinor representation is the 2-dimensional fundamental representation of $\su(2)$, often denoted by $\mathbf{2}$.
This is a real representation in the sense that $\bar{\mathbf{2}}=\mathbf{2}$.
In other words there is a conjugate-linear map $*:\mathbf{2}\to\mathbf{2}$ which commutes with the action of $\so(3)\simeq\su(2)$.
Explicitly, it is given by \begin{equation}
(u,v)\mapsto (-\bar v,\bar u).
\end{equation}
This map $*$ squares to minus one: $**=-1$.
Such a real representation is called a pseudoreal representation.
\item For $\so(4)$, we have $\so(4)\simeq \su(2)\times\su(2)'$. 
The two Weyl spinors of $\so(4)$ are the two-dimensional fundamental representations of $\su(2)$ and $\su(2)'$. 
This tells us that the complex conjugate of a Weyl spinor is the same Weyl spinor.
\item For $\so(5)$, we have $\so(5)\simeq \sp(2)$,
and the Dirac spinor of $\so(5)$ is the fundamental 4-dimensional representation of $\sp(2)$, often denoted by $\mathbf{4}$. 
This is a pseudoreal representation.
For more on the group $\Sp(n)$, see Section~\ref{sec:sp}.
\item For $\so(6)$, we have $\so(6)\simeq\su(4)$,
and the Weyl spinors of $\so(6)$ are simply the fundamental representation $\mathbf{4}$ and the anti-fundamental representation $\bar{\mathbf{4}}$ of $\su(4)$.
This means that the complex conjugate of a Weyl spinor is another Weyl spinor.
\item For $\so(7)$, we do not have a good accidental isomorphism to study the Dirac spinor of $\so(7)$, which is eight dimensional and often denoted by $\mathbf{8}$. 
Its reality can be understood by restricting a Weyl spinor of $\so(8)$ to $\so(7)$.
It can be also understood by using the octonion $\bO$, since $\Im \bO$ acting on $\bO$ provide gamma matrices for $\so(7)$.
Either way, one sees that the spinor is a strictly real representation, i.e.~there is a conjugate linear map $*:\mathbf{8}\to\mathbf{8}$ commuting with the $\so(7)$ action such that $**=+1$. 
In this case the subspace of $\mathbf{8}$ where $*$ acts by $+1$  is a 8-dimensional real vector space.
This real subspace is also known as a Majorana spinor representation.
\item The algebra $\so(8)$ has three eight-dimensional irreducible representations:
the 8-dimensional vector representation of $\so(8)$ is $\mathbf{8}_V$ 
and two Weyl spinors are $\mathbf{8}_S$ and $\mathbf{8}_C$.
These three are permuted by the outer automorphism $S_3$ fo $\so(8)$, which can be seen from its Dynkin diagram: \begin{equation}
\begin{tikzpicture}[baseline=(A)]
\node(A) at (0,0) {$\circ$};
\node(V) at (-1,0) {$\circ$};
\node(S) at (.5,.8) {$\circ$};
\node(C) at (.5,-.8) {$\circ$};
\draw[-] (A) to (V);
\draw[-] (A) to (S);
\draw[-] (A) to (C);
\end{tikzpicture}\, .
\end{equation}
Since $\mathbf{8}_V$ is clearly a strictly real representation, it follows that two Weyl spinors $\mathbf{8}_S$ and $\mathbf{8}_C$ are also strictly real.
In particular, this means that the complex conjugate of a Weyl spinor is the same Weyl spinor.
\end{itemize}

Since the spinors of $\so(8)$ are strictly real, it follows from \eqref{combine-gamma} that the reality properties of spinors of $\so(d+8)$ and those of $\so(d)$ are the same.
This is one example of a mathematical phenomenon called the Bott periodicity.

\subsubsection{Spinors of $\so(d-1,1)$}
To understand the spinors of $\so(d-1,1)$ instead of $\so(d)$, we need gamma matrices satisfying \begin{equation}
\{\Gamma^\mu,\Gamma^\nu\} = +2\eta^{\mu\nu}
\end{equation} where for concreteness we choose the mostly-plus convention for the metric $\eta^{\mu\nu}$ of the Minkowski space.
We note that the gamma matrices of $\so(1,1)$ and $\so(2,1)$ can be chosen so that \begin{equation}
\Gamma^0=\begin{pmatrix}
0 & -1 \\
1 & 0
\end{pmatrix},\quad
\Gamma^1=\begin{pmatrix}
0 & 1 \\
1 & 0
\end{pmatrix},\quad
\Gamma^2=\begin{pmatrix}
1 & 0 \\
0 & 1
\end{pmatrix}.
\label{so(2,1)-gamma}
\end{equation}
We can then combine the gamma matrices for $\so(1,1)$ and those for $\so(d-2)$ 
as in \eqref{combine-gamma} to form the required gamma matrices for $\so(d-1,1)$.
Since the gamma matrices for $\so(1,1)$ shown in \eqref{so(2,1)-gamma} are all real matrices,
it follows that the spinors of $\so(d-2)$ and $\so(d-1,1)$ behave the same way under the complex conjugation and therefore have the same reality conditions one can impose.

\exercise{We often perform the Wick rotation and study systems with $\so(d)$ spacetime symmetry
rather than $\so(d-1,1)$ symmetry. Should we use the reality properties of $\so(d)$  spinors rather than those of $\so(d-1,1)$?}

\answer No.  
In the Wick-rotated Euclidean theory, the reality property to be imposed does not demand that things are real in the naive sense of the word.
For example, consider the  Lagrangian of the gauge field \begin{equation}
\mathcal{L}_\text{L}:=\frac{1}{g^2}F_{\mu\nu}F^{\mu\nu} + \theta F_{\mu\nu}F_{\rho\sigma} \epsilon^{\mu\nu\rho\sigma}
\end{equation}
 in the Lorentzian signature.
 This is real, but is transformed to \begin{equation}
\mathcal{L}_\text{E}:=\frac{1}{g^2}F_{\mu\nu}F^{\mu\nu} + i\theta F_{\mu\nu}F_{\rho\sigma} \epsilon^{\mu\nu\rho\sigma} 
\end{equation}
in the Euclidean signature. 
This is no longer real, but has a modified reality condition:
\begin{equation}
\mathcal{L}_\text{E}= P \overline{\mathcal{L}_\text{E}}
\end{equation}
where $P$ is a parity action \begin{equation}
P:(x^1,x^2,\ldots, x^d)\mapsto (x^1,x^2,\ldots,-x^d).
\end{equation}
This extra action of $P$ in the reality condition comes from the fact that $x^d$ in the Wick-rotated spacetime originates from $ix^0$ in the original Lorentizan spacetime, and therefore multiplied by $-1$ when the complex conjugation is performed.

We can say that the Euclidean Lagrangian should be \emph{reflection real}.\footnote{%
This is not a standard terminology, but is a useful one when we discuss reality conditions in the Euclidean theory. 
In the end, we never impose the naive reality in the Euclidean theory. 
Therefore we can simply say \emph{real} instead of \emph{reflection real}.
The phrase \emph{reflection reality} was modeled after the concept called the reflection positivity, which is the Euclidean counterpart of the unitarity in the Lorentzian theory.
}
The fields are used to construct the Lagrangian.
Therefore, the reality conditions on fields in the Euclidean signature should make fields to be reflection real, and not real in the naive sense of the word.
Practically, this means that a real field should belong to a representation on which rotations act linearly and reflections act conjugate-linearly.\footnote{%
Note that the conjugate-linear action of the reflection comes for free, even in a parity non-symmetric theory.
If the original Lorentizan theory is parity symmetric, 
 we also have linear actions of reflections on the fields.
}

We can check that the reflection reality on  $\so(d)$ spinors can be imposed if and only if the standard reality condition can be put on $\so(d-1,1)$ spinors.
As an example, let us consider a Majorana-Weyl spinor $\psi$ for $\so(1,1)$.
In the Euclidean signature, $\SO(2)$ consists of the rotations by the angle $\varphi$,
and the parity $P$ sends $\varphi$ to $-\varphi$.
The Weyl spinor is indeed a reflection-real representation under which the $\varphi$ rotation acts by $e^{i\varphi/2}$ and the parity $P$ sends $\psi$ to $\psi^*$.

In a Euclidean theory, the correlators of reflection real fields are reflection real, since both the operator insertions and the weight factor $e^{-S_\text{E}}$ in the path integral are reflection real.
Again, let us take a 2d Majorana-Weyl fermion $\psi$. 
The two-point function of a free Majorana-Weyl spinor $\psi$ is given by \begin{equation}
\vev{\psi(z,\bar z) \psi(w,\bar w)} \propto \frac{1}{z-w}.
\end{equation}
This is complex, but is reflection real.
Indeed, let $z=x+iy$, and say that the parity $P$ sends $(x,y)$ to $(x,-y)$.
Then we indeed find that $z=P\overline{z}$.

\subsection{Possible types of supersymmetry}
We can now study the structure of supersymmetry in various dimensions.
In odd dimensions, we have supercharges $Q^i_\alpha$ in the Dirac spinor representation.
When the spinor is strictly real, we require $Q^i_\alpha$ to be Hermitean.
When the index $i$ runs from $1$ to $c$, it is called \Nequals{c} supersymmetric.

In even dimensions, we can have supercharges $Q^i_\alpha$ and $Q^j_{\dot\beta}$ in two Weyl representations.
When a complex conjugate of a Weyl representation is a different Weyl representation,
$Q^j_{\dot\beta}$ is simply the complex conjugate of $Q^i_\alpha$.
When the index $i$ then runs from $1$ to $c$,  it is again called \Nequals{c} supersymmetric.

Otherwise, one can choose the number of $Q^i_\alpha$ and $Q^j_{\dot\beta}$ independently.
Say the index $i$ runs from $1$ to $c_+$ and $j$ from $1$ to $c_-$.
Then the system is called \Nequals{(c_+,c_-)} supersymmetric.
When the Weyl spinor is strictly real, we require $Q^i_\alpha$ and $Q^j_{\dot\beta}$ to be Hermitean.

An R-symmetry is a symmetry which does not commute with the supersymmetry generators.
The maximal R-symmetry one can consider for \Nequals{c} supersymmetry is $\mathrm{U}(c)$, $\mathrm{O}(c)$, $\mathrm{Sp}(c)$, depending on whether the supercharges are complex, strictly real, or pseudoreal, respectively.
In even dimensions where we have \Nequals{(c_+,c_-)} supersymmetry,
the maximal R-symmetry is  $\mathrm{O}(c_+)\times \mathrm{O}(c_-)$
or $\mathrm{Sp}(c_+)\times\mathrm{Sp}(c_-)$, again depending on the reality of the supercharges.
Note that the actual R-symmetry group of a specific theory can be smaller than the maximal R-symmetry we just listed.

\begin{table}
\[
\begin{array}{r|cccc|ccccc}
\text{spacetime}& \text{type of} &  \text{dim$_\bC$ of} & \text{type of} & \text{dim$_\bR$ of} & \text{types of} & \text{maximal}\\
\text{dimension} & \text{spinor} & \text{spinor} & \text{reality} & \text{spinor} &  \text{susy} &\text{R symmetry}\\
\hline 
\so(1,1) & \text{Weyl} & 1 & \text{strictly real} &  1 &\cN{=} (c_+,c_-)  & \mathrm{O}(c_+)\times \mathrm{O}(c_-)\\
\so(2,1) & \text{Dirac} & 2 & \text{strictly real} &  2 &\cN{=} c & \mathrm{O}(c)\\
\so(3,1) & \text{Weyl} & 2 & \text{complex} &  4 & \cN{=}c & \mathrm{U}(c)\\
\so(4,1) & \text{Dirac} & 4 & \text{pseudoreal} &  8 & \cN{=}c & \mathrm{Sp}(c) \\
\so(5,1) & \text{Weyl} & 4 & \text{pseudoreal} &  8 & \cN{=}(c_+,c_-) & \mathrm{Sp}(c_+)\times \mathrm{Sp}(c_-)\\
\so(6,1) & \text{Dirac} & 8 & \text{pseudoreal} &  16 & \cN{=}c & \mathrm{Sp}(c) \\
\so(7,1) & \text{Weyl} & 8 & \text{complex} &  16 & \cN{=}c & \mathrm{U}(c)\\
\so(8,1) & \text{Dirac} & 16 & \text{strictly real} &  16 & \cN{=}c & \mathrm{O}(c)\\
\so(9,1) & \text{Weyl} & 16 & \text{strictly real} &  16 & \cN{=}(c_+,c_-) & \mathrm{O}(c_+)\times \mathrm{O}(c_-)\\
\so(10,1) & \text{Dirac} & 32 & \text{strictly real} &  32 & \cN{=}c & \mathrm{O}(c)\\
\so(11,1) & \text{Weyl} & 32 & \text{complex} &  64 & \cN{=}c & \mathrm{U}(c)
\end{array}
\]
\caption{Basic data of supersymmetry in various dimensions.\label{table:susytable}}
\end{table}

For an easy reference, various basic properties of the supersymmetry is tabulated in Table~\ref{table:susytable}.
There, for each spacetime dimension, we listed the type of the minimal spinor representation, its complex dimension, the type of reality condition which can be imposed on it, and the real dimension of the smallest spinor; 
then the types of supersymmetry, and finally the maximal R-symmetry allowed.

The concept of the total number of real component of supercharges is also useful;
this number is often just called the \emph{number of supercharges}.\footnote{%
Superconformal theories have $S$-supercharges in addition to $Q$-supercharges, as we will explain soon below.
For the number of supercharge, we do not usually count the $S$-supercharges.}
For example, a 3d \Nequals{6} theory would have $6\times 2=12$ supercharges.
The usefulness comes from the fact that supersymmetric theories with the same number of real component of supercharges behave quite similarly across spacetime dimensions.\footnote{%
The author has mainly been an 8-supercharge person.}

As an example of this behavior, let us consider the spin of particles contained in a massless representation of supersymmetry. 
Let us say a massless particle in $d\ge 4$ dimensions has the momentum \begin{equation}
P^\mu=(E,E,0,\ldots,0),
\end{equation} or in the lightcone coordinates, $P^+=E$, $P^-=0$, $P^i=0$ where $i=1,2,\ldots, d-2$.
In $d\ge 4$, the number of supercharges is automatically a multiple of $4$, so let us denote it by $4k$ where $k$ is an integer.
We denote the supercharges by $Q^{+i}$ and $Q^{-i}$, each Hermitean and $i=1,\ldots,2k$,
with the anticommutation relation \begin{equation}
\{Q^{+i},Q^{+j}\}=P^+ \delta^{ij},\qquad
\{Q^{-i},Q^{-j}\}=P^- \delta^{ij}.
\end{equation}
We see that $Q^{+i}$'s form $k$ pairs of fermionic oscillators,
whereas $Q^{-i}$'s are simply zero.
Since each fermionic oscillator change the helicity by $\pm1/2$,
any supermultiplet would contain a particle such that $|\text{helicity}| \ge k/4 $.

For example, when there are 16 supercharges (and therefore $k=4$), the  multiplet with the smallest $|\text{helicity}|$ has the helicity content \begin{equation}
-1,\quad
-\frac12,\quad
0,\quad
+\frac12,\quad
+1.
\end{equation}
Therefore, any massless theory in $d\ge 4$ with at least 16 supercharges
 will automatically have a massless vector field and will be a gauge theory.
This is true  in particular for all supersymmetric theories in $d\ge 7$.

Similarly, when there are 32 supercharges, any massless theory will automatically have a massless field of spin 2, and will contain dynamical gravity.
In particular, any supersymmetric theory in $d\ge 11$ will be a gravitational theory.
When we go one step further, we find that when there are 64 supercharges,
any massless theory will automatically have  massless higher-spin particles.
This is forced on us in $d\ge 12$, which is one reason why such theories are not  usually considered in the literature.

\subsection{Possible types of superconformal symmetries}
Let us next briefly study the superconformal symmetries.
In $d$ spacetime dimensions,
the Lorentz algebra is $\so(d-1,1)$.
The conformal algebra is then $\so(d,2)$.
Its generators are commonly denoted by $M_{\mu\nu}$ for $\so(d-1,1)$ generators,
$P_\mu$ and $K_\mu$ for translations and special conformal transformations,
and $D$ for the dilatation.

In the superconformal case, we first have the ordinary supercharges $Q_\alpha$.
Its commutator with $K_\mu$ is denoted by $S^\alpha$, and known as the superconformal supercharges.
The anti-commutator of $Q_\alpha$ and $S^\alpha$ will involve
not only $\so(d,2)$ generators but also R-symmetries.

The entire superconformal algebra is of the form \begin{equation}
\underbrace{\so(d,2)\oplus\text{R-symmetry}}_\text{bosonic} \oplus 
\underbrace{(Q_\alpha,S^\alpha)}_\text{fermionic}.
\end{equation}
All possible superconformal algebras were classified in \cite{Nahm:1977tg}.
Here a short summary will be presented.

We restrict attention to $d>2$, because the $d=2$ case is very special in that the non-supersymmetric conformal algebra itself is enhanced from $\so(2,2)$ to the infinite-dimensional $\text{Virasoro}\oplus\overline{\text{Virasoro}}$ algebra.
We demand that all R-symmetry generators do act nontrivially on $Q$ and $S$.
Then the superconformal algebra is simple.\footnote{%
In the technical sense of not having a nontrivial ideal.}

Now we can quote the classification of simple super Lie algebras \cite{Kac:1977qb}.
The bosonic part of the superconformal algebra is reductive\footnote{%
This means that it is  a direct sum of simple Lie algebras and $\mathfrak{u}(1)$'s. }
since it is a direct sum of $\so(d,2)$ and R-symmetries.
Therefore we only need to consider simple super Lie algebras whose bosonic part is reductive.
Their classification is taken from \cite{Kac:1977qb} and shown in Table~\ref{table:superLie}.

\begin{table}
\[
\begin{array}{r@{}l|r@{}l|r@{\,}l}
&&&\text{boson} &&\text{fermion} \\ 
\hline
\mathfrak{su}&(N|M) & \mathfrak{su}(N)&\oplus \mathfrak{su}(M) \oplus \mathfrak{u}(1) & \mathbf{N}&\otimes \bar{\mathbf{M}} 
\oplus \bar{\mathbf{M}}\otimes {\mathbf{N}}  \\
\mathfrak{osp}&(N|2M) & \mathfrak{so}(N)&\oplus\mathfrak{sp}(M) & \mathbf{N}&\otimes \mathbf{2M} \\
D&(2,1;\alpha)& \mathfrak{su}(2)&\oplus \mathfrak{su}(2)'\oplus \mathfrak{su}(2)''& \mathbf{2}&\otimes \mathbf{2}'\otimes \mathbf{2}'' \\
G&(3) & \mathfrak{g}_2&\oplus \mathfrak{su}(2) & \mathbf{7}&\otimes\mathbf{2}\\
F&(4)  &  \mathfrak{so}(7)&\oplus \mathfrak{su}(2) & \mathbf{8}&\otimes \mathbf{2} \\
\hline
P&(N) &\span{ \mathfrak{su}(N+1)} &   \mathbf{sym.}&\oplus\mathbf{antisym.}\\
Q&(N)& \span{ \mathfrak{su}(N+1)}& \mathbf{adj.}
\end{array}
\]
\caption{Data of simple super Lie algebras whose bosonic part is reductive. 
More precisely, $\su(N|N)$ needs to be replaced by $\mathfrak{psu}(N|N)$ which is obtained by removing the $\mathfrak{u}(1)$ part. 
$P(N)$ and $Q(N)$ are more exotic than the others in the sense that they have no invariant inner product.
\label{table:superLie}}
\end{table}

Now, we require for superconformal algebras that the fermion part is a spinor representation of  $\so(d-2,2)$ algebra. 
However, the Table~\ref{table:superLie} does not explicitly have spinors of $\so$ algebras, except for the exceptional case  of $F(4)$.
In other cases, we need to use accidental isomorphisms for low $d$ to convert a spinor of an $\so$ algebra into something else.
This means that the superconformal algebras only exists for low enough $d$.
We find only the possibilities listed in the Table~\ref{table:SC}.
In particular, the superconformal symmetries exist only up to $d=6$,
and one can only have \Nequals1 in $d=5$.

The definitive reference of superconformal multiplets is \cite{Cordova:2016emh}.
There, it is also shown that for $d=6$ and $d=4$, there are no superconformal theories with more than 16 supercharges,
while for $d=3$, there are superconformal theories with more than 16 supercharges but such theories are necessarily free.

\begin{table}
\[
\begin{array}{c|r@{}l|r@{}l|r@{}l|r@{}l}
\text{dimension} & \multicolumn{2}{c|}{\text{boson}} & \multicolumn{2}{c|}{\text{fermion}} & \multicolumn{2}{c|}{\text{total algebra}} & \multicolumn{2}{c}{\text{susy}}\\
\hline
3 & \so(3,2)&\oplus\so(N) & \mathbf{4}&\otimes \mathbf{N} & \mathfrak{osp}&(N|4)  & \cN&=N\\
4 & \so(4,2)&\oplus\su(N) & \mathbf{4}\otimes \mathbf{N} &\oplus \overline{\mathbf{4}}\otimes\overline{\mathbf{N}} & \mathfrak{su}&(4|N)  & \cN&=N\\
5 & \so(5,2)&\oplus\su(2) & \mathbf{8}&\otimes \mathbf{2} & F&(4) &\cN&=1\\
6 & \so(6,2)&\oplus\sp(N) & \mathbf{8}&\otimes \mathbf{2N}  & \mathfrak{osp}&(8|2N) &\cN&=(N,0)
\end{array}
\]
\caption{Data of superconformal algebras in $d\ge 3$.
Again, for $d=4$ and $N=4$, one needs to remove the $\mathfrak{u}(1)$ part.
\label{table:SC}}
\end{table}

\subsection{Summary}

The information obtained so far is summarized in the big Table~\ref{table:big}.
When referring to the table, it should be noted that:
\begin{itemize}
\item Only those supersymmetries which allow non-gravitational multiplets are listed.
\item In $3\le d\le 10$, all possibilities are listed, up to the exchange of $\cN{=}(c_+,c_-)\leftrightarrow \cN{=}(c_-,c_+)$.
\item In $d=2$ there are many more possibilities.
\item Those with {\bfseries\color{DarkGreen} superconformal algebras are shown in bold green}.
\item Those with \colorbox{blue!10}{16 supercharges have blue background}.
\item Those with  \colorbox{yellow!20}{8 supercharges have yellow background}.
\item Those with  \colorbox{red!10}{4 supercharges have red background}.
\end{itemize}

We also have a couple of comments:
\begin{itemize}
\item The dimensional reduction on $S^1$ of a theory in $d=D$ dimension with a supersymmetry 
given in an entry
will result in the theory in $d=D-1$ dimension in the entry just below the original entry,
with the understanding that both 6d \Nequals{(1,1)} and 6d \Nequals{(2,0)} 
reduce to \Nequals2 in 5d.
\item In the 5d supergravity literature, what we refer to $\cN{=}c$ supersymmetry is often called $\cN{=}2c$ supersymmetry.
\end{itemize}

\def\green#1{\mathbf{\color{DarkGreen}#1}}
\begin{table}
\[
\begin{array}{l|lcccccccc}
d=12 & \multicolumn{8}{c}{\text{automatically higher spin theory}}\\
d=11 & \multicolumn{8}{c}{\text{automatically contains dynamical gravity}} \\
d=10 & \cellcolor{blue!10} \cN=(1,0)\\
d=9 & \cellcolor{blue!10} {\cN=1} \\ 
d=8 & \cellcolor{blue!10} {\cN=1} \\
d=7 & \cellcolor{blue!10} {\cN=1}\\
d=6 &  \cellcolor{blue!10} \cN=(1,1)&  \cellcolor{blue!10} \green{  (2,0) }   & &&&  \cellcolor{yellow!20} \green{(1,0)} \\
d=5 & \cellcolor{blue!10} \cN=2  &  & &  & &   \cellcolor{yellow!20}\green1 \\
 d=4 & \cellcolor{blue!10}  \cN=\green4 &  && \green3  &&\cellcolor{yellow!20} \green2 &&  \cellcolor{red!10} \green1 &  \\
 d=3 &  \cellcolor{blue!10} \cN=\green8 & &\green7   & \green6& \green5  & \cellcolor{yellow!20}\green4 & \green3 &\cellcolor{red!10}  \green2 & \multicolumn{1}{l}{\green1} \\
 d=2 & \cellcolor{blue!10} \cN=\green{(8,8)}  && & &  & \cellcolor{yellow!20}\green{(4,4)} &  & \cellcolor{red!10} \green{(2,2)} & \\
\end{array}
\]
\caption{Supersymmetries in various dimensions. \label{table:big}}
\end{table}

\subsection{Minimal supersymmetric Yang-Mills theory}
\label{sec:minimalSYM}
Now let us ask in which dimensions one can have a supersymmetric Yang-Mills theory
which only contains the gauge field $A^a_\mu$  and a spinor $\lambda^a_\alpha$,
where $a$ is the index of the adjoint of the gauge group.
The on-shell degree of freedom in a vector field is $d-2$,
while the on-shell degree of freedom in a spinor is a power of $2$.
Therefore we need to have $d-2=2^p$ for some integer $p$.
Moreover, as we discussed, any supersymmetric theory in $d\ge 11$ contains dynamical gravity,
so we need to have $d\le 10$.

This leaves us with $d=2+2^0=3$, $d=2+2^1=4$, $d=2+2^2=6$, and $d=2+2^3=10$.
A priori, there is no guarantee that the reality conditions on spinors are such that 
there is a spinor field with the required number of on-shell degrees of freedom,
but it happens that the minimal spinor of each dimension does the job,
by consulting our Table~\ref{table:susytable};
note that the number of on-shell degrees of freedom of a spinor field is half the number of real off-shell components in the field.

Somewhat surprisingly,  the simplest Lagrangian one can think of, \begin{equation}
\tr F_{\mu\nu}F^{\mu\nu}+ \ii \tr \bar\lambda\slash D \lambda
\end{equation}
is actually supersymmetric \cite{Brink:1976bc}
and its supersymmetry can be checked uniformly \cite{Kugo:1982bn,Evans:1987tm}, 
by re-formulating  the $d=4$ case we already presented at the end of Sec.~\ref{sec:purelag}.

For this purpose, let us define $\sigma^\mu=\Gamma^\mu$ for $d=3$,
and $\Gamma^\mu=\begin{pmatrix}
0 & \sigma^\mu \\
\hat\sigma^\mu & 0
\end{pmatrix} $ for $d=4,6,10$.
The only supersymmetry variations allowed by the compatibility with Lorentz transformations are then
\begin{equation}
\delta A_\mu \propto \bar \epsilon \sigma_\mu \lambda + \cc ,\qquad
\delta\lambda \propto F_{\mu\nu}\hat\sigma^\mu \sigma^\nu \epsilon.
\label{generalsusyvar}
\end{equation} 
Then the supersymmetry transformation of the action  has the following structure:
\begin{equation}
\vcenter{\hbox{\begin{tikzpicture}[xscale=2.5,yscale=1.5]
\node(A) at (0,0) {$\tr FF$};
\node(B) at (2,0) {$\tr \bar\lambda \slash D \lambda$};
\node(P) at (1,-1) {$\overline\epsilon F_{\mu\nu}  \hat\sigma^\mu \sigma^\nu \lambda + \cc $};
\node(Q) at (3,-1){$f_{abc} (\bar{\epsilon} \sigma_\mu \lambda^a ) (\bar\lambda^b \sigma^\mu \lambda^c) +\cc $ };
\draw[->] (A) to node[midway,left]{$\delta A$} (P);
\draw[->] (B) to node[midway,left]{$\delta \lambda$} (P);
\draw[->] (B) to node[midway,right]{$\delta A$} (Q);
\end{tikzpicture}}}
\end{equation}

The variation linear in $\lambda$ can be canceled by appropriately choosing the proportionality coefficients in \eqref{generalsusyvar}.
The variation of the form $f_{abc} (\bar{\epsilon} \sigma_\mu \lambda^a ) (\bar\lambda^b \sigma^\mu \lambda^c) $ needs to vanish alone. 
This follows if \begin{equation}
\sigma_\mu{}^{(\alpha}_{(\hat\alpha}
\sigma^\mu{}^{\beta)}_{\hat\beta)}=0 \label{crucial-general}
\end{equation} where the parentheses in the subscripts and the superscripts mean separate symmetrization.

The essential relation \eqref{crucial-general} can be checked explicitly,
but a more conceptual explanation goes as follows, as in the case $d=4$.
Note that this identity follows if $p^\mu p_\mu=0$ for $p^\mu= \bar v \sigma^\mu v$ for arbitrary \emph{bosonic} spinors $v$.
Now, for $d=3,4,6,10$, the minimal spinor can be thought of a column 2-vector \begin{equation}
v=\begin{pmatrix}
v_1\\
v_2
\end{pmatrix}
\end{equation} where the entries $v_{1,2}$ are in the division algebra $\mathbb{K}$ given as follows: \begin{equation}
\begin{array}{c|ccccc}
d& 3 & 4 & 6 & 10\\
\hline
\mathbb{K} & \bR & \bC & \bH & \mathbb{O}
\end{array}
\end{equation}
where $\bH$ and $\mathbb{O}$ are the quaternion algebra
and the octonion algebra, briefly reviewed in Sec.~\ref{sec:sp}.
The sigma matrices can be chosen as \begin{equation}
\sigma^0=\begin{pmatrix}
1 & 0 \\
0 & 1
\end{pmatrix},\quad
\sigma^1=\begin{pmatrix}
1 & 0 \\
0 & -1
\end{pmatrix},\quad 
\sigma^{1+i}=\begin{pmatrix}
0 & e_i \\
\overline{e_i} & 0
\end{pmatrix}
\end{equation}
where $e_i$ with $i=1,\ldots, d-2=2^p$ are the real basis of $\mathbb{K}$.
The standard properties of the algebras $\mathbb{K}$ guarantee  that the $\sigma$ matrices defined above satisfy the required properties.
For details, see e.g.~\cite{Baez:2001dm}.

In particular, $\sigma^\mu$'s form a basis of 
Hermitean $2\times 2$ matrices with entries in $\mathbb{K}$, i.e. matrices of the form \begin{equation}
\begin{pmatrix}
a & w \\
\bar w & b
\end{pmatrix},\qquad a,b\in \bR,\ w\in \mathbb{K}.
\end{equation} 
We define for such a Hermitean matrix its determinant by \begin{equation}
\det \begin{pmatrix}
a & w \\
\bar w &b
\end{pmatrix} = ab-|w|^2.
\end{equation}
In particular, we have \begin{equation}
\det (p_\mu \sigma^\mu) = - p_\mu p^\mu.
\end{equation}

Now, for $p^\mu=\bar v\sigma^\mu v$, we have \begin{equation}
p^\mu \sigma_\mu \propto \begin{pmatrix}
\bar v_1 v_1 & \bar v_2 v_1 \\
\bar v_1 v_2 & \bar v_2 v_2 .
\end{pmatrix}
\end{equation}
We see that \begin{equation}
-p^\mu p_\mu = \det \begin{pmatrix}
\bar v_1 v_1 & \bar v_2 v_1 \\
\bar v_1 v_2 & \bar v_2 v_2 
\end{pmatrix}
=|v_1|^2 | v_2|^2 - |v_1 \bar v_2| ^2 = 0.
\end{equation}
This is what we wanted to show.

\subsection{Comments on theories with 16 supercharges}
Finally let us make some comments on theories with 16 supercharges.

\subsubsection{Super Yang-Mills theories and their R-symmetries}
We can start from the 10d minimal supersymmetric Yang-Mills theory.
As we saw already in Sec.~\ref{sec:minimalSYM}, 
it consists of the gauge field $A_\mu^a$ and a gaugino $\lambda^a_\alpha$.
To obtain the maximally supersymmetric Yang-Mills theory in dimension $D$,
one simply compactifies the 10d theory on $T^{10-D}$ and only keeps the modes which are constant along $T^{10-D}$.
This produces a $D$-dimensional gauge field $A^a_{0,1,\ldots, D-1}$
and $10-D$ real scalars $\phi^a_{D,\ldots,9}$.
The $\so(10-D)$ rotation of the scalars is the R-symmetry of the maximally supersymmetric Yang-Mills.
The results are explicitly listed in Table~\ref{table:max}.

\begin{table}
\[\begin{array}{r|l|rcl|l}
 & & \text{actual} && \text{maximal} \\
\text{dimension} & \text{scalars}& \text{R sym.}&& \text{R sym.}& \text{susy} \\
\hline
d=9 & \phi_9 & \so(1) &\simeq &  \so(1) & \cN=1 \\
d=8 & \phi_{8,9} & \so(2) &\simeq &  \mathfrak{u}(1) & \cN=1 \\
d=7 & \phi_{7,8,9} & \so(3) &\simeq &  \mathfrak{sp}(1) & \cN=1 \\
d=6 & \phi_{6,7,8,9} & \so(4) &\simeq &  \mathfrak{sp}(1)\oplus \mathfrak{sp}(1) & \cN=(1,1) \\
d=5 & \phi_{5,6,7,8,9} & \so(5) &\simeq &  \mathfrak{sp}(2) & \cN=2 \\
d=4 & \phi_{4,5,6,7,8,9} & \so(6) &\simeq &  \mathfrak{su}(4) & \cN=4 \\
d=3 & \phi_{3,4,5,6,7,8,9} & \so(7) &\subset &  \mathfrak{so}(8) & \cN=8
\end{array}\]
\caption{
R-symmetries of maximally supersymmetric Yang-Mills in various dimensions.
\label{table:max}}
\end{table}
Note that the actual R-symmetry of the maximal Yang-Mills theory is almost always the maximal R-symmetry allowed as tabulated in Table~\ref{table:susytable},
thanks to the accidental isomorphisms of $\so(10-d)$ with other algebras.
When $d=3$, the actual R-symmetry $\so(7)$ is smaller than the maximally allowed $\so(8)$.
But it  enhances to $\so(8)$ in the infrared superconformal limit.

The form of the Lagrangian of a maximally supersymmetric Yang-Mills theory is uniquely fixed once one fixes the spacetime dimension $d$ and the gauge algebra, up to the choice of the global form of the gauge group and the choice of the discrete theta angle.
It might be worth mentioning that \begin{itemize}
\item for high enough $d$, it is known that not all gauge algebra is allowed quantum mechanically.
\item for low enough $d$, there are known 16-supercharge theories which are \emph{not} super Yang-Mills theory.
\end{itemize}
As a final topic of this set of lecture notes, let us briefly review these points below.
\subsubsection{Allowed gauge algebras for super Yang-Mills}

In 10d, only  $E_8\times E_8$ and $\so(32)$ are allowed due to the anomaly cancellation,
assuming the presence of the gravity multiplet \cite{Green:1984sg}.
Both gauge algebras arise as part of the low-energy effective theory of heterotic string theories \cite{Gross:1984dd}.
In the older literature, a possibility of $E_8\times \mathfrak{u}(1)^{248}$ and  $  \mathfrak{u}(1)^{496}$ were often mentioned, but they are now known to be impossible \cite{Adams:2010zy}. 
The global structure of the gauge group will be interesting to study.
For $\so(32)$, the string theory chooses $\Spin(32)/\bZ_2$, where the $\bZ_2$ quotient is chosen so that it kills the vector representation while it keeps one of the spinor representation.
It would be interesting to study whether this choice is forced from a low-energy perspective.

In 9d, nobody seems to have studied in detail which gauge algebra is allowed.
By dimensionally reducing the 10d theory on $S^1$ with a holonomy,
one can realize subgroups of $E_8\times E_8$ or $\so(32)$ of rank 16.

In 8d, there is a known string/M theory realization except for $\so(\text{odd})$, $F_4$ and $G_2$.
Recently a field-theoretical argument was found for the non-existence of $\so(\text{odd})$ and $F_4$ \cite{Garcia-Etxebarria:2017crf}. 
The status of $G_2$ is not clear yet.

In 7d, a string/M theory realization is known for any simple gauge algebra.
For simply-laced cases, this is done by considering in M-theory the geometry $\bC^2/\Gamma$ where $\Gamma\subset \SU(2)$
is the finite subgroup of type ADE.
For non-simply-laced cases, one introduces a discrete torsion flux $\int_{S^3/\Gamma} C_3$ \cite{deBoer:2001px,Tachikawa:2015wka}.

Therefore, also in $d\le 6$,   the maximally supersymmetric Yang-Mills theory exists for any gauge algebra, by compactifying the $d=7$ theories.
Of course this conclusion can be obtained without the use of string/M-theory for $d\le 4$, 
since the Lagrangian defines a quantum field theory in these dimensions.

\subsubsection{16-supercharge theories which are not super Yang-Mills}
It does not seem likely that there are 16-supercharge theories which are not super Yang-Mills in $d\ge 7$.
This is because in these dimensions there are no superconformal group,
and therefore we expect any theory has a weakly-coupled Lagrangian description.
Then any 16-supercharge theory will be a theory of vector multiplets.

In 6d, in addition to the maximally supersymmetric Yang-Mills theory which has \Nequals{(1,1)} supersymmetry,
one can also consider \Nequals{(2,0)} theories. 
They are believed to follow an ADE classification, and can be realized e.g.~by considering type IIB string on $\bC^2/\Gamma$.
They form a very interesting class of superconformal theories in six dimensions.

In 5d, we expect to find only \Nequals2 supersymmetric Yang-Mills theory.
This follows from the same reasoning as in $d\ge 7$:
there is no \Nequals{2} superconformal symmetry in 5d,
and we expect any theory to be a Lagrangian theory.
This is in accord with the fact that 
the $S^1$ compactification of 6d \Nequals{(2,0)} superconformal theory
is believed to give the 5d maximally supersymmetric Yang-Mills theory of the corresponding gauge group.

In 4d, there is no known principle which says that all 16-supercharge theory is an \Nequals4 super Yang-Mills theory.
However, no explicit non-super-Yang-Mills example is known at present.
We mention in passing that there are known \Nequals3 theories, constructed first in \cite{Garcia-Etxebarria:2015wns}.

In 3d, there \emph{are} known 16-supercharge theory which is not  an \Nequals8 super Yang-Mills theory.\footnote{%
The author thanks G. Zafrir and D. Gang for the information contained in this paragraph.}
The most classic one is the Bagger-Lambert-Gustavsson (BLG) theory \cite{Bagger:2007jr,Gustavsson:2007vu} which is a Chern-Simons-matter theory which has \Nequals8 at the Lagrangian level.
To see that this is indeed not a super Yang-Mills, we use the supersymmetric moduli space of vacua.
The infrared limit of a \Nequals8 super Yang-Mills with gauge group $G$ has the moduli space of vacua given by $(\bR^8)^r/W$ where $r$ and $W$ are the rank and the Weyl group of $G$.
Now, the moduli space of vacua of the BLG theory  is known to be of the form $(\bR^8)^2/\mathbb{D}_{4k}$ or $(\bR^8)^2/\mathbb{D}_{2k}$, depending on whether the gauge group and the Chern-Simons level are either $\SU(2)_k\times \SU(2)_{-k}$ or its $\bZ_2$ quotient, where $\mathbb{D}_{2\ell}$ is the dihedral group of order $2\ell$ \cite{Distler:2008mk,Bagger:2012jb}.
A dihedral group $\mathbb{D}_{2\ell}$ is a Weyl group only when $\ell=1,2,3,4,6$.
Therefore, a generic BLG theory is \Nequals8 and is \emph{not} a super Yang-Mills theory.

Let us discuss next the Aharony-Bergman-Jafferis-Maldacena (ABJM) theory \cite{Aharony:2008ug}.
This is  a $\U(N)_k\times\U(N)_{-k}$ Chern-Simons-matter theory which generically has \Nequals6 at the Lagrangian level, which enhances to \Nequals8 in the infrared limit when $k=1$ or $2$.
The Aharony-Bergman-Jafferis (ABJ) theory \cite{Aharony:2008gk} of the special form $\U(N+1)_2 \times \U(N)_{-2}$ is also known to enhance to \Nequals8.
For the details of the enhancement, see \cite{Bashkirov:2010kz}.
We note that the $\U(N)_1\times \U(N)_{-1}$ ABJM theory is believed to be equal to \Nequals8 $\U(N)$ in the infrared limit \cite{Kapustin:2010xq},
the $\U(N)_2\times \U(N)_{-2}$ ABJM theory to \Nequals8 $\mathrm{O}(2N)$ theory,
and the $\U(N+1)_2\times \U(N)_{-2}$ ABJ theory to \Nequals8 $\mathrm{SO}(2N+1)$ theory \cite{Gang:2011xp}.
The BLG theories at $k=1,2,3,4$ were also believed to be equivalent to the infrared limit of \Nequals8 super Yang-Mills \cite{Lambert:2010ji,Bashkirov:2011pt,Agmon:2017lga},
Finally, we mention that all \Nequals7 theories in 3d are known to automatically enhance to \Nequals8 \cite{Bashkirov:2011fr,Bashkirov:2012rf}.

\newpage
\bibliographystyle{ytamsalpha}
\bibliography{ref}

\newcommand{\etalchar}[1]{$^{#1}$}
\providecommand{\bysame}{\leavevmode\hbox to3em{\hrulefill}\thinspace}
\providecommand{\MR}{\relax\ifhmode\unskip\space\fi MR }
\providecommand{\MRhref}[2]{%
  \href{http://www.ams.org/mathscinet-getitem?mr=#1}{#2}
}
\providecommand{\href}[2]{#2}
\providecommand{\doihref}[2]{\href{#1}{#2}}
\providecommand{\arxivfont}{\tt}
\begin{thebibliography}{BdlMMQ98}

\bibitem[AB69]{Adler:1969er}
S.~L. Adler and W.~A. Bardeen, \emph{{Absence of Higher Order Corrections in
  the Anomalous Axial Vector Divergence Equation}},
\doihref{http://dx.doi.org/10.1103/PhysRev.182.1517}{Phys. Rev. \textbf{182}
  (1969) 1517--1536}.

\bibitem[ABJ08]{Aharony:2008gk}
O.~Aharony, O.~Bergman, and D.~L. Jafferis, \emph{{Fractional M2-Branes}},
  \doihref{http://dx.doi.org/10.1088/1126-6708/2008/11/043}{JHEP \textbf{11}
  (2008) 043},
\href{http://arxiv.org/abs/0807.4924}{{\arxivfont arXiv:0807.4924 [hep-th]}}.

\bibitem[ABJM08]{Aharony:2008ug}
O.~Aharony, O.~Bergman, D.~L. Jafferis, and J.~Maldacena,
  \emph{{${\mathcal{N}}{=}6$ Superconformal Chern-Simons-Matter Theories,
  M2-Branes and Their Gravity Duals}},
  \doihref{http://dx.doi.org/10.1088/1126-6708/2008/10/091}{JHEP \textbf{10}
  (2008) 091},
\href{http://arxiv.org/abs/0806.1218}{{\arxivfont arXiv:0806.1218 [hep-th]}}.

\bibitem[ACP17]{Agmon:2017lga}
N.~B. Agmon, S.~M. Chester, and S.~S. Pufu, \emph{{A new duality between $
  \mathcal{N} $ = 8 superconformal field theories in three dimensions}},
  \doihref{http://dx.doi.org/10.1007/JHEP06(2018)005}{JHEP \textbf{06} (2018)
  005},
\href{http://arxiv.org/abs/1708.07861}{{\arxivfont arXiv:1708.07861 [hep-th]}}.

\bibitem[AD95]{Argyres:1995jj}
P.~C. Argyres and M.~R. Douglas, \emph{{New Phenomena in $SU(3)$ Supersymmetric
  Gauge Theory}},
  \doihref{http://dx.doi.org/10.1016/0550-3213(95)00281-V}{Nucl. Phys.
  \textbf{B448} (1995) 93--126},
\href{http://arxiv.org/abs/hep-th/9505062}{{\arxivfont arXiv:hep-th/9505062}}.

\bibitem[Adl69]{Adler:1969gk}
S.~L. Adler, \emph{{Axial Vector Vertex in Spinor Electrodynamics}},
\doihref{http://dx.doi.org/10.1103/PhysRev.177.2426}{Phys. Rev. \textbf{177}
  (1969) 2426--2438}.

\bibitem[ADS83]{Affleck:1983rr}
I.~Affleck, M.~Dine, and N.~Seiberg, \emph{{Supersymmetry Breaking by
  Instantons}},
\doihref{http://dx.doi.org/10.1103/PhysRevLett.51.1026}{Phys. Rev. Lett.
  \textbf{51} (1983) 1026}.

\bibitem[ADT10]{Adams:2010zy}
A.~Adams, O.~DeWolfe, and W.~Taylor, \emph{{String Universality in Ten
  Dimensions}},
  \doihref{http://dx.doi.org/10.1103/PhysRevLett.105.071601}{Phys. Rev. Lett.
  \textbf{105} (2010) 071601},
\href{http://arxiv.org/abs/1006.1352}{{\arxivfont arXiv:1006.1352 [hep-th]}}.

\bibitem[AEFJ97]{Anselmi:1997ys}
D.~Anselmi, J.~Erlich, D.~Z. Freedman, and A.~A. Johansen, \emph{{Positivity
  Constraints on Anomalies in Supersymmetric Gauge Theories}},
  \doihref{http://dx.doi.org/10.1103/PhysRevD.57.7570}{Phys. Rev. \textbf{D57}
  (1998) 7570--7588},
\href{http://arxiv.org/abs/hep-th/9711035}{{\arxivfont arXiv:hep-th/9711035}}.

\bibitem[AFGJ97]{Anselmi:1997am}
D.~Anselmi, D.~Z. Freedman, M.~T. Grisaru, and A.~A. Johansen,
  \emph{{Nonperturbative Formulas for Central Functions of Supersymmetric Gauge
  Theories}}, \doihref{http://dx.doi.org/10.1016/S0550-3213(98)00278-8}{Nucl.
  Phys. \textbf{B526} (1998) 543--571},
\href{http://arxiv.org/abs/hep-th/9708042}{{\arxivfont arXiv:hep-th/9708042}}.

\bibitem[AG80]{Abbott:1980jk}
L.~F. Abbott and M.~T. Grisaru, \emph{{The Three Loop Beta Function for the
  {Wess-Zumino} Model}},
\doihref{http://dx.doi.org/10.1016/0550-3213(80)90096-6}{Nucl. Phys.
  \textbf{B169} (1980) 415--429}.

\bibitem[AG83]{AlvarezGaume:1983at}
L.~\'Alvarez-Gaum\'e, \emph{{Supersymmetry and the Atiyah-Singer Index
  Theorem}},
\doihref{http://dx.doi.org/10.1007/BF01205500}{Commun. Math. Phys. \textbf{90}
  (1983) 161}.

\bibitem[Aga18]{Agarwal:2018oxb}
P.~Agarwal, \emph{{On Dimensional Reduction of 4D ${\mathcal{N}}\!=1$
  Lagrangians for Argyres-Douglas Theories}},
\href{http://arxiv.org/abs/1809.10534}{{\arxivfont arXiv:1809.10534 [hep-th]}}.

\bibitem[AGLV98]{SingularityTheory}
V.~I. Arnold, V.~V. Goryunov, O.~V. Lyashko, and V.~A. Vasil'ev,
  \doihref{http://dx.doi.org/10.1007/978-3-642-58009-3}{\emph{Singularity
  theory {I}}}, Encyclopedia of Mathematical Sciences, vol.~6, Springer, 1998.

\bibitem[AHM97a]{ArkaniHamed:1997ut}
N.~Arkani-Hamed and H.~Murayama, \emph{{Renormalization Group Invariance of
  Exact Results in Supersymmetric Gauge Theories}},
  \doihref{http://dx.doi.org/10.1103/PhysRevD.57.6638}{Phys. Rev. \textbf{D57}
  (1998) 6638--6648},
\href{http://arxiv.org/abs/hep-th/9705189}{{\arxivfont arXiv:hep-th/9705189}}.

\bibitem[AHM97b]{ArkaniHamed:1997mj}
\bysame, \emph{{Holomorphy, Rescaling Anomalies and Exact Beta Functions in
  Supersymmetric Gauge Theories}},
  \doihref{http://dx.doi.org/10.1088/1126-6708/2000/06/030}{JHEP \textbf{06}
  (2000) 030},
\href{http://arxiv.org/abs/hep-th/9707133}{{\arxivfont arXiv:hep-th/9707133}}.

\bibitem[AM18]{Argyres:2018urp}
P.~C. Argyres and M.~Martone, \emph{{Scaling Dimensions of Coulomb Branch
  Operators of 4D ${\mathcal{N}}{=}2$ Superconformal Field Theories}},
\href{http://arxiv.org/abs/1801.06554}{{\arxivfont arXiv:1801.06554 [hep-th]}}.

\bibitem[AMS16]{Agarwal:2016pjo}
P.~Agarwal, K.~Maruyoshi, and J.~Song, \emph{{$ \mathcal{N} $ =1 Deformations
  and RG flows of $ \mathcal{N} $ =2 SCFTs, part II: non-principal
  deformations}}, \doihref{http://dx.doi.org/10.1007/JHEP12(2016)103}{JHEP
  \textbf{12} (2016) 103}, \href{http://arxiv.org/abs/1610.05311}{{\arxivfont
  arXiv:1610.05311 [hep-th]}}.
[\doihref{10.1007/JHEP04(2017)113}{Addendum: JHEP04,113(2017)}].

\bibitem[Arg96]{ArgyresReview1996}
P.~C. Argyres, \emph{Introduction to supersymmetry}.
  \url{http://homepages.uc.edu/~argyrepc/cu661-gr-SUSY/susy1996.pdf}.

\bibitem[Arg01]{ArgyresReview2001}
\bysame, \emph{An introduction to global supersymmetry}.
  \url{http://homepages.uc.edu/~argyrepc/cu661-gr-SUSY/susy2001.pdf}.

\bibitem[ARPSW95]{Argyres:1995xn}
P.~C. Argyres, M.~R.~Plesser, N.~Seiberg, and E.~Witten, \emph{{New
  ${\mathcal{N}}{=}2$ Superconformal Field Theories in Four Dimensions}},
  \doihref{http://dx.doi.org/10.1016/0550-3213(95)00671-0}{Nucl. Phys.
  \textbf{B461} (1996) 71--84},
\href{http://arxiv.org/abs/hep-th/9511154}{{\arxivfont arXiv:hep-th/9511154}}.

\bibitem[ARSW13a]{Aharony:2013dha}
O.~Aharony, S.~S. Razamat, N.~Seiberg, and B.~Willett, \emph{{3D Dualities from
  4D Dualities}}, \doihref{http://dx.doi.org/10.1007/JHEP07(2013)149}{JHEP
  \textbf{1307} (2013) 149},
\href{http://arxiv.org/abs/1305.3924}{{\arxivfont arXiv:1305.3924 [hep-th]}}.

\bibitem[ARSW13b]{Aharony:2013kma}
\bysame, \emph{{3D dualities from 4D dualities for orthogonal groups}},
  \doihref{http://dx.doi.org/10.1007/JHEP08(2013)099}{JHEP \textbf{08} (2013)
  099},
\href{http://arxiv.org/abs/1307.0511}{{\arxivfont arXiv:1307.0511 [hep-th]}}.

\bibitem[ARW17]{Aharony:2017adm}
O.~Aharony, S.~S. Razamat, and B.~Willett, \emph{{From 3D Duality to 2D
  Duality}}, \doihref{http://dx.doi.org/10.1007/JHEP11(2017)090}{JHEP
  \textbf{11} (2017) 090},
\href{http://arxiv.org/abs/1710.00926}{{\arxivfont arXiv:1710.00926 [hep-th]}}.

\bibitem[Asp96]{Aspinwall:1996mn}
P.~S. Aspinwall, \emph{{K3 Surfaces and String Duality}},
  \href{http://arxiv.org/abs/hep-th/9611137}{{\arxivfont
  arXiv:hep-th/9611137}}.
(TASI 1996).

\bibitem[ASS17]{Agarwal:2017roi}
P.~Agarwal, A.~Sciarappa, and J.~Song, \emph{{$ \mathcal{N} $ =1 Lagrangians
  for generalized Argyres-Douglas theories}},
  \doihref{http://dx.doi.org/10.1007/JHEP10(2017)211}{JHEP \textbf{10} (2017)
  211},
\href{http://arxiv.org/abs/1707.04751}{{\arxivfont arXiv:1707.04751 [hep-th]}}.

\bibitem[AST13]{Aharony:2013hda}
O.~Aharony, N.~Seiberg, and Y.~Tachikawa, \emph{{Reading Between the Lines of
  Four-Dimensional Gauge Theories}},
  \doihref{http://dx.doi.org/10.1007/JHEP08(2013)115}{JHEP \textbf{1308} (2013)
  115},
\href{http://arxiv.org/abs/1305.0318}{{\arxivfont arXiv:1305.0318}}.

\bibitem[AT07]{Aharony:2007dj}
O.~Aharony and Y.~Tachikawa, \emph{{A Holographic Computation of the Central
  Charges of $d=4$, ${\mathcal{N}}{=}2$ SCFTs}},
  \doihref{http://dx.doi.org/10.1088/1126-6708/2008/01/037}{JHEP \textbf{0801}
  (2008) 037},
\href{http://arxiv.org/abs/0711.4532}{{\arxivfont arXiv:0711.4532 [hep-th]}}.

\bibitem[Bae01]{Baez:2001dm}
J.~C. Baez, \emph{{The Octonions}},
  \doihref{http://dx.doi.org/10.1090/S0273-0979-01-00934-X}{Bull. Am. Math.
  Soc. \textbf{39} (2002) 145--205},
  \href{http://arxiv.org/abs/math.RA/0105155}{{\arxivfont
  arXiv:math.RA/0105155}}.
\doihref{http://dx.doi.org/10.1090/S0273-0979-05-01052-9}{[Erratum: Bull. Am.
  Math. Soc. \textbf{42} (2005) 213]}.

\bibitem[Bas11]{Bashkirov:2011fr}
D.~Bashkirov, \emph{{A Note on ${\cal N}\ge 6$ Superconformal Quantum Field
  Theories in three dimensions}},
\href{http://arxiv.org/abs/1108.4081}{{\arxivfont arXiv:1108.4081 [hep-th]}}.

\bibitem[Bas12]{Bashkirov:2012rf}
\bysame, \emph{{BLG Theories at Low Values of Chern-Simons Coupling}},
\href{http://arxiv.org/abs/1211.4887}{{\arxivfont arXiv:1211.4887 [hep-th]}}.

\bibitem[BB12]{Benini:2012cz}
F.~Benini and N.~Bobev, \emph{{Exact Two-Dimensional Superconformal R-Symmetry
  and $c$-Extremization}},
  \doihref{http://dx.doi.org/10.1103/PhysRevLett.110.061601}{Phys.Rev.Lett.
  \textbf{110} (2013) 061601},
\href{http://arxiv.org/abs/1211.4030}{{\arxivfont arXiv:1211.4030 [hep-th]}}.

\bibitem[BBC04]{Bertolini:2004xf}
M.~Bertolini, F.~Bigazzi, and A.~L. Cotrone, \emph{{New Checks and Subtleties
  for {AdS/CFT} and $a$- Maximization}},
  \doihref{http://dx.doi.org/10.1088/1126-6708/2004/12/024}{JHEP \textbf{12}
  (2004) 024},
\href{http://arxiv.org/abs/hep-th/0411249}{{\arxivfont arXiv:hep-th/0411249}}.

\bibitem[BCI98]{Brodie:1998vv}
J.~H. Brodie, P.~L. Cho, and K.~A. Intriligator, \emph{{Misleading Anomaly
  Matchings?}}, \doihref{http://dx.doi.org/10.1016/S0370-2693(98)00353-0}{Phys.
  Lett. \textbf{B429} (1998) 319--326},
\href{http://arxiv.org/abs/hep-th/9802092}{{\arxivfont arXiv:hep-th/9802092}}.

\bibitem[BCKS97]{Berkooz:1997bb}
M.~Berkooz, P.~L. Cho, P.~Kraus, and M.~J. Strassler, \emph{{Dual Descriptions
  of SO(10) SUSY Gauge Theories with Arbitrary Numbers of Spinors and
  Vectors}}, \doihref{http://dx.doi.org/10.1103/PhysRevD.56.7166}{Phys. Rev.
  \textbf{D56} (1997) 7166--7182},
\href{http://arxiv.org/abs/hep-th/9705003}{{\arxivfont arXiv:hep-th/9705003}}.

\bibitem[BdlMMQ98]{Burgess:1998jh}
C.~P. Burgess, A.~de~la Macorra, I.~Maksymyk, and F.~Quevedo,
  \emph{{Supersymmetric Models with Product Groups and Field Dependent Gauge
  Couplings}}, JHEP \textbf{09} (1998) 007,
\href{http://arxiv.org/abs/hep-th/9808087}{{\arxivfont arXiv:hep-th/9808087}}.

\bibitem[BEHT13a]{Benini:2013nda}
F.~Benini, R.~Eager, K.~Hori, and Y.~Tachikawa, \emph{{Elliptic Genera of
  Two-Dimensional ${\mathcal{N}}{=}2$ Gauge Theories with Rank-One Gauge
  Groups}}, \doihref{http://dx.doi.org/10.1007/s11005-013-0673-y}{Lett. Math.
  Phys. \textbf{104} (2014) 465--493},
\href{http://arxiv.org/abs/1305.0533}{{\arxivfont arXiv:1305.0533 [hep-th]}}.

\bibitem[BEHT13b]{Benini:2013xpa}
\bysame, \emph{{Elliptic Genera of 2d ${\mathcal{N}}$ = 2 Gauge Theories}},
  \doihref{http://dx.doi.org/10.1007/s00220-014-2210-y}{Commun. Math. Phys.
  \textbf{333} (2015) 1241--1286},
\href{http://arxiv.org/abs/1308.4896}{{\arxivfont arXiv:1308.4896 [hep-th]}}.

\bibitem[BFH{\etalchar{+}}04]{Benvenuti:2004dy}
S.~Benvenuti, S.~Franco, A.~Hanany, D.~Martelli, and J.~Sparks, \emph{{An
  Infinite Family of Superconformal Quiver Gauge Theories with Sasaki-Einstein
  Duals}}, JHEP \textbf{06} (2005) 064,
\href{http://arxiv.org/abs/hep-th/0411264}{{\arxivfont arXiv:hep-th/0411264}}.

\bibitem[BFM99]{Borel:1999bx}
A.~Borel, R.~Friedman, and J.~W. Morgan, \emph{Almost commuting elements in
  compact {L}ie groups}, \doihref{http://dx.doi.org/10.1090/memo/0747}{Mem.
  Amer. Math. Soc. \textbf{747} (2002) x+136},
  \href{http://arxiv.org/abs/math.GR/9907007}{{\arxivfont
  arXiv:math.GR/9907007}}.

\bibitem[BG17a]{Benvenuti:2017kud}
S.~Benvenuti and S.~Giacomelli, \emph{{Abelianization and sequential
  confinement in $2+1$ dimensions}},
  \doihref{http://dx.doi.org/10.1007/JHEP10(2017)173}{JHEP \textbf{10} (2017)
  173},
\href{http://arxiv.org/abs/1706.04949}{{\arxivfont arXiv:1706.04949 [hep-th]}}.

\bibitem[BG17b]{Benvenuti:2017lle}
\bysame, \emph{{Supersymmetric Gauge Theories with Decoupled Operators and
  Chiral Ring Stability}},
  \doihref{http://dx.doi.org/10.1103/PhysRevLett.119.251601}{Phys. Rev. Lett.
  \textbf{119} (2017) 251601},
\href{http://arxiv.org/abs/1706.02225}{{\arxivfont arXiv:1706.02225 [hep-th]}}.

\bibitem[BG17c]{Benvenuti:2017bpg}
\bysame, \emph{{Lagrangians for Generalized Argyres-Douglas Theories}},
  \doihref{http://dx.doi.org/10.1007/JHEP10(2017)106}{JHEP \textbf{10} (2017)
  106},
\href{http://arxiv.org/abs/1707.05113}{{\arxivfont arXiv:1707.05113 [hep-th]}}.

\bibitem[BH94]{Berglund:1993fj}
P.~Berglund and M.~Henningson, \emph{{Landau-Ginzburg Orbifolds, Mirror
  Symmetry and the Elliptic Genus}},
  \doihref{http://dx.doi.org/10.1016/0550-3213(94)00389-V}{Nucl.Phys.
  \textbf{B433} (1995) 311--332},
\href{http://arxiv.org/abs/hep-th/9401029}{{\arxivfont arXiv:hep-th/9401029}}.

\bibitem[BJ69]{Bell:1969ts}
J.~S. Bell and R.~Jackiw, \emph{{A PCAC puzzle: $\pi^0 \to \gamma \gamma$ in
  the $\sigma$ model}},
\doihref{http://dx.doi.org/10.1007/BF02823296}{Nuovo Cim. \textbf{A60} (1969)
  47--61}.

\bibitem[BK10]{Bashkirov:2010kz}
D.~Bashkirov and A.~Kapustin, \emph{{Supersymmetry Enhancement by Monopole
  Operators}}, \doihref{http://dx.doi.org/10.1007/JHEP05(2011)015}{JHEP
  \textbf{05} (2011) 015},
\href{http://arxiv.org/abs/1007.4861}{{\arxivfont arXiv:1007.4861 [hep-th]}}.

\bibitem[BK11]{Bashkirov:2011pt}
\bysame, \emph{{Dualities Between ${\mathcal{N}}{=}8$ Superconformal Field
  Theories in Three Dimensions}},
  \doihref{http://dx.doi.org/10.1007/JHEP05(2011)074}{JHEP \textbf{05} (2011)
  074},
\href{http://arxiv.org/abs/1103.3548}{{\arxivfont arXiv:1103.3548 [hep-th]}}.

\bibitem[BL07]{Bagger:2007jr}
J.~Bagger and N.~Lambert, \emph{{Gauge Symmetry and Supersymmetry of Multiple
  M2-Branes}}, \doihref{http://dx.doi.org/10.1103/PhysRevD.77.065008}{Phys.
  Rev. \textbf{D77} (2007) 065008},
\href{http://arxiv.org/abs/0711.0955}{{\arxivfont arXiv:0711.0955 [hep-th]}}.

\bibitem[BLMP12]{Bagger:2012jb}
J.~Bagger, N.~Lambert, S.~Mukhi, and C.~Papageorgakis, \emph{{Multiple
  Membranes in M-theory}},
  \doihref{http://dx.doi.org/10.1016/j.physrep.2013.01.006}{Phys. Rept.
  \textbf{527} (2013) 1--100},
\href{http://arxiv.org/abs/1203.3546}{{\arxivfont arXiv:1203.3546 [hep-th]}}.

\bibitem[BMS08]{Buchel:2008vz}
A.~Buchel, R.~C. Myers, and A.~Sinha, \emph{{Beyond $\eta/s = 1/4\pi$}},
  \doihref{http://dx.doi.org/10.1088/1126-6708/2009/03/084}{JHEP \textbf{03}
  (2009) 084},
\href{http://arxiv.org/abs/0812.2521}{{\arxivfont arXiv:0812.2521 [hep-th]}}.

\bibitem[BPZT06]{Benvenuti:2006xg}
S.~Benvenuti, L.~A. Pando~Zayas, and Y.~Tachikawa, \emph{{Triangle anomalies
  from Einstein manifolds}},
  \doihref{http://dx.doi.org/10.4310/ATMP.2006.v10.n3.a4}{Adv. Theor. Math.
  Phys. \textbf{10} (2006) 395--432},
\href{http://arxiv.org/abs/hep-th/0601054}{{\arxivfont arXiv:hep-th/0601054}}.

\bibitem[Bro96]{Brodie:1996vx}
J.~H. Brodie, \emph{{Duality in Supersymmetric $SU(\hbox{$N_c$}$) Gauge Theory
  with Two Adjoint Chiral Superfields}},
  \doihref{http://dx.doi.org/10.1016/0550-3213(96)00416-6}{Nucl. Phys.
  \textbf{B478} (1996) 123--140},
\href{http://arxiv.org/abs/hep-th/9605232}{{\arxivfont arXiv:hep-th/9605232}}.

\bibitem[BS58]{BottSamelson}
R.~Bott and H.~Samelson, \emph{Applications of the theory of {M}orse to
  symmetric spaces}, \doihref{http://dx.doi.org/10.2307/2372843}{Amer. J. Math.
  \textbf{80} (1958) 964--1029}.

\bibitem[BSS77]{Brink:1976bc}
L.~Brink, J.~H. Schwarz, and J.~Scherk, \emph{{Supersymmetric Yang-Mills
  Theories}},
\doihref{http://dx.doi.org/10.1016/0550-3213(77)90328-5}{Nucl.Phys.
  \textbf{B121} (1977) 77}.

\bibitem[BW04]{Beasley:2004ys}
C.~Beasley and E.~Witten, \emph{{New Instanton Effects in Supersymmetric QCD}},
  \doihref{http://dx.doi.org/10.1088/1126-6708/2005/01/056}{JHEP \textbf{01}
  (2005) 056},
\href{http://arxiv.org/abs/hep-th/0409149}{{\arxivfont arXiv:hep-th/0409149}}.

\bibitem[BZ82]{Banks:1981nn}
T.~Banks and A.~Zaks, \emph{{On the Phase Structure of Vector-Like Gauge
  Theories with Massless Fermions}},
\doihref{http://dx.doi.org/10.1016/0550-3213(82)90035-9}{Nucl. Phys.
  \textbf{B196} (1982) 189--204}.

\bibitem[BZ05]{Butti:2005vn}
A.~Butti and A.~Zaffaroni, \emph{{R-Charges from Toric Diagrams and the
  Equivalence of $a$-maximization and $Z$-minimization}},
  \doihref{http://dx.doi.org/10.1088/1126-6708/2005/11/019}{JHEP \textbf{11}
  (2005) 019},
\href{http://arxiv.org/abs/hep-th/0506232}{{\arxivfont arXiv:hep-th/0506232}}.

\bibitem[Cap87]{Cappelli:1986ed}
A.~Cappelli, \emph{{Modular Invariant Partition Functions of Superconformal
  Theories}},
\doihref{http://dx.doi.org/10.1016/0370-2693(87)91532-2}{Phys. Lett.
  \textbf{B185} (1987) 82--88}.

\bibitem[Car88]{Cardy:1988cwa}
J.~L. Cardy, \emph{{Is There a $c$-Theorem in Four-Dimensions?}},
\doihref{http://dx.doi.org/10.1016/0370-2693(88)90054-8}{Phys. Lett.
  \textbf{B215} (1988) 749--752}.

\bibitem[Cas74]{Caswell:1974gg}
W.~E. Caswell, \emph{{Asymptotic Behavior of Nonabelian Gauge Theories to Two
  Loop Order}},
\doihref{http://dx.doi.org/10.1103/PhysRevLett.33.244}{Phys. Rev. Lett.
  \textbf{33} (1974) 244}.

\bibitem[CC18]{Caorsi:2018zsq}
M.~Caorsi and S.~Cecotti, \emph{{Geometric classification of 4d $\mathcal{N}=2$
  SCFTs}}, \doihref{http://dx.doi.org/10.1007/JHEP07(2018)138}{JHEP \textbf{07}
  (2018) 138},
\href{http://arxiv.org/abs/1801.04542}{{\arxivfont arXiv:1801.04542 [hep-th]}}.

\bibitem[CDFK14]{Closset:2014uda}
C.~Closset, T.~T. Dumitrescu, G.~Festuccia, and Z.~Komargodski, \emph{{From
  Rigid Supersymmetry to Twisted Holomorphic Theories}},
  \doihref{http://dx.doi.org/10.1103/PhysRevD.90.085006}{Phys. Rev.
  \textbf{D90} (2014) 085006},
\href{http://arxiv.org/abs/1407.2598}{{\arxivfont arXiv:1407.2598 [hep-th]}}.

\bibitem[CDI16]{Cordova:2016emh}
C.~Cordova, T.~T. Dumitrescu, and K.~Intriligator, \emph{{Multiplets of
  Superconformal Symmetry in Diverse Dimensions}},
\href{http://arxiv.org/abs/1612.00809}{{\arxivfont arXiv:1612.00809 [hep-th]}}.

\bibitem[CDI18]{Cordova:2018cvg}
C.~C{\'o}rdova, T.~T. Dumitrescu, and K.~Intriligator, \emph{{Exploring 2-Group
  Global Symmetries}},
\href{http://arxiv.org/abs/1802.04790}{{\arxivfont arXiv:1802.04790 [hep-th]}}.

\bibitem[CDSW02]{Cachazo:2002ry}
F.~Cachazo, M.~R. Douglas, N.~Seiberg, and E.~Witten, \emph{{Chiral rings and
  anomalies in supersymmetric gauge theory}},
  \doihref{http://dx.doi.org/10.1088/1126-6708/2002/12/071}{JHEP \textbf{12}
  (2002) 071},
\href{http://arxiv.org/abs/hep-th/0211170}{{\arxivfont arXiv:hep-th/0211170}}.

\bibitem[CEFS97]{Csaki:1997zg}
C.~Csaki, J.~Erlich, D.~Z. Freedman, and W.~Skiba, \emph{{${\mathcal{N}}\!=1$
  Supersymmetric Product Group Theories in the Coulomb Phase}},
  \doihref{http://dx.doi.org/10.1103/PhysRevD.56.5209}{Phys. Rev. \textbf{D56}
  (1997) 5209--5217},
\href{http://arxiv.org/abs/hep-th/9704067}{{\arxivfont arXiv:hep-th/9704067}}.

\bibitem[CG82]{Coleman:1982yg}
S.~R. Coleman and B.~Grossman, \emph{{'t Hooft's Consistency Condition as a
  Consequence of Analyticity and Unitarity}},
\doihref{http://dx.doi.org/10.1016/0550-3213(82)90028-1}{Nucl. Phys.
  \textbf{B203} (1982) 205--220}.

\bibitem[CGK97]{Cheung:1997id}
Y.-K.~E. Cheung, O.~J. Ganor, and M.~Krogh, \emph{{Correlators of the Global
  Symmetry Currents of 4D and 6D Superconformal Theories}},
  \doihref{http://dx.doi.org/10.1016/S0550-3213(98)00139-4}{Nucl. Phys.
  \textbf{B523} (1998) 171--192},
\href{http://arxiv.org/abs/hep-th/9710053}{{\arxivfont arXiv:hep-th/9710053}}.

\bibitem[Cho97a]{Cho:1997sa}
P.~L. Cho, \emph{{Exact Results in SO(11) SUSY Gauge Theories with Spinor and
  Vector Matter}},
  \doihref{http://dx.doi.org/10.1016/S0370-2693(97)00315-8}{Phys. Lett.
  \textbf{B400} (1997) 101--108},
\href{http://arxiv.org/abs/hep-th/9701020}{{\arxivfont arXiv:hep-th/9701020}}.

\bibitem[Cho97b]{Cho:1997kr}
\bysame, \emph{{More on Chiral - Nonchiral Dual Pairs}},
  \doihref{http://dx.doi.org/10.1103/PhysRevD.56.5260}{Phys. Rev. \textbf{D56}
  (1997) 5260--5271},
\href{http://arxiv.org/abs/hep-th/9702059}{{\arxivfont arXiv:hep-th/9702059}}.

\bibitem[CIZ87a]{Cappelli:1986hf}
A.~Cappelli, C.~Itzykson, and J.~B. Zuber, \emph{{Modular Invariant Partition
  Functions in Two-Dimensions}},
\doihref{http://dx.doi.org/10.1016/0550-3213(87)90155-6}{Nucl. Phys.
  \textbf{B280} (1987) 445--465}.

\bibitem[CIZ87b]{Cappelli:1987xt}
\bysame, \emph{{The ADE Classification of Minimal and $A_1^{(1)}$ Conformal
  Invariant Theories}},
\doihref{http://dx.doi.org/10.1007/BF01221394}{Commun. Math. Phys. \textbf{113}
  (1987) 1}.

\bibitem[Col88]{Coleman}
S.~Coleman, \emph{{Aspects of symmetry: selected Erice lectures of Sidney
  Coleman}}, Cambridge University Press, 1988.

\bibitem[Cre11]{Cremonini:2011iq}
S.~Cremonini, \emph{{The Shear Viscosity to Entropy Ratio: a Status Report}},
  \doihref{http://dx.doi.org/10.1142/S0217984911027315}{Mod. Phys. Lett.
  \textbf{B25} (2011) 1867--1888},
\href{http://arxiv.org/abs/1108.0677}{{\arxivfont arXiv:1108.0677 [hep-th]}}.

\bibitem[CS98]{Csaki:1998dp}
C.~Csaki and W.~Skiba, \emph{{Classification of the ${\mathcal{N}}\!=1$
  Seiberg-Witten Theories}},
  \doihref{http://dx.doi.org/10.1103/PhysRevD.58.045008}{Phys. Rev.
  \textbf{D58} (1998) 045008},
\href{http://arxiv.org/abs/hep-th/9801173}{{\arxivfont arXiv:hep-th/9801173}}.

\bibitem[CS13]{Closset:2013sxa}
C.~Closset and I.~Shamir, \emph{{The $\mathcal{N}=1$ Chiral Multiplet on
  $T^2\times S^2$ and Supersymmetric Localization}},
  \doihref{http://dx.doi.org/10.1007/JHEP03(2014)040}{JHEP \textbf{1403} (2014)
  040},
\href{http://arxiv.org/abs/1311.2430}{{\arxivfont arXiv:1311.2430 [hep-th]}}.

\bibitem[CSST97]{Csaki:1997cu}
C.~Csaki, M.~Schmaltz, W.~Skiba, and J.~Terning, \emph{{Selfdual
  ${\mathcal{N}}{=}1$ SUSY Gauge Theories}},
  \doihref{http://dx.doi.org/10.1103/PhysRevD.56.1228}{Phys.Rev. \textbf{D56}
  (1997) 1228--1238},
\href{http://arxiv.org/abs/hep-th/9701191}{{\arxivfont arXiv:hep-th/9701191}}.

\bibitem[Das71]{Dashen:1970et}
R.~F. Dashen, \emph{{Some Features of Chiral Symmetry Breaking}},
\doihref{http://dx.doi.org/10.1103/PhysRevD.3.1879}{Phys. Rev. \textbf{D3}
  (1971) 1879--1889}.

\bibitem[dBDH{\etalchar{+}}01]{deBoer:2001px}
J.~de~Boer, R.~Dijkgraaf, K.~Hori, A.~Keurentjes, J.~Morgan, D.~R. Morrison,
  and S.~Sethi, \emph{{Triples, Fluxes, and Strings}},
  \doihref{http://dx.doi.org/10.4310/ATMP.2000.v4.n5.a1}{Adv.Theor.Math.Phys.
  \textbf{4} (2002) 995--1186},
\href{http://arxiv.org/abs/hep-th/0103170}{{\arxivfont arXiv:hep-th/0103170}}.

\bibitem[dBHOO97]{deBoer:1997zy}
J.~de~Boer, K.~Hori, H.~Ooguri, and Y.~Oz, \emph{{K\"ahler Potential and Higher
  Derivative Terms from M Theory Five-Brane}},
  \doihref{http://dx.doi.org/10.1016/S0550-3213(98)00152-7}{Nucl. Phys.
  \textbf{B518} (1998) 173--211},
\href{http://arxiv.org/abs/hep-th/9711143}{{\arxivfont arXiv:hep-th/9711143}}.

\bibitem[Des96]{Deser:1996na}
S.~Deser, \emph{{Conformal Anomalies: Recent Progress}}, Helv. Phys. Acta
  \textbf{69} (1996) 570--581,
\href{http://arxiv.org/abs/hep-th/9609138}{{\arxivfont arXiv:hep-th/9609138}}.

\bibitem[DFY93]{DiFrancesco:1993dg}
P.~Di~Francesco and S.~Yankielowicz, \emph{{Ramond Sector Characters and
  ${\mathcal{N}}{=}2$ Landau-Ginzburg Models}},
  \doihref{http://dx.doi.org/10.1016/0550-3213(93)90452-U}{Nucl.Phys.
  \textbf{B409} (1993) 186--210},
\href{http://arxiv.org/abs/hep-th/9305037}{{\arxivfont arXiv:hep-th/9305037}}.

\bibitem[DG12]{Dimofte:2012pd}
T.~Dimofte and D.~Gaiotto, \emph{{An $E_7$ Surprise}},
  \doihref{http://dx.doi.org/10.1007/JHEP10(2012)129}{JHEP \textbf{1210} (2012)
  129},
\href{http://arxiv.org/abs/1209.1404}{{\arxivfont arXiv:1209.1404 [hep-th]}}.

\bibitem[Din15]{DineBook}
M.~Dine, \emph{Supersymmetry and string theory : beyond the standard model},
  2nd ed., Cambridge University Press, 2015.

\bibitem[DK96]{Distler:1996ub}
J.~Distler and A.~Karch, \emph{{${\mathcal{N}}\!=1$ Dualities for Exceptional
  Gauge Groups and Quantum Global Symmetries}},
  \doihref{http://dx.doi.org/10.1002/prop.2190450603}{Fortsch. Phys.
  \textbf{45} (1997) 517--533},
\href{http://arxiv.org/abs/hep-th/9611088}{{\arxivfont arXiv:hep-th/9611088}}.

\bibitem[DKM96]{Dorey:1996ez}
N.~Dorey, V.~V. Khoze, and M.~P. Mattis, \emph{{On Mass Deformed
  ${\mathcal{N}}{=}4$ Supersymmetric Yang-Mills Theory}},
  \doihref{http://dx.doi.org/10.1016/S0370-2693(97)00102-0}{Phys.Lett.
  \textbf{B396} (1997) 141--149},
\href{http://arxiv.org/abs/hep-th/9612231}{{\arxivfont arXiv:hep-th/9612231}}.

\bibitem[DMPVR08]{Distler:2008mk}
J.~Distler, S.~Mukhi, C.~Papageorgakis, and M.~Van~Raamsdonk, \emph{{M2-Branes
  on M-Folds}}, \doihref{http://dx.doi.org/10.1088/1126-6708/2008/05/038}{JHEP
  \textbf{05} (2008) 038},
\href{http://arxiv.org/abs/0804.1256}{{\arxivfont arXiv:0804.1256 [hep-th]}}.

\bibitem[DO08]{Dolan:2008qi}
F.~Dolan and H.~Osborn, \emph{{Applications of the Superconformal Index for
  Protected Operators and $q$-Hypergeometric Identities to ${\mathcal{N}}{=}1$
  Dual Theories}},
  \doihref{http://dx.doi.org/10.1016/j.nuclphysb.2009.01.028}{Nucl.Phys.
  \textbf{B818} (2009) 137--178},
\href{http://arxiv.org/abs/0801.4947}{{\arxivfont arXiv:0801.4947 [hep-th]}}.

\bibitem[DP85]{Dobrev:1985qv}
V.~K. Dobrev and V.~B. Petkova, \emph{{All Positive Energy Unitary Irreducible
  Representations of Extended Conformal Supersymmetry}},
\doihref{http://dx.doi.org/10.1016/0370-2693(85)91073-1}{Phys. Lett.
  \textbf{162B} (1985) 127--132}.

\bibitem[DS93]{Deser:1993yx}
S.~Deser and A.~Schwimmer, \emph{{Geometric Classification of Conformal
  Anomalies in Arbitrary Dimensions}},
  \doihref{http://dx.doi.org/10.1016/0370-2693(93)90934-A}{Phys. Lett.
  \textbf{B309} (1993) 279--284},
\href{http://arxiv.org/abs/hep-th/9302047}{{\arxivfont arXiv:hep-th/9302047}}.

\bibitem[DSWW86]{Dine:1986zy}
M.~Dine, N.~Seiberg, X.~G. Wen, and E.~Witten, \emph{{Nonperturbative Effects
  on the String World Sheet}},
\doihref{http://dx.doi.org/10.1016/0550-3213(86)90418-9}{Nucl. Phys.
  \textbf{B278} (1986) 769}.

\bibitem[Eag10]{Eager:2010yu}
R.~Eager, \emph{{Equivalence of A-Maximization and Volume Minimization}},
  \doihref{http://dx.doi.org/10.1007/JHEP01(2014)089}{JHEP \textbf{01} (2014)
  089},
\href{http://arxiv.org/abs/1011.1809}{{\arxivfont arXiv:1011.1809 [hep-th]}}.

\bibitem[EEH{\etalchar{+}}90]{Numbers}
H.-D. Ebbinghaus, J.~H. Ewing, H.~Hermes, F.~Hirzebruch, M.~Koecher,
  K.~Mainzer, J.~Neukirch, A.~Prestel, and R.~Remmert, \emph{Numbers}, Graduate
  Texts in Mathematics, vol. 123, Springer-Verlag, 1990.

\bibitem[EH13]{Elvang:2013cua}
H.~Elvang and Y.-t. Huang, \emph{{Scattering Amplitudes}},
\href{http://arxiv.org/abs/1308.1697}{{\arxivfont arXiv:1308.1697 [hep-th]}}.

\bibitem[EHS96]{Evans:1996hi}
N.~J. Evans, S.~D.~H. Hsu, and M.~Schwetz, \emph{{Phase Transitions in Softly
  Broken ${\mathcal{N}}{=}2$ SQCD at Nonzero Theta Angle}},
  \doihref{http://dx.doi.org/10.1016/S0550-3213(96)00595-0}{Nucl. Phys.
  \textbf{B484} (1997) 124--140},
\href{http://arxiv.org/abs/hep-th/9608135}{{\arxivfont arXiv:hep-th/9608135}}.

\bibitem[EN84]{Elitzur:1984kr}
S.~Elitzur and V.~P. Nair, \emph{{Nonperturbative Anomalies in Higher
  Dimensions}},
\doihref{http://dx.doi.org/10.1016/0550-3213(84)90024-5}{Nucl. Phys.
  \textbf{B243} (1984) 205}.

\bibitem[ES15]{EguchiSugawara}
T.~Eguchi and Y.~Sugawara, \emph{{\begin{uCJK}共形場理論\end{uCJK}}},
  Iwanami, 2015.

\bibitem[Eva88]{Evans:1987tm}
J.~M. Evans, \emph{{Supersymmetric {Yang-Mills} Theories and Division
  Algebras}},
\doihref{http://dx.doi.org/10.1016/0550-3213(88)90305-7}{Nucl. Phys.
  \textbf{B298} (1988) 92--108}.

\bibitem[EW09]{Elvang:2009gk}
H.~Elvang and B.~Wecht, \emph{{Semi-Direct Gauge Mediation with the 4-1
  Model}}, \doihref{http://dx.doi.org/10.1088/1126-6708/2009/06/026}{JHEP
  \textbf{06} (2009) 026},
\href{http://arxiv.org/abs/0904.4431}{{\arxivfont arXiv:0904.4431 [hep-ph]}}.

\bibitem[FHH00]{Feng:2000mi}
B.~Feng, A.~Hanany, and Y.-H. He, \emph{{D-Brane Gauge Theories from Toric
  Singularities and Toric Duality}},
  \doihref{http://dx.doi.org/10.1016/S0550-3213(00)00699-4}{Nucl. Phys.
  \textbf{B595} (2001) 165--200},
\href{http://arxiv.org/abs/hep-th/0003085}{{\arxivfont arXiv:hep-th/0003085}}.

\bibitem[FHM{\etalchar{+}}05]{Franco:2005sm}
S.~Franco, A.~Hanany, D.~Martelli, J.~Sparks, D.~Vegh, and B.~Wecht,
  \emph{{Gauge theories from toric geometry and brane tilings}},
  \doihref{http://dx.doi.org/10.1088/1126-6708/2006/01/128}{JHEP \textbf{01}
  (2006) 128},
\href{http://arxiv.org/abs/hep-th/0505211}{{\arxivfont arXiv:hep-th/0505211}}.

\bibitem[FI74]{Fayet:1974jb}
P.~Fayet and J.~Iliopoulos, \emph{{Spontaneously Broken Supergauge Symmetries
  and Goldstone Spinors}},
\doihref{http://dx.doi.org/10.1016/0370-2693(74)90310-4}{Phys. Lett.
  \textbf{51B} (1974) 461--464}.

\bibitem[Fri80]{Friedan:1980jf}
D.~H. Friedan, \emph{{Nonlinear Models in $2+\epsilon$ Dimensions}},
\doihref{http://dx.doi.org/10.1103/PhysRevLett.45.1057}{Phys. Rev. Lett.
  \textbf{45} (1980) 1057}.

\bibitem[Fri85]{Friedan:1980jm}
\bysame, \emph{{Nonlinear Models in $2+\epsilon$ Dimensions}},
\doihref{http://dx.doi.org/10.1016/0003-4916(85)90384-7}{Annals Phys.
  \textbf{163} (1985) 318}.

\bibitem[FS11]{Festuccia:2011ws}
G.~Festuccia and N.~Seiberg, \emph{{Rigid Supersymmetric Theories in Curved
  Superspace}}, \doihref{http://dx.doi.org/10.1007/JHEP06(2011)114}{JHEP
  \textbf{1106} (2011) 114},
\href{http://arxiv.org/abs/1105.0689}{{\arxivfont arXiv:1105.0689 [hep-th]}}.

\bibitem[FvNF76]{Freedman:1976xh}
D.~Z. Freedman, P.~van Nieuwenhuizen, and S.~Ferrara, \emph{{Progress Toward a
  Theory of Supergravity}},
\doihref{http://dx.doi.org/10.1103/PhysRevD.13.3214}{Phys. Rev. \textbf{D13}
  (1976) 3214--3218}.

\bibitem[Gai13]{Gaiotto:2013gwa}
D.~Gaiotto, \emph{{Kazama-Suzuki Models and BPS Domain Wall Junctions in
  ${\mathcal{N}}{=}1$ $SU(n)$ Super Yang-Mills}},
\href{http://arxiv.org/abs/1306.5661}{{\arxivfont arXiv:1306.5661 [hep-th]}}.

\bibitem[GEHO{\etalchar{+}}17]{Garcia-Etxebarria:2017crf}
I.~Garc{\'\i}a-Etxebarria, H.~Hayashi, K.~Ohmori, Y.~Tachikawa, and
  K.~Yonekura, \emph{{8D Gauge Anomalies and the Topological Green-Schwarz
  Mechanism}}, \doihref{http://dx.doi.org/10.1007/JHEP11(2017)177}{JHEP
  \textbf{11} (2017) 177},
\href{http://arxiv.org/abs/1710.04218}{{\arxivfont arXiv:1710.04218 [hep-th]}}.

\bibitem[Gep88]{Gepner:1987qi}
D.~Gepner, \emph{{Space-Time Supersymmetry in Compactified String Theory and
  Superconformal Models}},
\doihref{http://dx.doi.org/10.1016/0550-3213(88)90397-5}{Nucl. Phys.
  \textbf{B296} (1988) 757}.

\bibitem[GER15]{Garcia-Etxebarria:2015wns}
I.~Garc{\'\i}a-Etxebarria and D.~Regalado, \emph{{$ \mathcal{N}=3 $ four
  dimensional field theories}},
  \doihref{http://dx.doi.org/10.1007/JHEP03(2016)083}{JHEP \textbf{03} (2016)
  083},
\href{http://arxiv.org/abs/1512.06434}{{\arxivfont arXiv:1512.06434 [hep-th]}}.

\bibitem[GGP13]{Gadde:2013lxa}
A.~Gadde, S.~Gukov, and P.~Putrov, \emph{{(0,2) Trialities}},
  \doihref{http://dx.doi.org/10.1007/JHEP03(2014)076}{JHEP \textbf{1403} (2014)
  076},
\href{http://arxiv.org/abs/1310.0818}{{\arxivfont arXiv:1310.0818 [hep-th]}}.

\bibitem[GGP14]{Gadde:2014ppa}
\bysame, \emph{{Exact Solutions of 2D Supersymmetric Gauge Theories}},
\href{http://arxiv.org/abs/1404.5314}{{\arxivfont arXiv:1404.5314 [hep-th]}}.

\bibitem[GHMR85]{Gross:1984dd}
D.~J. Gross, J.~A. Harvey, E.~J. Martinec, and R.~Rohm, \emph{{The Heterotic
  String}},
\doihref{http://dx.doi.org/10.1103/PhysRevLett.54.502}{Phys. Rev. Lett.
  \textbf{54} (1985) 502--505}.

\bibitem[Gia17]{Giacomelli:2017ckh}
S.~Giacomelli, \emph{{RG Flows with Supersymmetry Enhancement and Geometric
  Engineering}}, \doihref{http://dx.doi.org/10.1007/JHEP06(2018)156}{JHEP
  \textbf{06} (2017) 156},
\href{http://arxiv.org/abs/1710.06469}{{\arxivfont arXiv:1710.06469 [hep-th]}}.

\bibitem[Gia18]{Giacomelli:2018ziv}
\bysame, \emph{{Infrared Enhancement of Supersymmetry in Four Dimensions}},
  \doihref{http://dx.doi.org/10.1007/JHEP10(2018)041}{JHEP \textbf{10} (2018)
  041},
\href{http://arxiv.org/abs/1808.00592}{{\arxivfont arXiv:1808.00592 [hep-th]}}.

\bibitem[GIR08]{Grinstein:2008qk}
B.~Grinstein, K.~A. Intriligator, and I.~Z. Rothstein, \emph{{Comments on
  Unparticles}},
  \doihref{http://dx.doi.org/10.1016/j.physletb.2008.03.020}{Phys. Lett.
  \textbf{B662} (2008) 367--374},
\href{http://arxiv.org/abs/0801.1140}{{\arxivfont arXiv:0801.1140 [hep-ph]}}.

\bibitem[GKKS17]{Gaiotto:2017yup}
D.~Gaiotto, A.~Kapustin, Z.~Komargodski, and N.~Seiberg, \emph{{Theta, Time
  Reversal, and Temperature}},
\href{http://arxiv.org/abs/1703.00501}{{\arxivfont arXiv:1703.00501 [hep-th]}}.

\bibitem[GKLP11]{Gang:2011xp}
D.~Gang, E.~Koh, K.~Lee, and J.~Park, \emph{{ABCD of 3d ${\cal N}=8$ and 4
  Superconformal Field Theories}},
\href{http://arxiv.org/abs/1108.3647}{{\arxivfont arXiv:1108.3647 [hep-th]}}.

\bibitem[GKO{\etalchar{+}}16]{Gomis:2016sab}
J.~Gomis, Z.~Komargodski, H.~Ooguri, N.~Seiberg, and Y.~Wang, \emph{{Shortening
  Anomalies in Supersymmetric Theories}},
  \doihref{http://dx.doi.org/10.1007/JHEP01(2017)067}{JHEP \textbf{01} (2017)
  067},
\href{http://arxiv.org/abs/1611.03101}{{\arxivfont arXiv:1611.03101 [hep-th]}}.

\bibitem[GKS{\etalchar{+}}10]{Green:2010da}
D.~Green, Z.~Komargodski, N.~Seiberg, Y.~Tachikawa, and B.~Wecht,
  \emph{{Exactly Marginal Deformations and Global Symmetries}},
  \doihref{http://dx.doi.org/10.1007/JHEP06(2010)106}{JHEP \textbf{06} (2010)
  106},
\href{http://arxiv.org/abs/1005.3546}{{\arxivfont arXiv:1005.3546 [hep-th]}}.

\bibitem[GKSW14]{Gaiotto:2014kfa}
D.~Gaiotto, A.~Kapustin, N.~Seiberg, and B.~Willett, \emph{{Generalized Global
  Symmetries}}, \doihref{http://dx.doi.org/10.1007/JHEP02(2015)172}{JHEP
  \textbf{02} (2015) 172},
\href{http://arxiv.org/abs/1412.5148}{{\arxivfont arXiv:1412.5148 [hep-th]}}.

\bibitem[GM09]{Gaiotto:2009gz}
D.~Gaiotto and J.~Maldacena, \emph{{The Gravity Duals of ${\mathcal{N}}\!=2$
  Superconformal Field Theories}},
\href{http://arxiv.org/abs/0904.4466}{{\arxivfont arXiv:0904.4466 [hep-th]}}.

\bibitem[GMSW04]{Gauntlett:2004yd}
J.~P. Gauntlett, D.~Martelli, J.~Sparks, and D.~Waldram, \emph{{Sasaki-Einstein
  Metrics on $S^2$ $\times$ $S^3$}}, Adv. Theor. Math. Phys. \textbf{8} (2004)
  711--734,
\href{http://arxiv.org/abs/hep-th/0403002}{{\arxivfont arXiv:hep-th/0403002}}.

\bibitem[GP95]{Giddings:1995ns}
S.~B. Giddings and J.~M. Pierre, \emph{{Some Exact Results in Supersymmetric
  Theories Based on Exceptional Groups}},
  \doihref{http://dx.doi.org/10.1103/PhysRevD.52.6065}{Phys. Rev. \textbf{D52}
  (1995) 6065--6073},
\href{http://arxiv.org/abs/hep-th/9506196}{{\arxivfont arXiv:hep-th/9506196}}.

\bibitem[GP97]{Giveon:1997sn}
A.~Giveon and O.~Pelc, \emph{{M Theory, Type IIA String and 4D
  ${\mathcal{N}}\!=1$ SUSY $SU(N_L)\times SU(N_R)$ Gauge Theory}},
  \doihref{http://dx.doi.org/10.1016/S0550-3213(97)00687-1}{Nucl. Phys.
  \textbf{B512} (1998) 103--147},
\href{http://arxiv.org/abs/hep-th/9708168}{{\arxivfont arXiv:hep-th/9708168}}.

\bibitem[GPR97]{Giveon:1997gr}
A.~Giveon, O.~Pelc, and E.~Rabinovici, \emph{{The Coulomb Phase in
  ${\mathcal{N}}\!=1$ Gauge Theories with an LG-Type Superpotential}},
  \doihref{http://dx.doi.org/10.1016/S0550-3213(97)00297-6}{Nucl. Phys.
  \textbf{B499} (1997) 100--124},
\href{http://arxiv.org/abs/hep-th/9701045}{{\arxivfont arXiv:hep-th/9701045}}.

\bibitem[GQ87]{Gepner:1986hr}
D.~Gepner and Z.-a. Qiu, \emph{{Modular Invariant Partition Functions for
  Parafermionic Field Theories}},
\doihref{http://dx.doi.org/10.1016/0550-3213(87)90348-8}{Nucl. Phys.
  \textbf{B285} (1987) 423}.

\bibitem[Gre97a]{Greene:1996cy}
B.~R. Greene, \emph{{String Theory on Calabi-Yau Manifolds}},
  \href{http://arxiv.org/abs/hep-th/9702155}{{\arxivfont
  arXiv:hep-th/9702155}}.
(TASI 1996).

\bibitem[Gre97b]{Gremm:1997sz}
M.~Gremm, \emph{{The Coulomb Branch of ${\mathcal{N}}\!=1$ Supersymmetric
  $SU(\hbox{$N_c$}$) $\times$ $SU(\hbox{$N_c$}$) Gauge Theories}},
  \doihref{http://dx.doi.org/10.1103/PhysRevD.57.2537}{Phys. Rev. \textbf{D57}
  (1998) 2537--2542},
\href{http://arxiv.org/abs/hep-th/9707071}{{\arxivfont arXiv:hep-th/9707071}}.

\bibitem[GRW15]{Gadde:2015wta}
A.~Gadde, S.~S. Razamat, and B.~Willett, \emph{{On the reduction of 4d $
  \mathcal{N}{=}1 $ theories on $ {\mathbb{S}}^2 $}},
  \doihref{http://dx.doi.org/10.1007/JHEP11(2015)163}{JHEP \textbf{11} (2015)
  163},
\href{http://arxiv.org/abs/1506.08795}{{\arxivfont arXiv:1506.08795 [hep-th]}}.

\bibitem[GS84]{Green:1984sg}
M.~B. Green and J.~H. Schwarz, \emph{{Anomaly Cancellation in Supersymmetric
  D=10 Gauge Theory and Superstring Theory}},
\doihref{http://dx.doi.org/10.1016/0370-2693(84)91565-X}{Phys. Lett.
  \textbf{149B} (1984) 117--122}.

\bibitem[Gus07]{Gustavsson:2007vu}
A.~Gustavsson, \emph{{Algebraic Structures on Parallel M2-Branes}},
  \doihref{http://dx.doi.org/10.1016/j.nuclphysb.2008.11.014}{Nucl. Phys.
  \textbf{B811} (2007) 66--76},
\href{http://arxiv.org/abs/0709.1260}{{\arxivfont arXiv:0709.1260 [hep-th]}}.

\bibitem[Hai02a]{Hailu:2002bg}
G.~Hailu, \emph{{${\mathcal{N}}\!=1$ Supersymmetric $SU(2)$ $^r$ Moose
  Theories}}, \doihref{http://dx.doi.org/10.1103/PhysRevD.67.085023}{Phys. Rev.
  \textbf{D67} (2003) 085023},
\href{http://arxiv.org/abs/hep-th/0209266}{{\arxivfont arXiv:hep-th/0209266}}.

\bibitem[Hai02b]{Hailu:2002bh}
\bysame, \emph{{Quantum Moduli Spaces of Linear and Ring Mooses}},
  \doihref{http://dx.doi.org/10.1016/S0370-2693(02)03160-X}{Phys. Lett.
  \textbf{B552} (2003) 265--272},
\href{http://arxiv.org/abs/hep-th/0209267}{{\arxivfont arXiv:hep-th/0209267}}.

\bibitem[Har05]{Harvey:2005it}
J.~A. Harvey, \emph{{TASI 2003 Lectures on Anomalies}},
\href{http://arxiv.org/abs/hep-th/0509097}{{\arxivfont arXiv:hep-th/0509097}}.

\bibitem[HBJ92]{Hirzebruch}
F.~Hirzebruch, T.~Berger, and R.~Jung, \emph{Manifolds and modular forms},
  Aspects of Mathematics, vol. E20, Vieweg, Braunschweig, 1992. With appendices
  by N.P. Skoruppa and by P. Baum, translated by P. S. Landweber.

\bibitem[HKLM99]{Hollowood:1999qn}
T.~J. Hollowood, V.~V. Khoze, W.-J. Lee, and M.~P. Mattis, \emph{{Breakdown of
  Cluster Decomposition in Instanton Calculations of the Gluino Condensate}},
  \doihref{http://dx.doi.org/10.1016/S0550-3213(99)00503-9}{Nucl. Phys.
  \textbf{B570} (2000) 241--266},
\href{http://arxiv.org/abs/hep-th/9904116}{{\arxivfont arXiv:hep-th/9904116}}.

\bibitem[HKP{\etalchar{+}}03]{MirrorBook}
K.~Hori, S.~Katz, R.~Pandharipande, R.~Thomas, C.~Vafa, R.~Vakil, and
  E.~Zaslow, \emph{Mirror symmetry}, Clay Mathematics Monographs, vol.~1,
  American Mathematical Society, Clay Mathematics Institute, 2003.

\bibitem[HKS10]{Hollands:2010xa}
L.~Hollands, C.~A. Keller, and J.~Song, \emph{{From SO/Sp Instantons to
  W-Algebra Blocks}}, \doihref{http://dx.doi.org/10.1007/JHEP03(2011)053}{JHEP
  \textbf{03} (2011) 053},
\href{http://arxiv.org/abs/1012.4468}{{\arxivfont arXiv:1012.4468 [hep-th]}}.

\bibitem[HLM{\etalchar{+}}16]{Hofman:2016awc}
D.~M. Hofman, D.~Li, D.~Meltzer, D.~Poland, and F.~Rejon-Barrera, \emph{{A
  Proof of the Conformal Collider Bounds}},
  \doihref{http://dx.doi.org/10.1007/JHEP06(2016)111}{JHEP \textbf{06} (2016)
  111},
\href{http://arxiv.org/abs/1603.03771}{{\arxivfont arXiv:1603.03771 [hep-th]}}.

\bibitem[Hor03]{Hori:2002fa}
K.~Hori, \emph{{Trieste Lectures on Mirror Symmetry}}, ICTP Lect. Notes Ser.
  \textbf{13} (2003) 109--202.
\url{http://users.ictp.it/~pub_off/lectures/lns013/Hori/Hori.pdf}.

\bibitem[HW95]{Horava:1995qa}
P.~Ho{\v r}ava and E.~Witten, \emph{{Heterotic and Type I String Dynamics from
  Eleven-Dimensions}},
  \doihref{http://dx.doi.org/10.1016/0550-3213(95)00621-4}{Nucl. Phys.
  \textbf{B460} (1995) 506--524},
\href{http://arxiv.org/abs/hep-th/9510209}{{\arxivfont arXiv:hep-th/9510209}}.

\bibitem[HW96]{Horava:1996ma}
P.~Ho{\v ra}va and E.~Witten, \emph{{Eleven-Dimensional Supergravity on a
  Manifold with Boundary}},
  \doihref{http://dx.doi.org/10.1016/0550-3213(96)00308-2}{Nucl. Phys.
  \textbf{B475} (1996) 94--114},
\href{http://arxiv.org/abs/hep-th/9603142}{{\arxivfont arXiv:hep-th/9603142}}.

\bibitem[HY15]{Honda:2015yha}
M.~Honda and Y.~Yoshida, \emph{{Supersymmetric Index on T$^2$ $\times$ S$^2$
  and Elliptic Genus}},
\href{http://arxiv.org/abs/1504.04355}{{\arxivfont arXiv:1504.04355 [hep-th]}}.

\bibitem[IN16]{Intriligator:2016sgx}
K.~Intriligator and E.~Nardoni, \emph{{Deformations of $W_{A,D,E}$ SCFTs}},
  \doihref{http://dx.doi.org/10.1007/JHEP09(2016)043}{JHEP \textbf{09} (2016)
  043},
\href{http://arxiv.org/abs/1604.04294}{{\arxivfont arXiv:1604.04294 [hep-th]}}.

\bibitem[Int95]{Intriligator:1995ff}
K.~A. Intriligator, \emph{{New RG Fixed Points and Duality in Supersymmetric
  $Sp(N_c)$ and $SO(N_c)$ Gauge Theories}},
  \doihref{http://dx.doi.org/10.1016/0550-3213(95)00296-5}{Nucl. Phys.
  \textbf{B448} (1995) 187--198},
\href{http://arxiv.org/abs/hep-th/9505051}{{\arxivfont arXiv:hep-th/9505051}}.

\bibitem[Int05]{Intriligator:2005if}
\bysame, \emph{{IR Free Or Interacting? a Proposed Diagnostic}},
  \doihref{http://dx.doi.org/10.1016/j.nuclphysb.2005.10.005}{Nucl. Phys.
  \textbf{B730} (2005) 239--251},
\href{http://arxiv.org/abs/hep-th/0509085}{{\arxivfont arXiv:hep-th/0509085}}.

\bibitem[IP95]{Intriligator:1995ne}
K.~A. Intriligator and P.~Pouliot, \emph{{Exact Superpotentials, Quantum Vacua
  and Duality in Supersymmetric Sp(\hbox{$N_c$}) Gauge Theories}},
  \doihref{http://dx.doi.org/10.1016/0370-2693(95)00618-U}{Phys.Lett.
  \textbf{B353} (1995) 471--476},
\href{http://arxiv.org/abs/hep-th/9505006}{{\arxivfont arXiv:hep-th/9505006}}.

\bibitem[IS94]{Intriligator:1994sm}
K.~A. Intriligator and N.~Seiberg, \emph{{Phases of ${\mathcal{N}}{=}1$
  Supersymmetric Gauge Theories in Four- Dimensions}},
  \doihref{http://dx.doi.org/10.1016/0550-3213(94)90215-1}{Nucl. Phys.
  \textbf{B431} (1994) 551--568},
\href{http://arxiv.org/abs/hep-th/9408155}{{\arxivfont arXiv:hep-th/9408155}}.

\bibitem[IS95a]{Intriligator:1995id}
\bysame, \emph{{Duality, Monopoles, Dyons, Confinement and Oblique Confinement
  in Supersymmetric SO(\hbox{$N_c$}) Gauge Theories}},
  \doihref{http://dx.doi.org/10.1016/0550-3213(95)00159-P}{Nucl.Phys.
  \textbf{B444} (1995) 125--160},
\href{http://arxiv.org/abs/hep-th/9503179}{{\arxivfont arXiv:hep-th/9503179}}.

\bibitem[IS95b]{Intriligator:1995au}
\bysame, \emph{{Lectures on Supersymmetric Gauge Theories and
  Electric--magnetic Duality}}, Nucl.Phys.Proc.Suppl. \textbf{45BC} (1996)
  1--28,
\href{http://arxiv.org/abs/hep-th/9509066}{{\arxivfont arXiv:hep-th/9509066}}.

\bibitem[ISS94]{Intriligator:1994rx}
K.~A. Intriligator, N.~Seiberg, and S.~H. Shenker, \emph{{Proposal for a Simple
  Model of Dynamical SUSY Breaking}},
  \doihref{http://dx.doi.org/10.1016/0370-2693(94)01336-B}{Phys. Lett.
  \textbf{B342} (1995) 152--154},
\href{http://arxiv.org/abs/hep-ph/9410203}{{\arxivfont arXiv:hep-ph/9410203}}.

\bibitem[IW03a]{Intriligator:2003jj}
K.~A. Intriligator and B.~Wecht, \emph{{The Exact Superconformal R-Symmetry
  Maximizes a}},
  \doihref{http://dx.doi.org/10.1016/S0550-3213(03)00459-0}{Nucl. Phys.
  \textbf{B667} (2003) 183--200},
\href{http://arxiv.org/abs/hep-th/0304128}{{\arxivfont arXiv:hep-th/0304128}}.

\bibitem[IW03b]{Intriligator:2003mi}
\bysame, \emph{{RG Fixed Points and Flows in SQCD with Adjoints}},
  \doihref{http://dx.doi.org/10.1016/j.nuclphysb.2003.10.033}{Nucl. Phys.
  \textbf{B677} (2004) 223--272},
\href{http://arxiv.org/abs/hep-th/0309201}{{\arxivfont arXiv:hep-th/0309201}}.

\bibitem[Kac77]{Kac:1977qb}
V.~G. Kac, \emph{{A Sketch of Lie Superalgebra Theory}},
\doihref{http://dx.doi.org/10.1007/BF01609166}{Commun. Math. Phys. \textbf{53}
  (1977) 31--64}.

\bibitem[Kap96]{Kapustin:1996nb}
A.~Kapustin, \emph{{The Coulomb Branch of ${\mathcal{N}}\!=1$ Supersymmetric
  Gauge Theory with Adjoint and Fundamental Matter}},
  \doihref{http://dx.doi.org/10.1016/S0370-2693(97)00209-8}{Phys. Lett.
  \textbf{B398} (1997) 104--109},
\href{http://arxiv.org/abs/hep-th/9611049}{{\arxivfont arXiv:hep-th/9611049}}.

\bibitem[Kar97]{Karch:1997jp}
A.~Karch, \emph{{More on ${\mathcal{N}}\!=1$ Selfdualities and Exceptional
  Gauge Groups}},
  \doihref{http://dx.doi.org/10.1016/S0370-2693(97)00604-7}{Phys. Lett.
  \textbf{B405} (1997) 280--286},
\href{http://arxiv.org/abs/hep-th/9702179}{{\arxivfont arXiv:hep-th/9702179}}.

\bibitem[Kat87]{Kato:1987td}
A.~Kato, \emph{{Classification of Modular Invariant Partition Functions in
  Two-Dimensions}},
\doihref{http://dx.doi.org/10.1142/S0217732387000732}{Mod. Phys. Lett.
  \textbf{A2} (1987) 585}.

\bibitem[Kaw96]{Kawano:1996bd}
T.~Kawano, \emph{{Duality of ${\mathcal{N}}\!=1$ Supersymmetric SO(10) Gauge
  Theory with Matter in the Spinorial Representation}},
  \doihref{http://dx.doi.org/10.1143/PTP.95.963}{Prog. Theor. Phys. \textbf{95}
  (1996) 963--968},
\href{http://arxiv.org/abs/hep-th/9602035}{{\arxivfont arXiv:hep-th/9602035}}.

\bibitem[Kaw09]{Kawai:2009ci}
T.~Kawai, \emph{{Twisted Elliptic Genera of ${\mathcal{N}}{=}2$ SCFTs in Two
  Dimensions}}, \doihref{http://dx.doi.org/10.1088/1751-8113/45/39/395401}{J.
  Phys. \textbf{A45} (2012) 395401},
\href{http://arxiv.org/abs/0909.1879}{{\arxivfont arXiv:0909.1879 [hep-th]}}.

\bibitem[KK10]{Kawai:2010yj}
H.~Kawai and Y.~Kikukawa, \emph{{A Lattice Study of ${\mathcal{N}}{=}2$
  Landau-Ginzburg Model Using a Nicolai Map}},
  \doihref{http://dx.doi.org/10.1103/PhysRevD.83.074502}{Phys. Rev.
  \textbf{D83} (2011) 074502},
\href{http://arxiv.org/abs/1005.4671}{{\arxivfont arXiv:1005.4671 [hep-lat]}}.

\bibitem[KL14a]{Kutasov:2014yqa}
D.~Kutasov and J.~Lin, \emph{{Exceptional ${\mathcal{N}}\!=1$ Duality}},
\href{http://arxiv.org/abs/1401.4168}{{\arxivfont arXiv:1401.4168 [hep-th]}}.

\bibitem[KL14b]{Kutasov:2014wwa}
\bysame, \emph{{${\mathcal{N}}\!=1$ Duality and the Superconformal Index}},
\href{http://arxiv.org/abs/1402.5411}{{\arxivfont arXiv:1402.5411 [hep-th]}}.

\bibitem[KM94]{Kawai:1994np}
T.~Kawai and K.~Mohri, \emph{{Geometry of (0,2) Landau-Ginzburg Orbifolds}},
  \doihref{http://dx.doi.org/10.1016/0550-3213(94)90178-3}{Nucl.Phys.
  \textbf{B425} (1994) 191--216},
\href{http://arxiv.org/abs/hep-th/9402148}{{\arxivfont arXiv:hep-th/9402148}}.

\bibitem[KMMR05]{Kinney:2005ej}
J.~Kinney, J.~M. Maldacena, S.~Minwalla, and S.~Raju, \emph{{An Index for 4
  Dimensional Super Conformal Theories}},
  \doihref{http://dx.doi.org/10.1007/s00220-007-0258-7}{Commun. Math. Phys.
  \textbf{275} (2007) 209--254},
\href{http://arxiv.org/abs/hep-th/0510251}{{\arxivfont arXiv:hep-th/0510251}}.

\bibitem[Kon84]{Konishi:1983hf}
K.~Konishi, \emph{{Anomalous Supersymmetry Transformation of Some Composite
  Operators in SQCD}},
\doihref{http://dx.doi.org/10.1016/0370-2693(84)90311-3}{Phys. Lett.
  \textbf{B135} (1984) 439--444}.

\bibitem[Kon96]{Konishi:1996iz}
\bysame, \emph{{Confinement, Supersymmetry Breaking and Theta Parameter
  Dependence in the Seiberg-Witten Model}},
  \doihref{http://dx.doi.org/10.1016/S0370-2693(96)01527-4}{Phys.Lett.
  \textbf{B392} (1997) 101--105},
\href{http://arxiv.org/abs/hep-th/9609021}{{\arxivfont arXiv:hep-th/9609021}}.

\bibitem[KOTY05]{Kawano:2005nc}
T.~Kawano, Y.~Ookouchi, Y.~Tachikawa, and F.~Yagi, \emph{{Pouliot Type Duality
  via $a$-maximization}},
  \doihref{http://dx.doi.org/10.1016/j.nuclphysb.2005.11.024}{Nucl. Phys.
  \textbf{B735} (2006) 1--16},
\href{http://arxiv.org/abs/hep-th/0509230}{{\arxivfont arXiv:hep-th/0509230}}.

\bibitem[KP07]{Kats:2007mq}
Y.~Kats and P.~Petrov, \emph{{Effect of Curvature Squared Corrections in AdS on
  the Viscosity of the Dual Gauge Theory}},
  \doihref{http://dx.doi.org/10.1088/1126-6708/2009/01/044}{JHEP \textbf{01}
  (2009) 044},
\href{http://arxiv.org/abs/0712.0743}{{\arxivfont arXiv:0712.0743 [hep-th]}}.

\bibitem[KPS03]{Kutasov:2003iy}
D.~Kutasov, A.~Parnachev, and D.~A. Sahakyan, \emph{{Central charges and
  U(1)(R) symmetries in N=1 superYang-Mills}},
  \doihref{http://dx.doi.org/10.1088/1126-6708/2003/11/013}{JHEP \textbf{11}
  (2003) 013},
\href{http://arxiv.org/abs/hep-th/0308071}{{\arxivfont arXiv:hep-th/0308071}}.

\bibitem[KS85]{Konishi:1985tu}
K.-i. Konishi and K.-i. Shizuya, \emph{{Functional Integral Approach to Chiral
  Anomalies in Supersymmetric Gauge Theories}},
\doihref{http://dx.doi.org/10.1007/BF02724227}{Nuovo Cim. \textbf{A90} (1985)
  111}.

\bibitem[KS89a]{Kazama:1988uz}
Y.~Kazama and H.~Suzuki, \emph{{Characterization of ${\mathcal{N}}\!=2$
  Superconformal Models Generated by Coset Space Method}},
\doihref{http://dx.doi.org/10.1016/0370-2693(89)91378-6}{Phys.Lett.
  \textbf{B216} (1989) 112}.

\bibitem[KS89b]{Kazama:1988qp}
\bysame, \emph{{New ${\mathcal{N}}\!=2$ Superconformal Field Theories and
  Superstring Compactification}},
\doihref{http://dx.doi.org/10.1016/0550-3213(89)90250-2}{Nucl.Phys.
  \textbf{B321} (1989) 232}.

\bibitem[KS95]{Kutasov:1995np}
D.~Kutasov and A.~Schwimmer, \emph{{On Duality in Supersymmetric Yang-Mills
  Theory}}, \doihref{http://dx.doi.org/10.1016/0370-2693(95)00676-C}{Phys.
  Lett. \textbf{B354} (1995) 315--321},
\href{http://arxiv.org/abs/hep-th/9505004}{{\arxivfont arXiv:hep-th/9505004}}.

\bibitem[KS97]{Kovner:1997im}
A.~Kovner and M.~A. Shifman, \emph{{Chirally Symmetric Phase of Supersymmetric
  Gluodynamics}}, \doihref{http://dx.doi.org/10.1103/PhysRevD.56.2396}{Phys.
  Rev. \textbf{D56} (1997) 2396--2402},
\href{http://arxiv.org/abs/hep-th/9702174}{{\arxivfont arXiv:hep-th/9702174}}.

\bibitem[KS99]{Kac:1999gw}
V.~G. Kac and A.~V. Smilga, \emph{{Vacuum structure in supersymmetric
  Yang-Mills theories with any gauge group}},
\href{http://arxiv.org/abs/hep-th/9902029}{{\arxivfont arXiv:hep-th/9902029}}.

\bibitem[KS11a]{Kamata:2011fr}
S.~Kamata and H.~Suzuki, \emph{{Numerical simulation of the $\mathcal{N}=(2,2)$
  Landau-Ginzburg model}},
  \doihref{http://dx.doi.org/10.1016/j.nuclphysb.2011.09.007}{Nucl. Phys.
  \textbf{B854} (2012) 552--574},
\href{http://arxiv.org/abs/1107.1367}{{\arxivfont arXiv:1107.1367 [hep-lat]}}.

\bibitem[KS11b]{Komargodski:2011vj}
Z.~Komargodski and A.~Schwimmer, \emph{{On Renormalization Group Flows in Four
  Dimensions}},
\href{http://arxiv.org/abs/1107.3987}{{\arxivfont arXiv:1107.3987 [hep-th]}}.

\bibitem[KS14]{Kapustin:2014gua}
A.~Kapustin and N.~Seiberg, \emph{{Coupling a QFT to a TQFT and Duality}},
  \doihref{http://dx.doi.org/10.1007/JHEP04(2014)001}{JHEP \textbf{1404} (2014)
  001},
\href{http://arxiv.org/abs/1401.0740}{{\arxivfont arXiv:1401.0740 [hep-th]}}.

\bibitem[KSS95]{Kutasov:1995ss}
D.~Kutasov, A.~Schwimmer, and N.~Seiberg, \emph{{Chiral Rings, Singularity
  Theory and Electric-Magnetic Duality}},
  \doihref{http://dx.doi.org/10.1016/0550-3213(95)00599-4}{Nucl. Phys.
  \textbf{B459} (1996) 455--496},
\href{http://arxiv.org/abs/hep-th/9510222}{{\arxivfont arXiv:hep-th/9510222}}.

\bibitem[KSS03]{Kovtun:2003wp}
P.~Kovtun, D.~T. Son, and A.~O. Starinets, \emph{{Holography and Hydrodynamics:
  Diffusion on Stretched Horizons}},
  \doihref{http://dx.doi.org/10.1088/1126-6708/2003/10/064}{JHEP \textbf{10}
  (2003) 064},
\href{http://arxiv.org/abs/hep-th/0309213}{{\arxivfont arXiv:hep-th/0309213}}.

\bibitem[KT83]{Kugo:1982bn}
T.~Kugo and P.~K. Townsend, \emph{{Supersymmetry and the Division Algebras}},
\doihref{http://dx.doi.org/10.1016/0550-3213(83)90584-9}{Nucl. Phys.
  \textbf{B221} (1983) 357--380}.

\bibitem[KTY97]{Kitao:1996mb}
T.~Kitao, S.~Terashima, and S.-K. Yang, \emph{{${\mathcal{N}}\!=2$ Curves and a
  Coulomb Phase in ${\mathcal{N}}\!=1$ SUSY Gauge Theories with Adjoint and
  Fundamental Matters}},
  \doihref{http://dx.doi.org/10.1016/S0370-2693(97)00261-X}{Phys. Lett.
  \textbf{B399} (1997) 75--82},
\href{http://arxiv.org/abs/hep-th/9701009}{{\arxivfont arXiv:hep-th/9701009}}.

\bibitem[Kut95]{Kutasov:1995ve}
D.~Kutasov, \emph{{A Comment on Duality in ${\mathcal{N}}{=}1$ Supersymmetric
  Nonabelian Gauge Theories}},
  \doihref{http://dx.doi.org/10.1016/0370-2693(95)00392-X}{Phys. Lett.
  \textbf{B351} (1995) 230--234},
\href{http://arxiv.org/abs/hep-th/9503086}{{\arxivfont arXiv:hep-th/9503086}}.

\bibitem[KWY10]{Kapustin:2010xq}
A.~Kapustin, B.~Willett, and I.~Yaakov, \emph{{Nonperturbative Tests of
  Three-Dimensional Dualities}},
  \doihref{http://dx.doi.org/10.1007/JHEP10(2010)013}{JHEP \textbf{10} (2010)
  013},
\href{http://arxiv.org/abs/1003.5694}{{\arxivfont arXiv:1003.5694 [hep-th]}}.

\bibitem[KY07]{Kawano:2007rz}
T.~Kawano and F.~Yagi, \emph{{Supersymmetric ${\mathcal{N}}{=}1$ Spin(10) Gauge
  Theory with Two Spinors via $a$-maximization}},
  \doihref{http://dx.doi.org/10.1016/j.nuclphysb.2007.07.007}{Nucl. Phys.
  \textbf{B786} (2007) 135--151},
\href{http://arxiv.org/abs/0705.4022}{{\arxivfont arXiv:0705.4022 [hep-th]}}.

\bibitem[KY10]{Kawano:2010hd}
\bysame, \emph{{$a$-maximization in ${\mathcal{N}}{=}1$ Supersymmetric Spin(10)
  Gauge Theories}}, \doihref{http://dx.doi.org/10.1142/S0217751X10051037}{Int.
  J. Mod. Phys. \textbf{A25} (2010) 5595--5645},
\href{http://arxiv.org/abs/1010.0065}{{\arxivfont arXiv:1010.0065 [hep-th]}}.

\bibitem[KYY93]{Kawai:1993jk}
T.~Kawai, Y.~Yamada, and S.-K. Yang, \emph{{Elliptic Genera and
  ${\mathcal{N}}{=}2$ Superconformal Field Theory}},
  \doihref{http://dx.doi.org/10.1016/0550-3213(94)90428-6}{Nucl.Phys.
  \textbf{B414} (1994) 191--212},
\href{http://arxiv.org/abs/hep-th/9306096}{{\arxivfont arXiv:hep-th/9306096}}.

\bibitem[KZ01]{KZ}
M.~Kontsevich and D.~Zagier, \emph{Periods}, Mathematics unlimited---2001 and
  beyond, Springer, Berlin, 2001, pp.~771--808.
  \url{http://www.ihes.fr/~maxim/TEXTS/Periods.pdf}.

\bibitem[LP10]{Lambert:2010ji}
N.~Lambert and C.~Papageorgakis, \emph{{Relating $U(N)\times U(N)$ to
  $SU(N)\times SU(N)$ Chern-Simons Membrane Theories}},
  \doihref{http://dx.doi.org/10.1007/JHEP04(2010)104}{JHEP \textbf{04} (2010)
  104},
\href{http://arxiv.org/abs/1001.4779}{{\arxivfont arXiv:1001.4779 [hep-th]}}.

\bibitem[LPT97]{Lykken:1997gy}
J.~D. Lykken, E.~Poppitz, and S.~P. Trivedi, \emph{{Chiral Gauge Theories from
  D-Branes}}, \doihref{http://dx.doi.org/10.1016/S0370-2693(97)01220-3}{Phys.
  Lett. \textbf{B416} (1998) 286--294},
\href{http://arxiv.org/abs/hep-th/9708134}{{\arxivfont arXiv:hep-th/9708134}}.

\bibitem[LR06]{Lee:2006ru}
S.~Lee and S.-J. Rey, \emph{{Comments on Anomalies and Charges of Toric-Quiver
  Duals}}, \doihref{http://dx.doi.org/10.1088/1126-6708/2006/03/068}{JHEP
  \textbf{03} (2006) 068},
\href{http://arxiv.org/abs/hep-th/0601223}{{\arxivfont arXiv:hep-th/0601223}}.

\bibitem[LRS15]{Liendo:2015ofa}
P.~Liendo, I.~Ramirez, and J.~Seo, \emph{{Stress-tensor OPE in $ \mathcal{N}=2
  $ superconformal theories}},
  \doihref{http://dx.doi.org/10.1007/JHEP02(2016)019}{JHEP \textbf{02} (2016)
  019},
\href{http://arxiv.org/abs/1509.00033}{{\arxivfont arXiv:1509.00033 [hep-th]}}.

\bibitem[LS95a]{Leigh:1995ep}
R.~G. Leigh and M.~J. Strassler, \emph{{Exactly Marginal Operators and Duality
  in Four-Dimensional ${\mathcal{N}}\!=1$ Supersymmetric Gauge Theory}},
  \doihref{http://dx.doi.org/10.1016/0550-3213(95)00261-P}{Nucl. Phys.
  \textbf{B447} (1995) 95--136},
\href{http://arxiv.org/abs/hep-th/9503121}{{\arxivfont arXiv:hep-th/9503121}}.

\bibitem[LS95b]{Leigh:1995qp}
R.~G. Leigh and M.~J. Strassler, \emph{{Duality of $Sp(2N_c)$ and $SO(N_c)$
  Supersymmetric Gauge Theories with Adjoint Matter}},
  \doihref{http://dx.doi.org/10.1016/0370-2693(95)00871-H}{Phys. Lett.
  \textbf{B356} (1995) 492--499},
\href{http://arxiv.org/abs/hep-th/9505088}{{\arxivfont arXiv:hep-th/9505088}}.

\bibitem[LS96]{Leigh:1996ds}
R.~G. Leigh and M.~J. Strassler, \emph{{Accidental Symmetries and
  ${\mathcal{N}}\!=1$ Duality in Supersymmetric Gauge Theory}},
  \doihref{http://dx.doi.org/10.1016/S0550-3213(97)00204-6}{Nucl. Phys.
  \textbf{B496} (1997) 132--148},
\href{http://arxiv.org/abs/hep-th/9611020}{{\arxivfont arXiv:hep-th/9611020}}.

\bibitem[LVW89]{Lerche:1989uy}
W.~Lerche, C.~Vafa, and N.~P. Warner, \emph{{Chiral Rings in
  ${\mathcal{N}}{=}2$ Superconformal Theories}},
\doihref{http://dx.doi.org/10.1016/0550-3213(89)90474-4}{Nucl.Phys.
  \textbf{B324} (1989) 427}.

\bibitem[Mac77]{Mack:1975je}
G.~Mack, \emph{{All Unitary Ray Representations of the Conformal Group
  $SU(2,2)$ with Positive Energy}},
\doihref{http://dx.doi.org/10.1007/BF01613145}{Commun. Math. Phys. \textbf{55}
  (1977) 1}.

\bibitem[Mal97]{Maldacena:1997re}
J.~M. Maldacena, \emph{{The Large N limit of superconformal field theories and
  supergravity}}, \doihref{http://dx.doi.org/10.1023/A:1026654312961}{Int. J.
  Theor. Phys. \textbf{38} (1999) 1113--1133},
\href{http://arxiv.org/abs/hep-th/9711200}{{\arxivfont arXiv:hep-th/9711200}}.

\bibitem[Man75]{Mandelstam:1974vf}
S.~Mandelstam, \emph{{Vortices and Quark Confinement in Nonabelian Gauge
  Theories}},
\doihref{http://dx.doi.org/10.1016/0370-2693(75)90221-X}{Phys. Lett.
  \textbf{53B} (1975) 476--478}.

\bibitem[Man76]{Mandelstam:1974pi}
\bysame, \emph{{Vortices and Quark Confinement in Nonabelian Gauge Theories}},
\doihref{http://dx.doi.org/10.1016/0370-1573(76)90043-0}{Phys. Rept.
  \textbf{23} (1976) 245--249}.

\bibitem[Mar98]{Maru:1998hp}
N.~Maru, \emph{{Confining Phase in SUSY SO(12) Gauge Theory with One Spinor and
  Several Vectors}}, \doihref{http://dx.doi.org/10.1142/S021773239800142X}{Mod.
  Phys. Lett. \textbf{A13} (1998) 1361--1370},
\href{http://arxiv.org/abs/hep-th/9801187}{{\arxivfont arXiv:hep-th/9801187}}.

\bibitem[Mar15]{Marino}
M.~Mari{\~n}o, \emph{Instantons and large {$N$} : an introduction to
  non-perturbative methods in quantum field theory}, Cambridge University
  Press, 2015.

\bibitem[Mar18]{MaruyoshiTalk}
K.~Maruyoshi, \emph{{Deformations of 4d SCFTs and infrared supersymmetry
  enhancement}}.
  \url{http://www2.yukawa.kyoto-u.ac.jp/~nfst2018/Slide/Maruyoshi.pdf}. {Talk
  at the workshop \it New Frontier in String Theory}.

\bibitem[Mat87]{Matsuo:1986cj}
Y.~Matsuo, \emph{{Character Formula of $\hat c< 1$ Unitary Representation of
  ${\mathcal{N}}{=}2$ Superconformal Algebra}},
\doihref{http://dx.doi.org/10.1143/PTP.77.793}{Prog. Theor. Phys. \textbf{77}
  (1987) 793}.

\bibitem[Min97]{Minwalla:1997ka}
S.~Minwalla, \emph{{Restrictions imposed by superconformal invariance on
  quantum field theories}},
  \doihref{http://dx.doi.org/10.4310/ATMP.1998.v2.n4.a4}{Adv. Theor. Math.
  Phys. \textbf{2} (1998) 783--851},
\href{http://arxiv.org/abs/hep-th/9712074}{{\arxivfont arXiv:hep-th/9712074}}.

\bibitem[MN96a]{Minahan:1996fg}
J.~A. Minahan and D.~Nemeschansky, \emph{{An ${\mathcal{N}}\!=2$ Superconformal
  Fixed Point with $E_{6}$ Global Symmetry}},
  \doihref{http://dx.doi.org/10.1016/S0550-3213(96)00552-4}{Nucl. Phys.
  \textbf{B482} (1996) 142--152},
\href{http://arxiv.org/abs/hep-th/9608047}{{\arxivfont arXiv:hep-th/9608047}}.

\bibitem[MN96b]{Minahan:1996cj}
\bysame, \emph{{Superconformal Fixed Points with $E_{N}$ Global Symmetry}},
  \doihref{http://dx.doi.org/10.1016/S0550-3213(97)00039-4}{Nucl. Phys.
  \textbf{B489} (1997) 24--46},
\href{http://arxiv.org/abs/hep-th/9610076}{{\arxivfont arXiv:hep-th/9610076}}.

\bibitem[MNS18]{Maruyoshi:2018nod}
K.~Maruyoshi, E.~Nardoni, and J.~Song, \emph{{Landscape of Simple
  Superconformal Field Theories in 4D}},
\href{http://arxiv.org/abs/1806.08353}{{\arxivfont arXiv:1806.08353 [hep-th]}}.

\bibitem[Mor18]{Morikawa:2018zys}
O.~Morikawa, \emph{{Numerical study of the $\mathcal{N}=2$ Landau--Ginzburg
  model with two superfields}},
\href{http://arxiv.org/abs/1810.02519}{{\arxivfont arXiv:1810.02519
  [hep-lat]}}.

\bibitem[MS16a]{Maruyoshi:2016tqk}
K.~Maruyoshi and J.~Song, \emph{{Enhancement of Supersymmetry via
  Renormalization Group Flow and the Superconformal Index}},
  \doihref{http://dx.doi.org/10.1103/PhysRevLett.118.151602}{Phys. Rev. Lett.
  \textbf{118} (2017) 151602},
\href{http://arxiv.org/abs/1606.05632}{{\arxivfont arXiv:1606.05632 [hep-th]}}.

\bibitem[MS16b]{Maruyoshi:2016aim}
\bysame, \emph{{$ \mathcal{N}=1 $ deformations and RG flows of $ \mathcal{N}=2
  $ SCFTs}}, \doihref{http://dx.doi.org/10.1007/JHEP02(2017)075}{JHEP
  \textbf{02} (2017) 075},
\href{http://arxiv.org/abs/1607.04281}{{\arxivfont arXiv:1607.04281 [hep-th]}}.

\bibitem[MS18]{Morikawa:2018ops}
O.~Morikawa and H.~Suzuki, \emph{{Numerical study of the $\mathcal{N}=2$
  Landau-Ginzburg model}}, \doihref{http://dx.doi.org/10.1093/ptep/pty088}{PTEP
  \textbf{2018} (2018) 083B05},
\href{http://arxiv.org/abs/1805.10735}{{\arxivfont arXiv:1805.10735
  [hep-lat]}}.

\bibitem[MSY05]{Martelli:2005tp}
D.~Martelli, J.~Sparks, and S.-T. Yau, \emph{{The Geometric dual of
  a-maximisation for Toric Sasaki-Einstein manifolds}},
  \doihref{http://dx.doi.org/10.1007/s00220-006-0087-0}{Commun. Math. Phys.
  \textbf{268} (2006) 39--65},
\href{http://arxiv.org/abs/hep-th/0503183}{{\arxivfont arXiv:hep-th/0503183}}.

\bibitem[MSY06]{Martelli:2006yb}
D.~Martelli, J.~Sparks, and S.-T. Yau, \emph{{Sasaki-Einstein Manifolds and
  Volume Minimisation}},
  \doihref{http://dx.doi.org/10.1007/s00220-008-0479-4}{Commun. Math. Phys.
  \textbf{280} (2008) 611--673},
\href{http://arxiv.org/abs/hep-th/0603021}{{\arxivfont arXiv:hep-th/0603021}}.

\bibitem[MZ04]{MaZhou}
X.~Ma and J.~Zhou, \emph{Elliptic genera of complete intersections},
  \doihref{http://dx.doi.org/10.1142/S0129167X05003259}{Internat. J. Math.
  \textbf{16} (2005) 1131--1155},
  \href{http://arxiv.org/abs/math.AG/0411081}{{\arxivfont
  arXiv:math.AG/0411081}}.

\bibitem[Nah78]{Nahm:1977tg}
W.~Nahm, \emph{{Supersymmetries and Their Representations}},
\doihref{http://dx.doi.org/10.1016/0550-3213(78)90218-3}{Nucl. Phys.
  \textbf{B135} (1978) 149}.

\bibitem[Nak13]{Nakayama:2013is}
Y.~Nakayama, \emph{{Scale Invariance vs Conformal Invariance}},
  \doihref{http://dx.doi.org/10.1016/j.physrep.2014.12.003}{Phys. Rept.
  \textbf{569} (2015) 1--93},
\href{http://arxiv.org/abs/1302.0884}{{\arxivfont arXiv:1302.0884 [hep-th]}}.

\bibitem[Naw11]{Nawata:2011un}
S.~Nawata, \emph{{Localization of ${\mathcal{N}}\!=4$ Superconformal Field
  Theory on S$^1$ $\times$ S$^3$ and Index}},
  \doihref{http://dx.doi.org/10.1007/JHEP11(2011)144}{JHEP \textbf{11} (2011)
  144},
\href{http://arxiv.org/abs/1104.4470}{{\arxivfont arXiv:1104.4470 [hep-th]}}.

\bibitem[NS86]{Nemeschansky:1986yx}
D.~Nemeschansky and A.~Sen, \emph{{Conformal Invariance of Supersymmetric
  $\sigma$ Models on Calabi-Yau Manifolds}},
\doihref{http://dx.doi.org/10.1016/0370-2693(86)91394-8}{Phys. Lett.
  \textbf{B178} (1986) 365--369}.

\bibitem[NSVZ83]{Novikov:1983uc}
V.~A. Novikov, M.~A. Shifman, A.~I. Vainshtein, and V.~I. Zakharov,
  \emph{{Exact Gell-Mann-Low Function of Supersymmetric Yang-Mills Theories
  from Instanton Calculus}},
\doihref{http://dx.doi.org/10.1016/0550-3213(83)90338-3}{Nucl. Phys.
  \textbf{B229} (1983) 381--393}.

\bibitem[NY14]{Nishioka:2014zpa}
T.~Nishioka and I.~Yaakov, \emph{{Generalized Indices for $ \mathcal{N} $ = 1
  Theories in Four-Dimensions}},
  \doihref{http://dx.doi.org/10.1007/JHEP12(2014)150}{JHEP \textbf{12} (2014)
  150},
\href{http://arxiv.org/abs/1407.8520}{{\arxivfont arXiv:1407.8520 [hep-th]}}.

\bibitem[Osb98]{Osborn:1998qu}
H.~Osborn, \emph{{${\mathcal{N}}{=}1$ Superconformal Symmetry in
  Four-Dimensional Quantum Field Theory}},
  \doihref{http://dx.doi.org/10.1006/aphy.1998.5893}{Annals Phys. \textbf{272}
  (1999) 243--294},
\href{http://arxiv.org/abs/hep-th/9808041}{{\arxivfont arXiv:hep-th/9808041}}.

\bibitem[Pes97]{Peskin:1997qi}
M.~E. Peskin, \emph{{Duality in Supersymmetric Yang-Mills Theory}},
  \href{http://arxiv.org/abs/hep-th/9702094}{{\arxivfont
  arXiv:hep-th/9702094}}.
(TASI 1996).

\bibitem[Pol05]{PolchinskiII}
J.~G. Polchinski, \emph{String theory}, vol.~2, Cambridge University Press,
  2005.

\bibitem[Pou95]{Pouliot:1995zc}
P.~Pouliot, \emph{{Chiral Duals of Nonchiral SUSY Gauge Theories}},
  \doihref{http://dx.doi.org/10.1016/0370-2693(95)01034-N}{Phys. Lett.
  \textbf{B359} (1995) 108--113},
\href{http://arxiv.org/abs/hep-th/9507018}{{\arxivfont arXiv:hep-th/9507018}}.

\bibitem[PS95]{Pouliot:1995sk}
P.~Pouliot and M.~J. Strassler, \emph{{A Chiral $SU(N)$ Gauge Theory and its
  Non-Chiral $Spin(8)$ Dual}},
  \doihref{http://dx.doi.org/10.1016/0370-2693(95)01554-X}{Phys. Lett.
  \textbf{B370} (1996) 76--82},
\href{http://arxiv.org/abs/hep-th/9510228}{{\arxivfont arXiv:hep-th/9510228}}.

\bibitem[PS96]{Pouliot:1996zh}
\bysame, \emph{{Duality and Dynamical Supersymmetry Breaking in Spin(10) with a
  Spinor}}, \doihref{http://dx.doi.org/10.1016/0370-2693(96)00241-9}{Phys.
  Lett. \textbf{B375} (1996) 175--180},
\href{http://arxiv.org/abs/hep-th/9602031}{{\arxivfont arXiv:hep-th/9602031}}.

\bibitem[PSY15]{Putrov:2015jpa}
P.~Putrov, J.~Song, and W.~Yan, \emph{{(0,4) Dualities}},
  \doihref{http://dx.doi.org/10.1007/JHEP03(2016)185}{JHEP \textbf{03} (2016)
  185},
\href{http://arxiv.org/abs/1505.07110}{{\arxivfont arXiv:1505.07110 [hep-th]}}.

\bibitem[Rai03]{Rains}
E.~M. Rains, \emph{Transformations of elliptic hypergeometric integrals},
  \doihref{http://dx.doi.org/10.4007/annals.2010.171.169}{Ann. of Math. 2nd
  ser. \textbf{171} (2010) 169--243},
  \href{http://arxiv.org/abs/math.QA/0309252}{{\arxivfont
  arXiv:math.QA/0309252}}.

\bibitem[R{\"o}m05]{Romelsberger:2005eg}
C.~R{\"o}melsberger, \emph{{Counting Chiral Primaries in ${\mathcal{N}}{=}1$,
  $d=4$ Superconformal Field Theories}},
  \doihref{http://dx.doi.org/10.1016/j.nuclphysb.2006.03.037}{Nucl.Phys.
  \textbf{B747} (2005) 329--353},
\href{http://arxiv.org/abs/hep-th/0510060}{{\arxivfont arXiv:hep-th/0510060}}.

\bibitem[R{\"o}m07]{Romelsberger:2007ec}
\bysame, \emph{{Calculating the Superconformal Index and Seiberg Duality}},
\href{http://arxiv.org/abs/0707.3702}{{\arxivfont arXiv:0707.3702 [hep-th]}}.

\bibitem[RSZ18]{Razamat:2018gbu}
S.~S. Razamat, O.~Sela, and G.~Zafrir, \emph{{Curious Patterns of IR Symmetry
  Enhancement}}, \doihref{http://dx.doi.org/10.1007/JHEP10(2018)163}{JHEP
  \textbf{10} (2018) 163},
\href{http://arxiv.org/abs/1809.00541}{{\arxivfont arXiv:1809.00541 [hep-th]}}.

\bibitem[RZ17]{Razamat:2017hda}
S.~S. Razamat and G.~Zafrir, \emph{{E$_{8}$ orbits of IR dualities}},
  \doihref{http://dx.doi.org/10.1007/JHEP11(2017)115}{JHEP \textbf{11} (2017)
  115},
\href{http://arxiv.org/abs/1709.06106}{{\arxivfont arXiv:1709.06106 [hep-th]}}.

\bibitem[Sai83]{saito1983}
K.~Saito, \doihref{http://dx.doi.org/10.2969/aspm/00110195}{\emph{The zeroes of
  characteristic function $\chi_f$ for the exponents of a hypersurface isolated
  singular point}}, Algebraic Varieties and Analytic Varieties, Mathematical
  Society of Japan, 1983, pp.~195--217.

\bibitem[Sei94a]{Seiberg:1994bz}
N.~Seiberg, \emph{{Exact Results on the Space of Vacua of Four-Dimensional SUSY
  Gauge Theories}}, \doihref{http://dx.doi.org/10.1103/PhysRevD.49.6857}{Phys.
  Rev. \textbf{D49} (1994) 6857--6863},
\href{http://arxiv.org/abs/hep-th/9402044}{{\arxivfont arXiv:hep-th/9402044}}.

\bibitem[Sei94b]{Seiberg:1994pq}
N.~Seiberg, \emph{{Electric--magnetic Duality in Supersymmetric Nonabelian
  Gauge Theories}},
  \doihref{http://dx.doi.org/10.1016/0550-3213(94)00023-8}{Nucl. Phys.
  \textbf{B435} (1995) 129--146},
\href{http://arxiv.org/abs/hep-th/9411149}{{\arxivfont arXiv:hep-th/9411149}}.

\bibitem[Shi97]{Shifman:1995ua}
M.~A. Shifman, \emph{{Nonperturbative Dynamics in Supersymmetric Gauge
  Theories}}, \doihref{http://dx.doi.org/10.1016/S0146-6410(97)00042-2}{Prog.
  Part. Nucl. Phys. \textbf{39} (1997) 1--116},
\href{http://arxiv.org/abs/hep-th/9704114}{{\arxivfont arXiv:hep-th/9704114}}.

\bibitem[Shi12]{Shifman}
\bysame, \emph{Advanced topics in quantum field theory : a lecture course},
  Cambridge University Press, 2012.

\bibitem[Son16]{SongTalk}
J.~Song, \emph{{N=1 Deformations and RG Flows of N=2 SCFTs}}.
  \url{http://home.kias.re.kr/MKG/upload/autumn/jaewon%20song_new.pdf}. {Talk
  at the \it Autumn Symposium on String Theory}.

\bibitem[Spi01]{Spiridonov1}
V.~P. Spiridonov, \emph{On the elliptic beta function},
  \doihref{http://dx.doi.org/10.1070/rm2001v056n01ABEH000374}{Russian Math.
  Surveys \textbf{56} (2001) 185--186}.

\bibitem[Spi03]{Spiridonov2}
\bysame, \emph{Theta hypergeometric integrals},
  \doihref{http://dx.doi.org/10.1090/S1061-0022-04-00839-8}{Algebra i Analiz
  \textbf{15} (2003) 161--215},
  \href{http://arxiv.org/abs/math.CA/0303205}{{\arxivfont math.CA/0303205}}.

\bibitem[SS81]{Sen:1981hk}
A.~Sen and M.~K. Sundaresan, \emph{{The Four Loop Beta Function for the
  {Wess-Zumino} Model}},
\doihref{http://dx.doi.org/10.1016/0370-2693(81)90489-5}{Phys. Lett.
  \textbf{101B} (1981) 61--63}.

\bibitem[SS89]{SugraDiverse}
A.~Salam and E.~Sezgin,
  \doihref{http://dx.doi.org/10.1142/0277}{\emph{Supergravities in diverse
  dimensions}}, World Scientific, 1989. Two volumes.

\bibitem[ST08]{Shapere:2008zf}
A.~D. Shapere and Y.~Tachikawa, \emph{{Central Charges of ${\mathcal{N}}{=}2$
  Superconformal Field Theories in Four Dimensions}},
  \doihref{http://dx.doi.org/10.1088/1126-6708/2008/09/109}{JHEP \textbf{0809}
  (2008) 109},
\href{http://arxiv.org/abs/0804.1957}{{\arxivfont arXiv:0804.1957 [hep-th]}}.

\bibitem[Str97]{Strassler:1997fe}
M.~J. Strassler, \emph{{Duality, Phases, Spinors and Monopoles in S$O(N)$ and
  Spin(N) Gauge Theories}},
  \doihref{http://dx.doi.org/10.1088/1126-6708/1998/09/017}{JHEP \textbf{09}
  (1998) 017},
\href{http://arxiv.org/abs/hep-th/9709081}{{\arxivfont arXiv:hep-th/9709081}}.

\bibitem[Str03]{Strassler:2003qg}
\bysame, \emph{{An Unorthodox Introduction to Supersymmetric Gauge Theory}},
  \href{http://arxiv.org/abs/hep-th/0309149}{{\arxivfont
  arXiv:hep-th/0309149}}.
{(TASI 2001)}.

\bibitem[SV86]{Shifman:1986zi}
M.~A. Shifman and A.~I. Vainshtein, \emph{{Solution of the Anomaly Puzzle in
  SUSY Gauge Theories and the Wilson Operator Expansion}},
  \doihref{http://dx.doi.org/10.1016/0550-3213(86)90451-7}{Nucl. Phys.
  \textbf{B277} (1986) 456}.
[Zh. Eksp. Teor. Fiz. \textbf{91} (1986) 723].

\bibitem[SV91]{Shifman:1991dz}
\bysame, \emph{{On Holomorphic Dependence and Infrared Effects in
  Supersymmetric Gauge Theories}},
\doihref{http://dx.doi.org/10.1016/0550-3213(91)90072-6}{Nucl. Phys.
  \textbf{B359} (1991) 571--580}.

\bibitem[SV08]{Spiridonov:2008zr}
V.~Spiridonov and G.~Vartanov, \emph{{Superconformal Indices for
  ${\mathcal{N}}\!=1$ Theories with Multiple Duals}},
  \doihref{http://dx.doi.org/10.1016/j.nuclphysb.2009.08.022}{Nucl.Phys.
  \textbf{B824} (2010) 192--216},
\href{http://arxiv.org/abs/0811.1909}{{\arxivfont arXiv:0811.1909 [hep-th]}}.

\bibitem[SV09]{Spiridonov:2009za}
\bysame, \emph{{Elliptic Hypergeometry of Supersymmetric Dualities}},
  \doihref{http://dx.doi.org/10.1007/s00220-011-1218-9}{Commun.Math.Phys.
  \textbf{304} (2011) 797--874},
\href{http://arxiv.org/abs/0910.5944}{{\arxivfont arXiv:0910.5944 [hep-th]}}.

\bibitem[SW94]{Seiberg:1994rs}
N.~Seiberg and E.~Witten, \emph{{Monopole Condensation, and Confinement in
  ${\mathcal{N}}{=}2$ Supersymmetric Yang-Mills Theory}},
  \doihref{http://dx.doi.org/10.1016/0550-3213(94)90124-4}{Nucl. Phys.
  \textbf{B426} (1994) 19--52},
\href{http://arxiv.org/abs/hep-th/9407087}{{\arxivfont arXiv:hep-th/9407087}}.

\bibitem[Tac06]{TachikawaThesis}
Y.~Tachikawa, \emph{{AdS/CFT} correspondence with eight supercharges}.
  \url{https://member.ipmu.jp/yuji.tachikawa/transp/thesis-corrected.pdf}.

\bibitem[Tac13]{Tachikawa:2013kta}
\bysame, \emph{{${\mathcal{N}}{=}2$ Supersymmetric Dynamics for Pedestrians}},
  \doihref{http://dx.doi.org/10.1007/978-3-319-08822-8}{Lect.Notes Phys.
  \textbf{890} (2013) 2014},
\href{http://arxiv.org/abs/1312.2684}{{\arxivfont arXiv:1312.2684 [hep-th]}}.

\bibitem[Tac14a]{Tachikawa:2014mna}
\bysame, \emph{{Magnetic Discrete Gauge Field in the Confining Vacua and the
  Supersymmetric Index}},
  \doihref{http://dx.doi.org/10.1007/JHEP03(2015)035}{JHEP \textbf{03} (2015)
  035},
\href{http://arxiv.org/abs/1412.2830}{{\arxivfont arXiv:1412.2830 [hep-th]}}.

\bibitem[Tac14b]{Tachikawa:2014dja}
\bysame, \doihref{http://dx.doi.org/10.1007/978-3-319-18769-3_4}{\emph{{A
  review on instanton counting and W-algebras}}}, New Dualities of
  Supersymmetric Gauge Theories (J.~Teschner, ed.), Springer, 2016,
  pp.~79--120.
\href{http://arxiv.org/abs/1412.7121}{{\arxivfont arXiv:1412.7121 [hep-th]}}.

\bibitem[Tac15]{Tachikawa:2015wka}
\bysame, \emph{{Frozen Singularities in M and F Theory}},
  \doihref{http://dx.doi.org/10.1007/JHEP06(2016)128}{JHEP \textbf{06} (2016)
  128},
\href{http://arxiv.org/abs/1508.06679}{{\arxivfont arXiv:1508.06679 [hep-th]}}.

\bibitem[Tan13]{TanedoReview}
F.~Tanedo, \emph{{Notes on Seibergology}}.
  \url{https://www.physics.uci.edu/~tanedo/files/notes/Seibergology.pdf}.

\bibitem[Ter09a]{TerningBook}
J.~Terning, \emph{Modern supersymmetry : dynamics and duality}, The
  international series of monographs on physics, no. 132, Oxford University
  Press, 2009.

\bibitem[Ter09b]{TerningSlides}
\bysame, \emph{{Slides for ``Modern supersymmetry : dynamics and duality''}}.
  \url{http://particle.physics.ucdavis.edu/modernsusy/slides.html}.

\bibitem[tH75]{tHooft:1975krp}
G.~'t~Hooft, \emph{{Gauge Fields with Unified Weak, Electromagnetic, and Strong
  Interactions}}, {High-Energy Physics: Proceedings, EPS International
  Conference, Palermo, Italy, 23-28 June 1975.}, 1975, p.~1225.
\url{http://www.staff.science.uu.nl/~hooft101/gthpub/GaugeFields_75.pdf}.

\bibitem[tH76]{tHooft:1976rip}
\bysame, \emph{{Symmetry Breaking Through Bell-Jackiw Anomalies}},
\doihref{http://dx.doi.org/10.1103/PhysRevLett.37.8}{Phys. Rev. Lett.
  \textbf{37} (1976) 8--11}.

\bibitem[tH79]{tHooft:1979uj}
\bysame, \emph{{A Property of Electric and Magnetic Flux in Nonabelian Gauge
  Theories}},
\doihref{http://dx.doi.org/10.1016/0550-3213(79)90595-9}{Nucl.Phys.
  \textbf{B153} (1979) 141}.

\bibitem[tH80]{tHooft:1979rat}
\bysame, \emph{{Naturalness, Chiral Symmetry, and Spontaneous Chiral Symmetry
  Breaking}}, \doihref{http://dx.doi.org/10.1007/978-1-4684-7571-5_9}{NATO Sci.
  Ser. B \textbf{59} (1980) 135--157}.
\url{http://www.staff.science.uu.nl/~hooft101/gthpub/Cargese_1997_III.pdf}.

\bibitem[Ton17]{Tong:2017oea}
D.~Tong, \emph{{Line Operators in the Standard Model}},
  \doihref{http://dx.doi.org/10.1007/JHEP07(2017)104}{JHEP \textbf{07} (2017)
  104},
\href{http://arxiv.org/abs/1705.01853}{{\arxivfont arXiv:1705.01853 [hep-th]}}.

\bibitem[TvN79]{Townsend:1979ha}
P.~K. Townsend and P.~van Nieuwenhuizen, \emph{{Dimensional Regularization and
  Supersymmetry at the Two Loop Level}},
\doihref{http://dx.doi.org/10.1103/PhysRevD.20.1832}{Phys. Rev. \textbf{D20}
  (1979) 1832}.

\bibitem[Var10]{Vartanov:2010xj}
G.~S. Vartanov, \emph{{On the ISS Model of Dynamical SUSY Breaking}},
  \doihref{http://dx.doi.org/10.1016/j.physletb.2010.12.040}{Phys. Lett.
  \textbf{B696} (2011) 288--290},
\href{http://arxiv.org/abs/1009.2153}{{\arxivfont arXiv:1009.2153 [hep-th]}}.

\bibitem[VW89]{Vafa:1988uu}
C.~Vafa and N.~P. Warner, \emph{{Catastrophes and the Classification of
  Conformal Theories}},
\doihref{http://dx.doi.org/10.1016/0370-2693(89)90473-5}{Phys.Lett.
  \textbf{B218} (1989) 51}.

\bibitem[VW93]{Veltman:1991xb}
M.~J.~G. Veltman and D.~N. Williams, \emph{{Schoonschip '91}},
\href{http://arxiv.org/abs/hep-ph/9306228}{{\arxivfont arXiv:hep-ph/9306228}}.

\bibitem[WB92]{WessBagger}
J.~Wess and J.~Bagger, \emph{Supersymmetry and supergravity}, 2nd ed., rev. and
  expanded ed., Princeton series in physics, Princeton University Press, 1992.

\bibitem[Wil09]{SchoonschipArchive}
D.~N. Williams, \emph{{Archive: Schoonschip}}.
  \url{http://www-personal.umich.edu/~williams/archive/schoonschip/index.html}.

\bibitem[Wit79]{Witten:1979ey}
E.~Witten, \emph{{Dyons of Charge $e\theta/2\pi$}},
\doihref{http://dx.doi.org/10.1016/0370-2693(79)90838-4}{Phys.Lett.
  \textbf{B86} (1979) 283--287}.

\bibitem[Wit82a]{Witten:1982fp}
\bysame, \emph{{An $SU(2)$ Anomaly}},
\doihref{http://dx.doi.org/10.1016/0370-2693(82)90728-6}{Phys. Lett.
  \textbf{B117} (1982) 324--328}.

\bibitem[Wit82b]{Witten:1982df}
\bysame, \emph{{Constraints on Supersymmetry Breaking}},
\doihref{http://dx.doi.org/10.1016/0550-3213(82)90071-2}{Nucl.Phys.
  \textbf{B202} (1982) 253}.

\bibitem[Wit93a]{Witten:1993yc}
\bysame, \emph{{Phases of ${\mathcal{N}}{=}2$ Theories in Two-Dimensions}},
  \doihref{http://dx.doi.org/10.1016/0550-3213(93)90033-L}{Nucl.Phys.
  \textbf{B403} (1993) 159--222},
\href{http://arxiv.org/abs/hep-th/9301042}{{\arxivfont arXiv:hep-th/9301042}}.

\bibitem[Wit93b]{Witten:1993jg}
\bysame, \emph{{On the Landau-Ginzburg Description of ${\mathcal{N}}{=}2$
  Minimal Models}},
  \doihref{http://dx.doi.org/10.1142/S0217751X9400193X}{Int.J.Mod.Phys.
  \textbf{A9} (1994) 4783--4800},
\href{http://arxiv.org/abs/hep-th/9304026}{{\arxivfont arXiv:hep-th/9304026}}.

\bibitem[Wit93c]{Witten:1993xi}
\bysame, \emph{{The Verlinde Algebra and the Cohomology of the Grassmannian}},
\href{http://arxiv.org/abs/hep-th/9312104}{{\arxivfont arXiv:hep-th/9312104}}.

\bibitem[Wit97]{Witten:1997bs}
\bysame, \emph{{Toroidal Compactification without Vector Structure}},
  \doihref{http://dx.doi.org/10.1088/1126-6708/1998/02/006}{JHEP \textbf{02}
  (1998) 006},
\href{http://arxiv.org/abs/hep-th/9712028}{{\arxivfont arXiv:hep-th/9712028}}.

\bibitem[Wit00]{Witten:2000nv}
\bysame, \emph{{Supersymmetric Index in Four-Dimensional Gauge Theories}}, Adv.
  Theor. Math. Phys. \textbf{5} (2002) 841--907,
\href{http://arxiv.org/abs/hep-th/0006010}{{\arxivfont arXiv:hep-th/0006010}}.

\bibitem[WZ74]{Wess:1974tw}
J.~Wess and B.~Zumino, \emph{{Supergauge Transformations in Four-Dimensions}},
\doihref{http://dx.doi.org/10.1016/0550-3213(74)90355-1}{Nucl. Phys.
  \textbf{B70} (1974) 39--50}.

\bibitem[Yam08]{Yamazaki:2008bt}
M.~Yamazaki, \emph{{Brane Tilings and Their Applications}},
  \doihref{http://dx.doi.org/10.1002/prop.200810536}{Fortsch. Phys. \textbf{56}
  (2008) 555--686},
\href{http://arxiv.org/abs/0803.4474}{{\arxivfont arXiv:0803.4474 [hep-th]}}.

\bibitem[Zam86]{Zamolodchikov:1986gt}
A.~B. Zamolodchikov, \emph{{Irreversibility of the Flux of the Renormalization
  Group in a 2D Field Theory}}, JETP Lett. \textbf{43} (1986) 730--732.
\url{http://www.jetpletters.ac.ru/ps/1413/article_21504.shtml}.

\end{thebibliography}

\end{document}